\documentclass[11pt]{book}

\usepackage{thesis}
\usepackage{chhead}
\usepackage{graphicx}
\usepackage{subeqn}
\usepackage[small,nooneline,bf]{caption2}
\usepackage[latin1]{inputenc}
\usepackage{german}
\usepackage{bm}

\def\Journal#1#2#3#4{{#1} {#2} (#4) #3 }

\def\NPA{{\em Nucl. Phys.} A}

\def\NPB{{\em Nucl. Phys.} B}

\def\PLB{{\em Phys. Lett.} B}

\def\PRL{\em Phys. Rev. Lett.}

\def\PREP{\em Phys. Rep.}

\def\PRD{{\em Phys. Rev.} D}

\def\RMP{{\em Rev. Mod. Phys.}}

\newcommand{\Diracslash}[1]{#1\llap{/\kern1pt}}
\newcommand{\diracslash}[1]{#1\llap{/\kern0pt}}
\renewcommand\a{\alpha}
\renewcommand\b{\beta}
\newcommand\g{\gamma}
\renewcommand\d{\delta}
\newcommand\e{\epsilon}

\newcommand\m{\mu}

\newcommand\vm{{\mbox{\boldmath${\mu}$}}}

\renewcommand\t{\tau}

\renewcommand\o{\omega}

\renewcommand\L{\Lambda}

\newcommand{\non}{\nonumber\\}

\newcommand{\trs}{\mbox{Tr}_s}

\newcommand{\be}{\begin{equation}}
\newcommand{\ee}{\end{equation}}
\newcommand{\bea}{\begin{eqnarray}}
\newcommand{\eea}{\end{eqnarray}}
\newcommand{\ba}[1]{\begin{array}{#1}}
\newcommand{\ea}{\end{array}}

\newcommand{\eqrf}[1]{Eq.\ (\ref{#1})}

\newcommand{\eqrftw}[2]{Eqs.\ (\ref{#1}) and (\ref{#2})}

\newcommand{\vs}{{\bm{s}}}

\newcommand{\ve}{\mathbf{e}}

\newcommand{\uk}{\hat{\mathbf{k}}}
\newcommand{\vk}{\mathbf{k}}

\newcommand{\vp}{\mathbf{p}}
\newcommand{\vq}{\mathbf{q}}

\newcommand{\vg}{{\mbox{\boldmath${\gamma}$}}}

\newcommand\ft{\tilde f}

\newcommand\Gt{\tilde G}

\def\agt{ \,\, \vcenter{\hbox{$\buildrel{\displaystyle >}\over\sim$}}
 \,\,}
\def\lesssim{ \,\, \vcenter{\hbox{$\buildrel{\displaystyle <}\over\sim$}}
 \,\,}
\def\alt{ \,\, \vcenter{\hbox{$\buildrel{\displaystyle <}\over\sim$}}
 \,\,}

\begin{document}

\originalTeX

\pagestyle{ch}

%%%%%%%%%%%%%%%%%%%%%%%%%%%%%%%%%%%%%%%%%%%%%%%%%%%%%%%%%%%%%%%%%%%%%%%
%
%                   TITELBLATT
% 
%%%%%%%%%%%%%%%%%%%%%%%%%%%%%%%%%%%%%%%%%%%%%%%%%%%%%%%%%%%%%%%%%%%%%%%
%
\thispagestyle{empty}
\vspace{14cm}
\begin{center}
{\Huge\bf{A General Effective Action for}} 

\bigskip
{\Huge\bf{Quark Matter}}
\bigskip
{\Huge\bf{and its Application to Color Superconductivity}}\\
\vspace{2.5cm}
{\Large
Dissertation \\
zur Erlangung des Doktorgrades \\
der Naturwissenschaften
\\[.8cm]
vorgelegt beim Fachbereich Physik \\
der Johann Wolfgang Goethe-Universit\"at \\
in Frankfurt am Main
\\[1.2cm]
von \\
Philipp Tim Reuter \\
aus Bonn
\\[1.2cm]
Frankfurt am Main, 2005
\\[0.3cm]
(D 30)
}\end{center}

\clearpage
%thispagestyle{empty}
\normalsize
\vspace*{8cm}
\noindent
vom Fachbereich Physik der Johann Wolfgang Goethe--Universit\"at\\
als Dissertation angenommen.\\[3cm]
Dekan: Prof.\ Dr.\ W.\ Aßmus\\[1cm]
Gutachter: Prof.\ Dr.\ D.-H.\ Rischke, HD Dr.\ J.\ Schaffner-Bielich\\[1cm]
Datum der Disputation: 30.11.2005
\clearpage

\newpage

\chapter*{Abstract}

I derive a general effective theory for hot and/or dense quark matter. 
After introducing general projection operators for hard
and soft quark and gluon degrees of freedom, I explicitly compute
the functional integral for the hard quark and gluon modes in the
QCD partition function. Upon appropriate choices for the projection 
operators one recovers various well-known effective theories
such as the Hard Thermal Loop/ Hard Dense Loop Effective Theories
as well as the High Density Effective Theory by Hong and Sch\"afer.
I then apply the effective theory to cold and dense quark matter and
show how it can be utilized to simplify the weak-coupling solution of 
the color-superconducting gap equation. In general, one considers as relevant 
quark degrees of freedom those within a thin layer of width 
$2\Lambda_{\rm q}$ around the Fermi surface and as relevant gluon 
degrees of freedom those with 3-momenta less than $\Lambda_{\rm gl}$. 
It turns out that it is necessary to choose $\Lambda_{\rm q} \ll
\Lambda_{\rm gl}$, i.e., scattering of quarks along the
Fermi surface is the dominant process. Moreover, this special choice 
of the two cut-off parameters $\Lambda_{\rm q}$ and $\Lambda_{\rm gl}$ 
facilitates the power-counting of the numerous contributions in 
the gap-equation. In addition, it is demonstrated that both the energy 
and the momentum dependence of the gap function has to be treated 
self-consistently in order to determine the imaginary part of the gap
function. For quarks close to the Fermi surface the imaginary part
is calculated explicitly and shown to be of sub-subleading order in
the gap equation.

\tableofcontents

\listoffigures
\listoftables

		\chapter{Introduction}
\section[Quark matter and strong interactions]{Quark matter and strong interactions} \label{QM}

In contrast to philosophers who search for  meaning in nature, physicists investigate the  properties of nature and search for the laws that inanimate nature obeys. 
Apparently, the properties of any given piece of (known) matter depend on its temperature and its density. It is therefore not at all a naive question to ask: ``What happens to some piece of matter if I heat and squeeze it\dots further and further?'' One correct answer could be: ``You may use QCD to describe it.'' At least this applies to ``normal'' matter made of atoms. Atoms have a nucleus, which is composed of neutrons and protons, which in turn consist of  quarks. Quarks are fermions and interact with each other by exchanging  gluons, which mediate the  strong interaction. The charge corresponding to the strong interaction is called  color and the quantum field theory describing this interaction  Quantum Chromodynamics (QCD). 
Generally, all particles that interact strongly are composed of quarks and called  hadrons. 
Those hadrons, which are composed of a quark and an antiquark, are called  mesons and those composed of three quarks, as the neutrons and protons mentioned above,  baryons. 
 While all quarks carry color charge, all hadrons are in total  color neutral. 

As long as the temperature and density of the considered system are not too large, i.e.\ below the scale $\Lambda_{\rm QCD}\simeq 200$ MeV, quarks are  strongly coupled and always  confined into these color neutral hadrons. As a consequence of this strong coupling perturbative approaches are impossible.
  The quarks are  deconfined at temperatures or densities above $\Lambda_{\rm QCD}$. Well above this scale the strong interaction exhibits the phenomenon of  asymptotic freedom \cite{GWP}, where QCD becomes  weakly coupled. In this regime all hadrons vanish and quarks and gluons  form a state called the  quark-gluon plasma or simply quark matter \cite{ColPer}. 

Quarks also feel the  electroweak  and the  gravitational force. The first one is described by a quantum field theory, which comprises electromagnetic and weak interactions. The theory for the latter,  general relativity, describes gravitation in terms of the curvature of space-time and is a classical field theory.
Due to their relative weakness as compared to the strong force it is often justified to neglect the gravitional and the electroweak against the strong interaction and describe the considered matter by QCD only.
As a counter example related to this work, where QCD alone is not sufficient, one may allude to neutron stars. These are the compact remnants of supernova explosions of type II \cite{phillips,weber,pons}. They have extremely large masses comparable to the mass of our sun, but only very small radii of several kilometers. The density in their inner cores are possibly sufficient to deconfine the quarks. Since these stellar objects have to be electrically neutral and $\beta-$equilibrated, electroweak processes have to be considered. Furthermore, neutron stars are held together gravitationally. In order to describe their bulk properties adequately, one therefore has to account for gravity as well.

In the present work, however, only the strong interaction will be accounted for and, moreover, only its weak coupling regime will be considered. The aim is to derive an  effective action for strongly interacting matter in the deconfined phase, i.e.\ for quark matter \cite{RRW}. This is motivated by the occurence of several well seperated momentum scales at high temperatures and/or baryonic densities, cf.\ Sec.\ \ref{EFT}. The derivation is performed in Sec.\ \ref{II}. The field of application comprises  low energy phenomena in weakly coupled quark matter. By construction this action will only contain low energy quark and gluon modes as explicit degrees of freedom. At large temperatures such modes are, e.g., long wave-length collective excitations \cite{braatenpisarski,Blaizot,LeBellac}. At large quark densities and low temperatures only those quarks, which are located close to the Fermi surface, are relevant degrees of freedom. The derived action can be adjusted to different high temperature and/or density regimes by a suitable choice of the projection operators, which separate high energy from low energy quark and gluon modes. These projection operators contain two independent cutoff parameters, $\Lambda_{\rm q}$ for quarks and $\Lambda_{\rm gl}$ for gluons. In Sec.\ \ref{III} it is shown that well-known effective theories for strongly interacting quark matter can be assigned to special choices of these cutoff parameters in this general effective action. In chapter \ref{chapterapp} the phenomenon of color superconductivity in quark matter will serve as a physical application to demonstrate the usefulness of the derived action in a semi-perturbative context.

Before going into more details, however, it is first necessary to provide an introductory overview of the basic principles of QCD and its phase diagram. This will be done in the remainder of this section. In Sec.\ \ref{CSC} the phenomenon of color superconductivity will be introduced. In Sec.\ \ref{EFT} the general concepts of effective theories are explained and some caveats with respect to quark matter are pointed out. Finally, the specific features of the effective approach developed in this work are discussed.

%As for any effective theory it is assumed that the { details} of the dynamics of the high energy modes do not affect the low energy physics. Therefore, they are not accounted for explicitly. 
%
%In fact, the typical size of a hadron can be estimated to be of the order of 1 fm. If one compresses the considered matter sufficiently hard, the hadrons in it will eventually overlap until one has one bulk piece of bulk deconfined { quark matter. 

\subsubsection{QCD: Asymptotic freedom, symmetries, and the phase diagram}

QCD is a non-Abelian gauge theory with the gauge group $SU(N_c)_c$, where the index $c$ refers to color and $N_c$ is the number of quark colors. The quarks correspond to the fundamental representation of  $SU(N_c)_c$, whereas the gluons are the gauge bosons and belong to the adjoint representation of  $SU(N_c)_c$. Apart from the color index, $1\leq c\leq N_c$, quarks also carry a flavor quantum number, $1\leq f\leq N_f$. 
 Suppressing all quark indices for simplicity 
the Lagrange density for QCD reads \cite{Peskin,khuang,introgauge,DHRreview}
\begin{equation} \label{LQCD}
{\cal L}_{\rm QCD} = \bar{\psi}\, \left( i \gamma^\mu D_\mu - m\right)
\psi - \frac{1}{4} \, F^{\mu \nu}_a \, F^a_{\mu \nu}
+ {\cal L}_{\rm gf}+ {\cal L}_{\rm ghost} \,\, .
\end{equation}
Since quarks are fermions with spin 1/2 their wave functions $\psi$ are $4  N_c  N_f$-dimensional spinors. The  Dirac conjugate spinor
is defined by $\bar{\psi} \equiv \psi^\dagger \gamma_0$, where 
$\gamma^\mu$ are the Dirac matrices and $m$ is the current quark mass, which is a diagonal matrix in flavor space, $m \equiv m_{ij}\,\delta_{ij}$. The six known flavors are called up ($u$), down ($d$), strange ($s$), charmed ($c$), bottom ($b$), and top ($t$). Their masses are ordered as $m_u\simeq m_d \ll m_s \ll m_c \ll m_b \ll m_t$.  Explicitly, the  current quark masses appearing in ${\cal L}_{\rm QCD}$ for the three lightest quark flavors are $m_u \simeq 5$ MeV,  $m_d \simeq 10$ MeV, and $m_s \simeq 100$ MeV \cite{Data}. The masses for the three remaining quark flavors are so much larger that these flavors will not play any role in the following.

The covariant derivative is defined as  $D_\mu = \partial_\mu - i g A_\mu^a \, T_a$, 
with the strong coupling constant $g$.
The vector fields $A^\mu_a$ represent the gluons, where the adjoint color index $a$ runs from 1 to $N_c^2-1$. The $N_c\times N_c$ matrices $T_a$ are the generators of the local $SU(N_c)_c$ gauge symmetry. Throughout this work I use
the representation $T_a \equiv \lambda_a/2$, where $\lambda_a$
are the Gell-Mann matrices. The gluonic field strength tensor is defined as
\begin{equation}\label{Fmunu}
F^{\mu \nu}_a = \partial^\mu A^\nu_a - \partial^\nu A^\mu_a +
g f_{abc} \, A^\mu_b \, A^\nu_c \,\, ,
\end{equation}
where $f_{abc}$ are the structure constants of $SU(N_c)_c$. They are  defined through $if_{abc}T_c\equiv [T_a,T_b]$. In the non-linear terms in Eq.\ (\ref{Fmunu}) proportional to  $f_{abc}$ the non-Abelian character of QCD manifests itself: the gluons carry color charge and couple to themselves. This ultimately gives rise to the numerous non-trivial features of this theory. 
The terms ${\cal L}_{\rm gf}$ and ${\cal L}_{\rm ghost}$ in Eq.~(\ref{LQCD}) contain gauge fixing terms 
and the contribution of the Faddeev-Popov ghosts. Since they are of no further relevance for this work I will not specify them here.

The QCD Lagrange density given in Eq.\ (\ref{LQCD}) is needed to calculate the  grand canonical partition function of QCD
\cite{LeBellac,FTFT}
\begin{equation} \label{Z}
{\cal Z}(T,V,\mu) = \int {\cal D}\bar{\psi} \,
{\cal D} \psi \, {\cal D}A^\mu_a \; \exp\{S_{\rm QCD}[\bar \psi, \psi, A] \}
\end{equation}
with the tree-level { action} of QCD at quark chemical potential $\mu$
\begin{equation}
S_{\rm QCD}[\bar \psi, \psi, A]\equiv\left[ \int_X
\left( {\cal L} + \mu \, {\cal N} \right) \right]\;.\label{SQCD}
\end{equation}
The quark chemical potential is associated with the net quark number conservation. 
The number density operator of the conserved net quark number is ${\cal N}\equiv \bar \psi \gamma_0 \psi$.
From Eq.\ (\ref{Z}) one can derive thermodynamical quantities such as pressure, $p(T,\mu)$, entropy density, $s(T,\mu)$, and particle number density, $n(T,\mu)$,
\begin{equation} \label{pressure}
p(T,\mu) =  \left.T \, \frac{\partial \ln {\cal Z}(T,\mu)}{\partial V}\right|_{T,\mu}\;,\;\;\;\;
s(T,\mu) = \left. \frac{\partial p(T,\mu)}{\partial T} \right|_\mu\;,
\;\;\;\;
n(T,\mu) =\left. \frac{\partial p(T,\mu)}{\partial \mu} \right|_T\;.
\end{equation}
In particular, the quark number density  $n(T,\mu)$ for massless quarks and at  $T\ll\mu$ is proportional to the third power of $\mu$, $n\sim \mu^3$. Note that in the thermodynamical limit, $V\rightarrow \infty$, $ \ln {\cal Z}(T,\mu)$ becomes an extensive quantity and as such is simply proportional to $V$. Therefore, the dependence of the pressure on the volume cancels out. In order to calculate the expectation value of any given operator ${\cal O}$ in the grand canonical ensemble one employs the following averaging prescription 
\begin{equation}
\langle {\cal O} \rangle \equiv \frac{1}{{\cal Z}}
\int {\cal D}\bar{\psi} \,
{\cal D} \psi \, {\cal D}A^\mu_a \; {\cal O} \; \exp\{S_{\rm QCD}[\bar \psi, \psi, A] \}\,\, .
\end{equation}

\begin{figure}[ht]
\centerline{
\includegraphics[width=7cm]{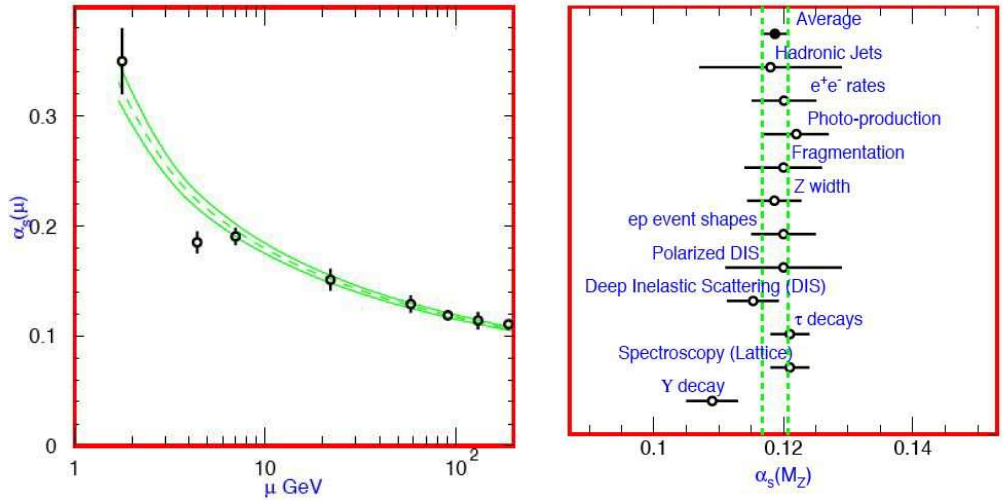}
\includegraphics[width=6.44cm]{strongcoupling.epsf}
}
\caption[The dependence of $\alpha_s$ on the considered energy scale $Q$.]{The experimental and theoretical dependence of $\alpha_s(Q)$ on the considered energy scale $Q\equiv\mu$ (left figure). Experimental values for $\alpha_s(M_Z)$ (right figure), \cite{Data}.}
\label{runningcoupling}
\end{figure}

Since QCD is a quantum field theory the vacuum is treated as a polarizable medium. Consequently, the result of any experiment that measures the effective color charge in the coupling constant $\alpha_s\equiv g^2/4\pi$ of two strongly interacting quarks will depend on the energy scale of the experiment. This dependence on the scale $Q$ is expressed by the $\beta-$function of QCD \cite{Peskin, khuang,introgauge}. In the so called $\overline{\rm MS}$-scheme  at the three loop level one finds for the scale $Q$ \cite{Chivukula:2004hw}
\begin{equation}
Q{\partial \alpha_s \over \partial Q} = 2 \beta(\alpha_s) =
-{\beta_0\over 2\pi} \alpha^2_s -{\beta_1 \over 4\pi^2} \alpha^3_s
-{\beta_2 \over 64\pi^3} \alpha^4_s - \ldots
\label{eq:betafunction}
\end{equation}
where
\begin{equation}
\beta_0 = 11-{2\over 3}N_f~, \qquad \beta_1 = 51-{19\over 3} N_f~,
\qquad \beta_2 = 2857-{5033\over 9} N_f +{325\over 27}N^2_f~.
\label{eq:betacoefficients}
\end{equation}
%QCD is a renormalizable theory and the physical { coupling constant} $g$ in Eq.\ (\ref{LQCD}) depends on the considered energy or momentum scale $Q$. It is related to the energy independent, { bare} coupling constant $g_B$ via
%\bea
%g_B = g \, Q^{\epsilon/2}Z_1\,Z_2^{-1}Z_3^{-1/2}\;,\label{gB}
%\eea
%where $\epsilon\rightarrow 0$. Here, $Z_1$ is the renormalization constant for the quark-gluon-vertex, $Z_2$ that of the quark wave function and $Z_3$ that of the gluon wave function. Calculating those functions to one loop order one finds
%\bea
%g_B \simeq g \,Q^{\epsilon/2} \left[1-\frac{g^2}{16\pi^2\epsilon} \left(11-\frac{2N_f}{3} \right)\right]\;.\label{gB1loop}
%\eea
%The running} of the coupling constant $g$ with $Q$ is generally determined by the $\b-$function
%\bea
%\beta(g) &\equiv& \frac{\partial g}{\partial \ln(Q)}\label{beta}\;.
%\eea
Since in nature the number of quark flavors is $N_f =6$ it follows that $\beta < 0 $ and QCD is indeed an asymptotically free theory as already mentioned above, cf.\ left diagram in \ Fig.\  \ref{runningcoupling}.
For a more quantitative analysis one has to solve the renormalization group equations. At three-loop order one finds  \cite{Chivukula:2004hw}
\begin{eqnarray}
\alpha_s(Q) &=& {4\pi \over \beta_0 \log(Q^2/\Lambda_{\rm QCD}^2)}
\left[1-{2\beta_1\over \beta^2_0}\,{\log\left[\log(Q^2/\Lambda_{\rm QCD}^2)\right] \over 
\log(Q^2/\Lambda_{\rm QCD}^2)}  +{4\beta^2_1 \over \beta^4_0 \log^2(Q^2/\Lambda_{\rm QCD}^2)}\right. \\
& \times & \left.\left(\left(\log\left[\log(Q^2/\Lambda^2_{\rm QCD})\right]-{1\over 2}\right)^2
+{\beta_2 \beta_0\over 8\beta^2_1}-{5\over 4}\right)\right]\;.
\label{eq:solution}
\end{eqnarray}
This solution may be used to compare the values of $\alpha_s(Q)$ and $\Lambda_{\rm QCD}$ from different experiments. With Eq.\ (\ref{eq:solution}) one may first  calculate the scale $\Lambda_{\rm QCD}$ by matching it to the measured $\alpha_s(Q)$ and then use it to extrapolate $\alpha_s$ to one standard reference scale, say to the mass of the $Z$ boson, $M_Z= 91.1876\pm0.0021$ GeV. This of course depends on the number of flavors used in Eq.\ (\ref{eq:solution}). For energies not much larger than $M_Z$ one may restrict oneself to $N_f=5$ active flavors and finds for $\Lambda_{\rm QCD}$ the value \cite{Data}
\bea
\Lambda_{\overline{\rm MS}}^{(5)} = 217^{+25}_{-23}~\mbox{MeV}\;.
\eea
For $\alpha_s(M_Z)$ one obtains \cite{Data}
\bea
\alpha_s(M_Z)= 0.1187\pm 0.002\;,
\eea
cf.\ right diagram in Fig.\ \ref{runningcoupling}. 
%The smallness of $\alpha_s$ in deep inelastic scattering experiments  substantiate the asymptotic freedom in QCD. 
Decreasing the scale Q, the coupling $\alpha_s(Q)$ grows very large for $Q\rightarrow \Lambda_{\rm QCD}$. Such a divergency of the coupling is sometimes called a Landau pole. In the case of quantum electrodynamics (QED) the Landau pole appears at very large energy scales, whereas the coupling is perturbatively small at low energies.%In fact for such theories the term Landau pole is used more commonly than in asymptotically free theories. 
%Generally, a Landau pole indicates the onset of a non-perturbative phenomenon. In QCD the formation of the chiral condensate or of the Cooper pair condensate is accompanied with a Landau pole.
%If one neglects all quark masses $\Lambda_{\rm QCD}$ remains as the only dimensionful quantity in QCD. However, considering the above arguments it is not really a parameter of QCD, but rather fixes the choice of units. In standard units one chooses $\Lambda_{\rm QCD}\equiv \Lambda_{\overline{\rm MS}}^{(5)}\simeq 200$ MeV $\simeq$ 1 fm$^{-1}$. In fact, the typical size of a hadron can be estimated to be of the order of 1 fm. 

Turning to the symmetries of the QCD Lagrangian given in Eq.\ (\ref{LQCD}) one observes that  ${\cal L}_{\rm QCD}$ is invariant under Lorentz transformations and translations in space-time. Furthermore, it respects charge conjugation, parity and time-reversal invariance. Another symmetry is its invariance under gauge transformations, which are generally local, i.e.\ dependent of space-time,  $U(x) \in SU(N_c)_c$
\bea
\psi(x) \rightarrow U(x)\psi(x)\;,~~~~A^\mu(x) \rightarrow U(x)A^\mu(x)U^\dagger(x) + iU(x)\partial^\mu U^\dagger(x)\;,
\eea
where $A^\mu\equiv A^\mu_aT_a$.
A further symmetry is the global (i.e.\ independent on space-time) quark flavor symmetry of ${\cal L}_{\rm QCD}$.
%: the dynamics of QCD are completely independent of flavor. 
To elaborate this, one may decompose the quark fields with respect to their chirality
\begin{equation}
\psi \equiv \psi_r + \psi_\ell
\;\; , \;\;\;\; 
\psi_{r,\ell} \equiv {\cal P}_{r,\ell} \, \psi
\;\; ,\;\;\;\;
{\cal P}_{r,\ell} \equiv \frac{1 \pm \gamma_5}{2} \,\, , \label{chiralbasis}
\end{equation}
where ${\cal P}_{r,\ell}$ are the chirality projectors. One observes that in ${\cal L}_{\rm QCD}$ only the mass term mixes quarks with different chirality
\begin{equation} \label{massterm}
\bar{\psi}^i \, m_{ij} \, \psi^j
\equiv \bar{\psi}^i_r \, m_{ij} \, \psi^j_\ell +
\bar{\psi}^i_{\ell} \, m_{ij} \, \psi^j_r \,\,,
\end{equation}
where use was made of the orthogonality of ${\cal P}_{r}$ and ${\cal P}_{\ell}$ and of ${\cal P}_{r,\ell} \gamma_0 = \gamma_0 {\cal P}_{\ell,r}$. Hence, assuming zero quark masses for all flavors, $m\equiv 0$, ${\cal L}_{\rm QCD}$ becomes chirally symmetric, i.e.\ invariant under global chiral $U(N_f)_r \otimes U(N_f)_\ell$ transformations given by
\begin{equation} \label{Urot}
\psi_{r,\ell} \rightarrow U_{r,\ell}\, \psi_{r,\ell}
\;\; ,\;\;\;\; 
U_{r,\ell} \equiv \exp\left( i \sum_{a=0}^{N_f^2-1} \alpha^a_{r,\ell}\,
T_a \right) \in U(N_f)_{r,\ell} \,\, .
\end{equation}
Here, the $\alpha^a_{r,\ell}$ are the parameters and $T_a$ the generators of $U(N_f)_{r,\ell}$ with $T_0 \sim \bm 1$. While the kinetic and interaction parts of ${\cal L}_{\rm QCD}$ are generally invariant under arbitrary transformations of $U(N_f)_r \otimes U(N_f)_\ell$, the mass term is only invariant in the case that the parameters $\alpha^a_{r}$ and $\alpha^a_\ell$ of the given transformations $U_r$ and $U_\ell$  are equal,  $\alpha^a_{r} = \alpha^a_\ell$. One therefore introduces the vectorial group $U(N_f)_V \subset U(N_f)_r \otimes U(N_f)_\ell$ with elements that fulfill $U_r = U_\ell$, which is often denoted as $U(N_f)_V \equiv U(N_f)_{r+\ell}$. These vectorial transformations leave  ${\cal L}_{\rm QCD}$ invariant if all masses are equal.  In nature, however, only the two lightest quarks have approximately equal masses, which reduces the vector symmetry to an approximative  $U(2)_V$ symmetry. %, the so-called isospin symmetry. 
Analogously one recovers an approximative $U(3)_V$ symmetry, if one considers energies much larger than $m_s$.
The subset $U(N_f)_A = U(N_f)/U(N_f)_V$ comprises all transformations fulfilling $U_r = U_\ell^\dagger$. Hence, the symmetry of these so-called {axial} transformations is always explicitly broken by the quark masses. 

%One may isomorphically map the chiral group  $U(N_f)_r \otimes U(N_f)_\ell$ onto the group of unitary vector and axial transformations,  $U(N_f)_V \otimes U(N_f)_A$. If $\alpha_{r}^a$ and $\alpha_{\ell}^a$ are the parameters of a given transformation in $U(N_f)_r  \otimes U(N_f)_\ell$ then the corresponding transformation in  $U(N_f)_V \otimes U(N_f)_A$ is given by the generators $\alpha_V\equiv (\alpha_r +\alpha_\ell)/2$ and $\alpha_A\equiv (\alpha_r -\alpha_\ell)/2$, respectively. 
Since any unitary transformation can be decomposed into the product of a special unitary transformation and a complex phase one may finally write $U(N_f)_r  \otimes U(N_f)_\ell$ as $SU(N_f)_r \otimes SU(N_f)_\ell \otimes U(1)_V\otimes U(1)_A$. The phase factors due to $ U(1)_V$ correspond to an exact symmetry (independent of the quark masses) of QCD, corresponding to the conservation of baryonic charge, i.e.\  of quark numbers, and is often denoted as $U(1)_B$. 
%Since it does not affect chiral dynamics it will be omitted in the following discussion. 
The $U(1)_A$ symmetry is explicitly broken by an anomaly at the quantum level of QCD \cite{tHooft}.
In hot and/or dense matter the instantons corresponding to this anomaly are screened  \cite{grosspisarskiyaffe} so that the $U(1)_A$ symmetry may become effectively restored again in matter. 

\begin{figure}[ht]
\centerline{\includegraphics[width=13cm]{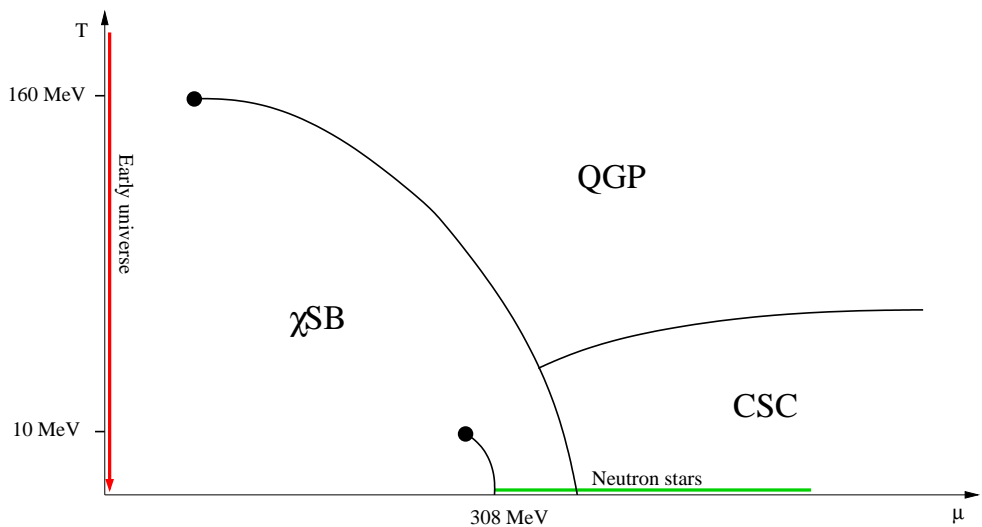}}
\caption[The phase diagram of strongly interacting matter.]{The phase diagram of strongly interacting matter.}
\label{phasediagram}
\end{figure}

Below some critical density and temperature of the order of magnitude of $\Lambda_{\rm QCD}$ the true ground state of the vacuum is populated by the so called chiral condensate. In Fig.\ \ref{phasediagram} the corresponding region is labelled by $\chi$SB. Often it is also referred to as the hadronic phase. It is described by the order parameter
\bea
\Phi^{ij}\sim\langle \bar{\psi}^i_\ell \, \psi^j_r \rangle +{\rm h.\,c.} \neq 0\;.\label{chiral}
\eea 
Its structure in flavor space connects right and left handed quarks. In the chiral limit of zero quark masses the difference among the quark flavors vanishes and $\Phi^{ij} = \delta^{ij}\Phi$, i.e.\ the order parameter of the chiral condensate becomes diagonal in flavor space just as the mass matrix discussed above. It follows analogosly that the chiral condensate spontaneously breaks the (approximate) axial part of the chiral symmetry. Since the broken symmetry is {global}, Goldstone's theorem \cite{Peskin,khuang,introgauge}  applies and {massless} Goldstone bosons are created. Their number is given by the number of broken generators. Since in the considered case the chiral condensate breaks the $SU(N_f)_A$ symmetry, $N_f^2-1$ Goldstone bosons are generated. In the case of $N_f=2$ (considering only up and down quarks) this corresponds to the generation of the three pseudoscalar pions. These are not exactly massless, since the symmetry was already explicitly broken by the up and down quark masses and therefore it was only an approximate symmetry. Generally, one calls the massive Goldstone bosons corresponding to the breaking of approximate global symmetries pseudo-Goldstone bosons. For $N_f=3$ also the heavier strange quark is involved, which makes the spontaneously broken symmetry even more approximate as in the case of $N_f=2$. Consequently, the created pseudo-Goldstone bosons are more massive. They correspond to the pseudoscalar meson octet made of pions, kaons, and the eta meson. Since all these mesons consist of a quark-antiquark pair, the quark number symmetry $U(1)_V$ and the gauge symmetry of electromagnetism $U(1)_{em}$ remain intact,
\bea
\chi{\rm SB}:~~ SU(3)_c\otimes SU(3)_V  \otimes U(1)_B \otimes U(1)_{em}\;. \label{chisym}
\eea
%The order magnitude of the masses of these dynamically generated bosons can be estimated as \cite{Chivukula:2004hw}
%\begin{eqnarray}
%m^2_\pi &\propto& (m_u + m_d) \Lambda_{\rm QCD}\;, \nonumber \\
%m^2_K &\propto& (m_s + m_{u,d})\Lambda_{\rm QCD}\;, \\
%m^2_{\eta} &\propto& {1\over 3} (m_u + m_d + 4 m_s) \Lambda_{\rm QCD}\;,
%\end{eqnarray}
%where the masses $m_i$ are the current quark masses given after Eq.\ (\ref{LQCD}). 
%\subsubsection{Exploring the QCD phase diagram}
Inside this hadronic phase the ground state of (infinite) nuclear matter with baryonic density $n_0\simeq 0.15$ fm$^{-3}$ is located at $T=0$ and $\mu = 308$ MeV.
% The value of the chemical potential can be obtained from the theorem of Hugenholz and Van Hove \cite{Hugenholz}, which states that the baryonic chemical potential equals the highest occupied single particle energy. In the considered case it is given by the mass of the neutron, $m_N = 939$ MeV, reduced by the average binding energy per nucleon in nuclear matter, $B_0=16$ MeV/A \cite{DasGupta}.
%of the phenomenological bag constant \cite{bag-model}
%which parametrizes the pressure of the vacuum. The latter is introduced to compensate the large kinetic energy of the quarks confined into the hadron. 
The value of the chemical potential is given by the mass of the neutron, $m_N = 939$ MeV, reduced by the average binding energy per nucleon in nuclear matter, $B_0=16$ MeV/A \cite{DasGupta}. One obtains $\mu_B\equiv$ 939 MeV -- 16 MeV = 923 MeV $= 3 \mu$. The factor 3 arises from the fact that a nucleon is a baryon with baryonic charge 1 and as such
%,  can be described by the non-relativistic quark model to 
consists of three (up and down) quarks each carrying baryonic charge $1/3$. These quarks are dressed by their strong interaction with the chiral condensate, so that the quark masses become proportional to the expectation value of the chiral condensate. This effectively gives them much larger constituent masses as compared to the respective current quark masses $m$ in ${\cal L}_{\rm QCD}$.

The line emerging from the point of infinite nuclear matter separates the gaseous from the liquid phase of hadronic matter and corresponds to a first order phase transition \cite{Jaq:83,cris:00}. At its endpoint at $T \approx 10$ MeV it becomes of second order exhibiting critical phenomena similarly to normal water \cite{Tra:96,Bo:95,Bug:00,Bug:01,Reuter,Bugaev}. 

The evolution of strongly interacting matter in the early universe corresponds to the red line at zero net baryonic density, $\mu=0$, in Fig.\ \ref{phasediagram}. For small values of $\mu$ the numerical lattice simulations of QCD \cite{Creutz,Owe} predict that the transition from the quark-gluon phase to the hadronic phase is a crossover \cite{Fodor,Ejiri}. These analyses fail so far in the regime $\mu> T$: lattice simulations are based on Monte Carlo importance sampling and rely on the probabilistic interpretation of the weight in the path integral. For $\mu> 0$ the fermion determinant becomes complex, which causes the famous sign problem \cite{Creutz,Owe}. For small values of $\mu$ a Taylor expansion technique has been proposed to investigate the critical endpoint of the chiral phase transition \cite{Stephanov}. For recent experimental and theoretical progress on the  properties of the confined and deconfined phases and the transitions among them see \cite{heavyion,Kogut}. As this field is not yet settled, the location of the line of the chiral phase transition and its critical endpoint in Fig. \ref{phasediagram} are only approximate. 

Also the boundaries around the color superconducting regime \cite{DHRreview,bailinlove,RWreview,shovy} labelled by CSC are essentially unknown.
Only in the limit of asymptotically large quark chemical potentials where QCD becomes weakly coupled it can be rigorously proven that the true ground state is color superconducting at sufficiently low temperatures. In this limit even the specific color superconducting phase is known. It is the so-called CFL phase, which breaks again the chiral symmetry, cf.\ Sec.\ \ref{CSC}. At chemical potentials and densities of physical relevance, however, the the situation becomes less transparent. The densities inside neutron stars are bounded by the condition of hydrostatic stability to be $n\lesssim 10 \, n_0$ \cite{shovy}. This corresponds to $\mu\lesssim 500$ MeV, cf.\ the green line in Fig.\ \ref{phasediagram} where QCD is not weakly coupled anymore. It is believed, however, that inside the cores of some neutron stars color superconductivity might occur and have observable effects on their properties \cite{svidzinsky,link,zhit}. The critical temperature and density for the onset of color superconductivity can only be estimated to be roughly of the order $\mu_c \agt 400$ and $T_c \lesssim 50$ MeV. Furthermore, the search for the most dominant color superconducting phase for given $T$ and $\mu$ is still ongoing wherefore the presumably complex structure inside the CSC region in Fig.\ \ref{phasediagram} is essentially unknown. These issues will be elaborated further in the following section.

\section{Color superconductivity} \label{CSC}

The discovery of color superconductivity goes back to the late 1970's \cite{bailinlove}. 
However, wider interest in the phenomenon of color superconductivity has only recently been
generated by the observation that, within a simple
Nambu--Jona-Lasinio (NJL) -- type model \cite{nambu} for 
the quark interaction, the so-called color-superconducting gap parameter $\phi$
assumes values of the order of 100 MeV \cite{ARWRSSV}.
The gap parameter $\phi$ appears in the dispersion relation of the gapped, i.e.\ Cooper paired quarks
in the color superconducting medium
\bea
\epsilon_{k,r} = \sqrt{(k-\mu)^2+\lambda_r\phi^2}\;,
\eea
where $k$ is the modulus of the 3-momentum of the quark. The index $r$ specifies the color and the flavor of the considered quark. In the special case that $\lambda_r=0$ the quark is ungapped. It follows that 
gapped quarks  carry a non-zero amount of energy $\epsilon_{{\mu},r} =\sqrt{\lambda_r}\, \phi$
even on the Fermi surface, $k=\mu$. Therefore, to excite (generate) a pair of gapped quarks
costs at least the energy $2\sqrt{\lambda_r}\,\phi$. Such an excitation process can be interpreted as the break-up of a Cooper pair and the required energy as the Cooper pair binding energy \cite{bcs,gala,fetter,tinkham,vollhardt}.

Gap parameters of $\sim$ 100 MeV would have important phenomenological
consequences for the physics of neutron stars \cite{svidzinsky,link,zhit}. 
%Some of them will be briefly mentioned in the following.
It is therefore of paramount importance to
put the estimates from NJL-type models \cite{4Fermi,4Fermi-rev,huang_2sc,Buballa2,Buballa3} on solid ground
and obtain a more reliable result for the magnitude of the gap
parameter based on first principles. To this end, the
color-superconducting gap parameter was also computed in QCD 
\cite{son,schaferwilczek,rdpdhr,shovkovy,Hsu}. In this section
a short review on color superconductivity is given including the
mechanism of Cooper pairing in QCD, the different symmetry breaking 
patterns for various color superconducting phases, and some implications on the 
observables of neutron stars. At the end of this section a schematic version of the QCD gap equation in the weak coupling limit will be presented. It is suitable to identify and power-count the various terms that arise from the different gluon sectors.

\begin{figure}[ht]
\centerline{\includegraphics[width=4cm]{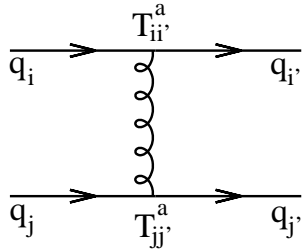}}
\caption{Feynman diagram for one gluon exchange among two quarks.}
\label{4fermi}
\end{figure}

As already explained, at high densities the quarks are deconfined and free to move individually. Since quarks are fermions, Pauli blocking forces them to build up a Fermi sea, where only quarks at the Fermi surface with Fermi momentum $k_F \equiv  \sqrt{\mu^2 - m^2}$ are able to interact and exchange momenta.  
For $\mu \gg m$ the typical momentum scale of these quarks is given by $\mu$. %which is much larger than the quark mass in the absence of any chiral condensate, $\mu \gg m_i$, so one finally can neglect the quark masses, $m_i\simeq 0$ (ultra relativistic limit). Then ${\cal L}_{\rm QCD}$ becomes chirally symmetric, i.e.\ invariant under $SU(N_f)_r\otimes SU(N_f)_\ell$ transformations.
Hence, for $\mu \gg \Lambda_{\rm QCD}$ the strong interaction becomes weakly coupled and single gluon exchange, cf.\ Fig.\ \ref{4fermi}, becomes dominant. The quark-quark scattering amplitude for one-gluon exchange is proportional to \cite{DHRreview,shovy}
\begin{eqnarray}\label{antitrip}
\sum\limits_{a=1}^{N_c^2-1}T^a_{ii^\prime}T^a_{jj^\prime}= 
-\frac{N_c +1}{4N_c}\left(\delta_{ii^\prime}\delta_{jj^\prime}-\delta_{ij^\prime}\delta_{i^\prime j}\right )+\frac{N_c-1}{4N_c}\left(\delta_{ii^\prime}\delta_{jj^\prime}+\delta_{ij^\prime}\delta_{i^\prime j}  \right)\;.\label{4fermiformula}
\end{eqnarray}
The indices $i,j$ are the fundamental colors of the two quarks in the incoming channel, and $i^\prime,j^\prime$ the respective colors in the outgoing channel. If one interchanges two color indices in the incoming or in the outgoing channel the sign of the first term changes while the second term remains intact.
The minus sign in front of the first, antisymmetric term indicates that this channel is attractive, whereas the other, symmetric term is repulsive. 
In terms of representation theory, Eq.\ (\ref{antitrip}) corresponds to the coupling of two fundamental color triplets to the antisymmetric color antitriplet and the symmetric color sextet
\begin{eqnarray}\label{antitriprep}
[3]_c \otimes [3]_c = [\bar 3 ]^a_c \oplus [6]^s_c\;.
\end{eqnarray}
Hereby, the 9-dimensional direct product of two 3-dimensional representations of $SU(3)_c$ has been uniquely decomposed into a direct sum of one  3-dimensional and one 6-dimensional representation. The term $[\bar 3 ]^a_c$ corresponds to the first term in Eq.\ (\ref{4fermiformula}) and $ [6]^s_c$ to the second.

The presence of an attraction among quarks is crucial. According to Cooper's theorem \cite{bcs} any attractive interaction destabilizes the Fermi surface in favor of the formation of Cooper pairs \cite{gala,fetter,tinkham,vollhardt}. Also for $\mu \sim 400 - 500$ MeV, where the theory is not yet in its weak coupling limit, $\alpha_s(\mu) \sim 1$, one expects the anti-triplet channel to be attractive. However, then it would be mediated by instantons \cite{ARWRSSV} rather than by one-gluon exchange. 

As in the case of the chiral condensate, the Cooper pairs spontaneously reduce the symmetry of the system. Normal quark matter without any condensate is invariant under arbitrary 
\bea
{\rm NQ}:~~ SU(3)_c\otimes SU(N_f)_V  \otimes SU(2)_J \otimes U(1)_B \otimes U(1)_{em} \label{Lsym}
\eea
transformations, where also the global spin group $SU(2)_J$ is included with the {total} quark spin $J=S+L$. $N_f$ is the number of quark flavors with approximately equal masses.
%Since quarks carry also  electric charge, the {electromagnetic} gauge group $U(1)_{em}$ has to be added. 
The residual symmetry of the system in the presence of a Cooper pair condensate is given by the symmetry of the order parameter $\Phi$, i.e.\ that of the Cooper pair wave function. Already from the symmetry breaking pattern of a specific color superconducting phase one can draw some important conclusions for its key properties. In this context, however, it is necessary to remember the Anderson-Higgs mechanism \cite{anderson,higgs}. It is roughly speaking the transfer of Goldstone's theorem onto {gauge} symmetries:
if a gauge symmetry is spontaneously broken then those gauge bosons, which correspond to the broken generators, acquire masses. In this way they absorb the degrees of freedom, which would have been occupied by the Goldstone bosons if the broken symmetries had been global. One jargonizes this by saying that these gauge bosons ``eat up the would-be-Goldstone-bosons and therefore become massive.'' (Strictly speaking, a gauge symmetry must first be fixed before it can be broken \cite{elitzur}. After its fixing the gauge the symmetry has become global and the occurring breaking is explicit. The corresponding Goldstone bosons depend on the choosen gauge and are therefore non-physical. The masses of the gauge bosons, however, are independent of the gauge and are physical.) 

One canonical example for the Anderson-Higgs mechanism is the Weinberg-Salam model for the electroweak interactions \cite{weinberg}, which has a $SU(2)_I \otimes U(1)_Y$ gauge symmetry. The index $I$ stands for the isospin and $Y$ for the hypercharge of the left-handed leptons and quarks. One introduces the so-called Higgs field \cite{higgs}, which spontaneously breaks the gauge symmetries to a residual $U(1)_{I+Y}$. Physically one interprets this residual group as the gauge group  $U(1)_{em}$ of electromagnetism with one massless gauge boson $A$, the photon. The three massive gauge bosons that emerge are identified with the $W^\pm$ and the $Z$ boson of the weak interaction. Since the remaining gauge group  $U(1)_{I+Y}$ consists of joint rotations in isospin and hypercharge space, the $Z$ and the $A$ bosons must be generated by a ``rotation'' in the space of the gauge boson of the original $U(1)_Y$ group with one of the gauge bosons of the original $U(1)_I$ group. Those ``original'' gauge bosons on the other hand become meaningless in the phase of the broken symmetries, for details see e.g.\ \cite{Peskin, khuang,introgauge}. %The corresponding rotation angle $\theta_W$ is called the Weinberg angle. 

In the case of color superconductivity the Cooper pairs always carry color charge. Depending on the considered colorsperconducting phase, they therefore reduce the color gauge symmetry to either  $SU(2)_c$ and give masses to five gluons, or break the color gauge symmetry completely. Then all eight gluons become massive. The respective gluon masses are called {color} Meissner masses and lead to the {color} Meissner effect: the respective magnetic gluon fields are expelled from the color superconducting medium. This is analogous to the more familiar electromagnetic Meissner effect, which occurs in electric superconductors. In that case the $U(1)_{em}$ gauge symmetry is spontaneously broken and the photons attain a Meissner mass. Since Cooper pairs of quarks are also electrically charged, a {color} superconductor may also be an {electromagnetic} superconductor depending on the breaking of the electromagnetic gauge symmetry in the considered phase \cite{spin-1-Meissner}. Analogously, the baryon number symmetry may be broken, since Cooper pairs have nonzero baryonic charge. A breaking of this symmetry is connected to the onset of {superfluidity}.

In the following, the requirement of antisymmetry of the Cooper pair wave function will be used to analyze different symmetry breaking patterns. This approach, however, cannot give any information on the breaking of the $U(1)_B$ and $U(1)_{em}$ symmetries. This as well as the values of the Meissner masses require more intensive calculations, which shall not be presented here. As shown above, the order parameter is an element of the representation $[\bar 3 ]^a_c$ of $SU(3)_c$ and is therefore {antisymmetric} in color space. In the case that the Cooper pair is a spin singlet, $J=0$, the wave function must be an element of the {antisymmetric} representation $[1 ]^a_J$ of $SU(2)_J$. If, however, it is a spin triplet, $J=1$, the corresponding representation of $SU(2)_J$ is $[3 ]^s_J$, which is {symmetric}. Therefore, its representation with respect to $SU(N_f)_V$ must be {antisymmetric} if $J=0$, and {symmetric} if $J=1$.

Starting with $J=0$ it follows that the Cooper pairs have to consist of two quarks with different flavors in order to fulfill antisymmetry. Assuming that only two flavors, i.e.\ up and down, participate in the pairing, $N_f=2$, one finds $[1]_f^a$ as the corresponding antisymmetric representation, since analogosly to the spin one has in flavor space
\begin{eqnarray}\label{Nf2}
[2]_V \otimes [2]_V = [1 ]^a_V \oplus [3]^s_V\;.
\end{eqnarray}
In color and flavor the most general ansatz for the order parameter reads
\bea
\Phi_{ij}^{fg} = \epsilon_{ijk}\epsilon^{fg}\, \Phi_k\;, \label{2SCansatz}
\eea
where the indices $i,j$ are the color and $f,g$ the flavor indices of the two quarks and the quantities $\epsilon_{ijk}$ and $\epsilon^{fg}$ are totally antisymmetric tensors of rank 3 in color and and of rank 2 in flavor space, respectively. It follows that the order parameter is a (complex) 3-vector in the fundamental color space. By a global color rotation one may transform the order parameter $\Phi_k$ to point into the (anti-)3 direction in color space. Then only those quarks carrying (transformed) colors 1 and 2 participate in the pairing. This demonstrates that in this phase the $SU(3)_c$ gauge symmetry is broken down to $SU(2)_c$. It follows that five of the eight gluons acquire a Meissner mass in this phase. The remaining three gluons do not interact with the anti-triplett Cooper pairs and remain massless. One finally finds that for this phase, which is called the 2SC phase (indicating that only two quark flavors participate in the pairing), the residual symmetry is given by
\bea
{\rm 2SC}:~~ SU(2)_c\otimes SU(2)_V  \otimes U(1)_{B+ c} \otimes U(1)_{em+c} \label{2SCsym}
\eea
Here, from the outset only the isospin symmetry $SU(2)_V$ was assumed, which is found unbroken. Furthermore, the condensate is invariant under joint rotations of $U(1)_{B}$ and $U(1)_{em}$ as well as under joint rotations of $SU(2)_{c}$ and $U(1)_{em}$ \cite{carter,rischke2SC}. Since the latter two are both gauge groups this implies that, as in the Weinberg-Salam model, the photon is only rotated but remains massless \cite{AlfordBergesRajagopal}. Therefore the 2SC phase is no electromagnetic superconductor. The presence of the symmetry group $U(1)_{B+c}$ suggests that this phase exhibits nontrivial features concerning superfluidity. In helium-3 one encounters similar symmetry breaking patterns where the particle number conservation group is locked with rotations in spin or angular momentum space \cite{vollhardt}. 

Assuming quark chemical potentials much larger than the strange quark mass one may consider the case that up, down, and strange quarks equally participate in the pairing. Then, for $N_f=3$, it is
\begin{eqnarray}\label{Nf3}
[3]_V \otimes [3]_V = [\bar 3 ]^a_V \oplus [6]^s_V\;,
\end{eqnarray}
wherefore the desired antisymmetric representation of $SU(3)_V$ is $[\bar 3 ]^a_J$. Consequently, in Dirac and flavor space the ansatz for the condensate reads
\bea
\Phi_{ij}^{fg} = \epsilon_{ijk}\epsilon^{fgh} \Phi_k^h\;.
\eea
It has a similar structure as the order parameter of superfluid helium-3 with spin $S=1$ and angular momentum $L=1$, which breaks the global $SO(3)_S$ and $SO(3)_L$ symmetries. There one encounters a multitude of different phases \cite{vollhardt}. In the present case of $SU(3)_c$ and  $SU(3)_V$ symmetry breaking, however, it turns out that the following ansatz for $\Phi_k^h$ is dominant \cite{pisarskirischke,arw2}
\bea
\Phi_{k}^{h} = \delta_k^h\, \Phi\;. 
\eea
This ansatz is similar to the order parameter of the chiral condensate, cf.\ Eq.\ (\ref{chiral}), in the chiral limit of zero quark masses, where the different flavors have identical properties and consequently $\Phi^{ij} = \delta^{ij}\Phi$. This order parameter simultaneously breaks the two global flavor symmetries $SU(3)_r \otimes SU(3)_\ell$ to the one global vectorial symmetry $SU(3)_V$. In the present case the global vectorial symmetry $SU(3)_V$ and the local symmetry $SU(3)_c$ are simultaneously broken to one global $SU(3)_{c+V}$ symmetry. 
Accordingly, the system is invariant under ``locked'' rotations in color and flavor space. This phase is therefore called the {color-flavor-locked} or CFL phase. All in all, it is invariant under
\bea
{\rm CFL}:~~ SU(3)_{c+V} \otimes SU(2)_J \otimes U(1)_{c+em}\;. \label{CFLsym}
\eea
In contrast to the 2SC phase all three colors and flavors participate in the pairing and one therefore expects all gluons to interact with the Cooper pair condensate and attain a Meissner mass. This is in fact true, since the remaining color symmetry is only global. Furthermore, since  $U(1)_{c+em}$ is a gauge group the photon remains massless. However, it is {mixed} with one of the gluons. Therefore, the CFL phase is no electromagnetic superconductor, i.e.\ the Cooper pairs are neutral with respect to the rotated electric charge \cite{AlfordBergesRajagopal}.  Due to the broken $U(1)_B$ it is a superfluid. In rotating systems as neutron stars superfluidity may result in vortices, through which magnetic fields can enter the superconducting phase in so called flux-tubes \cite{ruderman}. These vortices stick to the star's crust and and carry some amount of the stars total angular momentum. When the rotational velocity of the star gradually decreases due to electromagnetic radiation these vortices eventually rearrange and transfer some of the angular momentum to the star. As a consequence the rotational frequency is expected to suddenly increase leading to observable ``glitches'' \cite{gl-super}.

As mentioned above, the CFL phase is physically significant only for $\mu \gg m_s$. 
For such large chemical potentials one could in principle neglect all masses and assume invariance under chiral symmetry rotations, $SU(3)_{r} \otimes SU(3)_{\ell}$ instead of the vectorial $SU(3)_V$. In the CFL phase left and right handed quarks separately form Cooper pair condensates, which corresponds to the breaking pattern  $SU(3)_{r+c} \otimes SU(3)_{\ell+c}\,\hat = \,SU(3)_{r+\ell+c}$. Therefore, also in the limit of very large densities the chiral symmetry is broken, however, not by the quark masses or the chiral condensate but by the CFL phase \cite{Schafersympatterns}.

In the case that the Cooper pair carries spin 1, $J=1$, it may consist of two quarks of the {same} flavor \cite{bailinlove,sp1995,spin-1,spin-1-Meissner,bowers,andreas}. For simplicity one first considers the case that there is only one quark flavor in the system, $N_f=1$. As an element of $ [3]_J^s$ the order parameter is a 3-vector in momentum space with components $a=x,y,z$. The most general ansatz for the order parameter reads in spin and color space \cite{bailinlove,rdpdhr,schafer,asqwdhr}
\begin{equation}
\Phi_{ij}^a = \epsilon_{ijk} \, \Phi^a_k\;,
\end{equation}
which generally breaks the $SU(3)_c \otimes SU(2)_J$ symmetry and allows for a multitude of possible phases
\cite{bowers,andreas}. Here, only two special phases shall be analyzed, which are interesting due to their analogy to the 2SC and the CFL phases of spin-0 color superconductors. These are the {color-spin-locked} or CSL phase
\begin{equation} \label{CSL}
\Phi^a_k =  \delta^a_k \, \Phi\,\, ,
\end{equation}
and the {polar} phase, where
\begin{equation}
\Phi^a_k =  \delta_{k3}\, \delta^{az} \, \Phi \,\, .
\end{equation}
In the first the color gauge group $SU(3)_c$ and the global spin group $SU(3)_J$ are simultaneously broken to the global $SU(3)_{c+J}$ group, which ``locks'' color and spin rotations.  One has in total
\bea
{\rm CSL}:~~ SU(3)_{c+J} \otimes U(1)_V\;. \label{CSLsym}
\eea
As in the CFL phase all gluons and the photon become massive. In contrast to the CFL phase, however, the  $U(1)_B$ and the $U(1)_{em}$ symmetry are broken. Hence, the CSL phase is a superfluid and a electromagnetic superconductor. The implications of the electromagnetic superconductivity on the observables of a neutron star are discussed in \cite{spin-1-Meissner}.

The polar phase is similar to the 2SC phase with $SU(3)_c\otimes SU(2)_J$ breaking down to $SU(2)_c\otimes U(1)_J$. One finds for the residual symmetry group
\bea
{\rm polar}:~~ SU(2)_{c} \otimes U(1)_{J} \otimes U(1)_V\otimes U(1)_{B+c}\otimes U(1)_{em+c}\;. \label{polarsym}
\eea
Consequently, in the polar phase five gluons become massive. The condensate leaves rotational invariance in momentum space only with respect to one fixed axis. Indeed, the color superconducting gap parameter, see below, exhibits only an axial symmetry \cite{asqwdhr}. The $U(1)_{em}$ remains unbroken, wherefore the polar phase is no electromagnetic superconductor. However,  a many flavor system, where each flavor separately forms a polar spin-1 gap, can be shown to be electromagnetically superconducting \cite{asqwdhr}.

The physical significance of one-flavor color superconductors is that in neutron stars the Fermi momenta of quarks with different flavors may become sufficiently different to rule out the normal BCS pairing mechanism, which requires exactly opposite quark momenta. Such a mismatch of Fermi surfaces indeed occurs in neutron stars if one respects its electrical neutrality and $\beta-$equilibrium \cite{shovy,iida1,rajagopal,alford5,huang,neumann,ruester}. Total electric and color charge neutrality has to be fulfilled in order to allow for the gravitational binding of the star. Electrical charge neutrality for a two flavor system neglecting electrons, e.g., requires $n_d \approx 2n_u$, since the electrical charge of an up quark is $Q_u = 2/3$, whereas  $Q_d = -1/3$ for a down quark. With $n_{u,d} \sim \mu_{u,d}^3$ it therefore follows $\mu_d \approx 2^{1/3}\mu_u\approx 1.26 \mu_u$.
In this simple case of two flavors, the requirement of $\beta-$equilibrium will not change this estimate much, cf.\ \cite{shovy}. 

In spin-1 color superconductors quarks of the same flavor may pair, wherefore the BCS-pairing mechanism remains always intact. Alternatively, other color superconducting phases are proposed with different pairing mechanisms. In the so-called interior gap or breached pairing the quarks with the smaller Fermi momentum are excited to a higher energy state in order to form Cooper pairs with the quarks with larger Fermi momentum \cite{liu,gubankova}. Also displacements \cite{alford6,bowersphd} (LOFF phases) and deformations of the Fermi surfaces \cite{muether} are discussed to allow for Cooper pairing. The breached pairing mechanism has been shown to lead to Cooper pairs with gapless spectra \cite{shovkovy1,shovkovy2,shovkovy3,gCFL}, which however appear to be unstable \cite{Huang,honggapless,Casalbuoni,AlfordWang,Fukushima}.

To decide which of the many possible phases is favorable at given $T$ and $\mu$ one has to determine the phase with the largest pressure. Generally, the pressure grows with the gain of condensation energy of the Cooper pairs. This in turn depends on the number of quark degrees of freedom, which participate in the Cooper pairing, and on the magnitude of the respective color superconducting gap parameters. The gap parameter is determined from the gap equation. For a general derivation see \cite{DHRreview} and references therein. In Sec.\ \ref{IV} the gap equation is rederived within the new effective formalism. In the following it is sufficient to concentrate on the gap equation in the schematic form
\be \label{gapeq}
\phi = g^2\, \phi \left[ \zeta\, \ln^2 \left(\frac{\mu}{\phi}\right)
+ \beta\, \ln \left( \frac{\mu}{\phi}\right) + \alpha \right]
\ee
to discuss the various contributions from different gluon sectors and their respective orders of magnitude. To derive it one has to assume zero temperature, $T=0$, weak coupling, $g \ll 1$, 
and apply the mean-field approximation.
The solution is
\be \label{gapsol}
\phi = 2 \, b\, \mu \, \exp \left( - \frac{c}{g} \right) \left[ 1 +
O(g) \right]\;.
\ee
The first term in Eq.\ (\ref{gapeq}) is of {\em leading\/} order
since, according to Eq.\ (\ref{gapsol}), $g^2 \ln^2 (\mu/\phi) \sim 1$.
It originates from the exchange of almost static, long-range, Landau-damped
magnetic gluons \cite{son}. One factor $\ln (\mu/\phi)$ is the standard BCS
logarithm, which arises when integrating over quasiparticle modes 
from the bottom to the surface of the Fermi sea, 
$\int dq/\epsilon_q \sim \ln (\mu/\phi)$.
(In fact, as will be shown 
in Sec.\ \ref{estAB}, integrating over quasiparticle modes inside a layer around the Fermi surface of an 
 ``intermediate'' width is sufficient to build up a BCS log.)
The second factor $\ln (\mu/\phi)$ comes from a collinear enhancement $\sim
\ln(\mu/\epsilon_q)$ in the exchange of almost
static magnetic gluons. The coefficient $\zeta$ determines the constant
$c$ in the exponent in Eq.\ (\ref{gapsol}). As was 
first shown by Son \cite{son},
\be
c  \equiv \frac{3 \pi^2}{\sqrt{2}} \;.
\ee
The second term in Eq.\ (\ref{gapeq}) is of {\em subleading\/} order,
$g^2 \ln (\mu/\phi) \sim g \ll 1$. It originates from two sources. 
The first is the exchange of
electric and non-static magnetic gluons 
\cite{schaferwilczek,rdpdhr,shovkovy,Hsu}. In this case, the
single factor $\ln (\mu/\phi)$ is the standard BCS logarithm. The
second source 
is the quark wave-function renormalization factor in dense quark matter
\cite{rockefeller,qwdhr}. Here, the BCS logarithm does not arise, but
 the wave-function renormalization contains an additional $\ln
(\mu/\epsilon_q)$ which generates a $\ln(\mu/\phi)$. The coefficient
$\beta$ determines the prefactor $b$ of the exponent in Eq.\
(\ref{gapsol}).
For a two-flavor color superconductor,
\be
b \equiv 256\, \pi^4\, \left( \frac{2}{N_f \, g^2} \right)^{5/2} 
\exp\left( -\frac{\pi^2 + 4}{8} \right)\;,
\ee
where $N_f$ is the number of (massless) quark flavors participating in 
screening the gluon exchange interaction.
The third term in Eq.\ (\ref{gapeq}) is of {\em sub-subleading\/}
order, $\sim g^2$. The coefficient $\alpha$ determines the
$O(g)$ correction to the prefactor of the color-superconducting gap
parameter in Eq.\ (\ref{gapsol}). Since $\alpha$ has 
not yet been determined, the gap parameter can be reliably computed only
in weak coupling, i.e., when the $O(g)$ corrections to the prefactor
are small. 

Weak coupling, $g \ll 1$, however, requires {asymptotically}
large quark chemical potentials,  $\mu \gg \Lambda_{\rm QCD}$. 
The range of $\mu$ values of phenomenological
importance is, however, $\lesssim 1$ GeV.
Although the quark density $n$ is already quite large
at such values of $\mu$, $n \sim 10$ times
the nuclear matter ground state density, the coupling constant
is still not very small, $g \sim 1$. It is therefore of interest
to determine the coefficient of $g$ in the $O(g)$ corrections
to the prefactor in Eq.\ (\ref{gapsol}).
If it turns out to be small, one gains more confidence in
the extrapolation of the weak-coupling result
(\ref{gapsol}) to chemical potentials of order $\sim 1$ GeV.

It is worthwhile mentioning that an 
extrapolation of the weak-coupling result (\ref{gapsol}) for a two-flavor color
superconductor, neglecting sub-subleading
terms altogether and assuming
the standard running of $g$ with the chemical potential $\mu$,
yields values of $\phi$ of the order of $\sim 10$ MeV at chemical
potentials of order $\sim 1$ GeV, cf.\ Ref.\ \cite{DHRreview}.
This is within one order of magnitude of the predictions based on 
NJL-type models and thus might lead one to conjecture
that the true value of $\phi$ will lie somewhere in the
range $\sim 10 - 100$ MeV.

However, in order to confirm this and to obtain a more
reliable estimate of $\phi$ at values of $\mu$ of relevance in
nature, one ultimately has to
compute all terms contributing to sub-subleading order. In the following section the approach to this problem by an effective theory will be motivated.

%The effective action derived in this work may serve as an adequate tool to calculate the sub-subleading order corrections to the gap. As explicitly demonstrated in Chapter \ref{chapterapp} two cutoffs parameters can be effectively employed to identify various sub-subleading terms. In the following section the approach 

%An interesting consequence of a more realistic scenario with a rather strong coupling is that the spatial dimension of a Cooper pair becomes smaller allowing to interprete the pairs as one bosonic particles. These then may form a Bose condensate \cite{Itakura}. Due to the stronger coupling also quarks deeper in the Fermis sea could in principle participate in the coupling. This is on contrast to ordinary superconductivity of electrons in metals, where the relevant region around the Fermi surface is principally restriced by the Debye frequency $\omega_D$. 

\section{Search for an \mbox{\it effective} approach} \label{EFT}

Although possible in principle, the task to calculate the sub-subleading order contributions to the color
superconducting gap parameter is prohibitively 
difficult within the standard solution of the QCD gap
equation in weak coupling. So far, in the course of this solution
terms contributing at leading and subleading order have been identified, cf.\ discussion after Eq.\ (\ref{gapsol}). However, up to date it remained unclear which terms one would have to
keep at sub-subleading order. Moreover, 
additional contributions could in principle 
arise at any order from diagrams neglected in the
mean-field approximation \cite{rockefeller,DHQWDHR}.
Therefore, it would be {\it ideal} to have 
a computational scheme, which allows one
to determine {\em a priori\/}, i.e., at the outset of the
calculation, which terms contribute to the gap equation 
at a given order.

As a first step towards this goal, 
note that there are several scales in the problem. 
Besides the chemical potential $\mu$,
there is the inverse gluon screening length, which is of
the order of the gluon mass parameter $m_g$. At zero temperature
and for $N_f$ massless quark flavors \cite{LeBellac},
\be \label{gluonmass}
m_g^2 = N_f \, \frac{g^2 \mu^2}{6 \pi^2}\;,
\ee
i.e., $m_g \sim g \mu$.
Finally, there is the color-superconducting gap parameter $\phi$,
cf.\ Eq.\ (\ref{gapsol}). In weak coupling, $g \ll 1$, these three scales
are naturally ordered, $\phi \ll g \mu \ll \mu$.
This ordering of scales implies that the modes near the Fermi surface,
which participate in the formation of Cooper pairs and are therefore of
primary relevance in the gap equation, can be considered to be
independent of the detailed dynamics of the modes deep in the 
Fermi sea. This suggests that the most efficient way to compute 
properties such as the color-superconducting gap parameter is 
via an {\em effective theory for quark modes near the Fermi surface}.
Such an effective theory has been originally proposed by
Hong \cite{hong,hong2} and was subsequently refined by others
\cite{HLSLHDET,schaferefftheory,NFL,others}.

\begin{figure}[ht]
\centerline{\includegraphics[width=13cm]{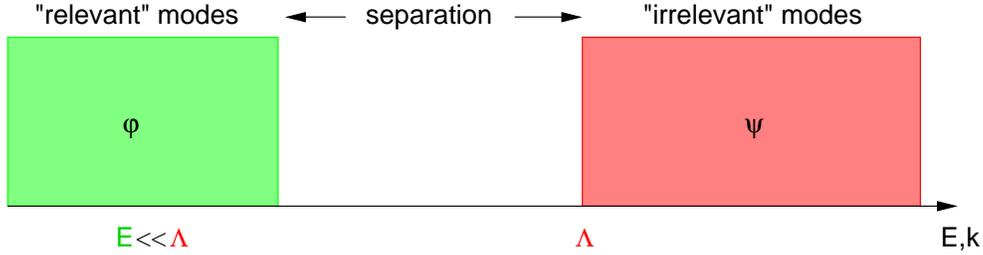}}
\caption[The separation of scales.]{The separation of scales.}
\label{scalesep}
\end{figure}

At this point it is worthwhile reviewing
the standard approach to derive an effective theory
\cite{polchinski,Kaplan,Manohar}. In the most simple case, one has a single 
scalar field, $\phi$, and a single momentum scale, $\Lambda$, which 
separates relevant modes, $\varphi$, from irrelevant modes, $\psi$, $\phi =
\varphi + \psi$. The relevant modes live on spatial scales 
$L \gg 1/\Lambda$, while the irrelevant modes 
live on scales $l \lesssim 1/ \Lambda \ll L$, cf.\ Fig.\ \ref{scalesep}.
In the derivation of the effective action, one is supposed to integrate out
the microscopic, irrelevant modes. 
Usually, however, this is not done explicitly. Instead, one
constructs all possible operators ${\cal O}_i$ composed of 
powers of the field $\varphi$ and its derivatives,
which are consistent with the symmetries of
the underlying theory, and writes the effective action as \cite{Kaplan}
\be \label{Seffstandard}
S_{\rm eff}[\varphi] = \int _X \sum_i g_i {\cal O}_i(\varphi)\;.
\ee
The coefficients, or {\em vertices\/}, $g_i$ determine the interactions of
the relevant modes $\varphi$. A priori, they are unknown 
functions of the single scale $\Lambda$, $g_i = g_i(\Lambda)$. 
All information about the microscopic scale $l$ is contained 
in these vertices. Since the microscopic scale 
$l \ll L$, the operators
${\cal O}_i$ are assumed to be {\em local\/} on the scale $L$.

The effective action (\ref{Seffstandard}) contains infinitely many terms.
In order to calculate physical observables within the effective
theory, one has to truncate the expansion after a finite number of
terms. One can determine the order of magnitude of various terms in the
expansion (\ref{Seffstandard}) via a dimensional scaling analysis, which allows
to classify the operators as {\em relevant\/} (they become
increasingly more important as the scale $L$ increases),
{\em marginal\/} (they do not change under scale transformations),
and {\em irrelevant\/} (they become increasingly less 
important as the scale $L$ increases).
To this end, one determines the naive scaling dimension of the
fields, ${\rm dim}(\varphi) \equiv \delta$, from the free
term in the effective action. Then, if the operator ${\cal O}_i$ consists of 
$M$ fields $\varphi$ and $N$ derivatives, its scaling dimension 
is ${\rm dim} ({\cal O}_i) \equiv \delta_i = M \delta + N$.
The operator ${\cal O}_i$ is then of order $\sim L^{-\delta_i}$. 
For dimensional reasons the constant
coefficients $g_i$ must then be of order
$\sim \Lambda^{d-\delta_i}$, where $d$ denotes the dimensionality of
space-time. Including the integration over space-time,
the terms in the expansion (\ref{Seffstandard}) are then of 
order $\sim (L \Lambda)^{d-\delta_i}$.
Consequently, relevant operators must have $\delta_i < d$,
marginal operators $\delta_i = d$, and irrelevant operators $\delta_i > d$.
At a given scale $L$, one has to take into account only relevant, 
or relevant and marginal, or all three types of operators,
depending on the desired accuracy of the calculation. To calculate
a process at energy $E =s\Lambda$ with $s\ll 1$ and error of order $s^{n+1}$ one
only needs operators ${\cal O}_i$ with dimension $\delta_i\leq D+n$.
The final result still depends on $\Lambda$ through the
coefficients $g_i(\Lambda)$. This dependence is 
eliminated by computing a physical observable in the
effective theory and in the underlying microscopic theory, and
matching the result at the scale $\Lambda$.

There are, however, cases where this naive dimensional scaling analysis fails
to identify the correct order of magnitude, and thus the relevance, 
of terms contributing to the effective action.
In the following, three examples shall be discussed briefly. For the first example, consider effective
theories where, in contrast to the above assumption, 
the vertices $g_i$ are in fact {\em non-local\/} functions. Such 
theories are, for instance, given by
the ``Hard Thermal Loop'' (HTL) or ``Hard Dense Loop'' (HDL) 
effective actions \cite{braatenpisarski,LeBellac}. 
In these effective theories, valid at length scales $L \sim 1/(gT)$ or
$\sim 1/(g \mu)$, respectively,
there are terms $g_n\, A^n$ in the effective action, which are
constructed from a quark or gluon (or ghost) loop
with $n$ external gluon legs;
$A$ is the external gluon field with $\delta = 1$. The 
coefficients $g_n$ are non-local and 
do not only depend on the scale $\Lambda \lesssim T$, or $\lesssim \mu$, 
but also on the relevant momentum scale $1/L \sim g T$, or $\sim g \mu$. 
Naively, one would expect
$g_n$ to belong to a local $n$-gluon operator and to
scale like $\Lambda^{4-n}$. Instead,
it scales like $L^{n-4}$ \cite{braatenpisarski}. 
For arbitrary $n$, the corresponding term $g_n\, A^n$
in the effective action then scales like $L^4$, independent
of the number $n$ of external gluon legs.

The second example pertains to the situation when there is
more than one single momentum scale $\Lambda$.
As explained above, for a single scale $\Lambda$ and a given length scale
$L$, the naive dimensional scaling analysis unambiguously determines
the order of magnitude of the terms in the expansion (\ref{Seffstandard}).
Now suppose that there are two scales, $\Lambda_1$ and $\Lambda_2$.
Then, the vertices $g_i$ may no longer be functions of a single scale,
say $\Lambda_1$, but could also depend on the ratio of
$\Lambda_2/\Lambda_1$. 
Two scenarios are possible: (a) two terms in the expansion
(\ref{Seffstandard}), say $g_n {\cal O}_n$ and
$g_m {\cal O}_m$, with the {\em same\/} scaling behavior may still be
of a {\em different\/} order of magnitude, or (b) the two terms can have
a {\em different\/} scaling behavior, but may still be of the {\em
same\/} order of magnitude.
In case (a), all that is required is that the operators ${\cal O}_n$
and ${\cal O}_m$ scale in the same manner, say $L^{-k}$, and that
$g_n \sim \Lambda_2^{d-k}$, but $g_m \sim \Lambda_1^{d-k}$.
If $\Lambda_1 \ll \Lambda_2$, $g_m \gg g_n$, and thus the two
terms are of different order of magnitude.
In case (b), assume $1/L \ll \Lambda_1 \ll \Lambda_2$,
with $\Lambda_1/ \Lambda_2 \sim 1/(\Lambda_1 L) \sim \epsilon \ll 1$ and let
 take the fields $\varphi$ to have naive scaling dimension $\delta =1$.
Then, at a given length scale $L$,
a term $g_n \varphi^n$, with a coefficient $g_n$ of order 
$\Lambda_2^{d-n}$, can be of the same order of magnitude 
as a term  $g_m \varphi^m$, $m \neq n$, if the coefficient
$g_m \sim \Lambda_1^{d-m} (\Lambda_2/ \Lambda_1)^k$ with $k= d+ m - 2n$.
Although the scaling behavior of the two terms is quite different as
$L$ increases, they can be of the same order of magnitude,
if the interesting scale $L$ happens to be
$\sim \Lambda_2/\Lambda_1^2$. In both cases (a) and (b) 
the naive dimensional scaling analysis fails to correctly sort the operators
${\cal O}_i$ with respect to their order of magnitude.

The third example where the naive dimensional scaling analysis fails
concerns quantities, which have to be calculated self-consistently.
Such a quantity is, for instance,
the color-superconducting gap parameter, which is computed
from a Dyson-Schwinger equation within a given many-body
approximation scheme. In this case, the self-consistent solution scheme
leads to large logarithms, like the BCS logarithm in Eq.\
(\ref{gapeq}). These logarithms cannot be identified {\em a priori\/}
on the level of the effective action, but only emerge {\em in the
course\/} of the calculation \cite{rdpdhr}.

In order to avoid these failures of the standard approach,
in this work I pursue a different venue to 
construct an effective theory. I introduce cut-offs in momentum space
for quarks, $\Lambda_{\rm q}$, and gluons, $\Lambda_{\rm gl}$. These cut-offs
separate relevant from irrelevant quark modes and soft from hard
gluon modes. I then explicitly integrate out irrelevant quark and
hard gluon modes and derive a {\it general} effective action for
hot and/or dense quark-gluon matter. One advantage of this approach is
that I do not have to guess the form of the possible operators
${\cal O}_i$ consistent with the symmetries of the underlying theory.
Instead, they are exactly derived from first principles. 
Simultaneously, the vertices $g_i$ are no longer unknown,
but are completely determined. 
Moreover, in this way I construct {\em all\/} possible operators and thus do
not run into the danger of missing a potentially important one.

I shall show that the standard HTL and HDL effective actions are
contained in the general effective action for a certain choice of the
quark and gluon cut-offs $\Lambda_{\rm q},\, \Lambda_{\rm gl}$. Therefore,
the new approach naturally generates non-local terms in the
effective action, including their correct scaling behavior, which, as
mentioned above, does not follow the rules of the naive 
dimensional scaling analysis. I also show that the action of
the high-density effective theory
derived by Hong and others 
\cite{hong,hong2,HLSLHDET,schaferefftheory,NFL,others}
is a special case of the general effective action. In this case, relevant
quark modes are located within a layer of width $2 \Lambda_{\rm q}$ around the
Fermi surface.

The two cut-offs, $\Lambda_{\rm q}$ and $\Lambda_{\rm gl}$, introduced in the present approach
are in principle different, $\Lambda_{\rm q} \neq \Lambda_{\rm gl}$. 
The situation is then as in the second example mentioned above, where the 
naive dimensional scaling analysis fails to
unambiguously estimate the order of magnitude of the various terms in the
effective action. Within the present approach, this problem does not
occur, since all terms, which may occur in the effective action, are 
automatically generated and can be explicitly kept in the further
consideration.
I shall show that in order to produce the correct result for the 
color-superconducting gap parameter to subleading order in weak
coupling, one has to demand $\Lambda_{\rm q} \lesssim g \mu \ll \Lambda_{\rm gl}
\lesssim \mu$, so that $\Lambda_{\rm q}/\Lambda_{\rm gl} \sim g \ll 1$.
Only in this case, the dominant contribution to the QCD gap equation
arises from almost static magnetic gluon exchange, while subleading
contributions are due to electric and non-static magnetic
gluon exchange. The fact that $\Lambda_{\rm q} \ll \Lambda_{\rm gl}$ is not
entirely unexpected: at asymptotically large densities, where
the scale hierarchy is $\phi \ll g\mu \sim m_g \ll \mu$,
the dominant contribution in the QCD gap equation
arises from gluons with momenta of order $\sim (m_g^2 \phi)^{1/3}$
\cite{son,schaferwilczek,rdpdhr}, while typical quark momenta lie in
a shell of thickness $\sim 2\, \phi \ll( m_g^2 \phi)^{1/3}$ around
the Fermi surface.

The color-superconducting gap parameter is computed 
from a Dyson-Schwinger equation for the quark propagator.
In general, this equation corresponds to a self-consistent resummation of
all one-particle irreducible (1PI) diagrams for the quark self-energy.
A particularly convenient way to derive Dyson-Schwinger
equations is via the Cornwall-Jackiw-Tomboulis (CJT) formalism \cite{CJT}. 
In this formalism, one constructs the set
of all two-particle irreducible (2PI) vacuum diagrams from the
vertices of a given tree-level action. The functional derivative of
this set with respect to the full propagator then defines the 
1PI self-energy entering the Dyson-Schwinger equation.
Since it is technically not feasible to include all possible
diagrams, and thus to solve the Dyson-Schwinger equation exactly,
one has to resort to a many-body approximation scheme, which
takes into account only particular classes of diagrams. 
The advantage of the CJT formalism is that such an approximation
scheme is simply defined by a truncation of the set of 2PI diagrams.
However, in principle there is no parameter, which controls the
accuracy of this truncation procedure.

The standard QCD gap equation in mean-field approximation studied in Refs.\
\cite{schaferwilczek,rdpdhr,shovkovy} follows from this approach
by including just the sunset-type diagram, which is constructed
from two quark-gluon vertices of the QCD tree-level action
(see, for instance, Fig.\ \ref{Gamma2eff} in Sec.\ \ref{IVa}).
I also employ the CJT formalism to derive the gap equation for
the color-superconducting gap parameter. However, I construct
all diagrams of sunset topology from the vertices of
the general {\em effective\/} action derived in this work.
The resulting gap equation is equivalent to the gap equation in 
QCD, and the result for the gap parameter to subleading order
in weak coupling is identical to that in QCD, provided 
$\Lambda_{\rm q} \lesssim g \mu \ll \Lambda_{\rm gl} \lesssim \mu$.
The advantage of using the effective
theory is that the appearance of the two scales $\Lambda_{\rm q}$ and
$\Lambda_{\rm gl}$ considerably facilitates the power counting
of various contributions to the gap equation as compared to full QCD.
I explicitly demonstrate this in the course of the calculation 
and suggest that, within this approach, 
it should be possible to 
identify the terms, which contribute beyond subleading order
to the gap equation. Of course, for a complete sub-subleading
order result one cannot restrict oneself to the sunset diagram,
but would have to investigate other 2PI diagrams as well.
This again shows that an {\em a priori\/} estimate of the relevance
of different contributions on the level of the effective action
does not appear to be feasible for quantities, which have to be computed
self-consistently.

This work is organized as follows. In Sec.\ \ref{II} I 
derive the general effective action by explicitly integrating out
irrelevant quark and hard gluon modes. In Sec.\ \ref{III} I 
show that the well-known HTL/HDL effective action, as well as the
high-density effective theory proposed by Hong and others,
are special cases of this general effective action for particular
choices of the quark and gluon cut-offs $\Lambda_{\rm q}$ and $\Lambda_{\rm gl}$,
respectively. Furthermore, I power count one special subset of diagrams, 
loops of irrelevant quark modes, analogously to the HTL power counting scheme
\cite{braatenpisarski} for the choice $\phi \ll\Lambda_{\rm q}\ll g\mu \ll \L_{\rm gl} \lesssim \mu$.

Section \ref{IV} contains the application of the
general effective action to the computation of the
real part of the color-superconducting gap parameter. 
In Sec.\ \ref{imsec} I consider the energy dependence of the gap function 
within the effective theory and compute its imaginary part for quarks 
near the Fermi surface.
In Sec.\ \ref{V} I conclude this work with a summary of the results and an outlook.

The units I use are $\hbar=c=k_B=1$. 4-vectors are denoted by
capital letters, $K^\mu = (k_0, {\bf k})$, with ${\bf k}$ being a
3-vector of modulus $|{\bf k}| \equiv k$ and direction
$\hat{\bf k}\equiv {\bf k}/k$. For the summation over Lorentz
indices, I use a notation familiar from Minkowski space, with metric
$g^{\mu \nu} = {\rm diag}(+,-,-,-)$, although 
I exclusively work in compact Euclidean space-time with 
volume $V/T$, where $V$
is the 3-volume and $T$ the temperature of the system. 
Space-time integrals 
are denoted as $\int_0^{1/T} d \tau \int_V d^3{\bf x} \equiv
\int_X$. Since space-time is compact, energy-momentum space is
discretized, with sums $(T/V)\sum_{K} \equiv T\sum_n (1/V) \sum_{\bf
k}$. For a large 3-volume $V$, the sum over 3-momenta
can be approximated by an integral, $(1/V)\sum_{\bf k} \simeq
\int d^3 {\bf k}/(2 \pi)^3$. For bosons, the sum over $n$ runs over
the bosonic Matsubara frequencies $\omega_n^{\rm b} = 2n \pi T$, while 
for fermions, it runs over the fermionic Matsubara frequencies
$\omega_n^{\rm f} = (2 n+1)\pi T$. In a Minkowski-like notation 
for four-vectors, $x_0 \equiv t \equiv -i \tau$, 
$k_0 \equiv -i \omega_n^{\rm b/f}$. The 4-dimensional delta-function
is conveniently defined as $\delta^{(4)}(X) \equiv \delta(\tau)\,
\delta^{(3)}({\bf x}) = -i \, \delta(x^0)\, \delta^{(3)}({\bf x})$.
		\chapter{A general effective action for quark matter}

\section{Deriving the effective action} \label{II}

In this section, I derive a general effective action for
hot and/or dense quark matter. I start from the QCD
partition function in the functional integral representation 
(Sec.\ \ref{IIa}). I first
integrate out irrelevant fermion degrees of freedom
(Sec.\ \ref{Intquarks}) and then hard gluon degrees of freedom
(Sec.\ \ref{Intgluons}). The final result is Eq.\ (\ref{Seff}) in
Sec.\ \ref{IId}. I remark that the same result could have been
obtained by first integrating out hard gluon modes and then
irrelevant fermion modes, but the intermediate steps leading to the 
final result are less transparent.

\subsection{Setting the stage} \label{IIa}

As discussed in the introduction, cf.\  Eq.\ (\ref{LQCD}-\ref{SQCD}),
the partition function for QCD in the absence of external
sources may be written as
\be \label{ZQCD}
{\cal Z} = \int {\cal D} A \, 
\exp \left\{ S_A [A]\right\}\, {\cal Z}_q[A] \,\, ,
\ee
where for convenience the pure (gauge-fixed) gluon action has been introduced
\be \label{SA}
S_A[A] = \int_X \left[ - \frac{1}{4} F^{\mu \nu}_a (X) \, F_{\mu \nu}^a
(X) \right] + S_{\rm gf}[A] + S_{\rm ghost}[A] \,\,.
\ee
The partition function for quarks in the presence of gluon fields is
\be \label{Zq}
{\cal Z}_q[A] = \int {\cal D} \bar{\psi} \, {\cal D}\psi\,
\exp \left\{ S_q[A,\bar{\psi},\psi] \right\}\,\, ,
\ee
where the quark action is
\be
S_q[A,\bar{\psi},\psi] = \int_{X} \bar{\psi}(X) \,
 \left( i \Diracslash{D}_X + \mu \gamma_0 - m \right) 
 \, \psi(X) \,\,.
\ee
In fermionic systems at nonzero density, it is advantageous to
additionally introduce charge-conjugate fermionic degrees of freedom,
\be \label{cc}
\psi_C(X) \equiv C \, \bar{\psi}^T(X) \;\;, \;\;\;\;
\bar{\psi}_C (X) \equiv \psi^T (X)\, C \;\; ,\;\;\;\;
\psi(X) \equiv C \, \bar{\psi}^T_C(X) \;\;, \;\;\;\;
\bar{\psi} (X) \equiv \psi^T_C (X)\, C \;\; ,
\ee
where $C \equiv i \gamma^2 \gamma_0$ is the charge-conjugation matrix,
$C^{-1} = C^\dagger = C^T = -C$, $C^{-1} \gamma_\mu^T C = - \gamma_\mu$;
a superscript $T$ denotes transposition. I may then rewrite the 
quark action in the form
\be \label{quarkaction}
S_q[A, \bar{\Psi}, \Psi] = 
\frac{1}{2} \int_{X,Y} \bar{\Psi}(X) \, {\cal G}_0^{-1} (X,Y)\,
\Psi(Y) + \frac{g}{2} \int_X \bar{\Psi}(X) \, \hat{\Gamma}^\mu_a
A_\mu^a(X) \, \Psi(X) \,\, ,
\ee
where I defined the Nambu-Gor'kov quark spinors
\be
\Psi \equiv \left( \begin{array}{c}
                    \psi \\
                    \psi_C \end{array} \right) \;\; , \;\;\;\;
\bar{\Psi} \equiv ( \bar{\psi} , \bar{\psi}_C )\,\, ,
\ee
and the free inverse quark propagator in the Nambu-Gor'kov basis
\be
{\cal G}_0^{-1}(X,Y)  \equiv \left( \begin{array}{cc} 
                          [G_0^+]^{-1}(X,Y) & 0 \\
                           0 & [G_0^-]^{-1}(X,Y) \end{array} \right)\,\, ,
\ee
with the free inverse propagator for quarks and charge-conjugate
quarks
\be
[G_0^\pm]^{-1}(X,Y) \equiv (i \Diracslash{\partial}_X \pm \mu
\gamma_0 - m )\, \delta^{(4)}(X-Y)\,\, .
\ee
The quark-gluon vertex in the Nambu-Gor'kov basis is defined as
\be \label{NGvertex}
\hat{\Gamma}^\mu_a \equiv \left( \begin{array}{cc}
                               \gamma^\mu T_a & 0 \\
                               0 & -\gamma^\mu T_a^T 
                              \end{array} \right) \,\, .
\ee
As I shall derive the effective action in momentum space,
I Fourier-transform all fields, as well as the free inverse
quark propagator,
\begin{subequations} \label{FT}
\bea
\Psi(X) & = & \frac{1}{\sqrt{V}} \sum _K e^{-i K \cdot X} \, \Psi(K)
\,\, , \\
\bar{\Psi}(X) & = & \frac{1}{\sqrt{V}} \sum_K e^{ i K \cdot X} \,
\bar{\Psi}(K)
\,\, , \\
{\cal G}_0^{-1}(X,Y) & = & \frac{T^2}{V} \sum_{K,Q} e^{-i K \cdot X}\,
e^{i Q \cdot Y} \, {\cal G}_0^{-1}(K,Q)
\,\, , \\
A^\mu_a(X) & = & \frac{1}{\sqrt{TV}} \sum_P e^{-i P \cdot X} \,
A^\mu_a(P) 
\,\, . \label{FTA}
\eea
\end{subequations}
The normalization factors are chosen such that the Fourier-transformed
fields are dimensionless quantities.
The Fourier-transformed free inverse quark propagator is diagonal
in momentum space, too,
\be 
\label{G0FT}
{\cal G}_0^{-1}(K,Q) = \frac{1}{T} \left( \begin{array}{cc}
                             [G_0^+]^{-1}(K) & 0 \\
                              0 & [G_0^-]^{-1}(K) \end{array} \right)
\delta^{(4)}_{K,Q} \,\, ,
\ee
where $[G_0^\pm]^{-1}(K) \equiv \Diracslash{K} \pm \mu \gamma_0 - m$.

Due to the relations (\ref{cc}), the Fourier-transformed
charge-conjugate quark fields are related to the original fields
via $\psi_C(K) = C \bar{\psi}^T(-K)$, $\bar{\psi}_C(K) = \psi^T(-K)C$.
The measure of the functional integration over quark fields
can then be rewritten in the form
\bea
{\cal D} \bar{\psi} \, {\cal D} \psi & \equiv & \prod_K d\bar{\psi}(K) 
\, d \psi(K) = {\cal N} \prod_{(K,-K)} d\bar{\psi}(K) \,
d \psi(K)\, d\bar{\psi}(-K) \, d \psi(-K) \nonumber \\
& = & {\cal N}' \prod_{(K,-K)} d\bar{\psi}(K) \,
d \psi(K)\, d\bar{\psi}_C(K) \, d \psi_C(K)
= {\cal N}'' \prod_{(K,-K)} d \bar{\Psi}(K) \, d\Psi(K)\non
&\equiv& {\cal D} \bar{\Psi}\, {\cal D}{\Psi} \,\, , \label{measure}
\eea
with the constant normalization factors ${\cal N},\, {\cal N}',
\, {\cal N}''$.
The last identity has to be considered as a definition for
the expression on the right-hand side.

Inserting Eqs.\ (\ref{FT}) -- (\ref{measure}) into Eq.\ (\ref{Zq}),
the partition function for quarks becomes
\be \label{Zq2}
{\cal Z}_q [A]= \int {\cal D} \bar{\Psi}\, 
{\cal D} \Psi \exp \left[ \frac{1}{2} \, \bar{\Psi} \left( 
{\cal G}_0^{-1} + g {\cal A} \right) \Psi \right]
\,\, .
\ee
Here, I employ a compact matrix notation,
\be \label{compact}
\bar{\Psi} \, \left( {\cal G}_0^{-1} + g {\cal A} \right)\,   \Psi \equiv
\sum_{K,Q} \bar{\Psi}(K) \, \left[ {\cal G}_0^{-1}(K,Q) 
+ g {\cal A}(K,Q) \right] \, \Psi(Q)\;\; ,
\ee
with the definition
\be \label{calA}
{\cal A}(K,Q) \equiv \frac{1}{\sqrt{VT^3}}\, \hat{\Gamma}^\mu_a
A_\mu^a(K-Q) \,\, .
\ee
The next step is to integrate out irrelevant quark modes.

\subsection{Integrating out irrelevant quark modes} \label{Intquarks}

Since I work in a finite volume $V$, 
the 3-momentum ${\bf k}$ is discretized.
Let us for the moment also assume that there is an ultraviolet
cut-off (such as in a lattice regularization) on the
3-momentum, i.e., the space of modes labelled by
3-momentum has dimension $D < \infty$.
I define projection operators ${\cal P}_1,\, {\cal P}_2$ for 
relevant and irrelevant quark modes, respectively,
\be \label{project}
\Psi_1 \equiv {\cal P}_1 \, \Psi\;\; , \;\;\;\;
\Psi_2 \equiv {\cal P}_2 \, \Psi \;\; , \;\;\;\;
\bar{\Psi}_1 \equiv \bar{\Psi} \, \gamma_0 {\cal P}_1 \gamma_0 
\;\; , \;\;\;\;
\bar{\Psi}_2 \equiv \bar{\Psi} \, \gamma_0 {\cal P}_2 \gamma_0 
\;\;.
\ee
The subspace of relevant quark modes has dimension $N_1$ in
the space of 3-momentum modes, 
the one for irrelevant modes dimension $N_2$, with $N_1 + N_2 = D$.

At this point, it is instructive to give an explicit
example for the projectors ${\cal P}_{1,2}$. In the effective theory 
for cold, dense quark matter, which contains
the high-density effective theory
\cite{hong,hong2,HLSLHDET,schaferefftheory,NFL,others} 
discussed in Sec.\ \ref{IIIB} as special case
and which I shall apply in Sec.\ \ref{IV} to the computation of 
the gap parameter, the projectors are chosen as
\begin{subequations} \label{P12}
\bea
{\cal P}_1(K,Q) & \equiv & \left( \begin{array}{cc}
 \Lambda_{\bf k}^+ & 0 \\
 0 & \Lambda_{\bf k}^- \end{array} \right) \, 
\Theta(\Lambda_{\rm q} - | k - k_F|) \, \delta^{(4)}_{K,Q} \;, \\
{\cal P}_2(K,Q) & \equiv & \left( \begin{array}{cc} 
\Lambda_{\bf k}^- + \Lambda_{\bf k}^+\, \Theta(| k - k_F| - \Lambda_{\rm q})
& 0 \\
0 &  \Lambda_{\bf k}^+ + \Lambda_{\bf k}^-\, \Theta(| k - k_F| -
\Lambda_{\rm q}) \end{array} \right)\,
\delta^{(4)}_{K,Q}\;.
\non
\eea
\end{subequations}
Here, 
\be
\Lambda^e_{\bf k} \equiv \frac{1}{2 E_{\bf k}} \, 
\left[ E_{\bf k} + e \gamma_0 \left(\vg \cdot
{\bf k} + m \right) \right] \,\, , 
\ee 
are projection operators onto states with positive ($e = +$)
or negative ($e=-$) energy, where $E_{\bf k} = \sqrt{{\bf k}^2 +m^2}$
is the relativistic single-particle energy.
The momentum cut-off $\Lambda_{\rm q}$ 
controls how many quark modes (with positive
energy) are integrated out. Thus, all quark modes within a layer of 
width $2 \Lambda_{\rm q}$ around the Fermi surface are considered as 
relevant, while all antiquark modes and quark modes
outside this layer are considered as irrelevant.
Note that,
for the Nambu-Gor'kov components corresponding to charge-conjugate
particles, the role of the projectors onto positive and negative energy
states is reversed with respect to the Nambu-Gor'kov components
corresponding to particles.
The reason is that, loosely speaking, a particle is actually a 
charge-conjugate antiparticle. For a more rigorous proof
compute, for instance, $\psi_{C,1}(K) \equiv C \, \bar{\psi}_1^T(-K)$ using 
$\bar{\psi}_1(-K) = \bar{\psi}(-K) \,\gamma_0 \Lambda^+_{-{\bf
k}}\gamma_0$ (for $|k-k_F| \leq \Lambda_{\rm q}$)
and $\gamma_0\, C [\Lambda_{-{\bf k}}^+]^T C^{-1}\,\gamma_0= 
\Lambda_{\bf k}^-$.
In Sec.\ \ref{III} I shall discuss other
choices for the projectors ${\cal P}_{1,2}$, pertaining to other
effective theories of hot and/or dense quark matter.
The following discussion in this section, however, will be completely
general and is not restricted to any particular choice for these projectors.

Employing Eq.\ (\ref{project}), the partition function (\ref{Zq2}) becomes
\be \label{Zq3}
{\cal Z}_q[A] = \int \prod_{n=1,2} {\cal D} \bar{\Psi}_n\, 
{\cal D} \Psi_n \, \exp \left( \frac{1}{2} \sum_{n,m=1,2} \bar{\Psi}_n \, 
{\cal G}^{-1}_{nm} \, \Psi_m \right)
\,\, .
\ee
{}From now on,  $\bar{\Psi}_{1,2}$, $\Psi_{1,2}$ are
considered as vectors restricted to the $N_{1,2}$-dimensional subspace of
relevant/irrelevant 3-momentum modes.
The matrices ${\cal G}_{nn}^{-1},\, n=1,2,$ are defined as
\be
{\cal G}_{nn}^{-1}(K,Q) = {\cal G}_{0,nn}^{-1}(K,Q) + g {\cal A}_{nn}(K,Q)\;,
\ee
where the indices indicate that, 
for a given pair of quark energies $k_0,\ q_0$, 
the 3-momenta ${\bf k}, \, {\bf q}$ belong to the subspace of
relevant ($n=1$) or irrelevant ($n=2$) quark modes, i.e.,
${\cal G}_{nn}^{-1}$ is an ($N_n \times N_n$)-dimensional matrix in
3-momentum space.
The matrices ${\cal G}_{nm}^{-1}$, $n \neq m$, reduce to
\be
{\cal G}_{nm}^{-1}(K,Q) = g \, {\cal A}_{nm}(K,Q)\;,
\ee
since ${\cal G}_0^{-1}$ is diagonal
in 3-momentum space, i.e. ${\cal G}_{0,nm}^{-1} \equiv 0$ for
$n \neq m$. For a given pair of quark energies
$k_0,\, q_0$, ${\cal G}_{nm}^{-1}$ is a $(N_n \times
N_m)$-dimensional matrix in 3-momentum space.

The Grassmann integration over the irrelevant quark fields $\bar{\Psi}_2,\,
\Psi_2$ can be done exactly, if one redefines them such that
the mixed terms $\sim {\cal G}_{nm}^{-1}$, $n \neq m$, are eliminated.
To this end, substitute 
\be
\Upsilon \equiv \Psi_2 + {\cal G}_{22}\, {\cal G}_{21}^{-1}\, \Psi_1\;\; ,
\;\;\;\; 
\bar{\Upsilon} \equiv \bar{\Psi}_2 + \bar{\Psi}_1\, 
{\cal G}_{12}^{-1}\, {\cal G}_{22}\; ,\label{shiftq}
\ee 
where ${\cal G}_{22}$ is the inverse of ${\cal G}_{22}^{-1}$, defined
on the subspace of irrelevant quark modes.
The result is
\be \label{Zq4}
{\cal Z}_q[A] 
= \int {\cal D} \bar{\Psi}_1\, 
{\cal D} \Psi_1 \, \exp \left[ \frac{1}{2} \, \bar{\Psi}_1 
\left( {\cal G}^{-1}_{11} - {\cal G}^{-1}_{12} \, {\cal G}_{22}
\, {\cal G}^{-1}_{21} \right) \Psi_1
+ \frac{1}{2}\, {\rm Tr}_q \ln {\cal G}_{22}^{-1} \right] \,\, .
\ee
The trace in the last term runs over all irrelevant quark momenta
$K$, and not only over pairs $(K,-K)$, as prescribed by 
the integration measure, Eq.\ (\ref{measure}).
This requires an additional factor $1/2$ in front of the trace. 
A more intuitive way of saying this is that this factor accounts 
for the doubling of the quark degrees of freedom in the Nambu-Gor'kov basis.
Of course, the trace runs not only over 4-momenta, but also 
over other quark indices, such as Nambu-Gor'kov, fundamental color, 
flavor, and Dirac indices. I indicated this by the subscript ``$q$''.

\begin{figure}[ht]
\centerline{\includegraphics[width=13cm]{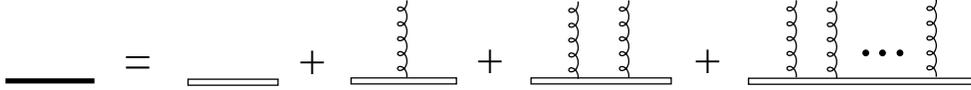}}
\caption[The full propagator for irrelevant quarks.]{The full propagator for irrelevant quarks. The right-hand
side symbolizes the expansion (\ref{expquark}). The free irrelevant
quark propagators ${\cal G}_{0,22}$ are denoted by double lines, the
gluon fields ${\cal A}_{22}$ by curly lines.}
\label{XXd}
\end{figure}

For a diagrammatic interpretation it is advantageous to rewrite
\be
{\cal G}^{-1}_{11} - {\cal G}^{-1}_{12} \, {\cal G}_{22}
\, {\cal G}^{-1}_{21} 
\equiv {\cal G}_{0,11}^{-1}  + g {\cal B}\;,
\ee
where
\be \label{B}
g{\cal B} \equiv g {\cal A}_{11} - g {\cal A}_{12} \, 
{\cal G}_{22} \, g {\cal A}_{21}\;.
\ee
The propagator for irrelevant quark modes,
${\cal G}_{22}$, has an expansion in powers of $g$ times the gluon field,
\be  \label{expquark}
{\cal G}_{22} = {\cal G}_{0,22} \sum_{n=0}^\infty 
(-1)^n g^n \left[ {\cal A}_{22}\, {\cal G}_{0,22} \right]^n\;.
\ee
This expansion is graphically depicted in Fig.\ \ref{XXd}.

\begin{figure}[ht]
\centerline{\includegraphics[width=11cm]{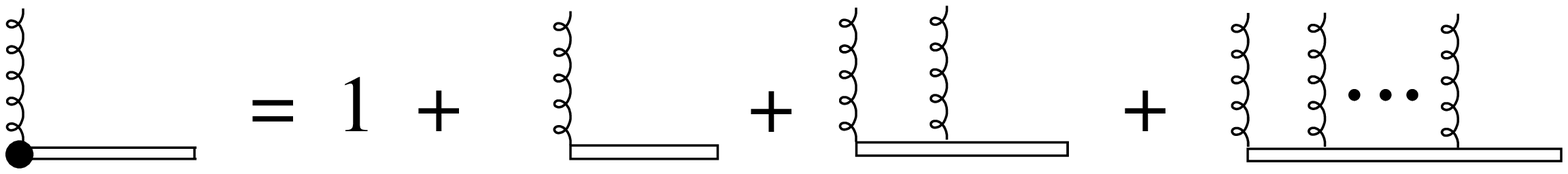}}
\caption{The diagrammatic symbol for the factor 
$\left(1 + g {\cal A}\, {\cal G}_{0,22} \right)^{-1}$.}
\label{XXb}
\end{figure}

\begin{figure}[ht]
\centerline{\includegraphics[width=13cm]{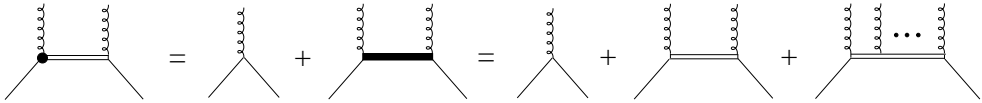}}
\caption[The term $\bar{\Psi}_1\, g {\cal B} \, \Psi_1$. ]{The term $\bar{\Psi}_1\, g {\cal B} \, \Psi_1$. A relevant
quark field is denoted by a single solid line.}
\label{XXa}
\end{figure}

\begin{figure}[ht]
\centerline{\includegraphics[width=12cm]{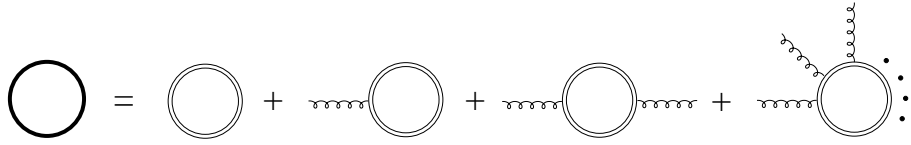}}
\caption{The graphical representation of the term ${\rm Tr}_q \ln
{\cal G}_{22}^{-1}$ in Eq.\ (\ref{Zq4}).}
\label{XXc}
\end{figure}

Using this expansion, and suppressing the indices on ${\cal A}$,
Eq.\ (\ref{B}) can be symbolically written as
\be
g {\cal B} = \left( 1 + g {\cal A}\, {\cal G}_{0,22} \right)^{-1}
g {\cal A}\;,\label{symbolB}
\ee
which suggests the interpretation of the field ${\cal B}$ as
a ``modified'' (non-local) gluon field. In the diagrams to
be discussed below, the factor
$\left(1 + g {\cal A}\, {\cal G}_{0,22} \right)^{-1}$ will
be denoted by the diagrammatical symbol shown in Fig.\ \ref{XXb}.
With this symbol, the expression $\bar{\Psi}_1\, g {\cal B} \,
\Psi_1$ can be graphically depicted as shown in Fig.\ \ref{XXa}.

Since
\be \label{explnquark}
\ln {\cal G}_{22}^{-1} = \ln {\cal G}_{0,22}^{-1} 
- \sum_{n=1}^\infty \frac{(-1)^n}{n} \, g^n \, \left[ {\cal G}_{0,22}
\, {\cal A}_{22} \right]^n\; ,
\ee
the last term in the exponent in Eq.\ (\ref{Zq4}) also has a
graphical interpretation, shown in Fig.\ \ref{XXc}.

This concludes the integration over irrelevant quark modes. Note that
the present treatment is (i) exact in the sense that no approximations have been
made and (ii) completely general, since it is independent of the
specific choice (\ref{P12}) for the projection operators.
The next step is to integrate out hard gluon modes.

\subsection{Integrating out hard gluon modes} \label{Intgluons}

Combining Eqs.\ (\ref{ZQCD}), (\ref{Zq4}), and (\ref{B}),
the partition function of QCD for relevant quark
modes and gluons reads
\begin{subequations} \label{ZQCD2}
\bea 
{\cal Z} & = &
\int {\cal D} \bar{\Psi}_1 \,  {\cal D} \Psi_1\, 
{\cal D} A \, \exp \left\{ S[A,\bar{\Psi}_1,\Psi_1] \right\}\,  \; \\
S[A,\bar{\Psi}_1,\Psi_1] & \equiv &
S_A[A] + \frac{1}{2} \, \bar{\Psi}_1 
\left\{ {\cal G}_{0,11}^{-1} +  g {\cal B}[A] \right\} \Psi_1
+ \frac{1}{2}\, {\rm Tr}_q \ln {\cal G}_{22}^{-1}[A] \;, \label{S}
\eea
\end{subequations}
where ${\cal D} A \equiv \prod_P d A(P)$. For
the sake of clarity, I restored
the functional dependence of the ``modified'' gluon field
${\cal B}$ and the inverse irrelevant quark propagator 
${\cal G}_{22}^{-1}$ on the gluon field $A$.

The gluon action in momentum space is 
\bea 
S_A [A] & = &
- \frac{1}{2} \sum_{P_1,P_2} A_\m^a(P_1)
\left[\Delta^{-1}_0\right]^{\mu\nu}_{ab}(P_1,P_2) A_\nu^b(P_2) \non
&  &
- \frac{1}{3!} \, \frac{g}{\sqrt{VT^3}} 
\sum_{P_1,P_2,P_3} \delta^{(4)}_{P_1+P_2+P_3,0} \,
{\cal V}_{\a \b \g}^{abc}(P_1,P_2,P_3) \,
A^{\a}_{a}(P_1) A^{\b}_{b}(P_2)A^{\g}_{c}(P_3)\non
&  & 
- \frac{1}{4!}\, \left(\frac{g}{\sqrt{VT^3}}\right)^2 
\sum_{P_1, \cdots ,P_4} \delta^{(4)}_{P_1+P_2+P_3+P_4,0}\, 
{\cal V}_{\a \b \g \d}^{abcd}\, 
A^{\a}_{a}(P_1) A^{\b}_{b}(P_2) A^{\g}_{c}(P_3) A^{\d}_{d}(P_4)  \non
&   &
+ {\rm Tr}_{gh} \ln {\cal W}^{-1} \;.
\label{Sgluon}
\eea
Here, $\Delta^{-1}_0(P_1,P_2)$ is the gauge-fixed inverse free gluon
propagator. 
To be specific, in general Coulomb gauge it reads
\begin{subequations}\label{D_0}
\bea 
\left[\Delta^{-1}_0 \right]^{\mu \nu}_{ab}(P_1,P_2) & \equiv& 
\frac{1}{T^2} \, \left[ \Delta^{-1}_0 \right]^{\mu \nu}_{ab}(P_1)\, 
\delta^{(4)}_{P_1, -P_2}\;, \\
\left[\Delta^{-1}_0\right]^{\mu \nu}_{ab}(P) & = & \delta_{ab}\, 
\left(P^2 g^{\mu \nu} -P^\mu P^\nu + \frac{1}{\xi_C}
\tilde{P}^\mu \tilde{P}^\nu \right)\;,
\eea
\end{subequations}
where $\xi_C$ is the Coulomb gauge parameter and $\tilde{P}^\mu \equiv
(0, {\bf p})$.
The vertex functions are
\begin{subequations}
\bea
\!\!\!\!\!\!\!\!\!\!\!\!\!\!\! {\cal V}_{\a \b \g}^{abc} (P_1,P_2,P_3) &\equiv &
\frac{i}{T}\, f^{abc} \, \left[ (P_1-P_2)_{\g}\, g_{\a \b}+
(P_2-P_3)_{\a}\, g_{\b \g}+ (P_3-P_1)_{\b}\, g_{\a \g} \right]\;, \\
{\cal V}^{abcd}_{\a \b \g \d} &\equiv&
 f^{abe} f^{ecd} \left( g_{\a\g} g_{\b \d} -
g_{\a \d} g_{\b \g} \right) 
+ f^{ace} f^{ebd}  \left( g_{\a\b} g_{\g \d} -
g_{\a \d} g_{\b \g} \right) \non
&&+ f^{ade} f^{ebc}  \left( g_{\a\b} g_{\g \d} -
g_{\a \g} g_{\b \d} \right) \;. 
\eea
\end{subequations}
The last term in Eq.\ (\ref{Sgluon}) is the trace of the
logarithm of the Faddeev-Popov
determinant, with the full inverse ghost propagator ${\cal W}^{-1}$.
The trace runs over ghost 4-momenta and adjoint color indices.
%\be
%\left[ {\cal W}^{-1} \right]^{ab}(P_1,P_2) \equiv
%\frac{p_12}{T2} \, \delta^{ab} \, \delta^{(4)}_{P_1,P_2}
%+ i \frac{g}{\sqrt{VT^3}} \, f^{abc} \, \frac{{\bf p}_1}{T} \cdot
%{\bf A}_c(P_1 - P_2) \; .
%\ee

Similar to the treatment of fermions in Sec.\ \ref{Intquarks}
I now define projectors ${\cal Q}_1,\, {\cal Q}_2$ for
soft and hard gluon modes, respectively,
\be
A_1  \equiv {\cal Q}_1 \, A \;\; , \;\;\;\; A_2 \equiv {\cal Q}_2 \,
A\;,
\ee
where
\begin{subequations} \label{Q12}
\bea
{\cal Q}_1(P_1,P_2) & \equiv & \Theta(\Lambda_{\rm gl} -p_1) \, \delta^{(4)}_{P_1,P_2}
\; , \\
{\cal Q}_2(P_1,P_2) & \equiv & \Theta(p_1-\Lambda_{\rm gl}) \, \delta^{(4)}_{P_1,P_2}
\; .
\eea
\end{subequations}
The gluon cut-off momentum $\Lambda_{\rm gl}$ defines which gluons are
considered to be soft or hard, respectively.

\begin{figure}[ht]
\centerline{\includegraphics[width=12cm]{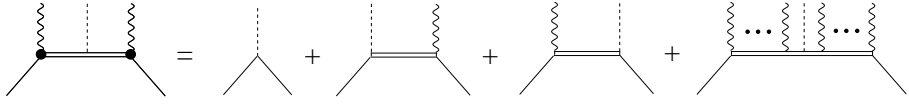}}
\caption[The term $A_2 {\cal J}_{{\cal B}}$.]{The term $A_2 {\cal J}_{{\cal B}}$. The hard gluon field
is denoted by a dashed line, the soft gluon fields by wavy lines.}
\label{YYa}
\end{figure}

\begin{figure}[ht]
\centerline{\includegraphics[width=12cm]{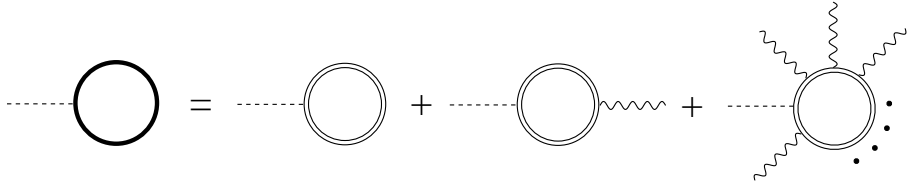}}
\caption[The fermionic contribution to the
term $A_2 {\cal J}_{\rm loop}$.]{The fermionic contribution to the
term $A_2 {\cal J}_{\rm loop}$. There is an additional contribution
from ghosts with similar topology.}
\label{YYb}
\end{figure}

\begin{figure}[ht]
\centerline{\includegraphics[width=3cm]{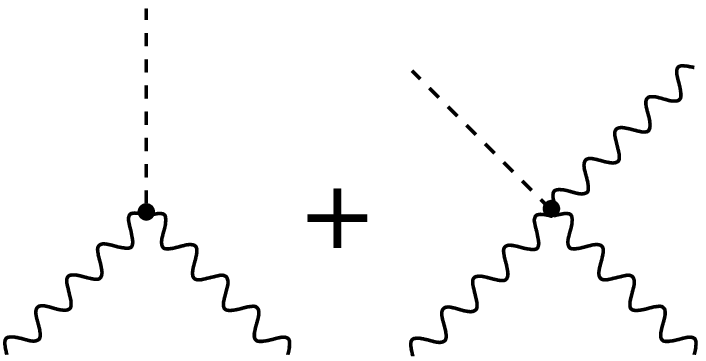}}
\caption{The term $A_2 {\cal J}_{{\cal V}}$.}
\label{YYc}
\end{figure}

I now insert $A \equiv A_1 + A_2$ into 
Eq.\ (\ref{ZQCD2}). The integration measure simply
factorizes, ${\cal D} A \equiv {\cal D} A_1\, {\cal D} A_2$.
The action $S[A,\bar{\Psi}_1,\Psi_1]$ can be sorted with respect to
powers of the hard gluon field,
\be \label{expansion}
S[A,\bar{\Psi}_1,\Psi_1] = S[A_1,\bar{\Psi}_1,\Psi_1]
+ A_2 {\cal J} [A_1,\bar{\Psi}_1,\Psi_1] - \frac{1}{2}\,
A_2 \, \Delta^{-1}_{22}[A_1, \bar{\Psi}_1,\Psi_1]\, A_2 
+ S_I[A_1,A_2,\bar{\Psi}_1,\Psi_1] \;.
\ee 
The first term in this expansion, containing no hard gluon fields
at all, is simply the action (\ref{S}), with $A$ replaced by
the relevant gluon field $A_1$.
The second term, $A_2 {\cal J}$,
contains a single power of the hard gluon field, where
\be \label{J}
{\cal J} [A_1,\bar{\Psi}_1,\Psi_1] \equiv 
\left. \frac{\delta S[A, \bar{\Psi}_1,\Psi_1]}{\delta A_2}
\right|_{A_2 =0} \equiv {\cal J}_{{\cal B}}[A_1, \bar{\Psi}_1, \Psi_1]
+ {\cal J}_{\rm loop}[A_1] + {\cal J}_{{\cal V}}[A_1]\,\, .
\ee
The first contribution, 
\be \label{J_B}
{\cal J}_{{\cal B}}[A_1,\bar{\Psi}_1,\Psi_1] 
= \frac{1}{2} \, \bar{\Psi}_1 \, \left( g \, \frac{\delta
{\cal B}}{\delta A_2}\right)_{A_2=0} \Psi_1 \;,
\ee 
arises from the coupling of the relevant fermions to the
``modified'' gluon field ${\cal B}$, i.e., from the second term in 
Eq.\ (\ref{S}). With the notation of Fig.\ \ref{XXb},
all diagrams corresponding to $A_2 {\cal J}_{{\cal B}}$ 
can be summarized into a
single one, cf.\ Fig.\ \ref{YYa}. It contains precisely two relevant
fermion fields, $\bar{\Psi}_1$ and $\Psi_1$.
The second contribution, ${\cal J}_{\rm loop}$, 
arises from the terms ${\rm Tr}_q \ln {\cal
G}_{22}^{-1}$ and ${\rm Tr}_{gh} \ln {\cal W}^{-1}$ in 
Eqs.\ (\ref{S}), (\ref{Sgluon}). The loop consisting of
irrelevant quark modes as internal lines, coupled to a single hard
and arbitrarily many soft gluons, is shown in Fig.\ \ref{YYb}.
Finally, the third contribution, ${\cal J}_{{\cal V}}$, 
arises from the non-Abelian vertices, cf.\ Fig.\ \ref{YYc}.

\begin{figure}[ht]
\centerline{\includegraphics[width=12cm]{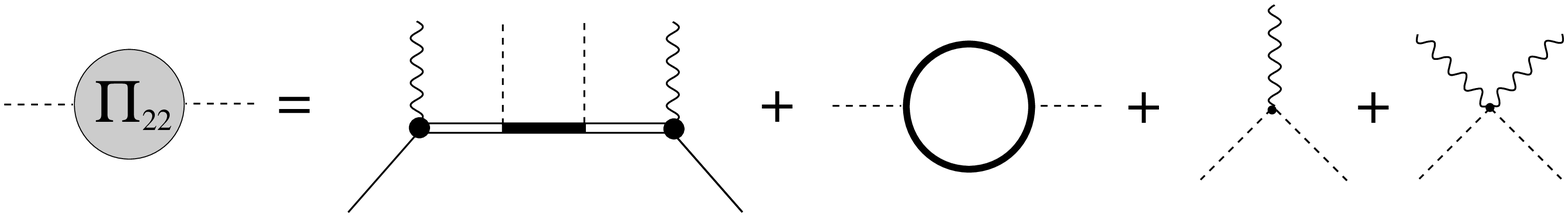}}
\caption[The term $A_2 \Pi_{22} A_2$ according to Eq.\ (\ref{Pi}). ]{The term $A_2 \Pi_{22} A_2$ according to Eq.\ (\ref{Pi}). 
The first diagram on the right-hand side corresponds to the term 
$A_2 \Pi_{\cal B}A_2 $. The second diagram is the fermion-loop contribution
to $A_2 \Pi_{\rm loop} A_2$; there is an analogous one from a
ghost loop. The last two diagrams correspond
to $A_2 \Pi_{\cal V} A_2$.}
\label{FigPi}
\end{figure}

\begin{figure}[ht]
\centerline{\includegraphics[width=13cm]{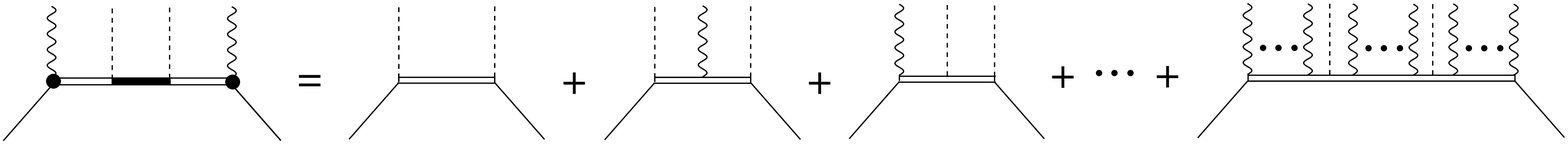}}
\caption{The term $A_2 \Pi_{{\cal B}}A_2$.}
\label{ZZa}
\end{figure}

\begin{figure}[ht]
\centerline{\includegraphics[width=12cm]{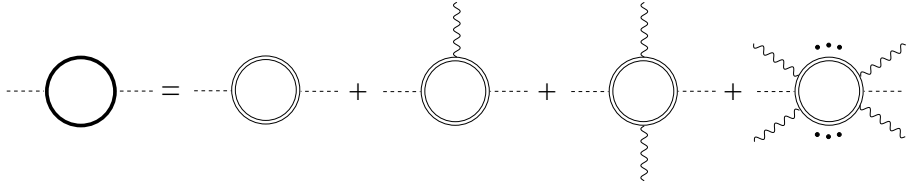}}
\caption{The fermionic contribution to the term $A_2 \Pi_{\rm loop}
A_2$.}
\label{ZZb}
\end{figure}

\begin{figure}[ht]
\centerline{\includegraphics[width=3cm]{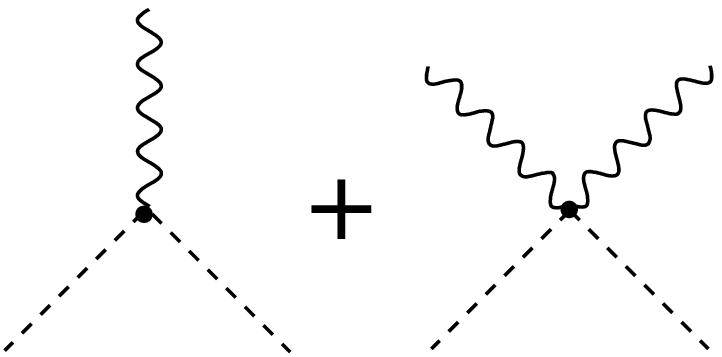}}
\caption{The term $A_2 \Pi_{{\cal V}} A_2$.}
\label{ZZc}
\end{figure}

The third term in Eq.\ (\ref{expansion}) is quadratic in $A_2$, 
where
\be
\Delta_{22}^{-1}[A_1,\bar{\Psi}_1,\Psi_1]\equiv 
- \left. \frac{\delta^2 S[A, \bar{\Psi}_1,\Psi_1]}{\delta A_2\,\delta A_2}
\right|_{A_2 =0} \equiv \Delta_{0,22}^{-1} + 
\Pi_{22}[A_1, \bar{\Psi}_1, \Psi_1]\,\, .
\ee
Here, $\Delta_{0,22}^{-1}$ is the free inverse propagator 
for hard gluons.
Similar to the ``current'' ${\cal J}$, cf.\ Eq.\ (\ref{J}),
the ``self-energy'' $\Pi_{22}$ of hard gluons consists of three
different contributions,
\be \label{Pi}
\Pi_{22}[A_1, \bar{\Psi}_1, \Psi_1] =
\Pi_{{\cal B}}[A_1, \bar{\Psi}_1, \Psi_1] +
\Pi_{\rm loop}[A_1] +
\Pi_{{\cal V}}[A_1] \;,
\ee
which has a diagrammatic representation as shown in Fig.\ \ref{FigPi}.
The first two contributions on the right-hand side of Eq.\ (\ref{Pi})
can be expanded as shown in Figs.\ \ref{ZZa} and \ref{ZZb}. 
Figure \ref{ZZc} depicts the three- and four-gluon vertices 
contained in the last term in Eq.\ (\ref{Pi}).
For further use, I explicitly give the first term,
\be
\Pi_{{\cal B}}[A_1, \bar{\Psi}_1, \Psi_1] =
- \frac{1}{2} \, \bar{\Psi}_1 \, \left( g \, \frac{\delta^2
{\cal B}}{\delta A_2 \delta A_2}\right)_{A_2=0} \Psi_1 \;.
\ee

Finally, I collect all terms with more than
two hard gluon fields $A_2$ in Eq.\ (\ref{expansion}) in the
``interaction action'' for hard gluons, 
$S_I[A_1,A_2,\bar{\Psi}_1,\Psi_1]$.
I then perform the functional integration over the hard gluon fields $A_2$.
Since functional integrals must be of Gaussian type in order to be 
exactly solvable, I resort to a method well known from perturbation theory. 
I add the source term $A_2 J_2$ to the action (\ref{S})
and may then replace the fields $A_2$ in $S_I$
by functional differentiation with respect to $J_2$, at
$J_2=0$. I then move
the factor $\exp\{ S_I[A_1, \delta/\delta J_2, \bar{\Psi}_1, \Psi_1 ]\}$
in front of the functional $A_2$-integral. Then, this functional
integral is Gaussian and can be readily
performed (after a suitable shift of $A_2$), with the result
\bea
{\cal Z} & = & \int {\cal D} \bar{\Psi}_1 \, {\cal D} \Psi_1 {\cal D} A_1
\, \exp\left\{ S[A_1,\bar{\Psi}_1, \Psi_1 ]- \frac{1}{2}\, {\rm Tr}_g \ln
\Delta_{22}^{-1} \right\} \non
&   & \times \left. 
\exp \left\{ S_I\left[A_1, \frac{\delta}{\delta J_2}, 
\bar{\Psi}_1, \Psi_1 \right]\right\}
\,  \exp \left[  \frac{1}{2} \, ({\cal J} + J_2) \,
\Delta_{22}\, ({\cal J} + J_2) \right] \right|_{J_2 = 0}\; . 
\label{Z3}
\eea
The trace over $\ln \Delta_{22}^{-1}$ runs over gluon 4-momenta, as
well as adjoint color and Lorentz indices. I indicate this with
a subscript ``$g$''.
Note that this result is still exact and completely general, since
so far the manipulations of the partition function were independent of
the specific choice (\ref{Q12}) 
for the projection operators ${\cal Q}_{1,2}$.
The next step is to derive the tree-level action
for the effective theory of relevant quark modes and soft gluons.

\subsection{Tree-level effective action} \label{IId}

In order to derive the tree-level effective action, I shall employ
two approximations. The first is based on the
principal assumption in the construction of any
effective theory, namely that soft and hard modes are well separated 
in momentum space. Consequently, momentum conservation
does not allow a hard gluon to
couple to any (finite) number of soft gluons. Under this assumption, the
diagrams generated by $A_2 ({\cal J}_{\rm loop} + {\cal J}_{{\cal V}})$,
cf.\ Fig.\ \ref{YYb}, \ref{YYc}, will not occur in the effective theory.
In the following, I shall therefore omit these terms, so that
${\cal J} \equiv {\cal J}_{{\cal B}}$. 
Note that similar arguments cannot be applied to
the diagrams generated by $A_2 (\Pi_{\rm loop} + \Pi_{{\cal V}}) A_2$,  
cf.\ Fig.\ \ref{ZZb}, \ref{ZZc}, 
since now there are two hard gluon legs which take care of 
momentum conservation.

My second approximation is that in the ``perturbative'' expansion
of the partition function (\ref{Z3}) with respect to powers
of the interaction action $S_I$, I only take the first term, i.e.,
I approximate $e^{S_I} \simeq 1$. 
At this point, this is simply a
matter of convenience, since I do not need
the higher-order terms in the expansion of $e^{S_I}$ for the 
effective theories to be discussed in Sec.\ \ref{III} or for the 
calculation of the gap parameter in Sec.\ \ref{IV}. However, one
can easily reinstall them if required by the particular problem at hand. 
I note in passing that the approximation $e^{S_I} \simeq 1$ becomes exact
in the derivation of the exact renormalization group \cite{Wegner}, where
one only integrates out modes in a shell of infinitesimal thickness.

Diagrams generated by the
higher-order terms in the expansion of $e^{S_I}$ are those with more
than one {\em resummed\/} hard gluon line.
Even with the approximation $e^{S_I} \simeq 1$, Eq.\ (\ref{Z3}) 
still contains diagrams with
arbitrarily many {\em bare\/} hard gluon lines, arising 
from the expansion of 
\be \label{expansion2}
\ln \Delta_{22}^{-1} = \ln \Delta_{0,22}^{-1}
- \sum_{n=1}^{\infty} \frac{(-1)^n}{n} \, \left(\Delta_{0,22}\, \Pi_{22}
\right)^n \;,
\ee
and from the term ${\cal J}_{{\cal B}} \Delta_{22} {\cal J}_{{\cal B}}$ in
Eq.\ (\ref{Z3}), when expanding
\be \label{expansion3}
\Delta_{22} = \Delta_{0,22} \sum_{n=0}^{\infty} (-1)^n \left( \Pi_{22} \,
\Delta_{0,22} \right)^n\;.
\ee

With these approximations, the partition function reads
\be \label{Z4}
{\cal Z} =  \int {\cal D} \bar{\Psi}_1 \, {\cal D} \Psi_1 {\cal D} A_1\,
\exp\{S_{\rm eff} [A_1,\bar{\Psi}_1, \Psi_1 ] \}\;,
\ee
where the effective action is defined as
\bea 
S_{\rm eff} [A_1,\bar{\Psi}_1, \Psi_1 ] & \equiv & 
S_A[A_1] + \frac{1}{2} \, \bar{\Psi}_1 
\left\{ {\cal G}_{0,11}^{-1} +  g {\cal B}[A_1] \right\} \Psi_1
+ \frac{1}{2}\, {\rm Tr}_q \ln {\cal G}_{22}^{-1}[A_1] \non
&& - \frac{1}{2}\,  {\rm Tr}_g \ln \Delta_{22}^{-1}[A_1,\bar{\Psi}_1,\Psi_1] \non
&  & + \; \frac{1}{2} \, {\cal J}_{{\cal B}}[A_1,\bar{\Psi}_1,\Psi_1]   \,
\Delta_{22}[A_1,\bar{\Psi}_1,\Psi_1] 
\, {\cal J}_{{\cal B}}[A_1,\bar{\Psi}_1,\Psi_1]  \; . 
\label{Seff}
\eea
This is the desired action for the effective theory describing the
interaction of relevant quark modes, $\bar{\Psi}_1, \Psi_1$, and
soft gluons, $A_1$. 
The functional dependence of the various terms on the right-hand
side on the fields $A_1,\bar{\Psi}_1, \Psi_1$
has been restored in order to facilitate the
following discussion of all possible interaction vertices occurring in this
effective theory.

\begin{figure}[ht]
\centerline{\includegraphics[width=3cm]{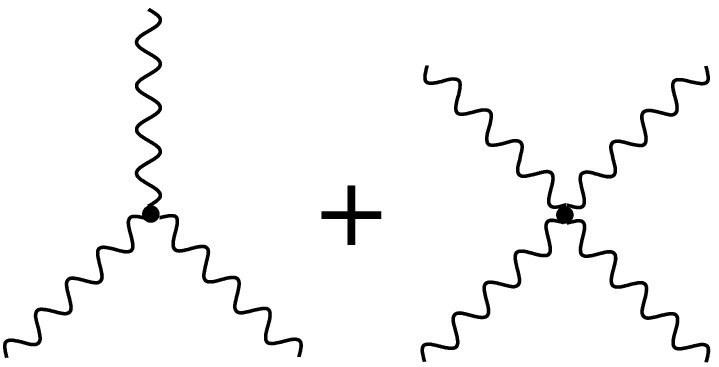}}
\caption[The three- and four-gluon vertices in \mbox{$S_A[A_1]$}.]{The three- and four-gluon vertices in $S_A[A_1]$, describing
the self-interaction of soft gluons in Eq.\ (\ref{Seff}).}
\label{AAa}
\end{figure}

\begin{figure}[ht]
\centerline{\includegraphics[width=14cm]{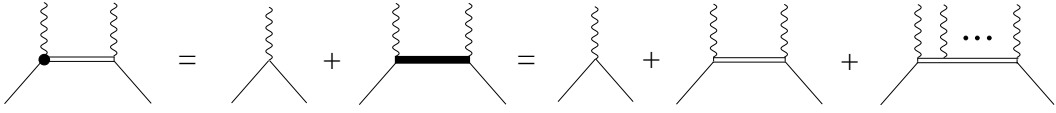}}
\caption{The term $\bar{\Psi}_1 \, g {\cal B}[A_1] \, \Psi_1$ in the
effective action (\ref{Seff}).}
\label{AAb}
\end{figure}

\begin{figure}[ht]
\centerline{\includegraphics[width=12cm]{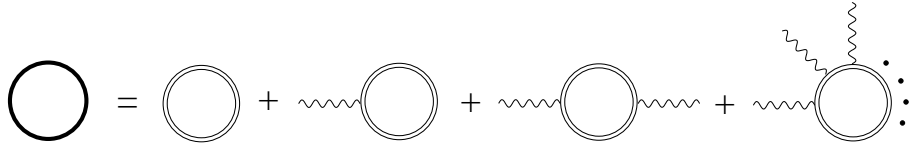}}
\caption{The term ${\rm Tr}_q \ln {\cal G}_{22}^{-1}[A_1]$ in
the effective action (\ref{Seff}).}
\label{AAc}
\end{figure}

\begin{figure}[ht]
\centerline{\includegraphics[width=14cm]{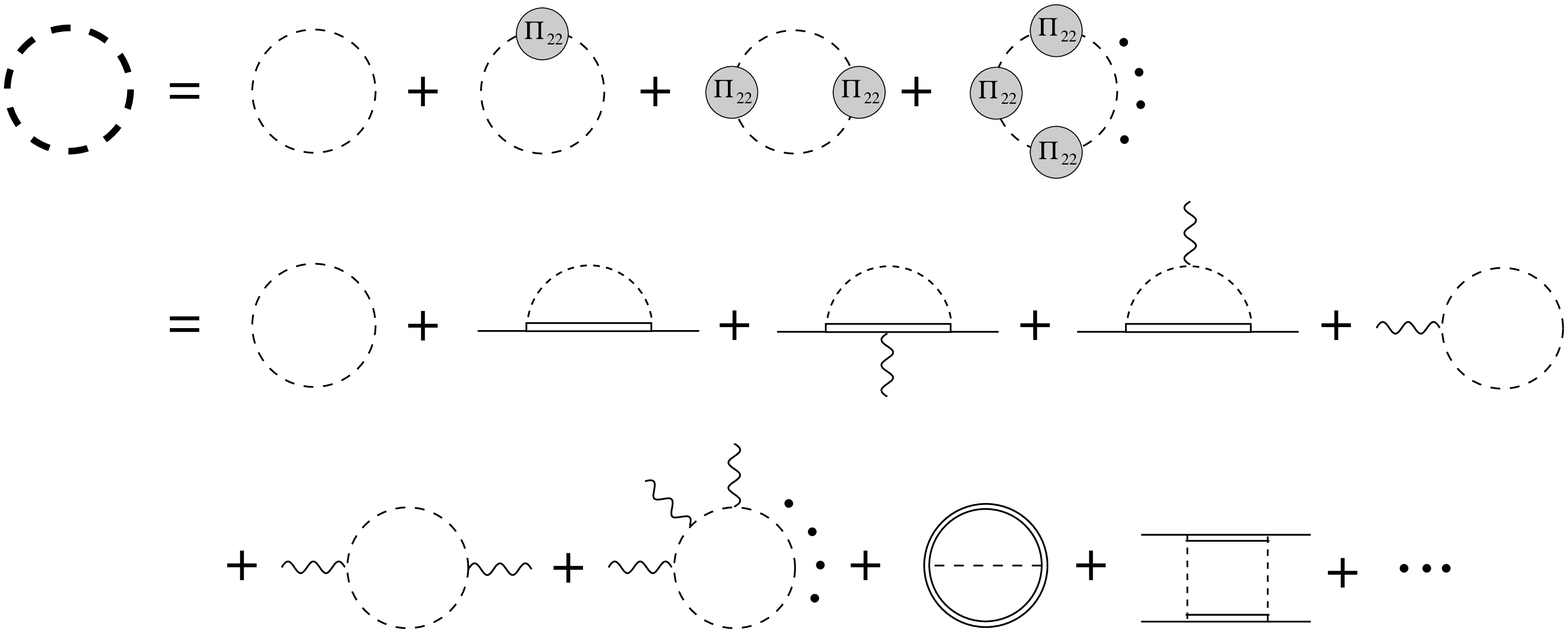}}
\caption[\mbox{The term ${\rm Tr}_g \ln
\Delta_{22}^{-1}[A_1,\bar{\Psi}_1,\Psi_1]$ 
in the effective action (\ref{Seff}).}]{The term ${\rm Tr}_g \ln
\Delta_{22}^{-1}[A_1,\bar{\Psi}_1,\Psi_1]$ 
in the effective action (\ref{Seff}). The first line corresponds to
the generic expansion (\ref{expansion2}), with ``self-energy'' insertions
$\Pi_{22}$, as shown in Fig.\ \ref{FigPi}. The second line contains
some examples
for diagrams generated when explicitly inserting the expression for 
$\Pi_{22}$.}
\label{AAd}
\end{figure}

\begin{figure}[ht]
\centerline{\includegraphics[width=12cm]{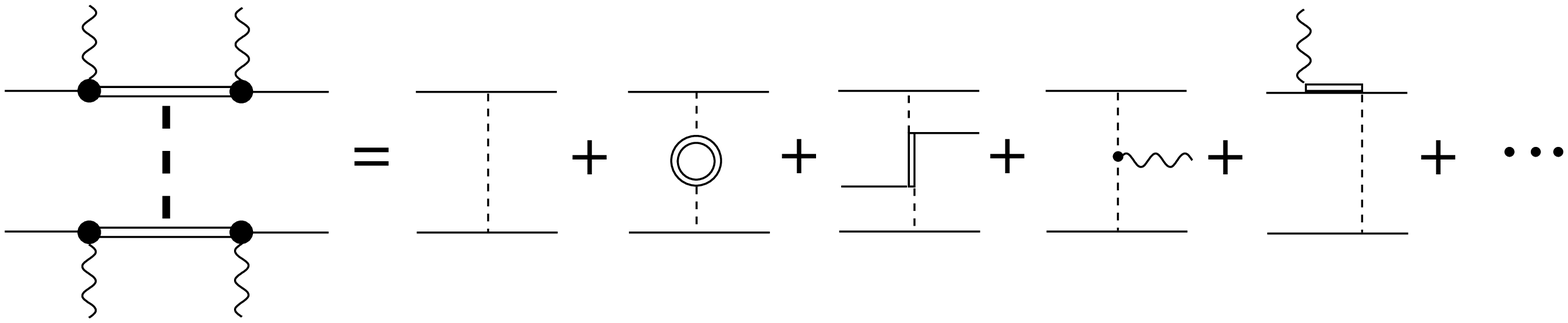}}
\caption[\mbox{The term ${\cal J}_{\cal B} \Delta_{22} {\cal J}_{\cal B}$ in
the effective action (\ref{Seff}).}]{The term ${\cal J}_{\cal B} \Delta_{22} {\cal J}_{\cal B}$ in
the effective action (\ref{Seff}). The thick dashed line is a full
hard gluon propagator, i.e., it has the expansion (\ref{expansion2}).
The first diagram on the right-hand side of this figure 
results from the $n=0$ term of this expansion, while the next
three diagrams originate from the $n=1$ term. Even a single insertion
of a hard gluon ``self-energy'' $\Pi_{22}$ gives rise to a variety of
diagrams. Here, I only show the contributions corresponding to
the first diagrams in Figs.\ \ref{ZZa}, \ref{ZZb}, and the three-gluon vertex.
The last diagram arises from the second term
of the expansion shown in Fig.\ \ref{XXb}. }
\label{AAe}
\end{figure}

The diagrams corresponding to these vertices are
shown in Figs.\ \ref{AAa}-\ref{AAe}. The three- and four-gluon 
vertices contained in
$S_A[A_1]$ are displayed in Fig.\ \ref{AAa}. In addition, $S_A[A_1]$ 
contains ghost loops with an arbitrary number of
attached soft gluon legs. The topology is equivalent to that
of the quark loops in Fig.\ \ref{AAc} and is therefore not shown
explicitly.
The interaction between two relevant quarks and the ``modified'' soft
gluon field, corresponding to $\bar{\Psi}_1 \, g {\cal B}[A_1]\, \Psi_1$, is
depicted in Fig.\ \ref{AAb}. 
This is similar to Fig.\ \ref{XXa}, except that now all gluon legs are soft.
Diagrams where an 
arbitrary number of soft gluon legs is attached
to an irrelevant quark loop are generated
by ${\rm Tr}_q \ln {\cal G}_{22}^{-1}$, cf.\ Fig.\ \ref{AAc}.
This is similar to Fig.\ \ref{XXc}, but now only soft gluon
legs are attached to the fermion loop.
The diagrams generated by the loop of a full hard gluon
propagator, ${\rm Tr}_g \ln \Delta_{22}^{-1}$,
are shown in Fig.\ \ref{AAd}. The first line in this figure
features the generic expansion of this term according to 
Eq.\ (\ref{expansion2}), where the hard gluon
``self-energy'' insertion $\Pi_{22}$, cf.\ Eq.\ (\ref{Pi}), is shown
in Fig.\ \ref{FigPi}.
The second line shows examples of diagrams generated by explicitly inserting
$\Pi_{22}$ in the generic expansion. Besides an arbitrary number of 
soft gluon legs, these diagrams also feature an arbitrary number
of relevant quark legs. If there are only two relevant 
quark legs, but
no soft gluon leg, one obtains the one-loop self-energy for relevant
quarks, cf.\ the second diagram in the second line of Fig.\ \ref{AAd}.
The next two diagrams are obtained by adding a soft gluon leg, 
resulting in vertex corrections for the bare vertex between
relevant quarks and soft gluons.
The first of these two diagrams arises from the 
$n=1$ term in Eq.\ (\ref{expansion2}), while the second originates
from the $n=2$ term.
Four relevant quark legs and no soft gluon leg
give rise to the scattering of two relevant quarks via
exchange of two hard gluons, contained in the $n=2$ term
in Eq.\ (\ref{expansion2}), cf.\ the last diagram in
Fig.\ \ref{AAd}. This diagram was also
discussed in the context of the effective theory presented
in Refs.\ \cite{hong,hong2}, cf.\ discussion in Sec.\ \ref{IIIB}.
Finally, the ``current-current'' interaction mediated by a full
hard gluon propagator, ${\cal J}_{{\cal B}} \,\Delta_{22} 
\, {\cal J}_{{\cal B}}$, Fig.\ \ref{AAe}, contains also a
multitude of quark-gluon vertices. The simplest one is the first on
the right-hand side in Fig.\ \ref{AAe}, corresponding to scattering of
two relevant fermions via exchange of a single hard gluon.

An important question is whether the introduction of the momentum
cut-offs $\Lambda_{\rm q},\, \Lambda_{\rm gl}$ could possibly spoil the gauge invariance
of the effective action (\ref{Seff}). This is not the case,
since gauge invariance is already explicitly broken
{\em from the very beginning\/} by the choice of gauge in the 
gauge-fixed gluon action (\ref{SA}). 
I do not perceive this to be a disadvantage of my approach, since
the computation of a physical quantity requires to fix the gauge
anyway. The final result should, of course, neither depend on the choice
of gauge, nor on $\Lambda_{\rm q}$ and $\Lambda_{\rm gl}$. 
Note that, up to this point, I was not required to specify the
gauge in the effective action (\ref{Seff}). 

The effective action (\ref{Seff}) is formally of the form
(\ref{Seffstandard}).
The difference is that Eq.\ (\ref{Seff}) contains more than one
relevant field: besides relevant quarks there are also soft gluons.
It is obvious that in this case there are many more possibilities
to construct operators ${\cal O}_i$ which occur in 
the expansion (\ref{Seffstandard}).
As pointed out in the introduction, it is therefore advantageous
to derive the effective action (\ref{Seff}) by explicitly integrating
out irrelevant quark and hard gluon modes, 
and not by simply guessing the form of the operators ${\cal O}_i$,
since then one is certain
that one has constructed {\em all\/} possible operators occurring in
the expansion (\ref{Seffstandard}).

As mentioned in the introduction, the standard approach
to derive an effective theory, namely guessing the form of the
operators ${\cal O}_i$ and performing a naive dimensional scaling analysis
to estimate their order of magnitude, fails precisely
when (a) there are non-local operators, or when (b) there is more than one
momentum scale. Both (a) and (b) apply here. As I shall
show below, the HTL/HDL effective action is one limiting
case of Eq.\ (\ref{Seff}), and it is well known that this action
is non-local. Moreover, as is obvious from the above derivation, there
are indeed several momentum scales occurring in Eq.\ (\ref{Seff}).
Let us focus on the case of zero temperature, $T=0$, and, for the
sake of simplicity, assume massless quarks, $m=0$, $\mu = k_F$.
To be explicit, I employ the choice (\ref{P12}) for the projectors
${\cal P}_{1,2}$.
In this case, the first momentum scale is defined by the Fermi energy
$\mu$. The propagator of antiquarks is $\sim 1/(k_0 + \mu + k)$.
If $\Lambda_{\rm q}, \Lambda_{\rm gl} \alt \mu$, the exchange of an antiquark 
can be approximated by a contact interaction with strength $\sim 1/\mu$,
on the scale of the relevant quarks, $L_{\rm q} \gg 1/\Lambda_{\rm q} \agt 1/ \mu$, 
or of the soft gluons, $L_{\rm gl} \gg 1/\Lambda_{\rm gl} \agt 1/ \mu$. 

The second momentum scale is defined by the quark cut-off momentum
$\Lambda_{\rm q}$. The propagator of irrelevant quark modes 
is $\sim 1/(k_0+\mu - k)$.
On the scale $L_{\rm q}$ of the relevant quarks, not only
the exchange of an antiquark, but also that of an irrelevant quark
with momentum ${\bf k}$ satisfying $|k- \mu| \geq \Lambda_{\rm q}$ 
is local, with strength $\sim 1/\Lambda_{\rm q}$.
However, suppose that the quark cut-off scale happens to be much smaller
than the chemical potential, $\Lambda_{\rm q} \ll \mu$. In this case,
antiquark exchange is ``much more
localized'' than the exchange of an irrelevant quark, 
$1/\mu \ll 1/ \Lambda_{\rm q}$.

The third momentum scale is defined by the gluon cut-off momentum
$\Lambda_{\rm gl}$. The propagator of a hard gluon is $\sim 1/P^2$.
On the scale $L_{\rm gl}$ of a soft gluon, the exchange of a hard gluon
with momentum $p \geq \Lambda_{\rm gl}$ can be considered local, with
strength $\sim 1/\Lambda_{\rm gl}^2$.
As I shall show below, in order to derive the value of the QCD gap
parameter in weak coupling and to subleading order, 
the ordering of the scales turns out to be $\Lambda_{\rm q} \alt g \mu
\ll \Lambda_{\rm gl} \alt \mu$.
Thus, antiquark exchange happens on a length scale of the same order as
hard gluon exchange, which in turn happens on a much smaller length scale
than the exchange of an irrelevant quark,
$1/\mu \alt 1/ \Lambda_{\rm gl} \ll 1/\Lambda_{\rm q}$.

		\section{Recovery of known effective theories} \label{III}

In this section I show that, for particular choices
of the projectors ${\cal P}_{1,2}$ in Eq.\ (\ref{project}),
several well-known and at first sight unrelated effective theories 
for hot and/or dense quark matter are in fact nothing but
special cases of the general effective theory defined by
the action (\ref{Seff}).
These are the HTL/HDL effective action for quarks and gluons, and
the high-density effective theory for cold, dense quark matter.

\subsection{HTL/HDL effective action} \label{IIIA}

Let us first focus on the HTL/HDL effective action.
This action defines an effective theory for massless
quarks and gluons with small momenta in a system
at high temperature $T$ (HTL), or large chemical potential $\mu$ (HDL). 
Consequently, the projectors ${\cal P}_{1,2}$ for quarks are given by
\begin{subequations} \label{PHTL}
\bea
{\cal P}_1 (K,Q) & = & \Theta(\Lambda_{\rm q} -k)\,\delta^{(4)}_{K,Q}\;, \\
{\cal P}_2 (K,Q) & = & \Theta(k - \Lambda_{\rm q})\,\delta^{(4)}_{K,Q}\;,
\eea
\end{subequations}
while the projectors for gluons are given by Eq.\ (\ref{Q12}).
(Note that, strictly speaking, the quarks and gluons
of the HTL/HDL effective action should also have small energies
in real time. Since the present effective action is defined in imaginary time,
one should constrain the energy only at the end
of a calculation, after analytically continuing the result 
to Minkowski space.)

The essential assumption to derive the HTL/HDL effective action is
that there is a single momentum scale, 
$\Lambda_{\rm q} = \Lambda_{\rm gl} \equiv \Lambda$,
which separates hard modes with momenta $\sim T$, or $\sim
\mu$, from soft modes with momenta $\sim gT$, or $\sim g\mu$.
In the presence of an additional energy scale $T$, or $\mu$,
naive perturbation theory in terms of powers of the coupling constant
fails. It was shown by Braaten and Pisarski \cite{braatenpisarski} that,
for the $n$-gluon scattering amplitude
the one-loop term, where $n$ soft gluon legs are attached to a
quark or gluon loop, is as important as the tree-level diagram.
The same holds for the scattering of $n-2$ gluons and 2 quarks.
At high $T$ and small $\mu$, the momenta of the quarks and gluons 
in the loop are of the order of the hard scale, $\sim T$. 
This gives rise to the name 
``Hard Thermal Loop'' effective action, and allows to simplify the 
calculation of the respective diagrams.
At large $\mu$ and small $T$, i.e., for the HDL effective action,
the situation is somewhat more involved.
As gluons do not have a Fermi surface, the only
physical scale which determines the order of magnitude of a loop
consisting exclusively of gluon propagators is the temperature. 
Therefore, at small $T$ and large $\mu$, such
pure gluon loops are negligible. On the other hand, 
the momenta of quarks in the loop are $\sim \mu$. Thus, only loops
with at least one quark line need to be considered in the HDL effective action.

In order to show that the HTL/HDL effective action is contained in the
effective action (\ref{Seff}), first note that a soft particle
cannot become hard by interacting with another soft particle.
This has the consequence that a soft quark cannot 
turn into a hard one by soft-gluon scattering. Therefore,
\be
g {\cal B}[A_1] \equiv g {\cal A}_{11}\;.
\ee
Another consequence is that the last term in Eq.\ (\ref{Seff}),
${\cal J}_{{\cal B}} \Delta_{22} {\cal J}_{{\cal B}}$, vanishes
since ${\cal J}_{{\cal B}}$
is identical to a vertex between a soft quark and a hard gluon, which
is kinematically forbidden. The resulting action then reads
\bea 
S_{\mbox{\scriptsize large}\,T/\mu} 
[A_1,\bar{\Psi}_1, \Psi_1 ] &\equiv&
S_A[A_1] + \frac{1}{2} \, \bar{\Psi}_1 
\left( {\cal G}_{0,11}^{-1} +  g {\cal A}_{11} \right) \Psi_1
+ \frac{1}{2}\, {\rm Tr}_q \ln {\cal G}_{22}^{-1}[A_1] \non
&&- \frac{1}{2}\,  {\rm Tr}_g \ln
\Delta_{22}^{-1}[A_1,\bar{\Psi}_1,\Psi_1] 
 \; . 
\label{SHTL}
\eea
Using the expansion (\ref{explnquark}) one realizes that the 
term ${\rm Tr}_q \ln {\cal G}_{22}^{-1}$ generates
all one-loop diagrams, where $n$ soft gluon legs 
are attached to a hard quark loop. This is precisely the quark-loop 
contribution to the HTL/HDL effective action.

For hard gluons with momentum $\sim T$ or $\sim \mu$,
the free inverse gluon propagator is $\Delta_{0,22}^{-1} \sim
T^2$ or $\sim \mu^2$, while the contribution $\Pi_{\rm loop}$ to the hard 
gluon ``self-energy'' (\ref{Pi}) is at most of the order $\sim g^2
T^2$ or $\sim g^2 \mu^2$. Consequently, $\Pi_{\rm loop}$ can be neglected and
$\Pi_{22}$ only contains tree-level diagrams, $\Pi_{22} \equiv \Pi_{{\cal B}}
+ \Pi_{{\cal V}}$.
Using the expansion (\ref{expansion2}) of
${\rm Tr}_g \ln \Delta_{22}^{-1}$, the terms which contain
only insertions of $\Pi_{{\cal V}}$ correspond to one-loop
diagrams where $n$ soft gluon legs
are attached to a hard gluon loop. As was shown in 
Ref.\ \cite{braatenpisarski}, with the exception of the two-gluon
amplitude, the loops with four-gluon vertices are suppressed.
Neglecting these, one is precisely left with 
the pure gluon loop contribution to the HTL effective action. 
As discussed above, for the HDL effective action, this contribution 
is negligible.

The ``self-energy'' $\Pi_{{\cal B}}$ contains only two soft
quark legs attached to a hard quark propagator (via emission and
absorption of hard gluons). Consequently, in 
the expansion (\ref{expansion2}) of ${\rm Tr}_g \ln \Delta_{22}^{-1}$, 
the terms which contain insertions of $\Pi_{{\cal V}}$ and $\Pi_{{\cal B}}$
correspond to one-loop diagrams where an arbitrary number of 
soft quark and gluon legs is attached to the loop. It was shown in Ref.\
\cite{braatenpisarski} that among these diagrams, only the ones with
two soft quark legs and no four-gluon vertices are kinematically
important and thus contribute to the HTL/HDL effective action.
I have thus shown that this effective action, $S_{\rm HTL/HDL}$,
is contained in the effective action (\ref{SHTL}), and constitutes its
leading contribution,
\be
S_{\mbox{\scriptsize large}\, T/\mu} = S_{\rm HTL/HDL} + \,
\mbox{higher orders}\;.
\ee
For the sake of completeness, let us briefly comment on possible
ghost contributions. Ghost loops arise from the term 
${\rm Tr}_{gh} \ln {\cal W}^{-1}$ in
$S_A[A_1]$. Their topology and consequently their properties
are completely analogous to those of the pure gluon loops discussed above.

I conclude with a remark regarding the HDL effective action.
According to Eq.\ (\ref{PHTL}), 
at zero temperature and large chemical potential,
a soft quark or antiquark has a momentum $k \sim g \mu$, i.e.,
it lies at the bottom of the Fermi sea, or at the top of the
Dirac sea, respectively.
These modes are, however, not that important in degenerate Fermi
systems, because it requires a large amount of energy $k_0 \sim \mu$
to excite them. The truly relevant modes are quark modes with large
momenta, $k \sim \mu$, close to the Fermi surface, because it costs
little energy to excite them.
A physically reasonable effective theory for cold, dense quark matter
should therefore feature no antiquark modes at all, and only
quark modes near the Fermi surface. Such a theory will be discussed
in Sec.\ \ref{IIIB}.

\subsection{High-density effective theory} \label{IIIB}

An effective theory
for high-density quark matter was first proposed
by Hong \cite{hong} and was further refined by Sch\"afer and others
\cite{HLSLHDET,schaferefftheory,NFL,others}. 
In the construction of this effective theory, one first proceeds
similarly to the discussion in Sec.\ \ref{II} and integrates
out antiquark modes. (From a technical point of view, this is
not done as in Sec.\ \ref{II} by functional integration, but by
employing the equations of motion for antiquarks. The result is
equivalent.) On the other hand, at first all quark modes in the Fermi
sea are considered as relevant. 
Consequently, in the notation of Sec.\ \ref{II},
the choice for the projectors ${\cal P}_{1,2}$ would be
\begin{subequations} \label{P12HDET}
\bea
{\cal P}_1(K,Q) & = & \left( \begin{array}{cc}
\Lambda_{\bf k}^+ & 0 \\
0 & \Lambda_{\bf k}^- \end{array} \right) \, \delta^{(4)}_{K,Q} \;, \\
{\cal P}_2(K,Q) & = &  \left( \begin{array}{cc}
\Lambda_{\bf k}^- & 0 \\
0 & \Lambda_{\bf k}^+ \end{array} \right)\, \delta^{(4)}_{K,Q} \;.
\eea
\end{subequations}
Also, at first gluons are not separated into soft and hard modes
either.
After this step, the partition function of the theory assumes the
form (\ref{ZQCD}) with ${\cal Z}_q$ given by Eq.\ (\ref{Zq4}).

In the next step, one departs from the rigorous approach of
integrating out modes, as done in Sec.\ \ref{II}, and follows
the standard way of constructing an effective theory, as
explained in the introduction.
One focusses exclusively on quark modes
close to the Fermi surface as well as on soft gluons. 
However, since quark modes far from the Fermi surface and hard gluons 
are not explicitly integrated out, the effective action does
not automatically contain the terms which
reflect the influence of these modes on the relevant quark and soft gluon
degrees of freedom. Instead, the corresponding terms
have to be written down ``by hand'' and the effective vertices
have to be determined via matching to the underlying microscopic
theory, i.e., QCD.

In order to further organize the terms occurring in the effective
action, one covers the Fermi surface with
``patches''. Each patch is labelled according to
the local Fermi velocity, ${\bf v}_F \equiv \hat{\bf k}\, 
k_F/\mu$ at its center.
A patch is supposed to have a typical size $\Lambda_\parallel$ in
radial ($\hat{\bf k}$) direction, and a size $\Lambda_\perp$
tangential to the Fermi surface.
The momentum of quark modes inside a patch is decomposed into a
large component in the direction of ${\bf v}_F$, the particular
Fermi velocity labelling the patch under consideration, and a small 
residual component, ${\bf l}$, residing exclusively inside the
patch,
\be \label{decomp}
{\bf k} = \mu \, {\bf v}_F + {\bf l}\;.
\ee
The residual component is further decomposed into a component 
pointing in radial direction, ${\bf l}_\parallel
\equiv {\bf v}_F ({\bf v}_F \cdot {\bf l})$, and the orthogonal one,
tangential to the Fermi surface, ${\bf l}_\perp \equiv {\bf l} - {\bf
l}_\parallel$.
The actual covering of the Fermi surface with such patches is not 
unique. One should, however, make sure that neighbouring
patches do not overlap, in order to avoid
double-counting of modes near the Fermi surface.
In this case the total number of patches on the Fermi surface is 
$\sim \mu^2/\Lambda_\perp^2$. 

\begin{figure}[ht]
\centerline{\includegraphics[width=10cm]{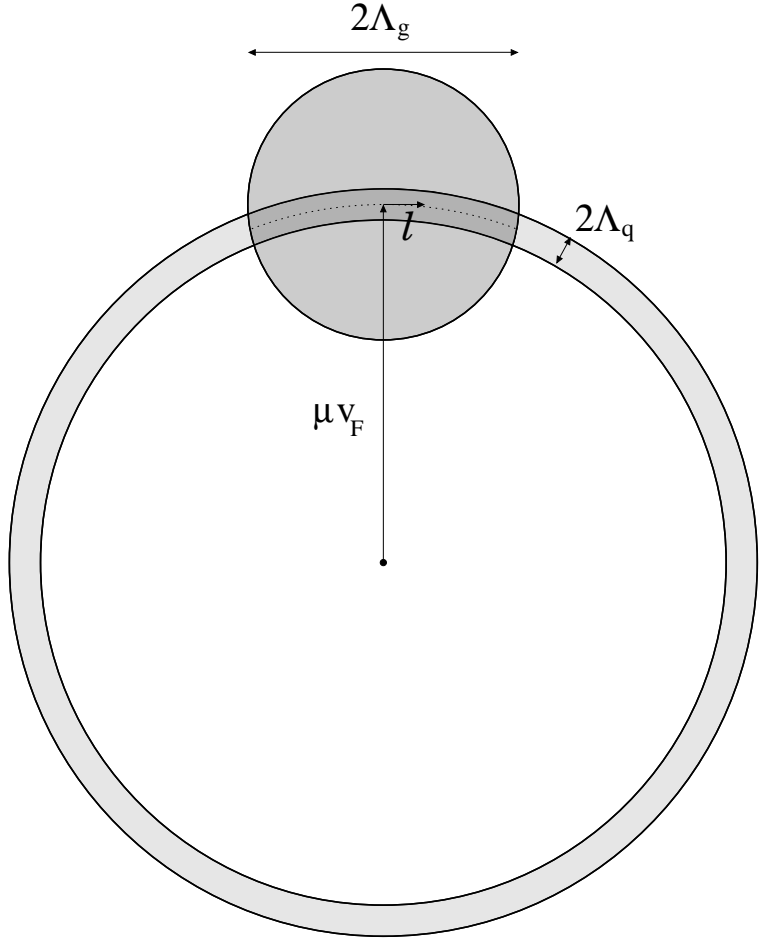}}
\caption[A particular patch covering the Fermi surface.]{A particular patch covering the Fermi surface.
The tangential dimension, $\Lambda_\perp$, is given by the
maximum momentum transferred via a soft gluon, $\Lambda_{\rm gl}$, while
the radial dimension, $\Lambda_\parallel$ is defined by the maximum 
distance of relevant quark modes from the Fermi surface, $\Lambda_{\rm q}$.
Also shown is a typical momentum transfer ${\bf l}$ via a soft gluon.}
\label{Sphere}
\end{figure}

In the following, I shall show that
the action of the high-density effective theory as discussed in Refs.\
\cite{hong,hong2,HLSLHDET,schaferefftheory,NFL,others} 
is contained in the effective action
(\ref{Seff}). To this end, however, I shall employ 
the choice (\ref{P12}) and (\ref{Q12}) for
the projectors for quark and gluon modes, and not Eq.\ (\ref{P12HDET})
for the quark projectors.
As in Refs.\ \cite{hong,hong2,HLSLHDET,schaferefftheory,NFL,others}, 
the quark mass will be set to zero, $m=0$. One also has to clarify how
the patches covering the Fermi surfaces introduced in Refs.\ 
\cite{hong,hong2,HLSLHDET,schaferefftheory,NFL,others} 
arise within the effective action Eq.\ (\ref{Seff}). 
It is obvious that the radial dimension $\Lambda_\parallel$
of a patch is related to the quark cut-off $\Lambda_{\rm q}$. One simply
chooses $ \Lambda_\parallel \equiv \Lambda_{\rm q}$. Similarly, since
soft-gluon exchange is not supposed to move a fermion from 
a particular patch to another, the dimension $\Lambda_\perp$ tangential
to the Fermi surface must be related to the gluon cut-off $\Lambda_{\rm gl}$.
Again, one adheres to the most simple choice $\Lambda_\perp \equiv 
\Lambda_{\rm gl}$. Since $\Lambda_{\rm gl} \alt \mu$, this is consistent 
with the matching procedure discussed in Ref.\ \cite{HLSLHDET}, 
where the matching scale is chosen as $\Lambda_\perp = \sqrt{2} \mu$
(which is only slightly larger than $\mu$).
The different scales $\Lambda_{\rm q}, \, \Lambda_{\rm gl}$, and $\mu$ are 
illustrated in Fig.\ \ref{Sphere}.
The modulus of the residual momentum ${\bf l}$ in Eq.\
(\ref{decomp}) is constrained to $l \leq {\rm max} \, 
(\Lambda_{\rm q},\, \Lambda_{\rm gl})$.

In Nambu-Gor'kov space, the leading, kinetic term in the Lagrangian
of the high-density effective theory reads
\be \label{kintermHDET}
{\cal L}_{\rm kin} = \frac{1}{2} \, \sum_{{\bf v}_F}
\bar{\Psi}_1(X,{\bf v}_F) \gamma_0 \, 
\left( \begin{array}{cc}
 i V \cdot D & 0 \\
 0 & i \bar{V} \cdot D_C \end{array} \right)\,
\Psi_1(X,{\bf v}_F)\; ,
\ee
cf.\ for instance Eq.\ (1) of Ref.\ \cite{schaferefftheory}.
Here, the 4-vectors
\be
V^\mu \equiv (1, {\bf v}_F)\;\;\;\; , \;\;\;\;\; 
\bar{V}^\mu \equiv (1, - {\bf v}_F)
\ee
are introduced. The covariant derivative for charge-conjugate fields is
defined as $D^\mu_{C} \equiv \partial^\mu + ig A^\mu_a T_a^T$.
The contribution (\ref{kintermHDET}) arises from
the term $\bar{\Psi}_1 \, \left( {\cal G}_{0,11}^{-1}\, + g {\cal
A}_{11} \right)\, \Psi_1$ in Eq.\ (\ref{Seff}). In order to see this,
use ${\cal P}_1^2 \equiv {\cal P}_1$ to write
\bea
\frac{1}{2} \, \bar{\Psi}_1 \, {\cal G}_{0,11}^{-1} \, \Psi_1 & = &
\frac{1}{2}\, \bar{\Psi}_1 \,\gamma_0 {\cal P}_1 \gamma_0 \, 
{\cal G}_{0,11}^{-1} \, {\cal P}_1\, \Psi_1 \non
& = & \frac{1}{2} \sum_{K,Q} \bar{\Psi}_1(K) \gamma_0 \, \frac{1}{T}\,
\left( \begin{array}{cc}
         k_0 + \mu -k & 0 \\
         0 & k_0 - \mu +k \end{array} \right) \, \delta^{(4)}_{K,Q}
\, \Psi_1(Q) \non
& \simeq & \frac{1}{2} \sum_{{\bf v}_F,L} \bar{\Psi}_1(L,{\bf v}_F) 
\gamma_0 \, \frac{1}{T}\,
\left( \begin{array}{cc}
         V \cdot L & 0 \\
         0 & \bar{V} \cdot L \end{array} \right) \, \Psi_1(L,{\bf
v}_F)\;.
\label{freekintermHDET}
\eea
In the last step, I have approximated $k \simeq \mu + {\bf v}_F
\cdot {\bf l}$, which holds up to terms of order $O(l^2/\mu)$, cf.\ 
Eq.\ (\ref{decomp}). This is a good approximation if the modulus of
a typical residual quark momentum in the effective theory
is $l \ll {\rm max}\, (\Lambda_{\rm q}, \, \Lambda_{\rm gl}) \alt \mu$.
I have also introduced the 4-vector $L^\mu \equiv (k_0, {\bf l})$
and, applying the decomposition (\ref{decomp}), I have written
the sum over ${\bf k}$ as a double sum over
${\bf v}_F$ and ${\bf l}$. The latter sum runs over all residual
momenta ${\bf l}$ inside a given patch, while the former runs over
all patches. With this decomposition, the spinors 
$\bar{\Psi}_1$, $\Psi_1$ are defined locally on a given patch
(labelled by the Fermi velocity ${\bf v}_F$), and depend on the
4-momentum $L$.
Note that a Fourier transformation to coordinate space
converts $V \cdot L \rightarrow i V \cdot \partial$.

Now consider the term $\bar{\Psi}_1 g {\cal A}_{11} \Psi_1$.
Since ${\cal A}_{11}$ is not diagonal in
momentum space, cf.\ Eq.\ (\ref{calA}), in principle the two
quark spinors $\bar{\Psi}_1$, $\Psi_1$ can belong to different
patches. However, I have chosen the tangential dimension of
a patch such that a (typical) soft gluon can by definition never move
a fermion across the border of a particular patch, 
$|{\bf k} - {\bf q}| \ll \Lambda_{\rm gl}$. Therefore,
both spinors reside in the same patch and, to leading order,
$\hat{\bf k} \simeq \hat{\bf q} \simeq {\bf v}_F$. With these
assumptions one may write
$\Lambda_{\bf k}^- \diracslash{A}_a(K-Q) \Lambda_{\bf q}^+
\simeq V \cdot A_a (K-Q) \, \gamma_0 \,\Lambda^+_{\bf k}$,
$\Lambda_{\bf k}^+ \diracslash{A}_a(K-Q) \Lambda_{\bf q}^-
\simeq \bar{V} \cdot A_a (K-Q) \, \gamma_0 \, \Lambda^-_{\bf k}$.
Then, introducing the residual momentum ${\bf l}'$
corresponding to the quark 3-momentum ${\bf q}$ and
defining ${L'}^\mu \equiv (q_0, {\bf l}')$, 
the respective term in the effective action becomes
\bea 
&&\frac{1}{2}\, \bar{\Psi}_1 \, g {\cal A}_{11} \,\Psi_1
\simeq \non
&&\frac{1}{2}\, \frac{g}{\sqrt{VT^3}} 
\sum_{{\bf v}_F, L, L'} 
\bar{\Psi}_1(L,{\bf v}_F) \gamma_0 \, 
\left( \begin{array}{cc}
 V \cdot A_a(L-L') \, T_a & 0 \\
 0 & - \bar{V} \cdot A_a(L-L') \, T_a^T \end{array} \right)\,
\Psi_1(L',{\bf v}_F) \;.\non
\label{gaugeHDET}
\eea
In coordinate space, the sum of Eqs.\ (\ref{freekintermHDET}) 
and (\ref{gaugeHDET}) becomes Eq.\ (\ref{kintermHDET}).

Subleading terms of order $O(1/\mu)$ in the high-density effective
theory are of the form
\bea
&&{\cal L}_{O(1/\mu)} =\non
&& - \frac{1}{2} \sum_{{\bf v}_F} \bar{\Psi}_1(X,{\bf
v}_F) \gamma_0 \, \frac{1}{2\mu} \left( \begin{array}{cc}
D_\perp^2 -\frac{g}{2} \, \sigma^{\mu \nu} F_{\perp \mu \nu}^aT_a & 0 \\
0 & - D_{C\perp}^2 - \frac{g}{2}\, \sigma^{\mu \nu} F_{\perp \mu
\nu}^a T_a^T \end{array} \right)\Psi_1(X,{\bf v}_F) \;,\non
\label{subleadingHDET}
\eea
cf.\ Eq.\ (2) of Ref.\ \cite{schaferefftheory}.
Here, $D_\perp^\mu \equiv 
\{0, ({\bf 1} - {\bf v}_F {\bf v}_F) \cdot {\bf D}\}$,
and similarly for $D_{C \perp}^\mu$. The commutator of two gamma matrices
is defined as usual,
$\sigma^{\mu \nu} \equiv (i/2) [\gamma^\mu , \gamma^\nu]$, and
$F_{\perp \mu \nu}^a T_a \equiv (i/g) [D_{\perp \mu}, D_{\perp \nu}]$.
As one will see in the following, this contribution arises from the term
$- g^2 \, \bar{\Psi}_1 \,{\cal A}_{12} \,{\cal G}_{22} \,{\cal
A}_{21}\, \Psi_1$ in Eq.\ (\ref{Seff}).

First, note that, with the projectors (\ref{P12}), the irrelevant quark
propagator ${\cal G}_{22}$ contains quark as well as antiquark modes.
In order to derive Eq.\ (\ref{subleadingHDET}), however, one has to
discard the quark and keep only the antiquark modes. In essence,
this is a consequence of the simpler choice (\ref{P12HDET}) for the 
projectors ${\cal P}_{1,2}$ in the high-density effective theory
of Refs.\ \cite{hong,hong2,HLSLHDET,schaferefftheory,NFL,others}.
In this case the propagator ${\cal G}_{22}$ may be simplified.
A calculation quite similar to that of Eqs.\
(\ref{freekintermHDET}) and (\ref{gaugeHDET}) now leads (in 
coordinate space) to
\be
{\cal G}_{22}^{-1} \equiv {\cal G}_{0,22}^{-1} + g{\cal A}_{22}
\simeq \gamma_0 \, \tau_3 \left( \begin{array}{cc}
2 \mu + i \bar{V} \cdot D & 0 \\
0 &  2\mu - i V \cdot D_C \end{array} \right) \; ,
\ee
where $\tau_3$ acts in Nambu-Gor'kov space.
This result may be readily inverted to yield
\be \label{G22}
{\cal G}_{22} \simeq \gamma_0 \, \tau_3\, \frac{1}{2 \mu}
\sum_{n=0}^\infty \frac{1}{(2 \mu)^n} \left( \begin{array}{cc}
- i \bar{V} \cdot D & 0 \\
0 & i V \cdot D_C \end{array} \right)^n\;.
\ee
Utilizing the projectors (\ref{P12HDET}),
one may also derive a simpler form for $g{\cal A}_{12}$ and
$g{\cal A}_{21}$. Consider, for instance, the term 
$\bar{\Psi}_1 \, g {\cal A}_{12}\, \Psi_2$. One follows the same steps
that led to Eqs.\ (\ref{freekintermHDET}), i.e., one assumes that the spinors 
$\bar{\Psi}_1$ and $\Psi_2$ reside in the same patch, such that
$\hat{\bf k} \simeq \hat{\bf q} \simeq {\bf v}_F$. This allows to derive
the identity $\Lambda_{\bf k}^\mp \, \diracslash{A}^a(K-Q) \Lambda_{\bf
q}^\mp \simeq \Lambda_{\bf k}^\mp \diracslash{A}^a_\perp(K-Q)$,
where $A_\perp^{\mu a} \equiv  
\{ 0, ({\bf 1} - {\bf v}_F {\bf v}_F) \cdot {\bf A}^a \}$. 
Now introduce the 4-vectors $L^\mu$, ${L'}^\mu$, as in Eq.\
(\ref{gaugeHDET}), which leads to
\be
\frac{1}{2} \, \bar{\Psi}_1 \, g {\cal A}_{12}\, \Psi_2
 \simeq  \frac{1}{2} \,\frac{g}{\sqrt{VT^3}}  \sum_{{\bf v}_F,L,L'} 
\bar{\Psi}_1 (L,{\bf v}_F)
\left( \begin{array}{cc}
\diracslash{A}^a_\perp(L-L')\, T_a & 0 \\
0 & - \diracslash{A}^a_\perp(L-L') \, T_a^T
\end{array} \right) \Psi_2(L',{\bf v}_F) \;.
\ee
One may add a term
$\diracslash{L}_\perp$ to the diagonal Nambu-Gor'kov
components, which trivially vanishes between spinors $\bar{\Psi}_1$ and
$\Psi_2$. This has the advantage that, in coordinate space,
\be \label{A12}
g {\cal A}_{12} \simeq \left( \begin{array}{cc}
i \Diracslash{D}_\perp & 0 \\
0 & i \Diracslash{D}_{C \perp} \end{array} \right)\;,
\ee
i.e., this term transforms covariantly under gauge transformations,
and no longer as a gauge field.
A similar calculation for $g{\cal A}_{21}$ gives the result
$g{\cal A}_{21} \equiv g {\cal A}_{12}$.
Combining Eqs.\ (\ref{G22}) and (\ref{A12}), the term
$- g^2 \, \bar{\Psi}_1 \,{\cal A}_{12} \,{\cal G}_{22} \,{\cal
A}_{21}\, \Psi_1$ corresponds to the following contribution in the
Lagrangian,
\bea \label{subleadingHDET2}
&&\!\!\!\!\!\!\!\!\!\!- \frac{1}{2} \sum_{{\bf v}_F}
\bar{\Psi}_1(X,{\bf v}_F) \gamma_0  \left( \begin{array}{cc}
\Diracslash{D}_\perp & 0 \\
0 & - \Diracslash{D}_{C \perp} \end{array} \right)
\frac{1}{2 \mu}
\sum_{n=0}^\infty \frac{1}{(2 \mu)^n} \left( \begin{array}{cc}
- i \bar{V} \cdot D & 0 \\
0 & i V \cdot D_C \end{array} \right)^n 
\left( \begin{array}{cc}
\Diracslash{D}_\perp & 0 \\
0 &  \Diracslash{D}_{C \perp} \end{array} \right)\non
&&\times \Psi_1(X,{\bf v}_F)\;. 
\eea
Taking only the $n=0$ term, and utilizing $\gamma^\mu \gamma^\nu
\equiv g^{\mu \nu} - i \sigma^{\mu\nu}$, one arrives at Eq.\ 
(\ref{subleadingHDET}). Note that my definition for transverse
quantities, e.g.\ $A_\perp^\mu \equiv 
\{ 0, ({\bf 1} - {\bf v}_F {\bf v}_F) \cdot {\bf A} \}$, 
slightly differs from that of Refs.\ \cite{hong,hong2},
where $A_\perp^\mu \equiv A^\mu - V^\mu V \cdot A$. However, both 
definitions agree when sandwiched between spinors 
$\bar{\Psi}_{1,2}$ and $\Psi_{2,1}$.

At order $O(1/\mu^2)$, besides the $n=1$ term in Eq.\ 
(\ref{subleadingHDET2}),
there are also four-fermion interaction terms,
cf.\ Eqs.\ (3-5) of Ref.\ \cite{schaferefftheory}. 
In the effective
action (\ref{Seff}), these contributions arise from 
the term ${\cal J}_{{\cal B}} \Delta_{22} {\cal J}_{{\cal B}}$
which originates from integrating out hard gluons.
(Since this is not done explicitly in the construction of the
high-density effective theory in Refs.\ 
\cite{hong,hong2,HLSLHDET,schaferefftheory,NFL,others}, this term
is not automatically generated, but has to be added ``by hand''.)
To leading order, this term corresponds to the exchange of a
hard gluon between two quarks, cf.\ the first diagram on the 
right-hand side of Fig.\ \ref{AAe}. 
If the quarks are close to the Fermi surface, the 
energy in the hard gluon propagator can be neglected, and
$\Delta_{0,22} \alt 1/\Lambda_{\rm gl}^2$. Since $1/\Lambda_{\rm gl}^2 \agt
1/\mu^2$, the contribution from hard-gluon exchange is of order
$O(1/\mu^2)$.
Four-fermion interactions also receive corrections at
one-loop order, cf.\ Fig.\ 5 of Ref.\ \cite{hong2}. In
Eq.\ (\ref{Seff}), they are contained in the
term ${\rm Tr} \ln \Delta_{22}^{-1}$, see the last diagram in Fig.\ \ref{AAd}.

Besides the quark terms in the Lagrangian of the high-density effective
theory \cite{hong,hong2,HLSLHDET,schaferefftheory,NFL,others}, 
there are also contributions from
gluons. The first is 
the standard Yang-Mills Lagrangian $-(1/4) F_{\mu \nu}^a F^{\mu
\nu}_a$, cf.\ Eq.\ (1) of Ref.\ \cite{schaferefftheory}.
This part is contained in the term $S_A[A_1]$ in Eq.\ (\ref{Seff}),
cf.\ Eq.\ (\ref{SA}).
The second contribution is a mass term for magnetic gluons,
\be
{\cal L}_{m_g} = - \frac{m_g^2}{2} \, {\bf A}^a \cdot {\bf A}^a \;,
\ee
cf.\ Eq. (19) of Ref.\ \cite{SonStephanov}, Eq.\ (18) of Ref.\ 
\cite{hong2}, or Eq.\ (27) of Ref.\ \cite{HLSLHDET}, 
where $m_g$ is
the gluon mass parameter (\ref{gluonmass}).
This term has to be added ``by hand'' in order to obtain the correct
value for the HDL gluon polarization tensor within the high-density
effective theory. In Eq.\ (\ref{Seff}) this contribution arises
from the $n=2$ term of the expansion (\ref{explnquark})
of ${\rm Tr} \ln {\cal G}_{22}^{-1}$. The gluon polarization tensor
has contributions
from particle-hole and particle-antiparticle excitations. 
The latter give rise to ${\cal L}_{m_g}$. 
While this term arises naturally within my derivation of the effective theory,
it does not in the high-density effective theory of Refs.\ 
\cite{hong,hong2,HLSLHDET,schaferefftheory,NFL,others}
because only antiquarks, but not irrelevant
quark modes, are explicitly integrated out. 
Irrelevant quark modes can then only be taken into account 
by adding the appropriate counter terms.

Sometimes, the full HDL action is added
to the Lagrangian of the high-density effective theory, cf.\ Eq.\ (8)
of Ref.\ \cite{schaferefftheory}.
This procedure requires a word of caution. For instance,
an important contribution to the HDL polarization tensor arises
from particle-hole excitations around the Fermi surface.
Such excitations are still relevant degrees of freedom in the
effective theory. However, in order for them to appear in the
gluon polarization tensor they would first have
to be integrated out. Therefore, strictly speaking 
such contributions cannot occur in the tree-level effective action. 
Of course, in an effective
theory one is free to add whatever contributions one deems necessary.
However, one has to be careful to avoid double counting.
As will be shown in Sec.\ \ref{IV}, the full HDL polarization tensor
will appear quite naturally in an approximate solution to the
Dyson-Schwinger equation for the gluon propagator, however, not at
tree-, but only at (one-)loop level.

It was claimed in Refs.\ \cite{hong2,HLSLHDET,schaferefftheory}
that a consistent power-counting scheme within the high-density
effective theory requires $\Lambda_\perp = \Lambda_\parallel$.
In contrast, I shall show in Sec.\ \ref{IV} 
that a computation of the gap parameter 
to subleading order requires $\Lambda_{\rm q} \equiv \Lambda_\parallel
\ll \Lambda_\perp \equiv \Lambda_{\rm gl}$.
This means that irrelevant quark modes become local on a scale
$l_q \gg 1/ \Lambda_{\rm q}$, while antiquark modes become local already on
a much smaller scale, $l_{\bar{q}} \gg 1/\mu$, cf.\ discussion at the
end of Sec.\ \ref{II}. As mentioned in the introduction,
for two different scales power counting of terms in
the effective action becomes a non-trivial problem.
While the high-density effective
theory of Refs.\
\cite{hong,hong2,HLSLHDET,schaferefftheory,NFL,others} 
contains effects
from integrating out antiquarks, i.e., from the scale $1/\mu$, the
effective action (\ref{Seff}) in addition keeps track of
the influence of irrelevant quark modes, i.e., from physics on the scale
$1/\Lambda_{\rm q} \gg 1/\mu$. Since all terms in the effective action
(\ref{Seff}) are kept, one can be certain not to miss any important
contribution just because the naive dimensional power-counting scheme
is invalidated by the occurrence of two vastly different length
scales.

\section{Towards a general effective \mbox{\it Theory}} \label{toeft}

As explained in Sec.\ \ref{EFT}, the calculation of any physical quantity within the framework of an effective theory requires the knowledge of the orders of magnitude of the operators in the action of the effective theory. Only after having determined all operators needed for the desired accuracy the actual calculation can be performed \cite{polchinski,Kaplan,Manohar}. Therefore, in principle, {\it all} operators in the general effective action (\ref{Z3}) yet await to be power-counted. This complete analysis is beyond the scope of this work and remains a formidable future project.

In the following, however, I take a first step towards this goal and begin with the class of quark loops that appear in the effective action after integrating out irrelevant quarks in the case of large $\mu$ and small $T$, cf.\ Eq.\ (\ref{Seff}) and Fig.\ \ref{AAc}.

\subsection{Power counting quark loops at large $\mu$ and small $T$}

In \cite{braatenpisarski} the power-counting rules for HTL/HDL were generally derived. Following this analysis, analogous rules are set up for loops of irrelevant quark modes at large $\mu$ and small $T$. However, I choose the projection operators (\ref{P12}) instead to the HTL/HDL projection operators (\ref{PHTL}). 
%In combination with the gluon projectors (\ref{Q12}) this corresponds to Fig.\ \ref{Sphere}.
Consequently, the internal {\it irrelevant} quark modes in the considered loops 
are antiquarks and quarks far from the Fermi surface with $|k-\mu|\geq \Lambda_{\rm q}$. The typical momenta of  {\it relevant} quark modes are located much closer to the Fermi surface,  $|k-\mu|\sim \Lambda_{\rm q}^n \ll \Lambda_{\rm q}\ll\mu$, where the upper index  $n$ abbreviates {\it near} to the Fermi surface. These {\it near} modes can be considered as the natural choice for the relevant degrees of freedom of a low-energy effective theory at large $\mu$ and small $T$ (in contrast to the low-momentum modes of the HDL theory, cf.\ the short discussion at the end of Sec.\ \ref{IIIA}). Therefore, a careful analysis of such loops seems worthwhile. 

The cutoff for {\it hard} gluons may be choosen as $\Lambda_{\rm gl} \lesssim \mu$, while the {\it soft} gluon modes in the effective action have typical momenta of order $\Lambda_{\rm gl}^s \ll \Lambda_{\rm gl}$, cf.\ Fig.\ \ref{Sphere}. It is clear from the outset that for $\Lambda^s_{\rm gl}\lesssim g\,\mu$ and $\Lambda_{\rm q}\rightarrow 0$ the considered loops must reproduce the corresponding HDLs discussed in Sec.\ \ref{IIIA}. Then, however, all quarks are integrated out and one ends up with a pure gauge theory for gluons. It would be more intriguing to keep a thin layer of quark modes around the Fermi surface as explicit degrees of freedom. If one chooses $\Lambda_{\rm q}$ sufficiently smaller than $\Lambda_{\rm gl}$ the difference between the considered loops and the HDLs will become arbitrarily small. 

An interesting ordering for all these scales could be 
\bea
\phi\sim\Lambda^n_{\rm q}\ll\Lambda_{\rm q} = \phi^yM^{1-y}\ll \Lambda^s_{\rm gl} \ll \Lambda_{\rm gl} \lesssim \mu \;,\label{naturalchoice}
\eea
where $1\agt y\gg g$ and $M^2\equiv (3\pi/4)\, m_g^2$, cf.\ Sec.\ \ref{estAB}. Soft gluons  being exchanged among relevant quarks would then have energies of order $\phi$. For the momentum scale $\Lambda^s_{\rm gl}= \phi^{1/3}M^{2/3}$ these gluons would feel the effect of Landau damping \cite{braatenpisarski,LeBellac}. 
For the choice $y=2/3$ the condition $\Lambda_{\rm q} \ll \Lambda^s_{\rm g}$ is fulfilled. Then the considered loops of irrelevant quark modes would contain Landau damping already on the tree-level of the effective action. On the other hand, the effect of the  Meissner masses would be suppressed by a factor $\phi/p\sim(\phi/M)^{2/3}\ll 1$ compared to the non-colorsuperconducting contributions in the loops (at least in the case of two external gluons) \cite{rischke2SC,shovy4}.

For $T\agt T_c$ one could use this theory for the investigation of precursory effects before the actual onset of Cooper pairing. In the strong coupling regime at $\mu \lesssim 500$ MeV numerical calculations based on the NJL-model (all gluons integrated out) showed the occurrence of low-energy collective modes in the form of strong diquark fluctuations with momenta close to the Fermi surface \cite{Koide,Kitazawa}. This is in agreement with the weak coupling analysis in \cite{Giannakis} where, after extrapolating down to $\mu \lesssim 500$ MeV, it was argued that color superconductivity could be a type-II superconductor with a second-order phase transition and strong pair fluctuations  at intermediate baryonic densities. On the other hand, at asymptotically large densities in weak coupling gauge field fluctuations are found to dominate and drive the CSC transition first-order \cite{bailinlove,Iida2,Giannakis2,Pisarski2}. The effective theory with the two different cutoffs proposed above might be useful in describing these fluctuations and estimating the relative order of magnitude of the diquark and the gauge field fluctuations.
 
%For $T<T_c$, howerver, the density of states of those relevant quark modes, which participate in the Cooper pairing, becomes very small. Ungapped quarks and also gappless modes, on the other hand, would remain and be particularly important for the cooling properties of neutron stars containing the respective phases \cite{shovy}.

The effect of the gap in the loops of irrelevant quarks certainly would have to be incorporated for $\Lambda^s_{\rm gl} \sim \phi$. This would require the introduction of a bilocal source term for irrelevant quark modes, cf.\ Sec.\ \ref{IVa}, leading to non-trivial off-diagonal  entries in Nambu-Gor'kov space and mixing quarks with charge-conjugate quarks. In the following, however, charge-conjugated quarks will not be used for simplicity.

\subsubsection{The quark propagator in the mixed representation}

The free propagator of massless quarks at chemical potential $\m$ and temperature $T$ is given by \cite{Meissner2f3f}
\be
\Gt_0(K) = \sum_{e=\pm}  \Lambda_{\vk}^e \g_0 \frac{k_0-(\m-ek)}{k_0^2-[\e_{\vk}^e]^2}
\ee
where 
$\e_{\vk}^e\equiv | \m - ek| $
and the projectors on positive ($e=+$) and negative ($e=-$) energy states defined as $\Lambda_{\vk}^e\equiv \frac{1}{2}\left(1+e\gamma_0 \vg \cdot \uk \right)$. Furthermore, $k_0\equiv-i\o_n$ with the fermionic Matsubara frequency \mbox{$\o_n=(2n+1)\pi T$}.
For the evaluation of the Matsubara sum in the loop it is advantageous \cite{loopease} to use 
the propagators in their mixed representation
\be
\Gt_0(\t , \vk )\equiv T\sum_{k_0} e^{-k_0 \t} \Gt_0(K)\;. \label{mixdef}
\ee
After performing the Matsubara sum in Eq.\ (\ref{mixdef}) in terms of a contour integral in the complex $k_0-$plane, one obtains for the range $0 \le \t \le \b$ \cite{Meissner2f3f}
\bea
\Gt_0(\t , \vk )&=& 
- \sum_{e=\pm} \Lambda_{\vk}^e \g_0 \left\{ 
\Theta(ek - \m) \left[1-N(\e_{\vk}^e)\right] e^{-\e_{\vk}^e \t}
+\Theta( \m -ek)N(\e_{\vk}^e) e^{\e_{\vk}^e \t}
 \right\}\\
&=& 
- \sum_{e,s=\pm} \Lambda_{\vk}^e \g_0  
\Theta(s[ek - \m]) \ft_s(\e_{\vk}^e) e^{-s\e_{\vk}^e \t}\;,             \label{G>}
\eea
where $N(x)\equiv (e^{x/T} + 1)^{-1}$ is the thermal distribution function for fermions. For the actual power counting the replacement 
\bea
\Theta(s[k-\mu])~  \rightarrow ~ \Theta(s[k - \m]-\Lambda_{\rm q}) \equiv\Theta^{ \Lambda_{\rm q}}_{s}  \label{ThetaLambda}
\eea
will be necessary. Before that, however, the form given by Eq.\ (\ref{G>}) may be kept to simplify the comparison with known results.
 Moreover, for the following calculations it is very convenient to introduce
\be
\ft_s(\e_{\vk}^e) \equiv \Theta(s) - sN(\e_{\vk}^e)\;,~~~~~~~~s=\pm\;. \label{ft}
\ee
In the following, the more compact notations
\begin{subequations}
\bea
\ft_{s}^e &\equiv& \ft_{s}(\e_{\vk}^e)\,,\\
\e_i&\equiv&\e_{\vk_i}^{e_i}= | \m - e_ik_i|\\
\Theta^{e}_{s} &\equiv& \Theta(s[ek - \m])
\eea
\end{subequations}
are used. The $\ft_s^e\,$'s obey the following relations
\bea
%\ft_{-s}(\e_{\vk}^e) &=& \ft_s(\e_{\vk}^e)\, e^{-s\e_{\vk}^e \b}\;, \\
% \ft_{-s}(\e_{\vk}^{e})+ \ft_{s}(\e_{\vk}^e)  &=&1\;, \\
%\ft_{-s_1}(\e_{\vk_1}^{e_1})\ft_{s_2}(\e_{\vk_2}^{e^2}) - \ft_{s_1}(\e_{\vk_1}^{e_1}) \ft_{-s_2}(\e_{\vk_2}^{e_2})&=&
%\frac{s+s}{2} -sN(\e_{\vk}^e) - s^\prime N(\e_{\vk^\prime}^{ e^\prime}) \;.
 \ft_{s}^{e}+ \ft_{-s}^e  &=&1\;, \label{rel2}\\
\ft_{-s}^e &=& \ft_s^e\, e^{-s\e_{\vk}^e \b}\;, \label{rel1}\\
\ft_{-s_1}^{e_1}\ft_{s_2}^{e_2} - \ft_{s_1}^{e_1} \ft_{-s_2}^{e_2}&=&
\frac{s_2-s_1}{2} + s_1 N(\e_1) -s_2N(\e_2) \label{rel3}\;.
\eea
While $e$ discriminates positive from negative energy states (quarks from antiquarks), the sign $s$ defines the direction of the evolution in imaginary time $\t$ (appearing in the exponent  $e^{-s\e_{\vk}^e \t}$ in the propagator): $s=+$ denotes propagation forward and  $s=-$ backwards in imaginary time. Simultaneously, for modes with positive energies $(e=+)$,  $s$ also splits the momentum space in one part above $(s=+)$ and one below $(s=-)$ the Fermi surface via $\Theta^+_s=\Theta(s[k-\mu])$. Hence, positive energy modes above the Fermi surface always propagate forward in imaginary time, whereas those below always propagate backwards. For antiquarks, however, one finds 
\bea
\Theta^-_s\equiv \Theta(-s)\;,\label{Thetaanti}
\eea
which means that negative energy modes always propagate backwards in imaginary time independently of their momentum.

Furthermore, $\ft^+_- =N(\e_{\vk}^+)$ is the {\it minimum} of the thermal distribution functions of quarks and quark-holes, whereas  $\ft^+_+ =\exp(|k-\mu|\b)\, N(\e_{\vk}^+)$ corresponds to their {\it maximum}.  Moreover, $\ft^-_- = N(\e_{\vk}^-) \simeq \exp[-(k+\mu)\b]$ is the thermal distribution of antiquarks (exponentially suppressed) and $\ft^-_+ = \exp[-(k+\mu)\b] N(\e_{\vk}^-) \simeq 1$ that of anti-quarkholes (abundant), respectively.
Eq.\ (\ref{rel2}) reflects the normalisation of  $\ft^e_\pm$, i.e.\ every state with positive (negative) energy and momentum $k$ is occupied, either by a(n) \mbox{(anti-)}quark or a(n) \mbox{(anti-)}quarkhole.
%Note that due to Eq.\ (\ref{rel1}) the lower index $s$ in $\ft^e_s$ can flip by multiplication with $e^{\mp s\e_{\vk}^e \b}$, while the sign $s$ in $e^{-s\e_{\vk}^e \b}$ and $\Theta^e_s$ is always the same. As a consequence, e.g., the there exist also quarkholes {\it above} the Fermi surface and anti-quarkholes. 

%Furthermore, $\ft^+_- =N(\e_{\vk}^+)$ is the thermal distribution function for quarks, whereas  $\ft^+_+ =\exp(|k-\mu|\b)\, N(\e_{\vk}^+)$ corresponds to quarkholes ($\ft^+_+$ being the reflection of $\ft^+_-$ at the Fermi surface and vice versa, $\ft^+_-(k-\mu)=\ft^+_+(\mu-k)$). Moreover, $\ft^-_- = N(\e_{\vk}^-) \simeq \exp[-(k+\mu)\b]$ is the distribution of antiquarks and $\ft^-_+ = \exp[-(k+\mu)\b] N(\e_{\vk}^-) \simeq 1$ that of anti-quarkholes, respectively.
%Eq.\ (\ref{rel2}) reflects the normalisation of  $\ft^e_\pm$, i.e.\ every state with positive (negative) energy and momentum $k$ is occupied, either by a(n) (anti-)quark or a(n) (anti-)quarkhole.
%Note that due to Eq.\ (\ref{rel1}) the lower index $s$ in $\ft^e_s$ can flip by multiplication with $e^{\mp s\e_{\vk}^e \b}$, while the sign $s$ in $e^{-s\e_{\vk}^e \b}$ and $\Theta^e_s$ is always the same. As a consequence, e.g., the there exist also quarkholes {\it above} the Fermi surface and anti-quarkholes. 

In the range $-\b \leq \t \leq 0$ one obtains similarly
\bea
\Gt_0(\t , \vk )&=& 
\sum_{e,s=\pm} \Lambda_{\vk}^e \g_0  
\Theta^{e}_{s} \ft_{-s}^e e^{-s\e_{\vk}^e \t}\;, \label{G<}
\eea
and therefore, for the whole range $-\b \leq \t \leq \b$
\bea
\Gt_0(\t , \vk )
&=&
-\mbox{sign}(\t)\sum_{e,s=\pm} \Lambda_{\vk}^e \g_0  
\Theta^{e}_{s} \ft_{\mbox{\tiny sign}(\t) s}^e e^{-s\e_{\vk}^e \t}\;.
\eea
If $-\b\leq \t \leq \b$ and  $-\b\leq \t\pm \b \leq \b$ are fulfilled, it follows with $\mbox{sign}(\t\pm \b)=\pm= -\mbox{sign}(\t)$
\bea
\Gt_0(\t \pm \b , \vk )&=&
\mbox{sign}(\t \pm \b )\sum_{e,s=\pm} \Lambda_{\vk}^e \g_0  
\Theta^{e}_{s} \ft_{\mbox{\tiny sign}(\t\pm \b  ) s}^e\, e^{-s\e_{\vk}^e (\t\pm \b)} \non
&=&
-\mbox{sign}(\t)\sum_{e,s=\pm} \Lambda_{\vk}^e \g_0  
\Theta^{e}_{s} \ft_{-\mbox{\tiny sign}(\t) s}^e\,e^{ \mbox{\tiny sign}(\t)  \e_{\vk}^e \b}\, e^{-s\e_{\vk}^e \t} \non
&=&
-\mbox{sign}(\t )\sum_{e,s=\pm} \Lambda_{\vk}^e \g_0  
\Theta^{e}_{s} \ft_{\mbox{\tiny sign}(\t) s}^e\, e^{-s\e_{\vk}^e \t} \non
&=&
-\Gt_0(\t , \vk )\;. \label{anti}
\eea
%where we used the compact notation $\ft_{s}^e \equiv \ft_{s}(\e_{\vk}^e)\,.$ So, 
Hence, as generally required, the mixed quark propagator $\Gt_0(\t , \vk )$ is antiperiodic with  period $\b$ within its range of definition.

%\subsubsection{Power counting quark loops with $N$ gloun legs at  $T\ll \Lambda_{\rm q} \ll \mu$}

%Before the actual power counting first the $\t-$integral for quark loops with 2, 3 and 4 gluon legs is explicitly performed. Then the remaining integrals over the internal 3-momentum $d^3k$ are power counted. 
The 4-momentum dependent part of the considered quark loops has the general form
\bea
{\cal J}^{\vm}(P_{1},\cdots ,P_{N})
%&=& T \sum_K \trs \g^\m \Gt_0(K) \g^{\m_1} \Gt_0(P_1-K) \g^{\m_2} \Gt_0(P_2-K)\cdots \g^{\m_{N-1}}\Gt_0(P_{N-1}-K) \non
&=&(-1)^{N+1} \frac{T}{V} \sum_K \trs \g^{\m_1} \Gt_0(K_1) \g^{\m_2} \Gt_0(K_2)\cdots \g^{\m_{N}}\Gt_0(K_{N}) \non
 & = & (-1)^{N+1}\int \frac{d^{3}\mathbf{k}}{(2\pi )^{3}}T\sum _{k^{0}}\mathrm{Tr}_s
\prod _{i=1}^{N}\gamma ^{\mu _{i}}\Gt_{0}(K_{i})\;,
%\nonumber \\
\eea
where $\vm\equiv (\m_1,\cdots \m_N)$ and $K_i\equiv K+P_i$. The prefactor $(-1)^{N+1}$ comes from the Feynman rule to take $-\Gt_0$ as the internal quark propagator and to write an overall $(-1)$ in front of each quark loop. The momenta of the external gluons are given by $P_{i+1}- P_i$ with $P_{N+1}= P_1$. 
With the inverse form of \eqrf{mixdef}
\bea
\Gt_0(K)= \int \limits _{0}^{\beta }d\tau e^{k^{0}\tau} \Gt_0(\tau,k)
\eea
it follows
\bea
\hspace*{-0.7cm}
{\cal J} ^{\vm}(P_{1},\cdots ,P_{N})
\!\!\!\! &=&\!\!\!\!(-1)^{N+1} \int \limits \frac{d^{3}\mathbf{k}}{(2\pi )^{3}}T\sum _{k^{0}}\mathrm{Tr}_s
\prod _{i=1}^{N}\gamma ^{\mu _{i}}\int \limits _{0}^{\beta }d\tau _{i} e^{ k_{i}^{0}\tau _{i}} \Gt_{0}(\tau _{i},k_{i})\nonumber \\
\!\!\!\! & = & \!\!\!\!(-1)^{N+1}\int \limits \frac{d^{3}\mathbf{k}}{(2\pi )^{3}}\mathrm{Tr}_s \prod _{i=1}^{N}\gamma ^{\mu _{i}}\int \limits _{0}^{\beta }d\tau _{i} e^{ p_{i}^{0}\tau _{i}} \Gt_{0}(\tau _{i},k_{i}) \;T\sum _{k^{0}} \;e^{ k^{0}\sum _{j=1}^{N}\tau _{j}}.
\end{eqnarray}
With the relation \cite{LeBellac}
\be
T\sum _{k^{0}}\exp \left( k^{0}\sum _{i=1}^{N}\tau _{i}\right) =\sum _{m=-\infty }^{\infty }(-1)^{m}\delta \left( {m}\b-\sum _{i=1}^{N}\tau _{i}\right)
\ee
one can perform the integral over one of the $\t$'s trivially, say over $\t_{N}$. The fact that in the integrals $0\leq \t_i \leq\b$ leads to
\be
0\leq \t_{N} = m\b - \sum _{i=1}^{N-1}\t_i \leq \b
\ee
and consequently the sum over $m$ runs from 1 to $N-1$ only. Using
$e^{p_im\beta}=1$ the integration over $\t_{N}$ yields 
%\mbox{
\bea
&&{\cal J} ^{\vm}(P_{1},\cdots ,P_{N})=\non
&=& 
(-1)^{N+1}\int \limits \frac{d^{3}\mathbf{k}}{(2\pi )^{3}}\mathrm{Tr}_s
\prod _{i=1}^{N-1}\left[ \gamma ^{\mu _{i}}
\int \limits _{0}^{\beta }d\tau _{i} e^{ (p_{i}^{0}-p_{N}^{0})\tau _{i}} 
%\non&&\times
\Gt_{0}(\tau _{i},\vk_{i})\right]\gamma^{\mu _{N}}
%\non&&
\non &&
\times
\sum_{m=1}^{N-1}(-1)^m
\Gt_{0}\left(m\b - \sum _{j=1}^{N-1}\t_j ,\vk_{N}\right)
\Theta \left( m\beta -\sum _{j=1}^{N-1}\tau _{j}\right) \Theta \left( \sum _{j=1}^{N-1}\tau _{j}-(m-1)\beta \right)
\non
&=&
- \int \limits \frac{d^{3}\mathbf{k}}{(2\pi )^{3}}
\sum_{\ve,\vs}\mathcal{T}_\ve^{\vm}
\Theta_\vs^\ve 
\sum_{m=1}^{N-1}(-1)^m
\prod _{i=1}^{N-1}\left[
\int \limits _{0}^{\beta }d\tau _{i} 
%\non&&\times
%\Theta^{e_i}_{s_i} 
\ft^{e_i}_{s_i}  e^{ (p_{i}^{0}-p_{N}^{0} -s_i\e_i)\tau _{i}}\right]
\times
\non&& \times
%\sum_{m=1}^{N-1}(-1)^m
%\Theta^{e_N}_{s_N} 
\ft^{e_N}_{s_N}
e^{ -s_N\e_N\left( m\beta -\sum _{j=1}^{N-1}\tau _{j}\right)} 
%\non &&
\Theta \left( m\beta -\sum _{j=1}^{N-1}\tau _{j}\right) \Theta \left( \sum _{j=1}^{N-1}\tau _{j}-(m-1)\beta \right)
\non
&=&
- \int \limits \frac{d^{3}\mathbf{k}}{(2\pi )^{3}}
\sum_{\ve,\vs}\mathcal{T}_\ve^{\vm}
\Theta_\vs^\ve
%\mathcal{F}_\vs^\ve
 \sum_{m=1}^{N-1}(-1)^m
\prod _{i=1}^{N-1}
\left[\int \limits _{0}^{\beta }d\tau _{i}
\ft^{e_i}_{s_i} e^{ \Omega_i \tau _{i}} \right]
\times\non&& \times
\ft^{e_N}_{s_N}e^{ -s_N\e_N m\beta}
%\times 
%\non &&
%\times
\Theta \left( m\beta -\sum _{j=1}^{N-1}\tau _{j}\right) \Theta \left( \sum _{j=1}^{N-1}\tau _{j}-(m-1)\beta \right)\;, \label{gentauint}
\eea 
where  $\vs \equiv (s_1,\cdots,s_N)$ and $\ve\equiv (e_1,\cdots,e_N)$, as well as 
\bea
\mathcal{T}_\ve^{\mu_1\cdots\mu_N} &\equiv &
\trs \left[ \prod_{j=1}^N \g^{\m_{j}} \Lambda_{\vk_j}^{e_j} \g_0 \right]\;,\\
\Omega_i &\equiv& p_{i}^{0}-p_{N}^{0} -s_i\e_i + s_N\e_N\;,\\
\Theta_\vs^\ve &\equiv& \prod_{i=1}^N \Theta_{s_i}^{e_i}
%,\\\mathcal{F}_\vs^\ve &\equiv& \prod_{i=1}^N \ft_{s_i}^{e_i}\;.
\eea
are  introduced. In the simplest case of two external gluons one has only $m = 1$ and therefore 
\bea
{\cal J}^{\mu_1\mu _{2}}(P_{1},P_2)&=& \int \limits \frac{d^{3}\mathbf{k}}{(2\pi )^{3}}
\sum_{\ve,\vs}\mathcal{T}_\ve^{\mu_1\mu_2}\Theta_\vs^\ve 
\int \limits _{0}^{\beta }d\tau _{1}
 \ft^{e_1}_{s_1}  e^{ \Omega_1 \tau _{1}}
 \ft^{e_2}_{s_2}
e^{ -s_2\e_2 \beta}\non
&=&
 \int \limits \frac{d^{3}\mathbf{k}}{(2\pi )^{3}}
\sum_{\ve,\vs}\mathcal{T}_\ve^{\mu_1\mu_2}\Theta_\vs^\ve 
\int \limits _{0}^{\beta }d\tau _1
\ft^{e_1}_{s_1} \ft^{e_2}_{-s_2}e^{ \Omega_1 \tau _{1}}\non
&=&
 \int \limits \frac{d^{3}\mathbf{k}}{(2\pi )^{3}}
\sum_{\ve,\vs}\mathcal{T}_\ve^{\mu_1\mu_2}
\Theta_\vs^\ve 
\frac{ \ft_{-s_1}^{e_1} \ft_{s_2}^{e_2} -  \ft_{s_1}^{e_1} \ft_{-s_2}^{e_2}}{\Omega_1}
\non
&=&
 \int \limits \frac{d^{3}\mathbf{k}}{(2\pi )^{3}}
\sum_{\ve,\vs}\mathcal{T}_\ve^{\mu_1\mu_2}
\Theta_\vs^\ve 
\frac{\frac{s_2-s_1}{2} +s_1N(\e_1) - s_2 N(\e_2)}{p_1^{0}-p_2^0 - s_1\e_{1} +s_2\e_{2}}
\;,~~~\label{2gluons}
\eea
 The above \eqrf{2gluons} is in accordance with the standard result for two external gluon fields Eq.(40) in \cite{Meissner2f3f}.
In order to simplify power counting arguments the energy with respect to the Fermi surface is introduced
\bea
E^e\equiv k-e\mu
\eea
Since $\Theta_s^e\equiv \Theta(s[ek-\mu])= \Theta(seE^e)$ one has $s=\mbox{sign} (eE^e)\,.$ Furthermore, 
$\e_{\vk}^e\equiv | \m - ek| = |eE^e|$ and therefore $s\e = eE^e = ek-\m$. So, in the denominators one can substitute
\bea
-s_1\e_1+s_2\e_2 = -e_1k_1+e_2k_2 \label{denom}
\eea
and obtains a form, which is suitable for a power-counting analysis
\bea
{\cal J}^{\mu_1\mu _{2}}(P_{1},P_2)=
\int \limits \frac{d^{3}\mathbf{k}}{(2\pi )^{3}}
\sum_{\ve,\vs}\mathcal{T}_\ve^{\mu_1\mu_2}
\Theta_\vs^\ve 
\frac{\frac{s_2-s_1}{2} +s_1N(\e_1) - s_2 N(\e_2)}{p_1^{0}-p_2^0  -e_1k_1+e_2k_2}
\;. \label{2gluonspower}
\eea
To recover a more familiar form one rewrites the numerators as
\bea
\frac{s_2-s_1}{2} +s_1N(\e_1) - s_2 N(\e_2) = \frac{e_2-e_1}{2} +e_1N(E^{e_1}) - e_2 N(E^{e_2}) \label{nom}
\eea
%and uses \eqrf{denom} and \eqrf{nom} 
and finds  after performing the sum over $\vs$
\bea
{\cal J}^{\mu_1\mu _{2}}(P_{1},P_2)=
 \int \limits \frac{d^{3}\mathbf{k}}{(2\pi )^{3}}
\sum_{\ve}\mathcal{T}_\ve^{\mu_1\mu_2}
\frac{\frac{e_2-e_1}{2} +e_1N(E^{e_1}) - e_2 N(E^{e_2})}{p_1^{0}-p_2^0  -e_1k_1+e_2k_2}
\;,~~~\label{2gluonsold}
\eea
which is in accordance with the standard result Eq.(44) in \cite{Meissner2f3f}. It is also in agreement with Eq.(5.77) in \cite{LeBellac}.
Moreover, in the limit ${\m \rightarrow 0} $ it reproduces the corresponding formular (A.11) in \cite{braatenpisarski} using 
\bea
\lim_{\m \rightarrow 0} \Theta_\vs^\ve  = \delta_{\vs,\ve} = \prod_{i=1}^N \delta_{s_i,e_i}\;. \label{limtheta}
\eea
In the cases of $N=3$ and 4 one obtains after rather lengthy calculations, cf.\ Sec.\ \ref{matsum}, Eqs.\ (\ref{3gluons}) and (\ref{4gluons})
\bea
{\cal J} ^{\mu_1\mu_2\mu _3}(P_{1},P_2,P_{3})
&=&
-\int \limits \frac{d^{3}\mathbf{k}}{(2\pi )^{3}}
\sum_{\ve,\vs}\mathcal{T}_\ve^{\mu_1\mu_2\mu_3}\Theta_\vs^\ve 
%\Theta^{e_1}_{s_1}\Theta^{e_2}_{s_2} \Theta^{e_3}_{s_3}
\;\frac{1} {p_2^{0}-p_3^0 - e_2k_2 +e_3k_3}
\times
\non
&&\times
\left[
\frac{\frac{s_3-s_1}{2} +s_1N(\e_1) - s_3 N(\e_3)} {p_1^{0}-p_3^0 - e_1k_1+s_3k_3}
-
\frac{\frac{s_2-s_1}{2} +s_1N(\e_1) - s_2 N(\e_2)} {p_1^{0}-p_2^0 - e_1k_1+e_2k_2 }
\right]\non\label{3gluonsb}
\eea
and
\bea
&&{\cal J} ^{\mu_1\cdots\mu _4}(P_{1},\cdots , P_{4})=\non
&=&
- \int \limits \frac{d^{3}\mathbf{k}}{(2\pi )^{3}}
\sum_{\ve,\vs}\mathcal{T}_\ve^{\mu_1\cdots\mu_4}\Theta_\vs^\ve 
\;\frac{1}{p_3^{0}-p_4^0 - e_3k_3+e_4k_4}\times \non
&&\left\{
\frac{1}{p_2^{0}-p_4^0 - e_2k_2+e_4k_4}
\left[ 
\frac{\frac{s_2-s_1}{2} +s_1N(\e_1) - s_2 N(\e_2)}{p_1^{0}-p_2^0 - e_1k_1+e_2k_2} -
\frac{\frac{s_4-s_1}{2} +s_1N(\e_1) - s_4 N(\e_4)}{p_1^{0}-p_4^0 - e_1k_1+e_4k_4}
\right]+ \right. \non
&&\left.
+\frac{1}{p_2^{0}-p_3^0 - s_2\e_{2} +s_3\e_{3}}\left[
\frac{\frac{s_3-s_1}{2} +s_1N(\e_1) - s_3 N(\e_3)}{p_1^{0}-p_3^0 - e_1k_1+e_3k_3}
-\frac{\frac{s_2-s_1}{2} +s_1N(\e_1) - s_2 N(\e_2)}{p_1^{0}-p_2^0 - e_1k_1+e_2k_2} \right]
\right\}\;.\non\label{4gluonsb}
\eea

\subsubsection{Power counting  of quark loops at large $\mu$ and small $T$}

%In order to perform a power counting analysis for $T\ll \Lambda_{\rm q} \ll \mu$ one replaces
%in the internal quark propagators (\ref{G>}) 

%As in the HTL case one can conjecture \cite{braatenpisarski} how to generalize the obtained results to $N>4$ external gluon legs. The generalisation to loops consisting of both quarks and gluons as internal lines is yet work under progress.
%In the following the order of magnitude of quark loops with 2, 3 and 4 external gluons is determined in terms of the fixed scales $\m \gg\L_{g}^{s} \gg\L_q^{n}$. The scaling properties of these diagramms for decreasing $\L_{g}^{s}$ and $\L_q^{n}$ but constant $\m$ will follow immediately.

In the case of two external gluons one has \eqrf{2gluonspower}
\bea
{\cal J}^{\mu_1\mu _{2}}(Q)=
 \int \limits \frac{d^{3}\mathbf{k}}{(2\pi )^{3}}
\sum_{\ve,\vs}\mathcal{T}_\ve^{\mu_1\mu_2}
\Theta_\vs^\ve 
\frac{\frac{s_2-s_1}{2} +s_1N(\e_1) - s_2 N(\e_2)}{-q^0  -e_1k_1+e_2k_2}\;,
\eea
where 
%$\Theta_\ve^\vs=\Theta(s_1[e_1k_1-\m])\Theta(s_2[e_2k_2-\m])$
$\Theta_\ve^\vs=\Theta_{s_1}^{e_1}\Theta_{s_2}^{e_2}$ and $\mathcal{T}_\ve^{\mu_1\mu_2}=\trs \left[ \g^{\m_{1}} \Lambda_{\vk_1}^{e_1} \g_0 \g^{\m_{2}} \Lambda_{\vk_2}^{e_2} \g_0 \right]$, which is
\begin{subequations}
\bea
\mathcal{T}_\ve^{00}&=&1+e_1e_2\, \uk_1\cdot \uk_2\;,\\
\mathcal{T}_\ve^{0i}&=&\mathcal{T}_\ve^{i0} = e_1\hat k_1^i+e_2\hat k_2^i\;,~~~i=x,y,z\;,\\
\mathcal{T}_\ve^{ij}&=&\d^{ij}\left(1-e_1e_2\, \uk_1\cdot \uk_2\right)+ e_1e_2\left(\hat k^i_1 \hat k^j_2 +\hat k^j_1 \hat k^i_2    \right)\;,~~~i,j=x,y,z\;.
\eea
\end{subequations}
First note that after the replacement given in Eq.\ (\ref{ThetaLambda}) with \mbox{$T\ll\Lambda_{\rm q}\ll\m $} one can estimate an upper boundary for the thermal distribution functions $N(\e_i)$. In the case $e_i=+$ one has $N(\e_i) <\exp(-\Lambda_{\rm q}/T)$ and for $e_i=-$ it is $N(\e_i) <\exp(-\m/T)$. Hence, all terms proportional to $N(\e_i)$ are {\it exponentially} suppressed compared to those proportional to $s_i$, and will therefore be neglected. Then, only terms with $s_1=-s_2$ contribute, where one mode in the loop propagates forward and one backwards in imaginary time. Making use of Eq.\ (\ref{Thetaanti}) one finds that all possible combinations are 
%Therefore, the only non-vanishing contributions are particle/anti-particle excitations (p-ap) and particle/particle-hole excitations (p-ph), as can be easily seen:
\begin{itemize}
\item $e_1 =- \Rightarrow s_1 =- \Rightarrow s_2 =+ \Rightarrow e_2 =+$ ~($\bar q /q^+$)
\item $e_1 =+,~ s_1=+ \Rightarrow s_2 = - \Rightarrow  e_2=+$ ~($q ^-/q^+$)  or~ $ e_2=-$ ~($\bar q /q^+$)
\item $e_1 =+,~ s_1=- \Rightarrow s_2 = + \Rightarrow  e_2=+$ ~($q ^-/q^+)\;,$
\end{itemize}
where $\bar q$ denotes an antiquark and $q^s$ a quark above ($s=+$) or below ($s=-$) the Fermi surface. After having neglected thermal excitations for irrelevant quarks, the latter two can be identified with quark and quarkhole modes, respectively.
Hence, one has
\bea
&&{\cal J}^{\mu_1\mu _{2}}(Q) =\non
&&= \int \limits \frac{d^{3}\mathbf{k}}{(2\pi )^{3}}
\left\{
\mathcal{T}_{+-}^{\mu_1\mu_2} \frac{\Theta(k_1-\m-\Lambda_{\rm q})}{q^0 + k_1+k_2}
+
\mathcal{T}_{-+}^{\mu_1\mu_2}\frac{\Theta(k_2-\m-\Lambda_{\rm q}) }{-q^0 +k_1 +k_2}
\right.\non
&&\left.+
\mathcal{T}_{++}^{\mu_1\mu_2}\frac{\Theta(k_2-\m-\Lambda_{\rm q})\Theta(\m-\Lambda_{\rm q}-k_1)-\Theta(k_1-\m-\Lambda_{\rm q})\Theta(\m-\Lambda_{\rm q}-k_2)  }{-q^0  -k_1+k_2}\right\}\non
&&=
 \int \limits \frac{d^{3}\mathbf{k}}{(2\pi )^{3}}
\left\{
\mathcal{T}_{+-}^{\mu_1\mu_2} \frac{\Theta(k-\m-\Lambda_{\rm q})}{q^0 + k+|\vk+\vq|}
+
\mathcal{T}_{-+}^{\mu_1\mu_2}\frac{\Theta(k-\m-\Lambda_{\rm q}) }{-q^0 + k+|\vk-\vq|}
\right.\non
&&\left.+
\mathcal{T}_{++}^{\mu_1\mu_2}\frac{\Theta(|\vk+\vq|-\m-\Lambda_{\rm q})\Theta(\m-\Lambda_{\rm q}-k)-\Theta(k-\m-\Lambda_{\rm q})\Theta(\m-\Lambda_{\rm q}-|\vk+\vq|)  }{-q^0  -k+|\vk+\vq|}\right\}.\non
\label{2quarkanalysis}
\eea
The first two terms correspond to the quark/antiquark loop. After writing $\Theta(k_2-\m-\Lambda_{\rm q})=1-\Theta(\m-\Lambda_{\rm q}-k_2)$ the 1 is identified as the ultraviolet-divergent vacuum contribution and has to be removed by renormalization. Since $\Lambda_{\rm q} \ll \mu$ one has $\Theta(\m-\Lambda_{\rm q}-k_i) \simeq \Theta(\m-k_i)$, which effectively restricts the integration region to the Fermi sphere with the volume $\m^3$. In the considered case of soft external gluons, $q^0,q\sim\Lambda_{\rm gl}^s\ll\m$, one can approximate $1/(\pm q^0 + k+|\vk+\vq|)\approx 1/2k$. Furthermore, one may expand $\mathcal{T}_\ve^{\mu_1\mu_2}$ in powers of $\frac{q}{k}$ yielding
\bea
\mathcal{T}_\ve^{00}&=&2\d_{e_1e_2} +\mathcal O\left(\frac{q^2}{k^2}\right)\;,\\
\mathcal{T}_\ve^{0i}=\mathcal{T}_\ve^{i0}&=& 2\d_{e_1e_2}\hat k^i +2e_1\d_{e_1,-e_2}(\d^{ij}-\hat k^i \hat k^j)\frac{q^j}{2k}+\mathcal O\left(\frac{q^2}{k^2}\right) \;,~~~i=x,y,z\;,\\
\mathcal{T}_\ve^{ij}
&=& 2\d^{ij}\d_{e_1,-e_2}+e_1e_2\hat k^i \hat k^j +\mathcal O\left(\frac{q^2}{k^2}\right)\;,~~~i,j=x,y,z\;.
\eea 
For $e_1=-e_2$ the dominant contribution arises from $\mathcal{T}_{+-}^{ij}=\mathcal{T}_{-+}^{ij}\approx 2(\d^{ij}-\hat k^i \hat k^j)$, whereas all others are suppressed by at least one power of $q/k$. Then the considered integral is of order $\mu^3/\mu\sim \mu^2$. After multiplication with $g^2$ from the vertices it is $\sim g^2\mu^2$. Choosing $\L_{\rm gl}^s\sim g\m \ll \m$ this would be of the order of the tree-level propagator, which is $Q^2 \sim (\L_{\rm gl}^s)^2$. For smaller values of $\L_{\rm gl}^s$, however, the considered term would be above tree-level, since its magnitude is independent of  the cutoff $\Lambda_{\rm gl}^s$. E.g.\ for $\Lambda_{\rm gl}^s\sim g^2\m$ this term would be one power of $1/g$ above tree-level.

\begin{figure}[ht]
\centerline{\includegraphics[width=14cm]{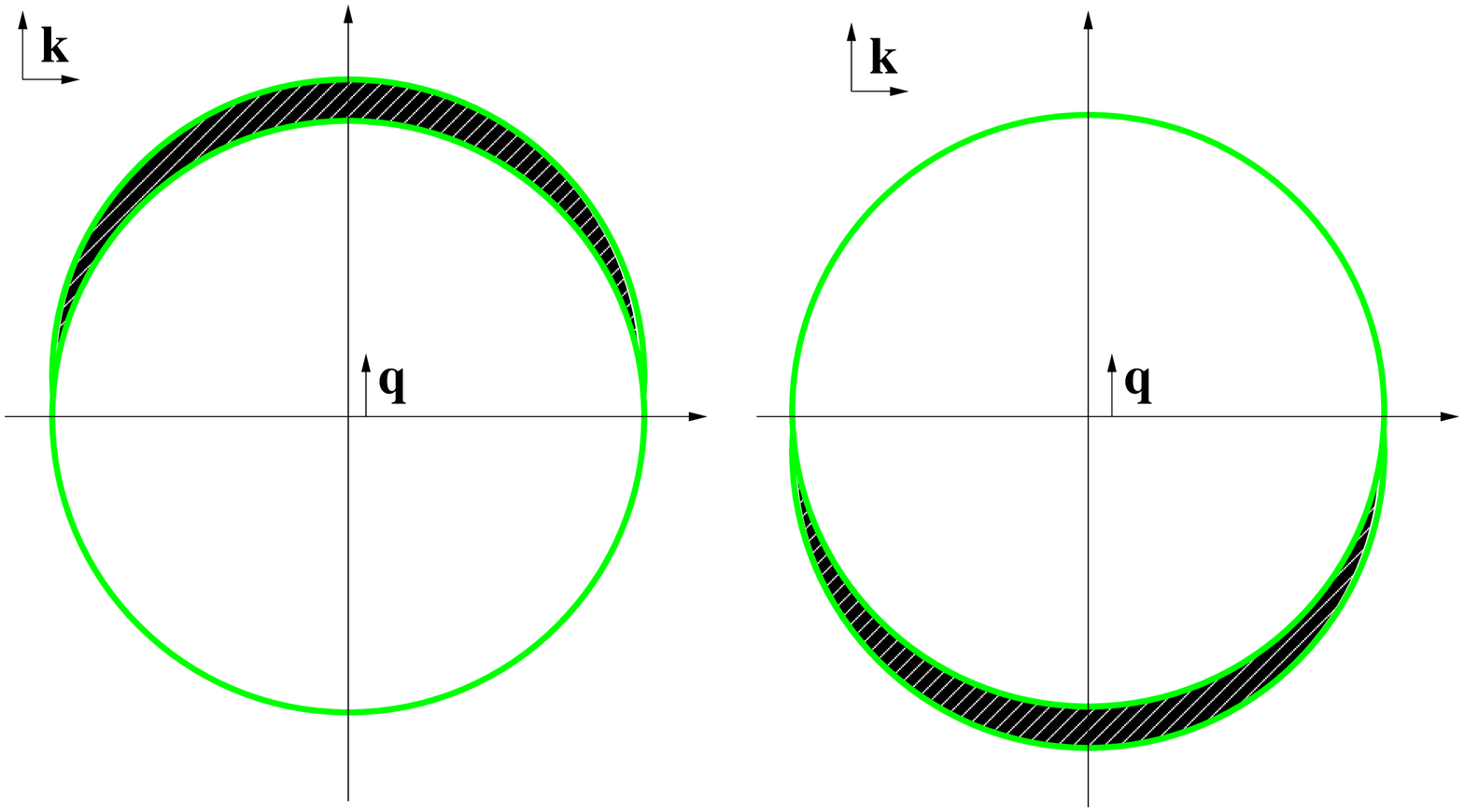}}
\caption[The integration region in the quark-quarkhole contribution in Eq.\ (\ref{2quarkanalysis}) ]{The integration region in the quark-quarkhole contribution in Eq.\ (\ref{2quarkanalysis}) is the dark shaded area for $\mu \gg q\sim \L_{\rm gl}^s \gg \L_{\rm q}$. The approximation in Eq.\ (\ref{app2}) is equivalent to including the light shaded areas corresponding to relevant quark modes with $|k-\mu|<\Lambda_{\rm q}$.}
\label{app2fig}
\end{figure}

The third term in \eqrf{2quarkanalysis} contains the particle/particle-hole excitations. Considering again the choice of cutoffs given in Eq.\ (\ref{naturalchoice}) with $\Lambda_{\rm q}\ll\Lambda^s_{\rm gl}$  one may approximate
\bea
|\vk+\vq|-k\ &\approx& \vq\cdot \uk\;,\label{app1}\\ 
\Theta(|\vk+\vq|-\m-\Lambda_{\rm q})\Theta(\m-\Lambda_{\rm q}-k)\non
-\Theta(k-\m-\Lambda_{\rm q})\Theta(\m-\Lambda_{\rm q}-|\vk+\vq|)
&\approx& \delta(k-\mu)\;\vq\cdot\uk\;.\label{app2}
\eea
To obtain this estimate note that the two quark modes in the loop must be located on opposite sides of the Fermi surface (above and below the Fermi surface) and they must be situated within a shell of width $\L_{\rm gl}^s$ around the Fermi surface since their relative momentum is given by the soft gluon momentum. Hence, $k\sim \mu \gg q \sim\L_{\rm gl}^s$. In Eq.\ (\ref{app2}) additionally use was made of tha fact that due to $\L_{\rm gl}^s \gg \Lambda_{\rm q}$ the cutoff $\Lambda_{\rm q}$ in the $\Theta$-functions on the l.h.s.\  can be neglected, cf.\ Fig.\ \ref{app2fig}. 
Note that in this way one will not recover the full result because the distribution functions $N(\e_i)$ as well as the gap $\phi$ have been dropped from the outset. Furthermore, for the choice $\Lambda_{\rm gl}^s \agt \Lambda_{\rm q}$ the approximation in Eq.\ (\ref{app2}) would become invalid and the calculation of the quark-quarkhole loop more complicated (but still feasible). Using Eqs.\ (\ref{app1},\ref{app2}) the quark/quarkhole excitation in \eqrf{2quarkanalysis} reads
\bea
\m^2 \int \frac{d\Omega}{(2\pi)^3} \mathcal{T}_{++}^{\m_1\m_2}\frac{\vq\cdot\uk}{q^0 - \vq\cdot \uk}\;, \label{2gluonslandau}
\eea
which is of order $\m^2$ for all $\m_1,\m_2$. i.e. it is at the same order as the quark-antiquark loop and as the tree-level gluon propagator. In fact,  Eq.\ (\ref{naturalchoice}) has turned out to be the right choice for the cutoffs to reproduce the HDL gluon self-energy, for details cf.\ \cite{Meissner2f3f}. For the choice $\m \gg \Lambda_{\rm gl}^s  \sim g\mu\agt \Lambda_{\rm q}$ the result would nontrivially depend on the fraction $x\equiv\Lambda_{\rm q}/\Lambda_{\rm gl}^s$ and converge to the HDL result in the limit $x \rightarrow 0$. 

Turning to the cases of three and four external gluons one first notes that, if not all internal lines correspond to particle and particle-hole modes the loop will be beyond tree-level. To show this one investigates the corresponding formulae \eqrftw{3gluonsb}{4gluonsb}. As in the two-gluon case the thermal distribution functions may be neglected. Assume that the line with label 1 is an antiparticle ($s_1=-$) and all other lines  are particles ($s_{j}=+\,,~j>1$). Then, using the approximation \eqrf{app1} all energy denominators are of order $\Lambda_{\rm gl}^s$ except for those inside the angular brackets, which are of order $\m$. Combining the latter, the energy denominators standing directly in front of the angular brackets are canceled. Then in the case of three external gluons the integral multiplied by $g^3$ is of order $g^3\m^3/\m^2\sim g^2 \Lambda_{\rm gl}^s$, which is suppressed by $g$ compared to the tree-level, which is $g\Lambda_{\rm gl}^s$. In the case of 4 external gluons the integral multiplied by $g^4$ is of order $g^4\m^3/(\m^2\Lambda_{\rm gl}^s)\sim g^3$ and thus is also suppressed by $g$ compared to tree-level, which is  $g^2$. Analogously one verifies that also {\it all} other integrals containing antiparticles are suppressed by at least one power of $g$ in comparison to tree-level. So, even though the integral region is of order $\m^3$ in the case that one mode corresponds to an antiparticle and all others are quarks above the Fermi surface no HDLs are produced.

In this context it seems worth mentioning that in \cite{braatenpisarski} it was claimed that loops with $N$ external gluons are HTLs if all but one of the $N-1$ occurring  energy denominators carry energies with opposite sign (and therefore are of order $\L_{\rm g}^s \sim gT$). It was argued there that  in this special case the distribution functions appear as a sum which is of order 1 in comparison to the difference of two distribution functions, which is only of order $g$, appearing when all energy denominators are of order $gT$. This does indeed compensate the one energy denominator which is of the order of the hard scale $T$. If one, however, sums up {\it all} terms of this type systematically as it has been done above in the present treatment, one finds that these terms will cancel, up to a remaining part which is suppressed by a factor of $g$ compared to tree-level. Hence, also in the regime of high temperatures {\it all} energy denominators must be of order of the soft scale $\L_{\rm gl}^s$ in order to produce a HTL.

The loops containing only quark and quarkhole modes can simplified in virtue of approximations \eqrftw{app1}{app2} to 
\bea
&&\hspace*{-1.8cm}{\cal J} ^{\mu_1\mu_2\mu _3}(Q_{1},Q_2)\approx\non
&\hspace*{-0.5cm}\approx&
\hspace*{-0.3cm}-\m^2\int \limits \frac{d \Omega}{(2\pi )^{3}}
\mathcal{T}_{+++}^{\mu_1\mu_2\mu_3} 
\;\frac{1} {(q_1^{0}+q_2^0)-(\vq_1+\vq_2)\cdot \uk}
%\times
%\non
%&&\times
\left[
\frac{\vq_2\cdot \uk } {q_2^0 -\vq_2\cdot \uk}
-
\frac{\vq_1\cdot \uk } {q_1^0 -\vq_1\cdot \uk}
\right]\label{3gluonsurface}
\eea
and
\bea
&&\hspace*{-3.cm}{\cal J} ^{\mu_1\cdots\mu _4}(Q_{1},Q_2 , Q_{3})\approx\non
&\approx&
-\m^2 \int \limits \frac{d \Omega}{(2\pi )^{3}}
\mathcal{T}_{++++}^{\mu_1\cdots\mu_4} 
\;\frac{1}{q_1^0+q_2^0+q_3^0 -(\vq_1+\vq_2+\vq_3)\cdot \uk}\times \non
&&\times\left\{
\frac{1}{q_1^{0}+q_2^0 - (\vq_1+\vq_2)\cdot \uk}
\left[ 
\frac{  \vq_1\cdot \uk }{q_1^{0}- \vq_1\cdot \uk  } -
\frac{ \vq_2\cdot \uk  }{q_2^{0}- \vq_2\cdot \uk}
\right] \right.+ \non
&&\left.+
\frac{1}{q_3^0  - \vq_3\cdot \uk}
\left[
\frac{ (\vq_2+\vq_3)\cdot \uk }{ q_2^{0}+q_3^0- (\vq_2+\vq_3)\cdot \uk}
-\frac{ \vq_2\cdot \uk  }{q_2^{0}- \vq_2\cdot \uk} \right]
\right\}
\eea
After multiplying the expression for the loop of three external gluons with $g^3$ it is of order $g^3\m^2/\Lambda_{\rm gl}^s\sim g \Lambda_{\rm gl}^s$, which is of the order of the corresponding tree-level diagram.
% It contains logarithmic divergencies for $q_1^0=\pm q_1\,,~q_2^0=\pm q_2\,,~q_1^0+q_2^0=q_1+q_2\,,$ which is analogous to the case of two external gluons \eqrf{2gluonslandau}. Moreover one can see, that 
Although in the angular brackets differences appear, no cancellations happen that effectively reduce the overall order of magnitude. E.g.\ the difference in the brackets in Eq.\ (\ref{3gluonsurface}) vanishes for $Q_2 \rightarrow -Q_1\;.$ In that case the energy denominator in front of the brackets disappears as well as preserving the overall order of magnitude of the loop.
%Note, that in the case of three external gluons this special configuration of external momenta means $Q_3 \rightarrow 0$ which violates the presumption that all external gluons are of order $\Lambda_{\rm gl}^s$ and therefore is of no practical interest.

Analogously, one finds for the loop with 4 external gluons $g^4\m^2/(\Lambda_{\rm gl}^s)^2\sim g^2$, i.e. it contributes also at tree-level. Again the various terms cannot cancel each other for any configuration of external gluon momenta $(Q_1,\cdots ,Q_4)$. If, e.g., again  $Q_2 \rightarrow -Q_1\;,$ the term in front of the first angular bracket will keep the order of magnitude of the term.
%Note, that this configuration has all momenta at $\Lambda_{\rm gl}^s$ and therefore is in accordance with the presumtion made.

One may conjecture as in \cite{braatenpisarski} how the above discussions of quark loops with 3 and 4 external gluons can be generalized to loops with any number $N \geq 3$ of external gluons.
Quark loops with $N$ external gluon legs with all internal lines corresponding to particle or particle-hole modes get support from a thin layer around the Fermi surface with volume proportional of $\m^2\Lambda_{\rm gl}^s$. The integrand consists of $N-1$ energy denominators which are all of order $1/\Lambda_{\rm gl}^s$ and the trace over Dirac space $\mathcal{T}^{\vm} _\ve $ which is dimensionless. Multiplying the loop with $g^N$ it is  of order 
\bea
g^N\frac{\Lambda_{\rm gl}^s \mu^2}{(\Lambda_{\rm gl}^s)^{N-1}} = g^{N-2}(\Lambda_{\rm gl}^s)^{4-N}\left(\frac{g\mu}{\Lambda_{\rm gl}^s}\right)^2 
=\mbox{tree-level} \times\left(\frac{g\mu}{\Lambda_{\rm gl}^s}\right)^2  \;.
\eea
Hence, if $\Lambda_{\rm gl}^s\sim g\mu$ the loop is at the order of the corresponding tree-level diagrams, $g^{N-2}(\Lambda_{\rm gl}^s)^{4-N}$, and therefore has to be resummed into the free gluon propagator or into a gluon vertex, respectively. Therefore, at this scale the free gluon propagation and the propagation via particle/hole and particle/antiparticle excitations are of equal importance. At scales $\Lambda_{\rm gl}^s<g\mu$ these loop diagrams even dominate the corresponding tree diagrams by a factor $\sim\left({g\mu}/{\Lambda_{\rm gl}^s}\right)^2 $. Gluons at scales $\Lambda_{\rm gl}^s < g\m$ prefer to propagate as quark/antiquark or as quark/quarkhole pair rather than to propagate freely. In other words, decreasing the scale $\Lambda_{\rm gl}^s$ {\it sufficiently  below} $g\m$ the free propagation becomes a {\it small correction} to the ``loop propagation'' and the tree-level gluon vertices become a small correction to the dominant 3 and 4 gluon loops. This is important when considering the static limit $P\rightarrow 0$ for gluons, for example. These loops of irrelevant quark modes are therefore {\it relevant} operators in the effective action, cf.\ Sec.\ \ref{EFT}.

At harder scales $\Lambda_{\rm gl}^s >g\m$ quark loops are  small corrections to the free propagation and the 3 and 4 gluon vertices as in vacuum perturbation theory. 
Note that in the case of hot quark matter with negligible chemical potential one has analogously HTL$\,\sim \mbox{tree-level}\times \left({gT}/{\Lambda_{\rm gl}^s}\right)^2 $.

All other types of quark loops will be suppressed by powers of $\Lambda_{\rm gl}^s/\m$. If $\Lambda_{\rm gl}^s\geq g\mu$ they are below tree-level and therefore part of the perturbative expansion. For $\Lambda_{\rm gl}^s< g\mu$, however, those terms which are suppressed by $\Lambda_{\rm gl}^s/\m$ still dominate the corresponding tree diagram by a factor of order ${g\mu}/{\Lambda_{\rm gl}^s}$. They are therefore still {\it relevant}. Those terms which are suppressed by $(\Lambda_{\rm gl}^s/\m)^2$ scale as tree-level and consequently are {\it marginal}. Diagrams, which are suppressed by $(\Lambda_{\rm gl}^s/\m)^3$ are always beyond tree-level and therefore perturbative.

Before applying the effective action Eq.\ (\ref{Seff}) to some perturbative quantity one first has to extend this analysis to all the remaining diagrams contained in the effective action. In the following Chapter \ref{chapterapp}, however, it will be demonstrated how the effective action can be combined with the CJT-formalism \cite{CJT} in order to facilitate the calculation of the non-perturbative color-superconducting gap-parameter.

		\chapter[Application to color superconductivity]{Application to \\ Color superconductivity}\label{chapterapp}

\section{Calculation of the QCD gap parameter} \label{IV}

In this section, I demonstrate how the effective action Eq.\ (\ref{Seff}) derived in
Sec.\ \ref{II} can be applied to compute the gap parameter of
color-superconducting quark matter to subleading order. To this end this effective action is combined with the CJT-formalism, which will be introduced below. 
The resulting subleading order gap equation, Eq.\ (\ref{gapequation5}) is well-known. However, as will be explicitly shown, the derivation given here is substantially more rigorous as the known derivations in full QCD. One of the most important advantages is that the self-consistency is required only for {\it relevant} fields and propagators. Furthermore, the expansion of the various terms in the gap equation in powers of $\Lambda_{\rm q}/\Lambda_{\rm gl}\sim g$ emerges as a powerful tool to rigorously identify leading, subleading and sub-subleading order contributions to the gap parameter. 
%For the sake of definiteness, a spin-zero, two-flavor color superconductor is considered (2SC phase, cf.\ Sec.\ \ref{CSC}). 

\subsection{CJT formalism for the effective theory} \label{IVa}

The gap parameter in superconducting systems is not accessible
by means of perturbation theory; one has to apply non-perturbative,
self-consistent, many-body resummation techniques to calculate it.
For this purpose it is convenient to employ the CJT formalism
\cite{CJT}. The first step is to add source terms 
to the effective action (\ref{Seff}),
\be \label{Seffsources}
S_{\rm eff}[A_1,\bar{\Psi}_1, \Psi_1]\; \longrightarrow\;
S_{\rm eff}[A_1,\bar{\Psi}_1, \Psi_1] + J_1 A_1 + 
\frac{1}{2}\, A_1 K_1 A_1  + \frac{1}{2}
\left( \bar{\Psi}_1 H_1 + \bar{H}_1 \Psi_1 + 
\bar{\Psi}_1 {\cal K }_1 \Psi_1 \right) \; ,
\ee
where I employed the compact matrix notation defined in Eq.\ (\ref{compact}).
$J_1$, $\bar{H}_1$, and $H_1$ 
are local source terms for the soft gluon and relevant quark fields, 
respectively, while $K_1$ and ${\cal K}_1$ are bilocal source terms.
The bilocal source ${\cal K}_1$ for quarks is also a matrix in
Nambu-Gor'kov space.
Its diagonal components are source terms which couple quarks to
antiquarks, while its off-diagonal components couple quarks
to quarks. The latter have to be introduced for systems which can become
superconducting, i.e., where the ground state has
a non-vanishing diquark expectation value,
$\langle \psi_1 \psi_1 \rangle \neq 0$.

One then performs a Legendre transformation with respect
to all sources and arrives at the CJT effective action \cite{CJT,kleinert}
\bea
\Gamma\left[A,\bar{\Psi},\Psi,\Delta, {\cal G}\right]
& = & S_{\rm eff} \left[A,\bar{\Psi},\Psi\right]
  -\frac{1}{2}\, {\rm Tr}_g \ln \Delta^{-1}
  -\frac{1}{2}\, {\rm Tr}_g \left( D^{-1} \Delta - 1 \right) \nonumber\\
& + & \frac{1}{2}\, {\rm Tr}_q\ln {\cal G}^{-1}
  +\frac{1}{2}\, {\rm Tr}_q\left(G^{-1}{\cal G}-1\right)
  +\Gamma_2\left[A,\bar{\Psi},\Psi,\Delta,{\cal G}\right]\; .   \label{Gamma}
\eea
Here, $S_{\rm eff}[A, \bar{\Psi}, \Psi]$ is the tree-level
action defined in Eq.\ (\ref{Seff}), which now depends on the
{\em expectation values\/} $A \equiv \langle A_1 \rangle$, $
\bar{\Psi} \equiv \langle \bar{\Psi}_1 \rangle$, and $\Psi \equiv
\langle \Psi_1 \rangle$ 
for the one-point functions of soft gluon and relevant quark fields.
In a slight abuse of notation, I use the same symbols for the
expectation values as for the original fields, prior to integrating
out modes. This should not lead to confusion, as the original fields
no longer occur in any of the following expressions.

The quantities $D^{-1}$ and $G^{-1}$ in Eq.\ (\ref{Gamma}) are
the inverse {\em tree-level\/} propagators for soft gluons and
relevant quarks, respectively, which are determined from 
the effective action $S_{\rm eff}$, see below.
The quantities $\Delta$ and ${\cal G}$ are the expectation values
for the two-point functions, i.e., the {\em full\/} propagators, of 
soft gluons and relevant quarks.
The functional $\Gamma_2$ is the sum of all two-particle irreducible
(2PI) diagrams. These diagrams are vacuum diagrams, i.e., they
have no external legs. They are constructed from the vertices
defined by the interaction part of $S_{\rm eff}$,
linked by full propagators $\Delta$, ${\cal G}$. 
The expectation values for the one- and two-point functions of
the theory are determined from the stationarity conditions
\be \label{statcond}
0 = \frac{\delta \Gamma}{\delta A} = \frac{\delta \Gamma}{\delta
\bar{\Psi}} = \frac{\delta \Gamma}{\delta \Psi} = \frac{\delta \Gamma}{\delta
\Delta } = \frac{\delta \Gamma}{\delta {\cal G}} \; .
\ee
The first condition yields the Yang-Mills equation for the 
expectation value $A$ of the soft gluon field. The second and third
condition correspond to the Dirac equation for $\Psi$ and
$\bar{\Psi}$, respectively. The effective action (\ref{Seff})
contains a multitude of terms which depend on $A, \bar{\Psi}, \Psi$,
and thus the Yang-Mills and Dirac equations are
rather complex, wherefore I refrain from explicitly presenting them here.
Nevertheless, for the Dirac equation the solution is trivial,
since $\bar{\Psi}_1,\, \Psi_1$ are Grassmann-valued fields, and
their expectation values must vanish identically,
$\bar{\Psi} = \langle \bar{\Psi}_1 \rangle = \Psi = \langle \Psi_1
\rangle \equiv 0$. On the other hand, for the Yang-Mills equation,
the solution $A$ is in general non-zero but, at least for
the two-flavor color superconductor considered here, it was shown 
\cite{Gerhold,DirkDennis} to be parametrically small, $A \sim \phi^2/(g^2
\mu)$, where $\phi$ is the color-superconducting gap parameter. 
Therefore, to subleading order in the gap equation it can be neglected.

The fourth and fifth condition (\ref{statcond}) are Dyson-Schwinger
equations for the soft gluon and relevant quark propagator,
respectively,
\begin{subequations} \label{DSE}
\bea 
\Delta^{-1} & = & D^{-1} + \Pi\;, \label{DSEgluon}\\
{\cal G}^{-1} & = & G^{-1} + \Sigma \;, \label{DSEquark}
\eea
\end{subequations}
where
\begin{subequations} \label{selfenergy}
\bea 
\Pi & \equiv & - 2 \, \frac{\delta \Gamma_2}{\delta \Delta^T}\; ,
\label{selfenergygluon}\\
\Sigma & \equiv & 2 \, \frac{\delta \Gamma_2}{\delta {\cal G}^T}
\label{selfenergyquark}
\eea
\end{subequations}
are the gluon and quark self-energies, respectively.
The Dyson-Schwinger equation for the relevant quark propagator is
a $2 \times 2$ matrix equation in Nambu-Gor'kov space,
\be \label{NGinvquark}
{\cal G}^{-1} =   \left( \begin{array}{cc}
                            [ G^+]^{-1} &  0 \\
                     0 & [ G^-]^{-1} \end{array} \right)
                + \left( \begin{array}{cc}
                            \Sigma^+ &  \Phi^- \\
                     \Phi^+ & \Sigma^- \end{array} \right) \;,
\ee
where $\Sigma^+$ is the regular self-energy for quarks and
$\Sigma^-$ the corresponding one for charge-conjugate quarks.
The off-diagonal self-energies $\Phi^\pm$, the so-called {\em gap
matrices}, connect regular with 
charge-conjugate quark degrees of freedom. A non-zero $\Phi^\pm$
corresponds to the condensation of quark Cooper pairs.
Only two of the four components of this matrix equation are
independent, say $[G^+]^{-1} + \Sigma^+$ and $\Phi^+$, 
the other two can be obtained via 
$[G^-]^{-1} + \Sigma^-= C \{[G^+]^{-1} + \Sigma^+\}^T C^{-1}$,
$\Phi^- \equiv \gamma_0 [\Phi^+]^\dagger \gamma_0$.
Equation (\ref{NGinvquark}) can be formally solved for ${\cal G}$
\cite{manuel2},
\be \label{NGquarkprop}
{\cal G} \equiv \left( \begin{array}{cc}
                            {\cal G}^+ &  \Xi^- \\
               \Xi^+ & {\cal G}^- \end{array} \right) \; ,
\ee
where
\be
{\cal G}^\pm \equiv \left\{ [G^\pm]^{-1} + \Sigma^\pm - 
\Phi^\mp \left( [G^\mp]^{-1} + \Sigma^\mp \right)^{-1} \Phi^\pm \right\}^{-1}
\ee
is the propagator describing normal propagation of quasiparticles
and their charge-conjugate counterpart, while
\be \label{Xi}
\Xi^\pm \equiv - \left( [G^\mp]^{-1} + \Sigma^\mp \right)^{-1}
\Phi^\pm {\cal G}^\pm
\ee
describes anomalous propagation of quasiparticles, which is possible
if the ground state is a color-superconducting quark-quark condensate,
for details, see Ref.\ \cite{DHRreview}.

The tree-level gluon propagator is defined as
\be
D^{-1} \equiv - \frac{\delta^2 S_{\rm eff} [A,\bar{\Psi}, \Psi]}{\delta 
A\,\delta A} \;.
\ee
Since I ultimately evaluate the tree-level propagator at
the stationary point of $\Gamma$, Eq.\ (\ref{statcond}), where
$\bar{\Psi} = \Psi =0$, I may omit all terms in $S_{\rm eff}$, 
Eq.\ (\ref{Seff}), which are proportional to the quark fields.
The only terms which contribute to the tree-level gluon propagator
are therefore 
\be
D^{-1} \equiv 
- \frac{\delta^2 }{\delta A\,\delta A}\left(
S_A+ \frac{1}{2}\, {\rm Tr}_q\ln {\cal G}_{22}^{-1}
- \frac{1}{2}\, {\rm Tr}_g\ln \Delta_{22}^{-1} \right) \;.
\ee
Using the expansions (\ref{expquark}),
(\ref{explnquark}), (\ref{expansion2}), and
(\ref{expansion3}), and exploiting the cyclic property of the trace, 
one finds
\be
D^{-1} = - \frac{\delta^2 S_A}{\delta A \, \delta A} 
- \frac{g}{2}\, 
{\rm Tr}_q\left( \frac{\delta {\cal G}_{22}}{\delta A} \, 
\frac{\delta {\cal A}_{22}}{\delta A} \right) 
+\frac{1}{2}{\rm Tr}_g \left(\frac{\delta \Delta_{22}}{ \delta A } \,
\frac{\delta \Pi_{22}}{\delta A} +\Delta_{22} \, 
\frac{\delta^2 \Pi_{22}}{\delta A\,\delta A}
\right)
\;.\label{D0noA}
\ee
In order to proceed, note
that the Dyson-Schwinger equations (\ref{DSE}) are evaluated
at the stationary point of the effective action, where 
$\bar{\Psi} = \Psi = 0, \, A \simeq 0$.
For $A=0$, the first term yields the free inverse propagator for
soft gluons, $\Delta_{0,11}^{-1}$, cf.\ Eq.\ (\ref{Sgluon}),
plus a contribution from the
Faddeev-Popov determinant, 
$(\delta^2 {\rm Tr}_{gh} \ln {\cal W}^{-1} / \delta A
\delta A)_{A=0}$. The contributions from the three- and four-gluon
vertex vanish for $A=0$. 
Furthermore, according to Eq.\ (\ref{calA}), 
\be \label{Gammatilde}
\frac{\delta {\cal A}_{22}(K,Q)}{\delta A(P)} =
\frac{1}{\sqrt{VT^3}} \, \hat{\Gamma} \, \delta^{(4)}_{K,Q+P}
\equiv \tilde{\Gamma}(K,Q;P)\; .
\ee
This is a matrix in fundamental color, flavor, and Nambu-Gor'kov space,
as well as in the space of quark 4-momenta $K,Q$. It is a vector in
Minkowski and adjoint color space ($\hat{\Gamma}$ carries a 
Lorentz-vector and a gluon color index), as well as in the space
of gluon 4-momenta $P$.
One evaluates $(\delta {\cal G}_{22}/\delta A)_{A=0}$ using the expansion
(\ref{expquark}). Only the term for $n=1$ survives when taking $A=0$.
For $\bar{\Psi} = \Psi = 0$, I have $\Pi_{\cal B}=0$, cf.\ Fig.\ \ref{ZZa},
and I only need to consider $\Pi_{22} = \Pi_{\rm loop} + \Pi_{{\cal V}}$.
Then, the term ${\cal V}^{(3)} \equiv 
(\delta \Pi_{{\cal V}} / \delta A )_{A=0}$ corresponds
to a triple-gluon vertex, cf.\ Fig.\ \ref{ZZc},
where two hard gluons couple to one soft gluon.
The term $(\delta \Pi_{\rm loop}/\delta A )_{A=0}$ is a correction to
this vertex: it couples two hard gluons to a soft one through an
(irrelevant) quark loop, cf.\ Fig.\ \ref{ZZb}.
According to arguments well known from the HTL/HDL effective theory,
this vertex correction can never be of the same order as
the tree-level vertex ${\cal V}^{(3)}$,
since the two incoming gluons are hard. I therefore neglect 
$(\delta \Pi_{\rm loop}/\delta A )_{A=0}$ in the following. 
Similarly, ${\cal V}^{(4)} \equiv (\delta^2 \Pi_{{\cal V}} / \delta A \delta
A )_{A=0}$ is a four-gluon vertex, cf.\ Fig.\ \ref{ZZc},
where two hard gluons couple to
two soft ones, and $(\delta^2 \Pi_{\rm loop} / \delta A \delta A )_{A=0}$ 
is the one-(quark-)loop correction to this vertex, cf.\ Fig.\
\ref{ZZb}. Applying
the same arguments as above I only keep ${\cal V}^{(4)}$.
Arguments from the HTL/HDL effective theory also show  that to leading order 
one may approximate $\Delta_{22} \simeq \Delta_{0,22}$.
Finally, utilizing the same arguments I approximate
$\delta \Delta_{22}/ \delta A  \simeq - \Delta_{0,22} {\cal V}^{(3)}
\Delta_{0,22}$.
Then, the inverse tree-level gluon propagator of Eq.\ (\ref{D0noA}) becomes
\bea
D^{-1} &=& \Delta_{0,11}^{-1} +\frac{g^2}{2}\, 
{\rm Tr}_q \left( {\cal G}_{0,22}\, \tilde{\Gamma} \, 
{\cal G}_{0,22}\, \tilde{\Gamma} \right) 
- \frac{1}{2}\, {\rm Tr}_g \left( \Delta_{0,22} \, {\cal V}^{(3)} \,
\Delta_{0,22}\,{\cal V}^{(3)} \right) + 
\frac{1}{2}\, {\rm Tr}_g \left(\Delta_{0,22}\,  {\cal V}^{(4)} 
\right)\non
&& - \left. \frac{\delta^2 {\rm Tr}_{gh} \ln {\cal W}^{-1} }{\delta A \, 
\delta A} \right|_{A = 0}
\;.\label{D0noA2}
\eea
The second term represents an (irrelevant) quark-loop, while
the third term is a hard gluon loop. The fourth term is a hard
gluon tadpole. Finally, the last term in Eq.\ (\ref{D0noA2}) 
corresponds to a ghost loop necessary to cancel loop contributions
from unphysical gluon degrees of freedom.
Note that in the effective theory loop
contributions involving irrelevant quarks and hard gluons 
occur already in the tree-level action (\ref{Seff}). Therefore,
such loops also arise in the inverse tree-level propagator 
(\ref{D0noA2}) for the soft gluons of the effective theory. 
For the projection operators (\ref{Q12}) and (\ref{PHTL})
the inverse tree-level propagator (\ref{D0noA2}) is precisely the
HTL/HDL-resummed inverse gluon propagator.
For small temperatures, $T \ll \mu$, 
the contribution from the gluon and ghost
loops is negligible as compared to that from the quark loop,
\be
D^{-1} \simeq  \Delta_{0,11}^{-1} +\frac{g^2}{2}\, 
{\rm Tr}_q \left( {\cal G}_{0,22}\, \tilde{\Gamma} \, 
{\cal G}_{0,22}\, \tilde{\Gamma} \right) 
\;.\label{D0noA3}
\ee

The inverse tree-level quark propagator is defined as
\be \label{S0}
G^{-1} \equiv
- 2 \,\frac{\delta^2 S_{\rm eff} [A,\bar{\Psi}, \Psi ]}{
\delta\bar{\Psi}\,\delta{\Psi}}\;.
\ee
For $\bar{\Psi} =\Psi =0$, the last term in Eq.\ (\ref{Seff}) does not
contribute to $G^{-1}$ because it has at least four external quark legs,
and the two functional derivatives $\delta/ \delta \bar{\Psi}$, 
$\delta/ \delta \Psi$ amputate only two of them. The first and the
third term in Eq.\ (\ref{Seff}) do not depend on $\bar{\Psi}, \Psi$ at
all, therefore
\be
G^{-1}  =   {\cal G}_{0,11}^{-1} +  g {\cal B}[A]
+ \frac{\delta^2 {\rm Tr}_g \ln \Delta_{22}^{-1} }{\delta\bar{\Psi}\,
\delta \Psi } \;.
\ee
Using the expansion formulae (\ref{expansion2}) and (\ref{expansion3})
and the fact that $\Pi_{22}$ depends
on $\bar{\Psi}, \Psi$ only through $\Pi_{\cal B}$, I obtain
\be
\frac{\delta^2{\rm Tr}_g \ln \Delta_{22}^{-1} }{\delta\bar{\Psi}\,
\delta{\Psi}}
= {\rm Tr}_g \left( \Delta_{22} 
\frac{\delta^2 \Pi_{{\cal B}}}{\delta\bar{\Psi}\,\delta{\Psi}}\right)\;.
\ee
I have exploited the fact that this expression is evaluated
at $\bar{\Psi} = \Psi = 0$, i.e., terms with external quark legs
will eventually vanish. The trace runs only over adjoint colors, Lorentz
indices, and (hard) gluon 4-momenta. 
Since $\Delta_{22}$ is a hard gluon propagator,
the contribution from $\Pi_{22}$ to $\Delta_{22}$ may be neglected
to the order I are computing, and I may set $\Delta_{22} 
\simeq \Delta_{0,22}$. Furthermore, 
$(\delta^2 \Pi_{{\cal B}}/ \delta \bar{\Psi} \delta \Psi)_{A=0}
\equiv - g^2 \,\tilde{\Gamma}\, {\cal G}_{0,22}\, \tilde{\Gamma}$, 
cf.\ Fig.\ \ref{ZZa}. At $\bar{\Psi}= \Psi = 0,\, A \simeq 0$ I are left with
\be \label{S0noA}
G^{-1}  =  {\cal G}_{0,11}^{-1}
- g^2 \, {\rm Tr}_g \left( \Delta_{0,22} \, \tilde{\Gamma} \,
{\cal G}_{0,22} \, \tilde{\Gamma} \right)\;.
\ee 
As was the case for the tree-level gluon propagator, also 
the tree-level quark propagator receives a loop contribution; here it
arises from a loop involving an irrelevant quark and a hard gluon
line. The term
$\tilde{\Gamma} {\cal G}_{0,22} \tilde{\Gamma}$ under the gluon trace remains
a matrix in the quark indices, i.e., fundamental color, flavor,
Dirac, and quark 4-momenta.

I now proceed to solve the Dyson-Schwinger equations (\ref{DSE}) for the
soft gluon and relevant quark propagator. To this end I have
to determine $\Gamma_2$. Of course it is not feasible to consider
{\em all\/} possible 2PI diagrams. The advantage of the CJT formalism is
that {\em any\/} truncation of $\Gamma_2$ defines a meaningful,
self-consistent many-body approximation for which one can solve
the Dyson-Schwinger equations (\ref{DSE}). In the truncation of
$\Gamma_2$ I only take into account 2-loop diagrams 
which are 2PI with respect to the soft gluon and
relevant quark propagators $\Delta, \, {\cal G}$,
\be 
\label{EqGamma2}
\Gamma_2 = - \frac{g^2}{4}\,{\rm Tr}_{q,g} \left( {\cal G} \,
\tilde{\Gamma} \, {\cal G} \, \tilde{\Gamma} \, \Delta \right)
- \frac{g^2}{2} \, {\rm Tr}_{q,g} \left( {\cal G} \, \tilde{\Gamma} \,
{\cal G}_{0,22} \, \tilde{\Gamma} \, \Delta \right)
- \frac{g^2}{4}\, {\rm Tr}_{q,g} \left( {\cal G}\, \tilde{\Gamma}\, {\cal
G}\, \tilde{\Gamma}\, \Delta_{0,22} \right) \; .
\ee
The traces now run over quark as well as over gluon indices. Consider,
for instance, the term ${\cal G} \, \tilde{\Gamma}\,  {\cal G}\,
\tilde{\Gamma}$. It is a matrix in the space of 
fundamental color, flavor, Dirac and quark 4-momenta, of which the
trace is taken through ${\rm Tr}_q$. In addition,
due to the two factors $\tilde{\Gamma}$ it carries two
Lorentz-vector, adjoint-color, and gluon-4-momenta indices.
The trace ${\rm Tr}_g$ contracts these indices with the corresponding
ones from the gluon propagator $\Delta$.

\begin{figure}[ht]
\centerline{\includegraphics[width=10cm]{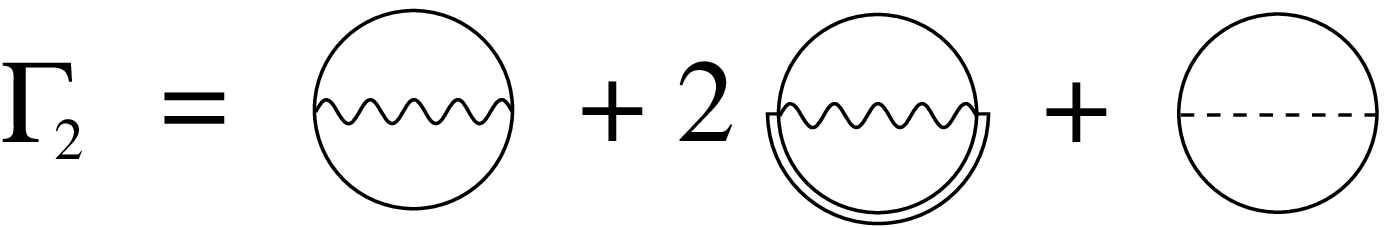}}
\caption{Diagrammatic representation of $\Gamma_2$, Eq.\
(\ref{EqGamma2}).}
\label{Gamma2eff}
\end{figure}

The diagrams corresponding to Eq.\ (\ref{EqGamma2}) are shown
in Fig.\ \ref{Gamma2eff}. 
The first two terms are constructed from the quark-gluon
coupling $\sim \bar{\Psi}\, g {\cal B}\, \Psi$. Using Eq.\ (\ref{B}),
one may either obtain an ordinary quark-gluon vertex 
$\sim g\, \bar{\Psi}\, {\cal A} \,\Psi$, 
involving one soft gluon and two relevant quark
legs, or a vertex $\sim g^2 \, \bar{\Psi}\, {\cal A} \, 
{\cal G}_{22}\, {\cal A}\, \Psi$, with (at least) two soft 
gluon legs and two relevant quark
legs. To lowest order, I approximate ${\cal G}_{22} \simeq {\cal
G}_{0,22}$, which neglects vertices with more than two soft gluon
legs. Taking two ordinary quark-gluon vertices and tying 
them together to obtain a 2PI 2-loop diagram, I arrive at the first
term in Eq.\ (\ref{EqGamma2}), or the first 
diagram in Fig.\ \ref{Gamma2eff}. Taking one of the two-gluon-two-quark
vertices and tying the legs together, one obtains the second term
in Eq.\ (\ref{EqGamma2}), or the second
diagram in Fig.\ \ref{Gamma2eff}, respectively. Finally, the third term/diagram
arises from the last term in Eq.\ (\ref{Seff}). To lowest order,
this corresponds to a four-quark vertex 
$\sim g^2 \, \bar{\Psi} \, \tilde{\Gamma}\, \Psi
\, \Delta_{0,22} \bar{\Psi} \, \tilde{\Gamma}\, \Psi$. Tying the quark
legs together to form a 2PI diagram, one obtains the corresponding
term/diagram in Eq.\ (\ref{EqGamma2})/Fig.\ \ref{Gamma2eff}. 

The combinatorial factors in front of the various terms in Eq.\ 
(\ref{EqGamma2}) are explained as follows. In the first diagram, there
are two ordinary quark-gluon vertices. According to Eq.\ (\ref{Seff}),
each diagram comes with a factor 1/2. Moreover, since there are two vertices,
the diagram is, in the perturbative sense, a diagram of second order,
which causes an additional factor 1/2 \cite{FTFT}. Finally, there are
two possibilities to connect the quark lines between the two
vertices. In total, I then have a prefactor 
$ -(1/2)^2 \times 1/2 \times 2 = -1/4$, where the minus sign arises
from the fermion loop. The second diagram arises from the
two-quark-two-gluon vertex, which already comes with a prefactor $-1/2$ in
Eq.\ (\ref{Seff}). It is perturbatively of first order, and 
there is only one possibility to tie the quark and gluon lines
together, so there is no additional combinatorial factor (and no
additional minus sign) for this diagram. Finally, the third diagram
arises from the four-quark vertex, $(1/2) {\cal J}_{{\cal B}}
\Delta_{0,22} {\cal J}_{{\cal B}}$, in Eq.\ (\ref{Seff}). 
This vertex comes with a factor 
1/2 and is perturbatively of first order. 
However, there are two additional factors 1/2 residing in ${\cal J}_{{\cal
B}}$, since ${\cal J}_{{\cal B}} \sim (1/2) \bar{\Psi} \, \hat{\Gamma}\,
\Psi$, cf.\ Eq.\ (\ref{J_B}). 
Again, there are two possibilities to tie the quark lines
together, so that, in total, I have a prefactor 
$-1/2 \times (1/2)^2 \times 2 = - 1/4$, where the minus sign 
again stands for the quark loop.

\begin{figure}[ht]
\centerline{\includegraphics[width=16cm]{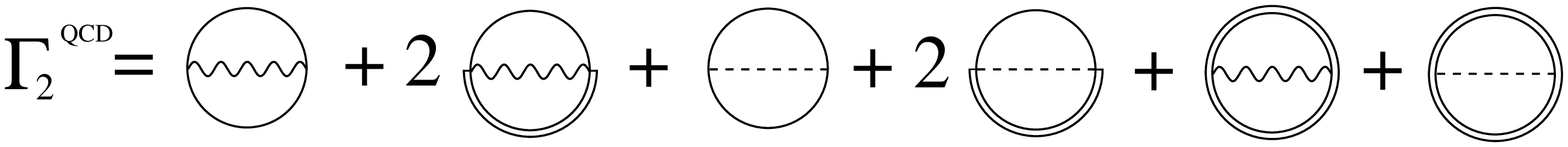}}
\caption[Diagrammatic representation of $\Gamma_2^{\rm QCD}$.]{Diagrammatic representation of $\Gamma_2^{\rm QCD}$, after
decomposing quark lines into relevant and
irrelevant, as well as gluon propagators into soft and hard
contributions.}
\label{Gamma2QCD}
\end{figure} 

At this point, it is instructive to compare $\Gamma_2$, Eq.\
(\ref{EqGamma2}), in the effective theory with 
$\Gamma_2^{\rm QCD}$ which one would have written
down in QCD at the same loop level.
$\Gamma_2^{\rm QCD}$ would be equivalent to the first diagram
of Fig.\ \ref{Gamma2eff}, but now the quark and gluon lines represent the
full propagators for {\em all\/} momentum modes, relevant {\em and\/}
irrelevant as well as soft {\em and\/} hard. In order to compare
with $\Gamma_2$ of the effective theory, I decompose the quark
propagators into relevant and irrelevant modes, and the gluon
propagator into soft and hard modes. One obtains the six diagrams
shown in Fig.\ \ref{Gamma2QCD}. The first three are precisely the
same that occur in $\Gamma_2$ of the effective theory, including
the combinatorial prefactors. The last three diagrams do not occur
in $\Gamma_2$ of the effective theory, because they are not 2PI
with respect to the relevant quark propagator ${\cal G}$ and
the soft gluon propagator $\Delta$. Nevertheless, they are
still included in the CJT effective action of the effective theory,
Eq.\ (\ref{Gamma}): opening the
relevant quark line of the fourth diagram, I recognize the loop
contribution to the tree-level quark propagator $G^{-1}$, cf.\
Eq.\ (\ref{S0noA}). Now consider the fifth term in Eq.\ (\ref{Gamma}):
here, this loop contribution to $G^{-1}$
is multiplied by ${\cal G}$ and traced over, which yields
the fourth diagram in $\Gamma_2^{\rm QCD}$.
Similarly, opening the soft gluon line
of the fifth diagram, I identify this diagram as 
the irrelevant quark-loop contribution
to the tree-level gluon propagator $D^{-1}$, cf.\ Eq.\ (\ref{D0noA3}).
The third term in Eq.\ (\ref{Gamma}), where this contribution is
multiplied by $\Delta$ and traced over, then yields the fifth diagram
of $\Gamma_2^{\rm QCD}$.
Finally, the sixth diagram resides in the term $\sim {\rm Tr}_g \ln 
\Delta_{22}^{-1}$ of the tree-level effective action $S_{\rm eff}$,
cf.\ Fig.\ \ref{AAd}.
Therefore, in principle, the CJT effective action (\ref{Gamma}) for
the effective theory contains the same information as the
corresponding one for QCD. However, while in QCD self-consistency
is maintained for {\em all\/} momentum modes via the solution of the
stationarity condition (\ref{statcond}), in the effective theory
self-consistency is only required for the {\em relevant\/} quark 
and {\em soft\/} gluon modes. These are the {\em only\/} 
dynamical degrees of freedom in the CJT effective action; the
irrelevant fermion and hard gluon modes, which were integrated out,
only appear in the vertices of the {\rm tree-level\/}
action (\ref{Seff}). In this sense the effective theory 
provides a simplification of the full problem.

\subsection{Dyson-Schwinger equations for relevant quarks and soft
gluons} \label{IVb}

After having specified $\Gamma_2$ in Eq.\ (\ref{EqGamma2})
I am now in the position to write down the Dyson-Schwinger equations
(\ref{DSE}) explicitly. For the full inverse propagator of soft gluons
I obtain with Eqs.\ (\ref{DSEgluon}), (\ref{selfenergygluon}), 
(\ref{D0noA3}), and (\ref{EqGamma2})
\be \label{DSEg}
\Delta^{-1} = \Delta_{0,11}^{-1} +\frac{g^2}{2}\,\left[ 
{\rm Tr}_q\left( {\cal G}_{0,22}\, \tilde{\Gamma} \, 
{\cal G}_{0,22}\, \tilde{\Gamma} \right) + 2\, 
{\rm Tr}_q\left( {\cal G}\, \tilde{\Gamma} \, 
{\cal G}_{0,22}\, \tilde{\Gamma} \right)+ 
{\rm Tr}_q\left( {\cal G}\, \tilde{\Gamma} \, 
{\cal G}\, \tilde{\Gamma} \right) \right] \;.
\ee
The first term in square brackets takes into account the effect of
quark-antiquark excitations as well as quark-hole excitations far
from the Fermi surface. The second term is the contribution from
excitations where one quark is close to the Fermi surface (a relevant
quark) while the second is far from the Fermi surface or an
antiquark (an irrelevant quark). The relevant quark propagator
${\cal G}$ can have diagonal elements in Nambu-Gor'kov space,
corresponding to normal propagation of quasiparticles, as well as
off-diagonal elements, corresponding to anomalous propagation of
quasiparticles, cf.\ Eq.\ (\ref{NGquarkprop}).
However, in the second term in square
brackets the latter contribution is absent, because ${\cal
G}_{0,22}$ is purely diagonal in Nambu-Gor'kov space, cf.\ Eq.\ (\ref{G0FT}).
This is different for the last term in square brackets, which
corresponds to quark-hole excitations close to the Fermi surface.
Both quark propagators have to be determined self-consistently and may
have off-diagonal elements in Nambu-Gor'kov space. Consequently, the
trace over Nambu-Gor'kov space gives two contributions, a loop where
both quarks propagate normally, and another one where they propagate
anomalously. Diagrams of this type have been evaluated in Ref.\
\cite{Meissner2f3f} and lead to the Meissner effect for gluons
in a color superconductor.

For the full inverse propagator of relevant quarks I obtain
with Eqs.\ (\ref{DSEquark}), (\ref{selfenergyquark}), (\ref{S0noA}),
and (\ref{EqGamma2})
\bea \label{DSEq}
&&\!\!\!\!\!\!\!\!{\cal G}^{-1} = \non
&&\!\!\!\!\!\!\!\!{\cal G}_{0,11}^{-1}
- g^2 \, \left[ {\rm Tr}_g\left( \Delta_{0,22} \, \tilde{\Gamma} \,
{\cal G}_{0,22} \, \tilde{\Gamma} \right)
+{\rm Tr}_g\left( \Delta \, \tilde{\Gamma} \,
{\cal G}_{0,22} \, \tilde{\Gamma} \right)
+  {\rm Tr}_g\left( \Delta_{0,22} \, \tilde{\Gamma} \,
{\cal G} \, \tilde{\Gamma} \right)
+  {\rm Tr}_g\left( \Delta \, \tilde{\Gamma} \,
{\cal G} \, \tilde{\Gamma} \right) \right] \; .\non
\eea
The first two terms in square brackets do not have off-diagonal
components in Nambu-Gor'kov space. They contribute only to the
regular quark self-energy. The other two terms in square brackets
have both diagonal and off-diagonal components in Nambu-Gor'kov
space. The diagonal components contribute to the regular quark self-energy,
in particular, the fourth term leads to the quark
wave-function renormalization factor computed first in Ref.\ 
\cite{manuel}. It gives rise to non-Fermi liquid behavior 
\cite{rockefeller}. The off-diagonal components enter the
gap equation for the color-superconducting gap parameter.

The system of Eqs.\ (\ref{DSEg}) and (\ref{DSEq}) has to be solved
self-consistently for the full propagators of quarks and gluons.
However, as was shown in Ref.\ \cite{dirkselfenergy}, 
in order to extract the color-superconducting gap parameter to
subleading order it is sufficient to consider the gluon propagator 
in HDL approximation;
corrections arising from the color-superconducting gap in the
quasiparticle spectrum are of sub-subleading order in the gap
equation. For the present purpose this means that it is not necessary to 
self-consistently solve Eq.\ (\ref{DSEg}) together with Eq.\
(\ref{DSEq}); one may approximate ${\cal G}$ on the right-hand
side of Eq.\ (\ref{DSEg}) by ${\cal G}_{0,11}$. 
In essence, this is equivalent to
considering only the first term
on the right-hand side of Eq.\ (\ref{DSEq}) when solving
Eq.\ (\ref{DSEg}). Of course, under this approximation
the effect of the regular quark self-energy (leading to wave-function
renormalization) and of the anomalous quark self-energy
(which accounts for the gap in the quasiparticle excitation spectrum)
are neglected.

With this approximation, and using ${\cal G}_0 \equiv {\cal G}_{0,11}
\oplus {\cal G}_{0,22}$, one may combine the terms in Eq.\ (\ref{DSEg})
to give
\be \label{DeltaHDL}
\Delta^{-1} \simeq \Delta_{0,11}^{-1} + \frac{g^2}{2}\,
{\rm Tr}_q\left( {\cal G}_0\, \tilde{\Gamma} \, 
{\cal G}_0\, \tilde{\Gamma} \right) \;.
\ee
Taking the gluon cut-off scale $\Lambda_{\rm gl}$ to fulfill 
$g \mu \ll \Lambda_{\rm gl} \lesssim \mu$,
soft gluons are defined to have momenta of order $g \mu$. 
I compute the fermion loop in Eq.\ (\ref{DeltaHDL}) under this
assumption (taking the soft gluon energy to be of the same
order of magnitude as the gluon momentum). I then realize
that the soft gluon propagator determined by Eq.\ (\ref{DeltaHDL}) 
is just the gluon propagator in HDL approximation. I indicate this
fact in the following by a subscript, $\Delta \equiv \Delta_{\rm HDL}$.
Armed with this (approximate) solution of the Dyson-Schwinger equation
(\ref{DSEg}) I now proceed to solve Eq.\ (\ref{DSEq}).
I consider the two independent components $[G^+]^{-1} + \Sigma^+$
and $\Phi^+$ in Nambu-Gor'kov space separately. Due to translational
invariance it is convenient to define
$[G^+]^{-1}(K,Q) \equiv (1/T) [G^+]^{-1}(K)\,
\delta^{(4)}_{K,Q}$, $\Sigma^+(K,Q) \equiv (1/T) \Sigma^+(K)\,
\delta^{(4)}_{K,Q}$
and using Eqs.\ (\ref{NGvertex}),
(\ref{G0FT}), (\ref{D_0}), 
(\ref{Gammatilde}), I obtain 
the Dyson-Schwinger equation for $[G^+]^{-1} + \Sigma^+$,  
\bea 
&&[G^+]^{-1}(K) + \Sigma^+(K)  =  \non
&&[ G_{0,11}^+]^{-1}(K) -  g^2 \, \frac{T}{V} \sum_Q \left\{ \frac{}{} 
\left[ \Delta_{0,22}\right]^{\mu \nu}_{ab}(K-Q) +  
\left[ \Delta_{\rm HDL}\right]^{\mu \nu}_{ab}(K-Q) \right\} 
\gamma_\mu T^a \, G_{0,22}(Q) \, \gamma_\nu T^b  \non 
&&-  g^2 \, \frac{T}{V} \sum_Q \left\{ \frac{}{}
\left[ \Delta_{0,22}\right]^{\mu \nu}_{ab}(K-Q) +  
\left[ \Delta_{\rm HDL}\right]^{\mu \nu}_{ab}(K-Q) \right\}
\gamma_\mu T^a \, {\cal G}^+(Q) \, \gamma_\nu T^b \;.
\label{DSEq2}
\eea
Note that the first sum over $Q$ runs over irrelevant quark momenta,
$0 \leq q < \mu - \Lambda_{\rm q}$ and $\mu+ \Lambda_{\rm q} < q < \infty$, while
the second sum runs over relevant quark momenta,
$\mu - \Lambda_{\rm q} \leq q \leq \mu + \Lambda_{\rm q}$.
There is no double counting of gluon exchange contributions, since
the hard gluon propagator $\Delta_{0,22}$ has support only for gluon momenta
$|{\bf k} - {\bf q}| > \Lambda_{\rm gl}$, while the HDL propagator 
is restricted to gluon momenta $|{\bf k} - {\bf q}| \leq \Lambda_{\rm gl}$.
To subleading order in the gap equation, I do not have to solve
this Dyson-Schwinger equation self-consistently. It is sufficient
to use the approximation ${\cal G}^+ \simeq G_{0,11}^+$ on the
right-hand side of Eq.\ (\ref{DSEq2}) and to keep only the last term
which, as discussed above, is responsible for non-Fermi liquid
behavior in cold, dense quark matter. The net result is then simply
a wave-function renormalization for the free quark propagator
$G_{0,11}^+$ \cite{manuel},
\be \label{fullinversequarkprop}
[G^+]^{-1}(K) + \Sigma^+(K) \simeq [ G_{0,11}^+]^{-1}(K)
+ \bar{g}^2  \, k_0 \, \gamma_0 \, \ln \frac{M^2}{k_0^2} 
\equiv \left[ Z^{-1}(k_0)\, k_0 + \mu \right] \,
\gamma_0 - \vg \cdot {\bf k}\; ,
\ee
where $\bar{g} \equiv g/ (3 \sqrt{2} \pi)$ and $M^2 = (3 \pi/4) m_g^2$,
with the gluon mass parameter $m_g$ defined in Eq.\ (\ref{gluonmass}). 
Neglecting effects from the finite life-time of
quasi-particles \cite{manuel2}, which are of sub-subleading
order in the gap equation, the wave-function renormalization factor is
\be \label{wavefunc}
Z(k_0) = \left( 1 + \bar{g}^2  \,\ln \frac{M^2}{k_0^2} \right)^{-1}\;.
\ee
Due to translational invariance, it is convenient to define
$\Phi^+(K,Q) \equiv (1/T) \, 
\Phi^+(K) \, \delta^{(4)}_{K,Q}$ and $\Xi^+(K,Q) \equiv T \, \Xi^+(K)\,
\delta^{(4)}_{K,Q}$, and the Dyson-Schwinger equation for $\Phi^+(K)$ becomes
\be
\Phi^+ (K)  =  g^2 \, \frac{T}{V} \sum_Q \left\{ \frac{}{}
\left[ \Delta_{0,22}\right]^{\mu \nu}_{ab}(K-Q) 
+  \left[ \Delta_{\rm HDL}\right]^{\mu \nu}_{ab}(K-Q)\frac{}{}
\right\} \, \gamma_\mu (T^a)^T \, \Xi^+(Q) \, \gamma_\nu T^b   \;.
\label{Phi}
\ee
Here, the sum runs only over relevant quark momenta, $\mu - \Lambda_{\rm q} 
\leq q \leq \mu + \Lambda_{\rm q}$. This is the gap equation for the 
color-superconducting gap parameter within my effective theory.
There is no contribution from irrelevant fermions, since their
propagator is diagonal in Nambu-Gor'kov space.

While the gluon cut-off was taken to be $\Lambda_{\rm gl} \lesssim \mu$, so
that soft gluons have typical momenta of order $g \mu$,
so far I have not specified the magnitude of $\Lambda_{\rm q}$.
In weak coupling, the color-superconducting gap function is strongly
peaked around the Fermi surface \cite{son,rdpdhr,schaferwilczek}. 
For a subleading-order calculation of the gap parameter, it is
therefore sufficient to consider as relevant quark modes those within
a thin layer of width $2 \Lambda_{\rm q}$ around the Fermi surface.
For the following, my principal assumption is
$\Lambda_{\rm q} \lesssim g \mu \ll \Lambda_{\rm gl} \lesssim \mu$.
As I shall see below, this assumption is crucial to 
identify sub-subleading corrections to the gap equation (\ref{Phi}), which
arise, for instance, from the pole of the gluon propagator.
Note that this assumption is different from that of
Refs.\ \cite{hong2,schaferefftheory}, where it is assumed that 
$\Lambda_{\rm q} \simeq \Lambda_{\rm gl}$.

For a two-flavor color superconductor, the color-flavor-spin
structure of the gap matrix is \cite{DHRreview}
\be \label{gapmatrix}
\Phi^+(K) = J_3 \tau_2 \gamma_5\, \Lambda_{\bf k}^+ \, 
\Theta(\Lambda_{\rm q} -|k-\mu|)\, \phi(K)\;, 
\ee
where $(J_3)_{ij} \equiv -i \epsilon_{ij3}$ and $(\tau_2)_{fg} \equiv
-i \epsilon_{fg}$ represent the fact
that quark pairs condense in the color-antitriplet, flavor-singlet channel.
The Dirac matrix $\gamma_5$ restricts quark pairing to the even-parity
channel (which is the preferred one due to the $U(1)_A$ anomaly of
QCD). In the effective action (\ref{Seff}), antiquark and irrelevant
quark degrees of freedom are integrated out. The condensation of
antiquark or irrelevant quark pairs, while in principle possible,  
is thus not taken into account;
the bilocal source terms in Eq.\ (\ref{Seffsources}) only allow for
the condensation of relevant quark degrees of freedom.
The condensation of antiquarks or irrelevant quarks
could also be accounted for, if one introduces bilocal
source terms already in Eq.\ (\ref{quarkaction}), 
i.e., {\em prior\/} to integrating out any of the quark degrees of freedom.
While there is in principle no obstacle in following this course of
action, it is, however, not really necessary if one is interested in 
a calculation of the color-superconducting gap parameter to subleading
order in weak coupling: antiquarks contribute to the gap
equation beyond subleading order \cite{aqgap}, and the gap function for quarks
falls off rapidly away from the Fermi surface, i.e., in the region
of irrelevant quark modes, and thus also contributes at most to
sub-subleading order to the gap equation.
Consequently, the Dirac structure of the gap matrix (\ref{Phi})
contains only the projector $\Lambda_{\bf k}^+$ onto positive energy 
states. The theta function accounts for the fact that
the gap function $\phi(K)$ pertains only to
relevant quark modes.

Inserting Eq.\ (\ref{fullinversequarkprop}) and the corresponding
one for $[G^-]^{-1} +\Sigma^-$, as well as Eq.\ (\ref{gapmatrix}),
into the definition (\ref{Xi}) for the
anomalous quark propagator, one obtains
\be
\Xi^+(Q) = J_3 \tau_2 \gamma_5 \, \Lambda_{\bf q}^- \,
\Theta(\Lambda_{\rm q} - |q-\mu|) \, \frac{\phi(Q)}{[q_0/Z(q_0)]^2 - \epsilon_q^2}\;.
\ee
One now plugs this expression into the gap equation (\ref{Phi}), 
multiplies both sides with $J_3 \tau_2 \gamma_5 \Lambda_{\bf k}^+$,
and traces over color, flavor, and Dirac degrees of freedom. These
traces simplify considerably since both hard 
and HDL gluon propagators are diagonal in adjoint color space,
$[\Delta_{0,22}]^{\mu \nu}_{ab} \equiv \delta_{ab}\, \Delta_{0,22}^{\mu
\nu}$, $[\Delta_{\rm HDL}]^{\mu \nu}_{ab} \equiv \delta_{ab}\, 
\Delta_{\rm HDL}^{\mu \nu}$. The
result is an integral equation for the gap function $\phi(K)$,
\be \label{gapequation}
\phi(K) = \frac{g^2}{3} \, \frac{T}{V} \sum_Q \left[ \frac{}{}
\Delta_{0,22}^{\mu \nu}(K-Q) 
+  \Delta_{\rm HDL}^{\mu \nu}(K-Q)\frac{}{}
\right] \, {\rm Tr}_s \left( \Lambda_{\bf k}^+ \gamma_\mu 
\Lambda_{\bf q}^- \gamma_\nu \right) \, 
\frac{\phi(Q)}{[q_0/Z(q_0)]^2 - \epsilon_q^2}\;.
\ee
The sum over $Q$ runs only over relevant quark momenta, 
$|q- \mu| \leq \Lambda_{\rm q}$. Also, the 3-momentum ${\bf k}$ is relevant,
$|k-\mu| \leq \Lambda_{\rm q}$.

\subsection{Solution of the gap equation to sub-leading order} \label{IVc}

In pure Coulomb gauge, both the hard gluon and the HDL propagators
have the form
\be \label{propgen}
\Delta^{00}(P) = \Delta^{\ell}(P) \;\;\;\; , \;\;\;\;\;
\Delta^{0i}(P) = 0 \;\;\;\;, \;\;\;\;\;
\Delta^{ij}(P) =  (\delta^{ij} - \hat{p}^i \hat{p}^j) \,
\Delta^{t}(P)\;,
\ee
where $\Delta^{\ell,t}$ are the propagators for
longitudinal and transverse gluon degrees of freedom.
For hard gluons 
\begin{subequations} \label{prop022}
\bea
\Delta^{\ell}_{0,22}(P) & = & -\frac{1}{p^2}\;,  \\
\Delta^{t}_{0,22}(P) & = &  - \frac{1}{P^2}\; ,
\eea
\end{subequations}
while for soft, HDL-resummed gluons
\begin{subequations} \label{propHDL}
\bea 
\Delta^{\ell}_{\rm HDL}(P) & = & - \frac{1}{p^2 - \Pi^{\ell}_{\rm HDL}(P)} 
\; , \\
\Delta^t_{\rm HDL}(P) & = & - \frac{1}{P^2 - \Pi^t_{\rm HDL}(P)} \;,
\eea
\end{subequations}
with the HDL self-energies \cite{LeBellac}
\begin{subequations} \label{HDLselfenergies}
\bea
\Pi^{\ell}_{\rm HDL} (p_0,p) & = & - 3 \, m_g^2 \left[ 1 - \frac{p_0}{2p}\,
\ln \left( \frac{p_0 + p}{p_0-p} \right) \right]\; , \\
\Pi^t_{\rm HDL} (p_0,p) & = & \frac{3}{2}\, m_g^2 \left[
\frac{p_0^2}{p^2}  + \left( 1 - \frac{p_0^2}{p^2} \right)
\, \frac{p_0}{2p} \,\ln \left( \frac{p_0 + p}{p_0-p} \right) \right]\;.
\eea
\end{subequations}
The HDL propagators (\ref{propHDL}) have quasiparticle poles at
$p_0 = \pm \omega_{\ell,t}(p)$, and a cut between
$p_0 = - p$ and $p_0 = p$ \cite{LeBellac}. 
The gluon energy on the quasiparticle mass-shell is always
larger than the gluon mass parameter, $\omega_{\ell,t}(p) \geq m_g$,
where the equality holds for zero momentum, $p=0$.

\begin{figure}[ht]
\centerline{\includegraphics[width=16cm]{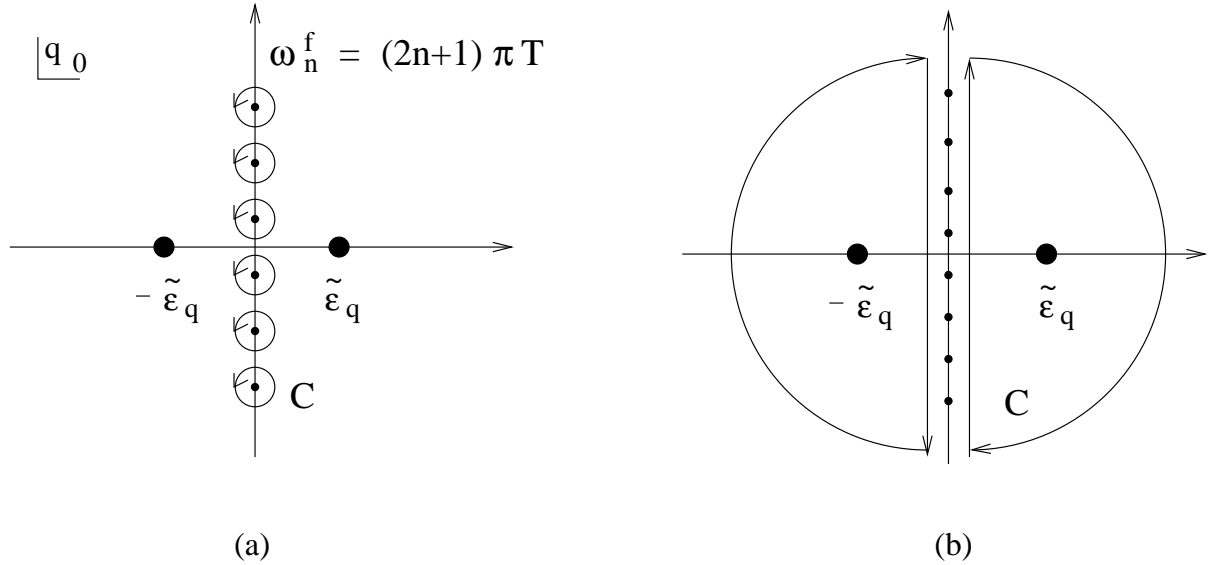}}
\caption[The contour ${\cal C}$ in Eq.\ (\ref{genMsum}).]{(a) The contour ${\cal C}$ in Eq.\ (\ref{genMsum})
encloses the poles of $\tanh [q_0/(2T)]$ on the
imaginary $q_0$ axis. (b) Deforming the contour ${\cal C}$ and
adding semicircles at infinity to enclose the poles
of the quark propagator on the real $q_0$ axis.}
\label{contour1}
\end{figure} 

I first perform the Matsubara sum, using the method of contour
integration in the complex $q_0$ plane \cite{LeBellac,FTFT}, 
\be \label{genMsum}
T \sum_n f(q_0) \equiv \frac{1}{2 \pi i } \oint_{\cal C} dq_0 \, \frac{1}{2} \,
\tanh\left( \frac{q_0}{2T} \right) \, f(q_0)\;
\ee
where the contour ${\cal C}$ consists of circles running around
the poles $\omega_n^{\rm f}= (2 n+1) \pi T $ of 
$\tanh [q_0/((2T)]$ on the imaginary $q_0$ axis, cf.\ Fig.\
\ref{contour1} (a).
Inserting the propagators
(\ref{prop022}) and (\ref{propHDL}) into Eq.\ (\ref{gapequation}), 
I have to compute four distinct
terms. The first one arises from the exchange of static electric hard gluons.
Since $\Delta^{\ell}_{0,22}(P)$ does not depend on $p_0=k_0 - q_0$, 
only the quark propagator gives rise to a pole of $f(q_0)$, cf.\ 
Fig.\ \ref{contour1} (b).
After deforming the contour and closing it at infinity
as shown in Fig.\ \ref{contour1} (b),
one employs the residue theorem to pick up the
poles of the quark propagator, 
\be \label{appEHGE}
T \sum_n \Delta_{0,22}^{\ell}(P) \, 
\frac{\phi(Q)}{[q_0/Z(q_0)]^2 - \epsilon_q^2} = 
\frac{1}{p^2}\, \tanh \left( \frac{\tilde{\epsilon}_q}{2T} \right) \,
\frac{Z^2(\tilde{\epsilon}_q)}{4 \, \tilde{\epsilon}_q} \,
\left[ \phi(\tilde{\epsilon}_q,{\bf q}) + \phi(-\tilde{\epsilon}_q,{\bf
q})\right]\;,
\ee
with $\tilde{\epsilon}_q \equiv \epsilon_q \, Z(\tilde{\epsilon}_q)$.
Here, I have used the fact that
the quark wave-function renormalization factor is an even function
of its argument, $Z(q_0) \equiv Z(-q_0)$, cf.\ Eq.\ (\ref{wavefunc}).
An essential assumption in order to derive Eq.\ (\ref{appEHGE}) 
is that the gap function $\phi(Q)$ is analytic in the complex $q_0$
plane. This assumption will also be made in all subsequent
considerations.

\begin{figure}[ht]
\centerline{\includegraphics[width=14cm]{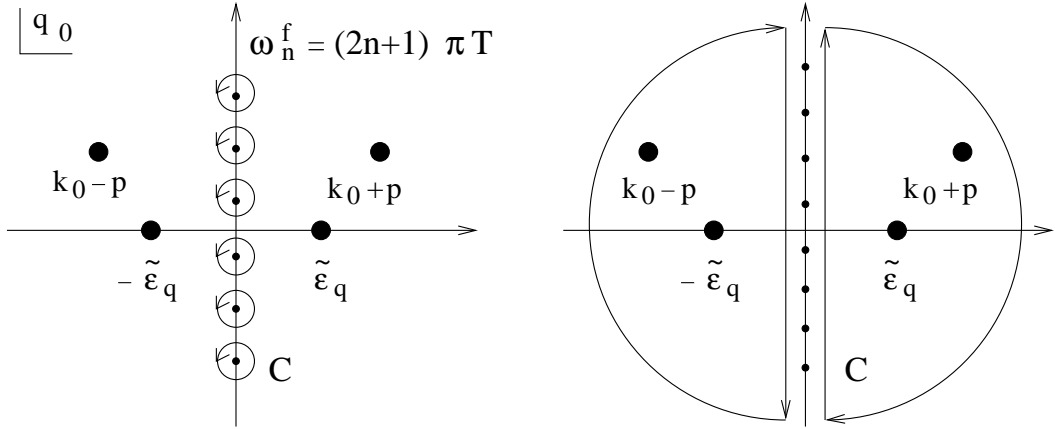}}
\caption[Same as in Fig.\ \ref{contour1}, but for magnetic
hard gluon exchange.]{Same as in Fig.\ \ref{contour1}, but for magnetic
hard gluon exchange. Now also the gluon propagator has 
poles at $k_0 \pm p$ in the complex $q_0$ plane. 
These are further away from the imaginary axis than the poles
$\tilde{\epsilon}_q$ of the quark propagator, because for
my choice of quark and gluon cut-offs, $\Lambda_{\rm q} \ll \Lambda_{\rm gl}$,
I have $\tilde{\epsilon}_q \alt \Lambda_{\rm q} \ll \Lambda_{\rm gl} \leq p$.}
\label{contour2}
\end{figure} 

By the same method one computes the second term in Eq.\
(\ref{gapequation}), corresponding to
magnetic hard gluon exchange. This is slightly more
complicated, since not only the quark propagator but
also $\Delta^{t}_{0,22}(P)$ has poles in the
complex $q_0$ plane. The latter are located at $p_0= \pm p$, 
i.e., $q_0 = k_0 \pm p$ , cf.\ Fig.\ \ref{contour2}. The external
quark energy $k_0$ is fixed and, prior to analytic continuation
$k_0 \rightarrow \tilde{\epsilon}_k + i \eta$
to the quasiparticle mass-shell, is equal to one particular
fermionic Matsubara frequency, cf.\ Fig.\ \ref{contour2}. The residue theorem
now yields four contributions, two from the quark and two from the
gluon poles. Using $\tanh [(k_0 \pm p)/(2T)] \equiv \pm \coth(p/2T)$
and analytically continuing $k_0 \rightarrow \tilde{\epsilon}_k + i
\eta$ I find
\bea
&&T \sum_n \Delta_{0,22}^{t}(P) \, 
\frac{\phi(Q)}{[q_0/Z(q_0)]^2 - \epsilon_q^2} = \non 
&&\hspace*{-1cm} \tanh \left( \frac{\tilde{\epsilon}_q}{2T} \right) \,
\frac{Z^2(\tilde{\epsilon}_q)}{4 \, \tilde{\epsilon}_q}\,
\left[\frac{\phi(\tilde{\epsilon}_q,{\bf q})}{(\tilde{\epsilon}_k
- \tilde{\epsilon}_q + i \eta)^2 - p^2}
+ \frac{\phi(-\tilde{\epsilon}_q,{\bf q})}{(\tilde{\epsilon}_k
+ \tilde{\epsilon}_q + i \eta)^2 - p^2}\right]  \non
&&\hspace*{-1cm} + \, \coth \left( \frac{p}{2T} \right) \, \frac{1}{4p} \,
\left[ \frac{Z^2(p+\tilde{\epsilon}_k) \,\phi(p+\tilde{\epsilon}_k,{\bf q})}{
(p+\tilde{\epsilon}_k  + i \eta)^2 - \epsilon_q^2
Z^2(p+\tilde{\epsilon}_k)}
+ \frac{Z^2(p- \tilde{\epsilon}_k) \, \phi(\tilde{\epsilon}_k-p,{\bf q})}{
(p-\tilde{\epsilon}_k - i \eta)^2 - \epsilon_q^2
Z^2(p-\tilde{\epsilon}_k)} \right] \;.
\eea
Since the gluon momentum is hard, $p \geq \Lambda_{\rm gl}$, and thus much
larger than the quasiparticle energies $\tilde{\epsilon}_k, \tilde{\epsilon}_q$
which are at most of the order of the quark cut-off $\Lambda_{\rm q} \ll
\Lambda_{\rm gl}$, to order $O(\Lambda_{\rm q}/\Lambda_{\rm gl})$ one may neglect the
terms $(\tilde{\epsilon}_k \pm \tilde{\epsilon}_q + i \eta)^2$ in the
energy denominators of the first term. Furthermore, in the second term
one may approximate $Z(p \pm \tilde{\epsilon}_k) \simeq Z(p) = 1 +
O(g^2)$ and $\phi(p \pm \tilde{\epsilon}_k, {\bf q})
\simeq \phi(p,{\bf q})$. Note that the gap function is far off-shell
for $p \geq \Lambda_{\rm gl} \gg \Lambda_{\rm q} \geq |\mu -q|$. Then, to order
$O(\Lambda_{\rm q}/\Lambda_{\rm gl})$, one may also neglect 
$\tilde{\epsilon}_k, \tilde{\epsilon}_q$ in the energy denominators
of the second term. I obtain
\bea
&&T \sum_n \Delta_{0,22}^{t}(P) \, 
\frac{\phi(Q)}{[q_0/Z(q_0)]^2 - \epsilon_q^2} =\non
&&- \frac{1}{p^2} \, \tanh \left( \frac{\tilde{\epsilon}_q}{2T} \right) \,
\frac{Z^2(\tilde{\epsilon}_q)}{4 \, \tilde{\epsilon}_q}\,
\left[ \phi(\tilde{\epsilon}_q,{\bf q}) + \phi(-\tilde{\epsilon}_q,{\bf
q})\right]\, \left[ 1 + O\left( \frac{\Lambda_{\rm q}^2}{\Lambda_{\rm gl}^2}\right) 
\right]
 \non&   &  
+ \coth \left( \frac{p}{2T} \right) \,
\frac{\phi(p,{\bf q})}{2 \, p^3}\,
\left[ 1 + O\left( \frac{\Lambda_{\rm q}^2}{\Lambda_{\rm gl}^2} \right)\right]\,
\;. \label{appMHGE}
\eea
As a next step one estimates to which order the two remaining terms contribute to
the gap equation (\ref{gapequation}).
At $T=0$, one may set the hyperbolic functions to one. 
I shall also ignore the difference
between the on-shell and off-shell gap functions, and take
$\phi(p,{\bf q}) \simeq \phi(\pm \tilde{\epsilon}_q,{\bf q}) \equiv \phi
 = const.$. For the purpose of power counting, one may restrict
oneself to the leading contribution of the Dirac traces in Eq.\
(\ref{gapequation}), which is of order one, cf.\ Eqs.\ (\ref{traces})
below. In order to obtain the leading contribution of the first term in
Eq.\ (\ref{appMHGE}), one may also set $Z^2(\tilde{\epsilon}_q) \simeq
1$. The integral over the absolute magnitude of the quark
momentum is $\int dq \, q^2$, while the angular integration is 
$\int d \cos \theta \equiv \int d p \, p / (kq)$.
Then, the first term in Eq.\ (\ref{appMHGE}) leads to the
following contribution in the gap equation
\be
 g^2\, \frac{\phi}{k} \int_{\mu-\Lambda_{\rm q}}^{\mu+ \Lambda_{\rm q}} 
dq \, \frac{q}{\epsilon_q} \int_{\Lambda_{\rm gl}}^{k+q} \frac{dp}{p}
\simeq g^2 \, \phi \, \ln \left( \frac{2\Lambda_{\rm q}}{\phi} 
\right) \, \ln \left( \frac{ 2\mu}{\Lambda_{\rm gl}} \right) 
\sim g^2 \, \phi \, \frac{1}{g} = g\, \phi\;,
\ee
where I approximated $k \simeq q \simeq \mu$ and
employed the weak-coupling solution (\ref{gapsol})
to estimate $\ln ( 2 \Lambda_{\rm q}/\phi ) \sim 1/g$. Furthermore, for
$\Lambda_{\rm gl} \alt \mu$, the angular logarithm is 
$\ln( 2\mu/\Lambda_{\rm gl}) \sim O(1)$. According to the discussion
presented in the introduction, the contribution
from hard magnetic gluon exchange is thus of subleading order 
in the gap equation. Note that the term arising from hard electric
gluon exchange, Eq.\ (\ref{appEHGE}), is of the same order
as the first term in Eq.\ (\ref{appMHGE}), and thus also contributes
to subleading order. The way I estimated the first term
on the right-hand side of Eq.\ (\ref{appMHGE}) is equivalent to just
taking the hard magnetic gluon propagator in the static limit,
$\Delta^t_{0,22}(P) \simeq  1/p^2$, which is correct
up to terms of order $O(\Lambda_{\rm q}^2/\Lambda_{\rm gl}^2)$. To this order,
the propagator for hard magnetic gluons is thus (up to a sign)
identical to the one
for hard electric gluons. Since the ratio $\Lambda_{\rm q}/\Lambda_{\rm gl}
\simeq g \mu/ \mu \equiv g$, this approximation introduces corrections
at order $O(g^3 \phi)$ in the gap equation, which is {\em beyond\/}
sub-subleading order, $O(g^2 \phi)$. 

Similarly, I estimate the contribution of the second term
in Eq.\ (\ref{appMHGE}) to the gap equation (\ref{gapequation}),
\be
 g^2\, \frac{\phi}{k} \int_{\mu-\Lambda_{\rm q}}^{\mu+ \Lambda_{\rm q}} 
dq \, q \int_{\Lambda_{\rm gl}}^{k+q} \frac{dp}{p^2}
\sim g^2 \, \phi \,  \frac{\Lambda_{\rm q}}{\Lambda_{\rm gl}} 
\sim g^3 \phi\;,
\ee
i.e., for my choice $\Lambda_{\rm q}/\Lambda_{\rm gl} \sim g$,
this term contributes beyond sub-subleading order.
Note that this estimate is conservative, as I assumed the
off-shell gap function to be of the same
order as the gap at the Fermi surface, $\phi(p,{\bf q}) \sim \phi$.
However, I know \cite{rdpdhr} that, for energies far from the
Fermi surface, $\tilde{\epsilon}_q \sim \Lambda_{\rm q} \alt g \mu$,
even the on-shell gap function
is suppressed by one power of $g$ compared to the value
of the gap at the Fermi surface, $\phi(\Lambda_{\rm q},{\bf q}) \sim g \phi$.
The off-shell gap function at $q_0 = p \agt \Lambda_{\rm gl} \gg \Lambda_{\rm q}$ 
may be even smaller. In order to decide this issue, one would have
to perform a computation of the gap function for arbitrary
values of the energy $q_0$, and not just on the
quasiparticle mass-shell, $q_0 \equiv \tilde{\epsilon}_q$.
I note that for the choice $\Lambda_{\rm q} \simeq \Lambda_{\rm gl}$ for the cut-offs 
\cite{hong2,schaferefftheory}, the ratio
$\Lambda_{\rm q}/\Lambda_{\rm gl}$ is of order one and cannot be used as a 
parameter to sort the various contributions according to
their order of magnitude. The expansion of the denominators
in powers of $\Lambda_{\rm q}/\Lambda_{\rm gl}$ as seen on the right-hand side
of Eq.\ (\ref{appMHGE}) is then inapplicable. 

\begin{figure}[ht]
\centerline{\includegraphics[width=14cm]{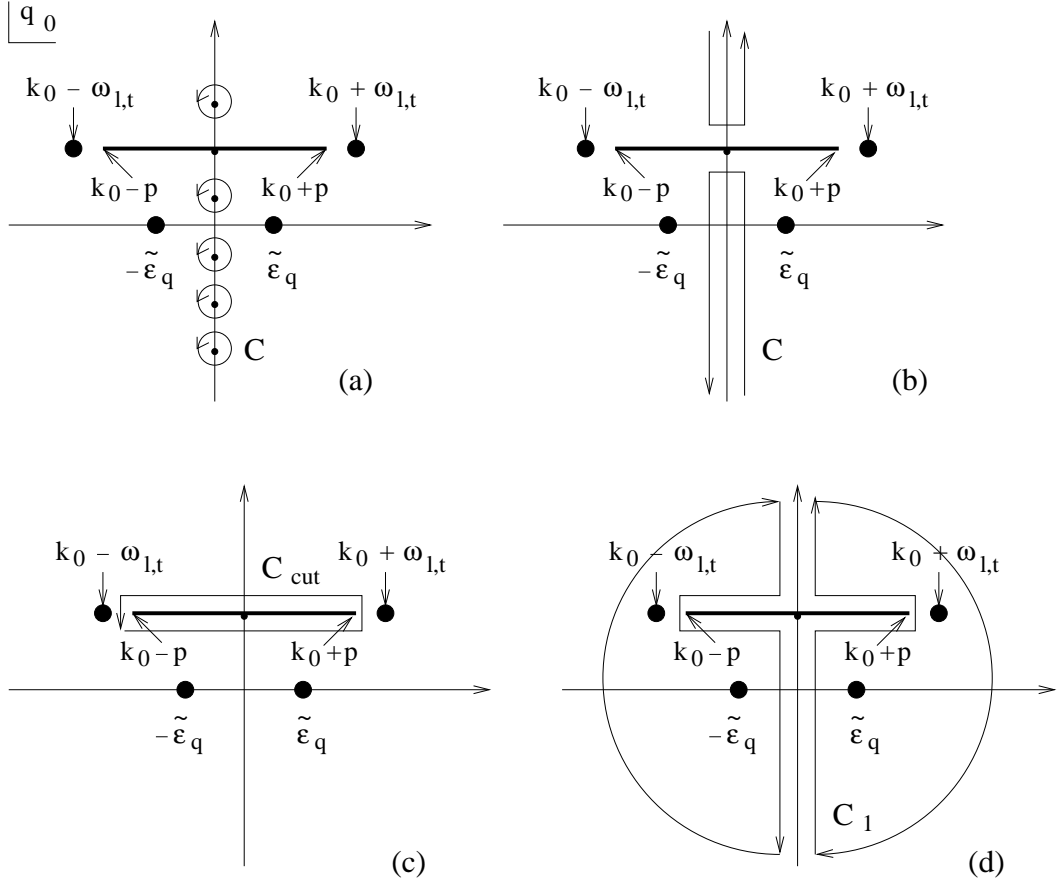}}
\caption[Evaluating the Matsubara sum for HDL-resummed gluon
propagators.]{Evaluating the Matsubara sum for HDL-resummed gluon
propagators. (a) The original contour ${\cal C}$ in Eq.\
(\ref{genMsum}). There is no circle around the point 
$k_0 = q_0$, where the corresponding term in the Matsubara sum 
has a cut arising from the HDL gluon propagator.
(b) Deforming the contour ${\cal C}$.
(c) The contour ${\cal C}_{\rm cut}$ running
around the cut. (d) The contour ${\cal C}_1 = {\cal C} + {\cal C}_{\rm
cut}$ which is closed at infinity.}
\label{contour3}
\end{figure} 

The third and fourth terms in the gap equation (\ref{gapequation})
arise from soft, HDL-resummed electric and magnetic gluon exchange.
Evaluating the Matsubara sum via contour integration in the complex
$q_0$ plane is considerably more difficult than in the
previous cases, because the HDL gluon propagators
$\Delta^{\ell,t}_{\rm HDL}$ do not only have poles but also cuts.
The analytic structure is shown in Fig.\ \ref{contour3} (a).
Besides the poles of the quark propagator at $q_0 = \pm \tilde{\epsilon}_q$,
there are also those from the gluon propagator at 
$q_0 = k_0 \pm \omega_{\ell,t}(p)$. The cut of the gluon propagator
between $- p \leq p_0 \leq p$ translates into a cut
between $k_0 -p \leq q_0 \leq k_0 + p$. Prior to analytic
continuation, the gluon poles and the cut are shifted away from
the real axis and located at the 
(imaginary) external Matsubara frequency $k_0$.

The Matsubara sum over $q_0$ is evaluated in the standard way, 
cf.\ Eq.\ (\ref{genMsum}), with
the caveat that the contribution at $q_0 = k_0$,
where the cut of the gluon propagator is located, has to be omitted. 
This is similar to the zero-temperature case where the Matsubara sum
becomes a continuous integral along the imaginary $q_0$ axis and
where one has to avoid integrating over the cut.
Alternatively, the term $q_0 = k_0$ can be included in the
Matsubara sum if one shifts the cut by some small amount
$\pm i \epsilon$ along the imaginary $q_0$ axis.
The final result will be the same, as one still has to circumvent
the cut by a proper choice of the integration contour.

I now deform the contour as shown in Fig.\ \ref{contour3} (b), and
add and subtract a contour integral running around the cut, Fig.\
\ref{contour3} (c). The integral over the contour ${\cal C} + {\cal
C}_{\rm cut}$ can be closed at infinity, yielding the contour
${\cal C}_1$ shown in Fig.\ \ref{contour3} (d). One obtains 
\bea \label{gapequation2}
&&T \sum_n \Delta_{\rm HDL}^{\ell, t}(P) \, 
\frac{\phi(Q)}{[q_0/Z(q_0)]^2 - \epsilon_q^2} = \non
&&\frac{1}{2 \pi i} \left[ \oint_{{\cal C}_1} - \oint_{{\cal C}_{\rm
cut}} \right] d q_0\, \frac{1}{2} \,
\tanh\left( \frac{q_0}{2T} \right) \, \Delta_{\rm HDL}^{\ell,t}(P)\,
\frac{\phi(Q)}{[q_0/Z(q_0)]^2 - \epsilon_q^2}\;.
\eea
Evaluating the integral over ${\cal C}_1$ is rather similar
to the case of hard gluon exchange: one just picks up the
poles of the quark and gluon propagators inside the contour ${\cal
C}_1$. 
After analytic continuation $k_0 \rightarrow \tilde{\epsilon}_k + i \eta$
one obtains
\bea
\lefteqn{\frac{1}{2 \pi i} \oint_{{\cal C}_1} d q_0\, \frac{1}{2} \,
\tanh\left( \frac{q_0}{2T} \right) \, \Delta_{\rm HDL}^{\ell,t}(P)\,
\frac{\phi(Q)}{[q_0/Z(q_0)]^2 - \epsilon_q^2} \simeq } \non
&&\hspace*{-0.8cm} -  \tanh \left( \frac{\tilde{\epsilon}_q}{2T} \right) \,
\frac{Z^2(\tilde{\epsilon}_q)}{4 \, \tilde{\epsilon}_q}\,
\left[ \Delta_{\rm HDL}^{\ell,t}(\tilde{\epsilon}_k - \tilde{\epsilon}_q
+ i \eta, {\bf p})\, \phi(\tilde{\epsilon}_q,{\bf q}) 
+ \Delta_{\rm HDL}^{\ell,t}(\tilde{\epsilon}_k + \tilde{\epsilon}_q
+ i \eta, {\bf p})\, \phi(-\tilde{\epsilon}_q,{\bf q})\right] \non
&&\hspace*{-0.8cm}  - \coth \left( \frac{\omega_{\ell,t}}{2T} \right) \,
\frac{1}{2  \, \omega_{\ell,t}^2}\,
\left[ \phi(\omega_{\ell,t} + \tilde{\epsilon}_k,{\bf q})\,
Z_{\ell,t}(- \omega_{\ell,t},p)
- \phi(\tilde{\epsilon}_k-\omega_{\ell,t} ,{\bf q})\,
Z_{\ell,t}( \omega_{\ell,t},p) \right]\,\non
&&\times\left[ 1 + O\left( \frac{\epsilon_q^2}{\omega_{\ell,t}^2} 
\right)\right]
\;. \label{C_1}
\eea
Here, I approximated the quark wave-function renormalization
factor $Z(\omega_{\ell,t}\pm \tilde{\epsilon}_k) \simeq 1 + O(g^2)$.
I also expanded the denominators of the quark propagator 
$(\tilde{\epsilon}_k \pm \omega_{\ell,t} + i \eta)^2 - \epsilon_q^2
\simeq \omega_{\ell,t}^2 \, [ 1 +
O(\epsilon_q^2/\omega_{\ell,t}^2)]$.
For my  choice of the cut-off $\Lambda_{\rm q} \alt g \mu \sim m_g$, I may estimate
$\omega_{\ell,t} \geq m_g \agt \Lambda_{\rm q} \geq \epsilon_q$, i.e.,
the corrections of order 
$O(\epsilon_q^2/\omega_{\ell,t}^2)$ are small everywhere
except for a small region of phase space where $p \simeq 0$ and
$\epsilon_q \simeq \Lambda_{\rm q}$. 
(In principle, in the expansion of the denominators 
there are also linear terms, $\sim \pm \tilde{\epsilon}_k/
\omega_{\ell,t}$, but these are very small everywhere for
external momenta close to the Fermi surface, $k \simeq \mu$.)
Note that the gap function
is again off-shell at the gluon pole, although not as far as
in the case of hard gluon exchange, cf.\ Eq.\ (\ref{appMHGE}).
The residues of the HDL gluon propagators at the respective poles
are \cite{LeBellac}
\begin{subequations}\label{gluonresidues}
\bea
Z_{\ell}(\omega_\ell,p) & =&  \frac{\omega_\ell (\omega_\ell^2 - p^2)}{
p^2 (p^2 + 3 m_g^2 - \omega_\ell^2)} \; , \\
Z_t (\omega_t,p) & = & \frac{\omega_t (\omega_t^2 - p^2)}{
3 m_g^2 \omega_t^2 - (\omega_t^2 - p^2)^2}\;.
\eea
\end{subequations}
To very good approximation, one finds that
$Z_t(\omega_t,p) \simeq  1/(2 \omega_t)$ for all momenta $p$. 
In the longitudinal case, the residue is very well approximated
by $Z_l(\omega_l,p) \simeq  \omega_l/(2\, p^2)$ for small momenta
$p \alt m_g$, while for large momenta, $m_g \ll
p$, $Z_l(\omega_l,p) \sim  \exp [-2 p^2/(3 m_g^2)]/p$, 
i.e., it is exponentially suppressed \cite{RDPphysicaA}.

These approximate forms allow for a simple power counting of the
gluon-pole contribution in Eq.\ (\ref{C_1}) to the gap equation 
(\ref{gapequation}). To this end, I approximate the gap function
by its value at the Fermi surface, 
$\phi(\pm \omega_{\ell,t} + \tilde{\epsilon}_k, {\bf
q}) \simeq \phi$, and consider the limiting case $T=0$ where
$\coth[\omega_{\ell,t}/(2T)] = 1$.
Then, the contribution from the longitudinal gluon pole is 
\be \label{appESGE}
 g^2\, \frac{\phi}{k} \int_{\mu-\Lambda_{\rm q}}^{\mu+ \Lambda_{\rm q}} 
dq \, q \left[ \int_{|k-q|}^{m_g} 
\frac{dp}{2 \, p\, \omega_\ell} + \int_{m_g}^{\Lambda_{\rm gl}}
\frac{dp}{ \omega_\ell^2}\, 
\exp \left(-\frac{2 p^2}{3 m_g^2}\right) \right]
\sim g^2 \, \phi \, \frac{\Lambda_{\rm q}}{m_g} 
\sim g^2 \phi \;.
\ee
In the first $p$ integral,
which only runs up to the scale $m_g$,
one may approximate $\omega_\ell \simeq m_g$, while
in the second $p$ integral, which runs from $m_g$ to
$\Lambda_{\rm gl} \alt \mu$, one may take $\omega_\ell \simeq p$.
To obtain the right-hand side of Eq.\ (\ref{appESGE}) I have 
set $k \simeq q \simeq \mu$, and I have employed
my  choice $\Lambda_{\rm q} \alt g \mu$ for the quark cut-off. This also
allowed me to approximate logarithms of $\Lambda_{\rm q}/m_g$ by numbers
of order $O(1)$. With this choice for the quark cut-off,
the contribution (\ref{appESGE}) is of
sub-subleading order, $\sim O(g^2 \phi)$, to the gap
equation. 

With a more careful evaluation of 
the integrals, one could extract the precise numerical prefactor
of the sub-subleading contribution (\ref{appESGE}). 
Note, however, that further suppression factors
may arise from the off-shellness of the gap function at 
$\phi(\pm \omega_{\ell,t} + \tilde{\epsilon}_k, {\bf q})$,
which consequently would render this
contribution beyond sub-subleading order. As noted previously,
this issue can only be decided if $\phi(q_0,{\bf q})$ is known
also off the quasiparticle mass-shell, and not only on-shell.
I also note that the $1/p^2$
factor in the residue $Z_\ell$ is an artifact of the Coulomb gauge
\cite{RDPphysicaA}, and does not appear in e.g.\ covariant gauge.
One would have to collect all other terms
of sub-subleading order to make sure that the complete 
sub-subleading contribution is gauge invariant and the term
(\ref{appESGE}) is not cancelled by some other terms.

Similarly, I estimate the contribution from the transverse gluon pole,
\be \label{appMSGE}
 g^2\, \frac{\phi}{k} \int_{\mu-\Lambda_{\rm q}}^{\mu+ \Lambda_{\rm q}} 
dq \, q \int_{|k-q|}^{\Lambda_{\rm gl}} \frac{dp\, p}{2 \, \omega_t^3}
\sim g^2 \, \phi \int_0^{\Lambda_{\rm q}} d \xi\,
\int_{m_g}^{\Lambda_{\rm gl}} \frac{d \omega_t}{\omega_t^2}
\sim g^2 \, \phi \, \frac{\Lambda_{\rm q}}{m_g} \sim g^2 \, \phi\;,
\ee
where I defined $\xi \equiv q - \mu$. I
approximated $dp \, p \simeq d \omega_t \, \omega_t$ since,
for the purpose of power counting, to very good approximation one
may take the dispersion relation of the transverse gluon 
equal to that of a relativistic particle with mass
$m_g$, $\omega_t(p) \simeq (p^2 + m_g^2)^{1/2}$. 
I also used  $\Lambda_{\rm q} \alt m_g \ll \Lambda_{\rm gl} $ and $k \simeq q \simeq \mu$.
In conclusion, also the
transverse gluon pole possibly contributes to sub-subleading order in the
gap equation, with the same caveats concerning the off-shellness of
the gap function as mentioned previously. 

Let us now focus on the integral around the cut of the gluon
propagator in Eq.\ (\ref{gapequation2}). 
I substitute $q_0$ by $p_0 = k_0 - q_0 \equiv \omega$ and
use the fact that $\tanh[q_0/(2T)] \equiv - \coth [\omega/(2T)]$.
Since the gluon propagator is the only part of the integrand 
which is discontinuous across the cut, I obtain after analytic continuation 
$k_0 \rightarrow \tilde{\epsilon}_k + i \eta$
\bea
\lefteqn{- \frac{1}{2 \pi i} \oint_{{\cal C}_{\rm cut}} d q_0\, \frac{1}{2} 
\tanh\left( \frac{q_0}{2T} \right)  \Delta_{\rm HDL}^{\ell,t}(P)\,
\frac{\phi(Q)}{[q_0/Z(q_0)]^2 - \epsilon_q^2}} \non
 & = & 
\int_{-p}^p d\omega\, \frac{1}{2} \coth \left( \frac{\omega}{2T}
\right) \, \frac{Z^2(\tilde{\epsilon}_k - \omega)\, 
\phi(\tilde{\epsilon}_k - \omega,{\bf q})}{(\tilde{\epsilon}_k -
\omega + i \eta)^2 - [Z(\tilde{\epsilon}_k - \omega)\, \epsilon_q]^2} \, 
\rho^{\ell,t}_{\rm cut}(\omega,{\bf p})\;,
\label{C_cut}
\eea
where $\rho^{\ell,t}_{\rm cut}(\omega,p) \equiv {\rm Im} \Delta^{\ell,t}_{\rm
HDL} (\omega+i \eta,p)/ \pi$ is the spectral density of the
HDL propagator arising from the cut. Explicitly,
\begin{subequations}\label{rhocuts}
\bea
\rho^{\ell}_{\rm cut}(\omega,{\bf p}) & = & 
\frac{2 M^2}{\pi}\, \frac{\omega}{p}\,
\left\{ \left[ p^2 + 3\, m_g^2 \left(
1 - \frac{\omega}{2p}\, \ln
\left| \frac{ p+ \omega}{p-\omega} \right| \right) \right]^2
+ \left( 2 M^2\, \frac{\omega}{p} \right)^2
\right\}^{-1} \;,\\
\rho^t_{\rm cut} (\omega,{\bf p}) & = & 
\frac{M^2}{\pi}\, \frac{\omega}{p}\,
\frac{p^2}{p^2-\omega^2} \, 
\left\{ \left[ p^2 + \frac{3}{2}\, m_g^2 \left(
\frac{\omega^2}{p^2-\omega^2} + \frac{\omega}{2p}\, \ln
\left| \frac{ p+ \omega}{p-\omega} \right| \right) \right]^2
+ \left( M^2\, \frac{\omega}{p} \right)^2
\right\}^{-1}
\!\!. \non\label{rhotcut}
\eea
\end{subequations}
In order to power count the contribution from the cut of
$\Delta^{\ell}_{\rm HDL}$ to the gap equation, it is sufficient
to approximate the spectral density by \cite{rdpdhr}
\be
\rho^{\ell}_{\rm cut}(\omega,{\bf p}) \simeq
\frac{2 M^2}{\pi}\, \frac{\omega}{p}\,\frac{1}{
( p^2 + 3\, m_g^2 )^2} \;.
\ee
This form reproduces the correct behavior for $\omega \ll p$.
For $\omega \alt p$, it overestimates the spectral density when
$p\alt m_g$, while it slightly underestimates 
it for $p \agt m_g$. For the gap equation, however,
this region is unimportant, since the respective
contribution is suppressed by the
large energy denominator $(\tilde{\epsilon}_k -
\omega + i \eta)^2 - [Z(\tilde{\epsilon}_k - \omega)\, \epsilon_q]^2
\simeq p^2$ in Eq.\ (\ref{C_cut}). To leading order,
one may set $Z(\tilde{\epsilon}_k - \omega) \simeq 1$. I also
approximate $\phi(\tilde{\epsilon}_k - \omega,{\bf
q}) \simeq \phi$. 
Then, the $\omega$ integral can be performed analytically. (One may compute
this integral with the principal value prescription; the
contribution from the complex pole contributes to the imaginary part
of the gap function, which I neglect throughout this computation to subleading order, cf.\ Sec.\ \ref{imsec}.)
This produces at most logarithmic singularities, which
are integrable. I therefore simply approximate
the $\omega$ integral by a number of order $O(1)$.
Consequently, the contribution from Eq.\ (\ref{C_cut}) to
the gap equation is of order
\bea \label{appESGEcut}
g^2 \, \frac{\phi}{k} \int_{\mu-\Lambda_{\rm q}}^{\mu+ \Lambda_{\rm q}} 
dq \, q \int_{|k-q|}^{\Lambda_{\rm gl}} dp\, \frac{m_g^2}{(p^2 + 3\, m_g^2)^2}
&\sim&
 g^2 \, \phi \int_0^{\Lambda_{\rm q}} d \xi \left( \int_\xi^{m_g}
\frac{dp}{m_g^2} +  m_g^2 \int_{m_g}^{\Lambda_{\rm gl}}
\frac{dp}{p^4} \right) \non
&\sim& g^2 \, \phi\, \frac{\Lambda_{\rm q}}{m_g}
\sim g^2 \, \phi\;,
\eea
where I approximated the $p$ integral by a method similar 
to the one employed in Eq.\ (\ref{appESGE}). For my choice
$\Lambda_{\rm q} \alt g \mu$, Eq.\ (\ref{appESGEcut}) constitutes another 
(potential) contribution of sub-subleading order to the gap equation.

Finally, I estimate the contribution from the cut of the transverse
gluon propagator. For all momenta $p$ and energies $-p \leq \omega
\leq p$, a very good approximation for the spectral density
(\ref{rhotcut}) is given by the formula
\be
\rho_{\rm cut}^t (\omega, {\bf p}) \simeq \frac{M^2}{\pi} \,
\frac{\omega \, p}{p^6 + (M^2\, \omega)^2}\;.
\ee
This approximate result constitutes an upper bound for
the full result (\ref{rhotcut}). The advantage of using this
approximate form is that, interchanging the order of the
$p$ and $\omega$ integration in the gap equation, 
the former may immediately be performed.
Approximating $Z(\tilde{\epsilon}_k - \omega) \simeq 1$,
neglecting the dependence of the gap function on the
direction of ${\bf q}$, and defining $\lambda \equiv {\rm max}(
|k-q|, \omega)$, at $T=0$ the contribution to the gap equation is
\bea \label{appMSGEcut}
%\lefteqn{
&&g^2 \int_{\mu-\Lambda_{\rm q}}^{\mu + \Lambda_{\rm q}}
d q \, \frac{q}{k} \int_0^{\Lambda_{\rm gl}} d \omega
\, 
\sum\limits_{\sigma=\pm}
\frac{\phi(\tilde{\epsilon}_k- \sigma\omega,q)}{
(\tilde{\epsilon}_k -\sigma\omega)^2 - \epsilon_q^2} 
\left[ {\rm arctan} \left( \frac{\Lambda_{\rm gl}^3}{\omega M^2} \right)
- {\rm arctan} \left(\frac{\lambda^3}{\omega M^2} \right) \right]
%} 
\non
&& \sim g^2 \int_0^{\Lambda_{\rm q}} \frac{d \xi }{\epsilon_q}
\int_0^M d \omega 
\sum\limits_{\sigma_1\!,\sigma_2=\pm}
\frac{\sigma_1\, \phi(\tilde{\epsilon}_k- \sigma_2\,\omega,q)}
{\tilde \epsilon_k -\sigma_2\omega-\sigma_1\epsilon_q}
%\left[ \phi(\tilde{\epsilon}_k- \omega,q)
%\left( \frac{1}{\tilde{\epsilon}_k - \omega -  \epsilon_q} -
%\frac{1}{\tilde{\epsilon}_k - \omega +  \epsilon_q} \right)\right.\non
%&&\left.+ \phi(\tilde{\epsilon}_k +\omega,q)
%\left( \frac{1}{\tilde{\epsilon}_k + \omega -  \epsilon_q} -
%\frac{1}{\tilde{\epsilon}_k + \omega +  \epsilon_q} \right)\right]\;.
\eea
Here, I have used the fact that the particular combination
of arctan's in the first line effectively cuts off the $\omega$
integral at the scale $\omega \sim M$. As usual, I have 
set $k \simeq q \simeq \mu$.
After estimating $\phi(\tilde{\epsilon}_k \pm \omega,q) \simeq
\phi$ one first performs the integral over $\omega$
\bea
&&g^2 \phi \int_0^{\Lambda_{\rm q}} \frac{d \xi }{\epsilon_q}
\int_0^M d \omega \!\sum\limits_{\sigma_1\!,\sigma_2=\pm}
\frac{\sigma_1}{\tilde \epsilon_k -\sigma_2\omega-\sigma_1\epsilon_q} \non
&&=
-g^2 \phi \int_0^{\Lambda_{\rm q}} \frac{d \xi }{\epsilon_q} 
\sum\limits_{\sigma_1\!,\sigma_2=\pm}\!\! \sigma_1\sigma_2
\ln\left|\frac{\tilde\epsilon_k-\sigma_1\epsilon_q-\sigma_2 M}
{\tilde\epsilon_k-\sigma_1\epsilon_q}\right|\;.\label{appMSGEcut2}
\eea
The denominator under the logarithm in Eq.\ (\ref{appMSGEcut2}) does not contribute 
because of the sum over $\sigma_2$. In the region $\xi > \phi$ one may neglect $\tilde\epsilon_k$ in the numerator under the logarithm and obtains
\bea
2 g^2 \phi \int_\phi^{\Lambda_{\rm q}} \frac{d \xi }{\epsilon_q} 
\ln\left|\frac{\epsilon_q+M}
{\epsilon_q-M}\right| \sim g^2 \phi \;,
\eea
where the estimate on the r.h.s\ results after expanding the logarithm 
in powers of $\epsilon_q/M$. In the region $\xi <  \phi$ one has
\bea\label{region2}
g^2 \phi \int_0^{\phi} \frac{d \xi }{\epsilon_q} 
\sum\limits_{\sigma_1\!,\sigma_2=\pm}\!\! \sigma_1\sigma_2 
\ln\left|1-\sigma_2\frac{\tilde\epsilon_k-\sigma_1\epsilon_q}{M}\right|
\sim g^2 \phi \,\frac{\phi}{M}\;.
\eea
Consequently, the contribution (\ref{appMSGEcut}) is at most of sub-subleading order.
%If we simply neglect the off-shell behavior of the gap function
%and approximate $\phi(\tilde{\epsilon}_k \pm \omega,q) \simeq
%\phi$, this contribution would (at least) be of subleading order.
%Note that the corresponding contribution in previous treatments of the
%QCD gap equation, cf.\ for instance Eq.\ (67) of Ref.\ \cite{rdpdhr},
%was discarded as being of higher order. 
%At this point, we refrain from a more careful evaluation
%of the contribution (\ref{appMSGEcut}), because this 
%requires a calculation of the gap function off the quasiparticle
%mass-shell. Since the purpose of the present work is to
%show that our method reproduces previous results, we follow
%Ref.\ \cite{rdpdhr} and also
%discard the contribution (\ref{appMSGEcut}) in the following.

The remaining term from the evaluation of the Matsubara sum
in Eq.\ (\ref{gapequation2}) is the contribution from
the quark pole, i.e., the first line of Eq.\ (\ref{C_1}).
This has to be combined with the subleading-order terms 
from hard-gluon exchange, i.e., from Eq.\ (\ref{appEHGE}) and
from the first line of Eq.\ (\ref{appMHGE}), in order to
obtain the gap equation which contains all contributions of
leading and subleading order. Before doing so, however, I
also evaluate the Dirac traces in Eq.\ (\ref{gapequation}).
In pure Coulomb gauge, I only require
\begin{subequations} \label{traces}
\bea
{\rm Tr}_s \left( \Lambda_{\bf k}^+ \gamma_0 
\Lambda_{\bf q}^- \gamma_0 \right) & = & \frac{(k+q)^2 - p^2}{2\, k\,q} \;
, \\
(\delta^{ij} - \hat{p}^i \hat{p}^j)\, 
{\rm Tr}_s \left( \Lambda_{\bf k}^+ \gamma_i \Lambda_{\bf q}^-
\gamma_j \right) & = & -2 - \frac{p^2}{2\, k\,q} + 
\frac{(k^2-q^2)^2}{2\, k\, q\, p^2}\;,
\eea
\end{subequations}
where I used $p^2 \equiv ({\bf k}-{\bf q})^2 = k^2 + q^2 - 2\, k\, q \,
\hat{\bf k} \cdot \hat{\bf q}$ to eliminate $\hat{\bf k} \cdot
\hat{\bf q}$ in favor of $p^2$. Let us estimate the order of magnitude
of the terms arising from the traces at the Fermi surface, $k \equiv
\mu$. Setting $q \equiv \mu + \xi$, where 
$-\Lambda_{\rm q} \leq \xi \leq \Lambda_{\rm q}$, one obtains
\begin{subequations} \label{traces2}
\bea
{\rm Tr}_s \left( \Lambda_{\bf k}^+ \gamma_0 
\Lambda_{\bf q}^- \gamma_0 \right) & = & 
2 - \frac{p^2}{2\, k \, q} + O\left( \frac{\xi^2}{\mu^2} \right) \;, 
\label{traces2a} \\
(\delta^{ij} - \hat{p}^i \hat{p}^j)\, 
{\rm Tr}_s \left( \Lambda_{\bf k}^+ \gamma_i \Lambda_{\bf q}^-
\gamma_j \right) & = & -2 - \frac{p^2}{2\, k \, q} 
+ O\left( \frac{\xi^2}{\mu^2} \right)  \;. \label{traces2b}
\eea
\end{subequations}
As shown above, the contribution from hard-gluon exchange is at most
of subleading order. Thus, for this contribution
it is sufficient to keep only the leading terms in Eq.\
(\ref{traces2}), i.e., one may safely neglect terms
of order $O(\xi^2/\mu^2) \alt O(\Lambda_{\rm q}^2/\Lambda_{\rm gl}^2) \sim O(g^2)$ 
or higher.
Note that, since for hard gluon exchange $p \sim \mu \agt \Lambda_{\rm gl}$,
the terms $p^2/(2 k q)$ cannot be omitted. However, since
the magnetic gluon propagator is effectively $\sim 1/p^2$, cf.\
Eq.\ (\ref{appMHGE}), i.e., (up to a sign) identical to the electric
propagator, these terms will ultimately cancel
between the electric and the magnetic contribution.
This cancellation is well known, see for instance Ref.\ \cite{asqwdhr}, and 
is special to the spin-zero case. It does not occur in spin-one
color superconductors where there is an additional exponential
prefactor which suppresses the magnitude of the spin-one gap relative
to the spin-zero case \cite{asqwdhr}.

As is well known, electric soft-gluon exchange also
contributes to subleading order in the gap equation.
Thus, as in the case of hard-gluon exchange, one may drop the terms
of order $O(\xi^2/\mu^2)$ in Eq.\ (\ref{traces2a}).
On the other hand, magnetic soft-gluon exchange constitutes the
leading order contribution to the gap equation. I therefore would have to
keep all terms up to {\em sub\/}leading order, i.e., $\sim
O(\xi/\mu)$. Fortunately, the corrections to the result
(\ref{traces2b}) are of order $O(\xi^2/\mu^2) \sim O(g^2)$, i.e., they
are of {\em sub-sub\/}leading order and thus can also be omitted.

I combine Eqs.\ (\ref{appEHGE}), (\ref{appMHGE}), and the first
line of Eq.\ (\ref{C_1}), and assume that the gap function is
even in its energy argument, $\phi(-\tilde{\epsilon}_q, {\bf q})
= \phi(\tilde{\epsilon}_q, {\bf q})$, and isotropic,
$\phi(\tilde{\epsilon}_q, {\bf q}) \equiv \phi(\tilde{\epsilon}_q,q)
\equiv \phi_q$.
Then, on the quasiparticle mass-shell
$k_0 = \tilde{\epsilon}_k$ the gap equation (\ref{gapequation})
becomes
\bea
\phi_k & = & \frac{g^2}{24 \pi^2} \, 
\int_{\mu - \Lambda_{\rm q}}^{\mu + \Lambda_{\rm q}}  dq\, 
\frac{q}{k}\,\frac{Z^2(\tilde{\epsilon}_q)}{\tilde{\epsilon}_q}\,
\tanh \left( \frac{\tilde{\epsilon}_q}{2T} \right) \, \phi_q
\int_{|k-q|}^{k+q} dp\, p \, 
\left\{ \Theta(p - \Lambda_{\rm gl}) \, \frac{4}{p^2} + 
\Theta(\Lambda_{\rm gl} - p)\right. \non
& &\hspace*{-1cm} \times \left. \!\!\sum_{s = \pm} \left[
 \Delta_{\rm HDL}^{\ell}(\tilde{\epsilon}_k - s \tilde{\epsilon}_q
+ i \eta, p) \, \left( - 1 +  \frac{p^2}{4\,k\,q} \right) 
+  \Delta_{\rm HDL}^{t}(\tilde{\epsilon}_k - s \tilde{\epsilon}_q
+ i \eta, p) \, \left(  1 +  \frac{p^2}{4\,k\,q} \right) \right]
\right\}\;.\non
\label{gapequation3}
\eea
The next step is to divide the integration region in the $p-q$ plane
into two parts, separated by the gluon ``light cone'' 
$|\tilde{\epsilon}_k - s \tilde{\epsilon}_q| = p$. For my  choice
$\Lambda_{\rm q} \ll \Lambda_{\rm gl}$ the region,
where $|\tilde{\epsilon}_k - s \tilde{\epsilon}_q| < p$, is very large,
while its complement is rather small.
In order to estimate the contribution from the latter 
to the gap equation,  one may approximate the HDL gluon propagators by their
limiting forms for large gluon energies, cf.\ Eqs.\ (\ref{propHDL}), 
(\ref{HDLselfenergies}),
\be
p_0 \gg p \; : \;\;\;\;\;\;\;
\Delta^\ell_{\rm HDL} (P) \simeq \frac{p_0^2}{m_g^2\,p^2} \;\;\;\; ,
\;\;\;\;\;
\Delta^t_{\rm HDL} (P) \simeq \frac{1}{m_g^2} \;.
\ee
Following the power-counting scheme employed previously, 
the contribution from the electric sector is of order
\be
g^2 \, \frac{\phi}{k} \int_{\mu-\Lambda_{\rm q}}^{\mu+\Lambda_{\rm q}}
dq\, \frac{q}{\epsilon_q} \, 
\frac{(\tilde{\epsilon}_k - s \tilde{\epsilon}_q)^2}{m_g^2}
\int_{|k-q|}^{|\tilde{\epsilon}_k - s \tilde{\epsilon}_q|}
\frac{dp}{p}
\sim g^2 \, \frac{\phi}{m_g^2} \int_0^{\Lambda_{\rm q}} d \xi \, \epsilon_q
\sim g^2 \, \phi\, \frac{\Lambda_{\rm q}^2}{m_g^2} \sim g^2 \, \phi\;.
\ee
This is a contribution of sub-subleading order, as long as one
adheres to the choice $\Lambda_{\rm q} \alt g \mu$.
Analogously, I estimate the contribution from the magnetic
sector to be
\be
g^2 \, \frac{\phi}{k} \int_{\mu-\Lambda_{\rm q}}^{\mu+\Lambda_{\rm q}}
dq\, \frac{q}{\epsilon_q} \, 
\int_{|k-q|}^{|\tilde{\epsilon}_k - s \tilde{\epsilon}_q|}
dp \,p \, \frac{1}{m_g^2}
\sim g^2 \, \frac{\phi}{m_g^2} \int_0^{\Lambda_{\rm q}} \frac{d
\xi}{\epsilon_q} \, \xi^2
\sim g^2 \, \phi\, \frac{\Lambda_{\rm q}^2}{m_g^2} \sim g^2 \, \phi\;.
\ee
Consequently, all contributions from the region
$|\tilde{\epsilon}_k - s \tilde{\epsilon}_q| \geq p$ are of
sub-subleading order, and the further analysis can be restricted
to the region $|\tilde{\epsilon}_k - s \tilde{\epsilon}_q| < p$.
In this region, it is permissible to use the low-energy limit 
of the HDL gluon propagator, which follows from Eqs.\ (\ref{propHDL}),
(\ref{HDLselfenergies})
keeping only the leading terms in the gluon energy,
\be \label{lel}
p_0 \ll p \; : \;\;\;\;\;\;\;
\Delta_{\rm HDL}^{\ell} (P) 
 \simeq  - \frac{1}{p^2 + 3\, m_g^2} \;\;\;\;  , \;\;\;\;\;
\Delta_{\rm HDL}^t (P)  \simeq  
\frac{p^4}{p^6 + M^4 \,p_0^2} \; .
\ee
Here, I only retained the real part of the transverse gluon propagator,
since the imaginary part contributes to
the imaginary part of the gap function, which is usually ignored.
(In Ref.\ \cite{rdpdhr} it was argued that, at least close
to the Fermi surface, the contribution of the
imaginary part is of sub-subleading order in the gap equation.)
With the approximation (\ref{lel}), the gap equation 
(\ref{gapequation3}) becomes
\bea
\phi_k & = & \frac{g^2}{24 \pi^2} \, 
\int_{\mu - \Lambda_{\rm q}}^{\mu + \Lambda_{\rm q}}  dq\, 
\frac{q}{k}\,\frac{Z^2(\tilde{\epsilon}_q)}{\tilde{\epsilon}_q}\,
\tanh \left( \frac{\tilde{\epsilon}_q}{2T} \right) \, \phi_q
\left\{  4 \, \ln \left( \frac{k+q}{\Lambda_{\rm gl}}\right) \right. \non
&   & + \left.
\sum_{s=\pm}
\int_{|\tilde{\epsilon}_k - s \tilde{\epsilon}_q|}^{\Lambda_{\rm gl}} 
dp\,\left[ \frac{p}{p^2 + 3\, m_g^2} \, 
\left(  1 -  \frac{p^2}{4\,k\,q} \right) 
+ \frac{p^5}{p^6 + M^4 (\tilde{\epsilon}_k - s \tilde{\epsilon}_q)^2}
 \, \left(  1 +  \frac{p^2}{4\,k\,q} \right) \right] \right\},\non
\label{gapequation4}
\eea
where I already performed the integration over hard gluon momenta 
$p\geq \Lambda_{\rm gl}$. The integration over soft gluon momenta can 
also be performed analytically. Formally, the terms $\sim p^2/(4kq)$ 
give rise to subleading-order contributions, $\sim \Lambda_{\rm gl}^2/(8kq)$,
but they ultimately cancel, since they come with different signs in
the electric and the magnetic part. Other contributions from these
terms are at most of sub-subleading order. 
Exploiting the symmetry of the integrand around
the Fermi surface and setting $k \simeq \mu$, I arrive at
\bea
&&\phi_k = \non
&&
\hspace*{-0.8cm}
\frac{g^2}{12 \pi^2} \int_{0}^{\Lambda_{\rm q}}
d(q-\mu) \, \frac{Z^2(\tilde{\epsilon}_q)}{\tilde{\epsilon}_q} \,
\tanh \left( \frac{\tilde{\epsilon}_q}{2T} \right) \, \phi_q\,
\left[ 2\, \ln \left( \frac{4 \mu^2}{\Lambda_{\rm gl}^2} \right)
+ \ln \left( \frac{\Lambda_{\rm gl}^2}{3\, m_g^2} \right)
+ \frac{1}{3} \, \ln \left( 
\frac{\Lambda_{\rm gl}^{6}}{M^4 |\tilde{\epsilon}_k^2 -
\tilde{\epsilon}_q^2|} \right) \right].\non
\eea
Here, I have neglected terms $\sim \tilde{\epsilon}_k -
s\tilde{\epsilon}_q$ against $3\, m_g^2$ under the logarithm
arising from soft electric gluons, and terms
$\sim (\tilde{\epsilon}_k - s\tilde{\epsilon}_q)^6$ against 
$M^4 (\tilde{\epsilon}_k - s\tilde{\epsilon}_q)^2$ under the
logarithm from soft magnetic gluons.

Now observe that the gluon cut-off $\Lambda_{\rm gl}$ cancels in the final
result,
\be
\phi_k = \frac{g^2}{18 \pi^2} \int_{0}^{\Lambda_{\rm q}}
d(q-\mu) \, \frac{Z^2(\tilde{\epsilon}_q)}{\tilde{\epsilon}_q} \,
\tanh \left( \frac{\tilde{\epsilon}_q}{2T} \right) \, \phi_q\,
\frac{1}{2}\, \ln  \left( \frac{\tilde{b}^2 \mu^2}{|\tilde{\epsilon}_k^2 -
\tilde{\epsilon}_q^2|} \right) \;,
\label{gapequation5}
\ee
where $\tilde{b} \equiv 256 \pi^4 [2/(N_f g^2)]^{5/2}$.
This is Eq.\ (19) of Ref.\ \cite{qwdhr}, since
$\bar{g}^2 \equiv g^2/(18 \pi^2)$, with the upper limit
of the $(q-\mu)$ integration, $\delta$, replaced by the
quark cut-off $\Lambda_{\rm q}$.

The solution of the gap equation (\ref{gapequation5}) is well known,
and given by Eq.\ (\ref{gapsol}).
As was shown in Ref.\ \cite{rdpdhr}, the dependence on the cut-off
$\Lambda_{\rm q}$ enters only at sub-subleading order, i.e., it
constitutes an $O(g)$ correction to the prefactor in 
Eq.\ (\ref{gapsol}). Therefore, to subleading order I do not need
a matching calculation to eliminate $\Lambda_{\rm q}$.

The result (\ref{gapequation5}) shows that the standard gap equation 
of QCD can be obtained from the effective action (\ref{Seff}). The
above, rather elaborate derivation of Eq.\ (\ref{gapequation5}) 
demonstrates that, in order
to obtain this result, it is mandatory to choose $\Lambda_{\rm q} \ll 
\Lambda_{\rm gl}$. This also enabled me to identify potential 
sub-subleading order contributions. However, I argued that,
at this order, the off-shell behavior of the gap function 
has to be taken into account. This requires a complex ansatz for the gap function, 
which will be considered in the following Sec.\ \ref{imsec}.

		\section{The imaginary part of the gap function} \label{imsec}

The scalar gap function $\phi(K)$ contains the energy and momentum dependence of the (2SC) gap matrix 
$
\Phi^+(K) = J_3 \tau_2 \gamma_5\, \Lambda_{\bf k}^+ \, 
\Theta(\Lambda_{\rm q} -|k-\mu|)\, \phi(K)$, which solves the gap equation Eq.\ (\ref{gapequation})
\be\label{gapequation1}
\Phi^+ (K)  =  g^2 \, \frac{T}{V} \sum_Q  \Delta^{\mu \nu}_{ab}(K-Q)\, \gamma_\mu (T^a)^T \, \Xi^+(Q) \, \gamma_\nu T^b   \;.
\ee
Due to the dependence of the gluon propagator $\Delta(K-Q)$ on the external energy $k_0= -\omega_n=-i(2n+1)\pi T$ the solution $\phi(K)$ must be energy-dependent itself. Hence, solving the gap equation {\it self-consistently} requires an {\it energy dependent} ansatz for the gap function.

\begin{figure}[ht]
\centerline{\includegraphics[width=8cm]{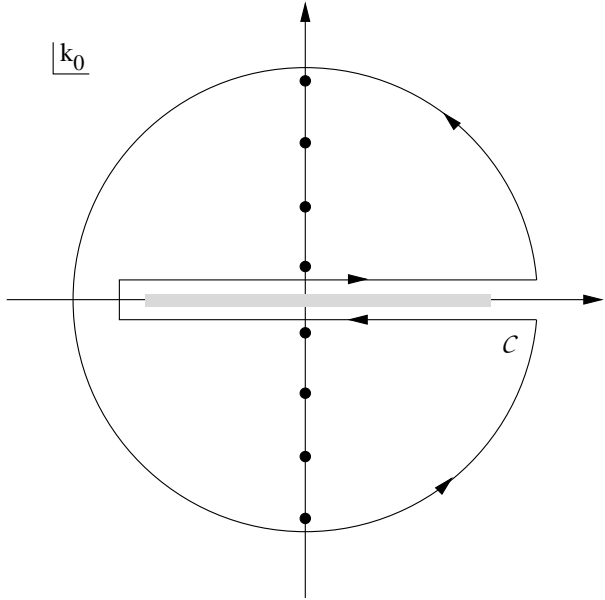}}
\caption[A contour circumventing all non-analyticities of $\phi(K)$.]{The contour ${\mathcal C}$ circumvents all non-analyticities of $\phi(K)$ on the real $k_0-$axis in the shaded area. The dots indicate the fermionic Matsubara frequencies, $k_0=-i(2n+1)\pi T$.}
\label{phiC}
\end{figure}

The function  $\phi(K)$ solving Eq.\ (\ref{gapequation1}) for some Matsubara frequency $k_0$ can be considered as the {\it analytic continuation} of the gap function $\phi(\omega,\vk)$ of physical and therefore real frequencies $\omega$. Then the physical gap function $\phi(\omega,\vk)$ has to have singularities along the real $\omega$-axis so that the analytic continuation $\phi(K)$ is, as required, {\it not} a constant in complex energy $k_0$. To see this write without loss of generality
\bea
\phi(K)\equiv \tilde\phi(K) +\phi_0(\vk)\;,
\eea
where the energy dependence of $\phi(K)$ is contained in $ \tilde\phi(K)$. It is physical to assume, and it will be confirmed when solving the gap equation, that $\phi(K)$ becomes local, i.e.\, independent of energy, for asymptotically large energies, i.e.\ that $\tilde\phi(K) \rightarrow 0 $ and $\phi(K) \rightarrow \phi_0(\vk)$ for $k_0\rightarrow \infty$. Cauchy's theorem gives for any $k_0$ off the real axis
\bea
\phi(k_0,\vk) &=& \oint\limits_{\mathcal C}\frac{dz}{2\pi i}\, \frac{\phi(z,\vk)}{z-k_0}\non
&=& \int\limits_{-\infty}^\infty \frac{d\omega^\prime}{2\pi i}\,\frac{\phi(\omega^\prime+i\eta,\vk)-\phi(\omega^\prime-i\eta,\vk)}{\omega^\prime-k_0}
 +\oint\limits_{\mathcal C^\prime}\frac{dz}{2\pi i}\, \frac{\phi(z,\vk)}{z}\;,
\eea
where the contour ${\mathcal C}$ circumvents all non-analyticities of $\phi(K)$ on the real $k_0-$axis  as depicted in Fig.\ \ref{phiC} and ${\mathcal C^\prime}$  is a circle whose radius tends to infinity.
I define the spectral density of $\phi(K)$ by
\bea \label{specphi3}
\rho_\phi(\omega,\vk) &\equiv &  \frac{1}{2\pi i}\,\left[\phi(\omega+i\eta,\vk)-\phi(\omega-i\eta,\vk)\right]\;,
%\equiv \frac{i}{2\pi}\,{\rm Disc}\; \phi(\omega+i\eta,\vk)
\eea
which contains all non-analyticities of $\phi(\omega,\vk)$.
Using $\phi(z,\vk)|_{z\in \mathcal C^\prime} \equiv \phi_0(\vk)$ one obtains
\bea\label{specphi}
\phi(k_0,\vk)
&=& \int\limits_{-\infty}^\infty d\omega\,\frac{\rho_\phi(\omega,\vk)}{\omega-k_0}+\phi_0(\vk)\;,
\eea
where the first term is identified as $\tilde\phi(K)$.
Eq.\ (\ref{specphi}) holds for all $k_0$ surrounded by contour ${\mathcal C}$, in particular for frequencies infinitesimally close to the real axis, $k_0 = \omega +i\epsilon$ with $\epsilon >\eta >0$. For such frequencies one can employ the Dirac identity $1/(x-i\epsilon) \equiv \mathcal P(1/x) +i\pi \delta(x)$ to obtain
\bea\label{phi3}
&&\phi(\omega +i\epsilon,\vk)=\non
&=& \int\limits_{-\infty}^\infty d\omega^\prime\,\frac{\rho_\phi(\omega^\prime,\vk)}{\omega^\prime-\omega -i\epsilon}+\phi_0(\vk)=\mathcal P\int\limits _{-\infty}^\infty d\omega^\prime \,\frac{\rho_\phi(\omega^\prime,\vk)}{\omega^\prime -\omega}+\phi_0(\vk) + i\pi\rho_\phi(\omega,\vk)\;.
\eea
Hence, for real $\phi_0(\vk)$ and $\rho_\phi(\omega,\vk)$
\bea\label{realphi2}
{\rm Re}\,\phi(\omega+i\epsilon,\vk)&=& \mathcal P\int\limits _{-\infty}^\infty d\omega^\prime \,\frac{\rho_\phi(\omega^\prime,\vk)}{\omega -\omega^\prime}+\phi_0(\vk)\;,\\
{\rm Im}\,\phi(\omega+i\epsilon,\vk)&=& {\pi}\,\rho_\phi(\omega,\vk)\;.\label{imphi2}
\eea
Combining Eq.\ (\ref{specphi3}) with Eq.\ (\ref{imphi2}) yields the identity
\bea\label{imphispec}
{\rm Im}\,\phi(\omega+i\epsilon,\vk) = \frac{1}{2i}\,\left[\phi(\omega+i\epsilon,\vk)-\phi(\omega-i\epsilon,\vk)\right]\;,
\eea
i.e.\, non-analyticities along the real $k_0$-axis are equivalent to a nonzero imaginary part of the gap function. Furthermore, one finds the dispersion relations for $\tilde\phi(\omega+i\eta,\vk)$
\begin{subequations}
\bea\label{disp1}
{\rm Re}\,\tilde\phi(\omega+i\epsilon,\vk)&=& \frac{1}{\pi}\,\mathcal P\int\limits _{-\infty}^\infty d\omega^\prime \,\frac{{\rm Im}\,\tilde\phi(\omega^\prime+i\epsilon,\vk)}{\omega^\prime -\omega}\;,\\
{\rm Im}\,\tilde\phi(\omega+i\epsilon,\vk)&=& -\frac{1}{\pi} \,\mathcal P\int\limits _{-\infty}^\infty d\omega^\prime \,\frac{{\rm Re}\,\tilde\phi(\omega^\prime+i\epsilon,\vk)}{\omega^\prime -\omega}\;,\label{disp2}
\eea
\end{subequations}
i.e.\ $\rm{Im}\,\tilde\phi(\omega+i\epsilon,\vk)$ and $\rm{Re}\,\tilde\phi(\omega+i\epsilon,\vk)$ are Hilbert transforms of each other, 
${\rm Re}\,\tilde\phi= {\cal H}[{\rm Im}\,\tilde\phi]$.
It follows that $\tilde \phi \neq 0$ only if both Re$\, \tilde\phi$ and Im$\,\tilde\phi$ are nonzero. Consequently, the gap function $\phi$ is energy-dependent only if it has a nonzero imaginary part, Im$\,\phi \equiv {\rm Im}\,\tilde\phi\neq 0$, which in turn is generated by its non-analyticities along the real $k_0$ axis, cf.\ Eq. (\ref{imphispec}).

This investigation accounts for the energy dependence of the gap function by including the non-analyticities of $\phi$ in the solution of the gap equation. This finally leads to a {\it complex gap equation}, whose real and imaginary parts are coupled integral equations
\begin{subequations}\label{coupled}
\bea\label{coupledRe}
{\rm Re}\,\phi={\rm Re}\,\phi\,[\phi,\Delta,\Sigma]\;,\\
{\rm Im}\,\phi={\rm Im}\,\phi\,[\phi,\Delta,\Sigma]\;,\label{coupledIm}
\eea
\end{subequations}
which have to be solved self-consistently. Of course, there are analogous DSEs for $\Delta$ and $\Sigma$, which in principle would have to be solved self-consistently simultaneously with Eqs.\ (\ref{coupled}). However, it has been shown \cite{dirkselfenergy} that to subleading order one may replace $\Delta= \Delta_{\rm HDL}$ in Eq.\ (\ref{coupledRe}). (This can be refined by taking  $\Delta= \Delta_{\rm HDL}$ for soft gluon momenta and the free propagator $\Delta= \Delta_{0,22}$ for hard momenta \cite{Reuter}.) Also $\Sigma$ may be replaced by its one-loop approximation in Eq.\ (\ref{coupledRe}) to subleading order \cite{manuel,qwdhr}. %Consequently, in the following we can restrict to Eqs.\ (\ref{coupled}) and may abandon the DSEs of $\Delta$ and $\Sigma$.
Furthermore, the contribution of Im$\,\Sigma$ to ${\rm Re}\,\phi$ in Eq.\ (\ref{coupledRe})  has been shown to be of sub-subleading order \cite{manuel} and therefore will be neglected in the following. However, the contribution of ${\rm Im}\,\phi$  to ${\rm Re}\,\phi$ in Eq.\ (\ref{coupledRe}) has not been systematically accounted for, yet.
In the following it will be checked if Im$\,\phi$ contributes leading or subleading order corrections to the real part of the gap function or not, i.e. whether the common simplification of Eq.\ (\ref{coupledRe})
\bea\label{assumption1}
{\rm Re}\,\phi={\rm Re}\,\phi\,[{\rm Re}\,\phi,\Delta,\Sigma]
\eea
is justified to subleading order or not. To this end it is admissible to make the {\it same} approximations in Eq.\ (\ref{coupledIm}) as in Eq.\ (\ref{coupledRe}) yielding
\begin{subequations}\label{coupled2}
\bea\label{coupled2Re}
{\rm Re}\,\phi={\rm Re}\,\phi\,[\phi,\Delta_{\rm HDL},\Delta_{0,22},{\rm Re}\,\Sigma_{\rm 1\,loop}]\;,\\
{\rm Im}\,\phi={\rm Im}\,\phi\,[\phi,\Delta_{\rm HDL},\Delta_{0,22},{\rm Re}\,\Sigma_{\rm 1\,loop}]\;.\label{coupled2Im}
\eea
\end{subequations}
This can be justified when considering Im$\,\phi \equiv {\rm Im}\,\tilde\phi$ as the Hilbert transform of Re$\,\tilde\phi$ in its subleading order approximation. As such it must be insensitive to any sub-subleading order corrections to Re$\,\tilde\phi$. In other words, one assumes that contributions, which do not enter Re$\,\phi$ directly via Eq.\ (\ref{coupledRe}) do not enter Re$\,\phi$ indirectly via  Eq.\ (\ref{coupledIm}). As stated in the beginning the main contributions to Im$\,\phi$ are expected to arise from the energy dependence of the gluon propagator $\Delta(K-Q)$.

The gap in the quasiparticle excitation spectrum is given by the modulus of the complex gap function
\bea\label{gapexcite}
|\phi|= \sqrt{({\rm Re}\,\phi)^2+({\rm Im}\,\phi)^2}\;.
\eea
Hence, there are two possibilities how the physical quantity $|\phi|$ is affected by ${\rm Im}\,\phi$ to subleading order. The first is that  ${\rm Im}\,\phi$  enters ${\rm Re}\,\phi$ to subleading order through Eq.\ (\ref{coupled2Re}) as already discussed. The second is that  ${\rm Im}\,\phi$ appearing on the r.h.s.\ of Eq.\ (\ref{gapexcite}) is of order ${\rm Re}\,\phi$. In \cite{rdpdhr} it is estimated that Im$\,\phi = 0$ on the Fermi surface and Im$\,\phi\sim g\,{\rm Re}\,\phi$ exponentially close to the Fermi surface. This will be confirmed in the following. It follows that in {\it these} momentum regimes Im$\,\phi$ contributes at most at order $g^2{\rm Re}\,\phi$ to $|\phi|$, i.e.\ at sub-subleading order. Hence, the second of the two possibilities mentioned above can be discarded in this momentum regime. However, at momenta $|k-\mu|\sim g\,\mu$ it is Im$\,\phi\sim {\rm Re}\,\phi$ and both the real and the imaginary part contribute at the same order of magnitude to $|\phi|$. Also this will be confirmed in the following. The modulus appears in the complex gap equation only in the on-shell energy dispersion relation
\bea
{\epsilon_q} = \sqrt{(k-\mu)^2 + |\phi|^2}\,.
\eea
In the region $\zeta=|k-\mu|\sim g\,\mu$, where Im$\phi\sim$ Re$\phi$, it is $\zeta\gg \phi$ and therefore 
${\epsilon_q}\simeq \zeta$. Hence, in the gap equation it is self-consistent to subleading-order to approximate $|\phi|\simeq{\rm Re}\,\phi$ and write
\bea
{\epsilon_q} \simeq \sqrt{(k-\mu)^2 + ({\rm Re}\,\phi_q)^2}\;.\label{modulusisreal}
\eea
Consequently, in contrast to $|\phi|$ the on-shell energy ${\epsilon_q}$ can be affected by Im$\,\phi_q$ only through Re$\,\phi_q$ for all $\zeta$.

Before actually solving the complex gap equation it is worthwhile investigate how the symmetry properties of the gap matrix $\Phi^+(K)$ affect the gap function $\phi(K)$. Remembering $\psi_C(K) = C \bar{\psi}^T(-K)$, $\bar{\psi}_C(K) = \psi^T(-K)C$, and $C=-C^{-1} = -C^T$ it follows from the antisymmetry of the quark fields that
\bea
\sum\limits_K \bar \psi_C (K) \Phi^+(K)\psi(K) &\equiv& 
-\left(\sum\limits_K \bar \psi_C (K) \Phi^+(K)\psi(K) \right)^T \non
&=&-\sum\limits_K  \psi^T (K) \left[\Phi^+(K)\right]^T\bar\psi^T_C(K)\non
&=& \sum\limits_K \bar \psi_C(K)\, C^{-1}\left[  \Phi^+(-K)\right]^T C\,\psi(K)\;.
\eea
Hence, the gap matrix must fulfill
\bea
C\,\Phi^+(K)\,C^{-1} =\left[  \Phi^+(-K)\right]^T\;.
\eea
Since $C\gamma_5\Lambda^+_{\vk}C^{-1}= [\gamma_5\Lambda_{-\vk}^+]^T$ and
in the 2SC case $[J_3\tau_2]^T=J_3\tau_2$ it follows for the gap function
\bea 
\phi(K) &=& \phi(-K)\;.
\eea
Assuming that the gap function is symmetric under reflection of 3-momentum $\vk$, $\phi(k_0,\vk)=\phi(k_0,-\vk)$ one obtains with Eqs.\ (\ref{realphi2},\ref{imphi2})
\begin{subequations}
\bea
{\rm Re}\,\phi(\omega+i\eta,\vk)&=&{\rm Re}\,\phi(-\omega+i\eta,\vk)\;,\\
{\rm Im}\,\phi(\omega+i\eta,\vk)&=&-{\rm Im}\,\phi(-\omega+i\eta,\vk)\;,\label{oddIm}\\
\rho_{\phi}(\omega,\vk)&=&-\rho_{\phi}(-\omega,\vk)\;.\label{odd}
\eea
\end{subequations}
Hence, one found that ${\rm Re}\,\phi$ is an even function in $\omega$ while ${\rm Im}\,\phi$ and $\rho_\phi$ are odd. 

Furthermore, one introduces the spectral representations for
the gluon propagators. In the case of soft, HDL-resummed electric and magnetic
 gluon propagators, cf.\ Eqs.\ (\ref{propHDL}), one has \cite{LeBellac,RDPphysicaA}
\bea\label{specdens}
\!\!\!\!\!\!\!\!\!\!\!\!\!\!\! \rho_{\ell,t}(\omega,{\bf p}) & = & \rho^{\rm pole}_{\ell,t} 
(\omega,{\bf p})\, \left\{\delta \left[\omega - \omega_{\ell,t}({\bf p})\right] -\delta \left[\omega - \omega_{\ell,t}({\bf p})\right]\right\}
+\rho^{\rm cut}_{\ell,t}(\omega,{\bf p}) \, \theta(p-|\omega|)\;,
\eea
where $\rho^{\rm cut}_{\ell,t}$ are the contributions from the cut of the logarithm in the HDL self-energies $\Pi^{\ell,t}_{\rm HDL}$, cf.\ (\ref{HDLselfenergies}), and already given in Eqs.\ (\ref{rhocuts}). The contributions from the gluon poles at $\omega =\pm\omega_{\ell,t}$ are given by the respective residues of the propagator, $\rho^{\rm pole}_{\ell,t}(\omega,{\bf p})\equiv Z_{\ell,t}(\omega,{\bf p})$,  cf.\ Eqs.\ (\ref{gluonresidues}).
For the respective hard gluon propagators, cf.\ Eqs.\ (\ref{prop022}), one has $\Delta_{0,22}^\ell(\omega\pm i\eta,\vp) = -1/p^2$ and $\Delta_{0,22}^t(\omega\pm i\eta,\vp)= -1/[(\omega\pm i\eta)^2-p^2]$ and therefore with $\rho_{0,22}^{\ell,t}(\omega,\vp)\equiv[\Delta_{0,22}^{\ell,t}(\omega+i\eta,\vp)-\Delta_{0,22}^{\ell,t}(\omega-i\eta,\vp)]/(2\pi i)$
\begin{subequations}
\bea \label{hardspecl}
\rho_{0,22}^\ell(\omega,\vp) &=& 0\;,\\
\rho_{0,22}^t(\omega,\vp) &=& {\rm sign}(\omega)\delta(\omega^2-p^2)\;.\label{hardspect}
\eea
\end{subequations}
With these spectral densitis,  $\rho_{\rm HDL}^{\ell,t}$ and $\rho_{0,22}^{\ell,t}$, one may express the respective gluon propagators in their spectral representations
\bea\label{longtransprops4}
\Delta^\ell(P) = - \frac{1}{p^2} + \int\limits_{-\infty}^{\infty} d \omega \, 
\frac{\rho^\ell(\omega,\vp)}{\omega-p_0}
 \,\,\,\,\,\,  ,  \,\,\,\,\,\,
\Delta^t(P) =  \int\limits_{-\infty}^{\infty} d \omega \, 
\frac{\rho^t(\omega,\vp)}{\omega-p_0}\;.
\eea

\subsection{Solving the complex gap equation}\label{solving}

In the following solution of the complex gap equation (\ref{coupled2}) it will be shown that its imaginary part, Eq.\  (\ref{coupled2Im}), can be decomposed as
\bea \label{AB}
{\rm Im}\,\phi ={\cal A}+ {\cal B}\;,
%{\cal A}[{\rm Re}\,\phi]+ {\cal B}[{\rm Im}\,\phi]\;,
\eea
%where the dependency of $\cal A$ and $\cal B$ on $\Delta_{\rm HDL},\Delta_{0,22}$ and ${\rm Re}\,\Sigma_{\rm 1\,loop}$ has been suppressed for simplicity. The term ${\cal A}[{\rm Re}\,\phi]$ will be shown to depend only on the %real part of 
where $\cal A$ contains only contributions, where the gap function is on the quasi-particle mass-shell, and $\cal B$ contains the rest.
%the gap function on the mass-shell $\phi(\tilde\epsilon_{\vq},\vq)$, 
Then all contributions in the imaginary part of the gap function, which are generated by the energy dependence, i.e.\ by the non-analyticities of $\phi$, are collected in ${\cal B}$. Therefore, all terms in $\cal B$ have not been considered so far and neglecting $\cal B$ must lead to the known subleading order result for Re$\phi(\tilde\epsilon_q,q)$. In the following it will checked, whether this result is self-consistent. Self-consistency is fulfilled only if the contributions contained in $\cal B$ enter  Re$\phi(\tilde\epsilon_q,q)$ beyond subeading order, i.e. at sub-subleading order. To this end it is first assumed that self-consistency is indeed fulfilled and then analyzed, if $\cal B$ really contributes beyond subleading order. This procedure comprises of the following steps.

In order to obtain $\cal A$ and $\cal B$, first the Matsubara sum over the internal frequency $q_0$ is perfomed, cf.\ Sec.\ \ref{Matsu}, and then all imaginary contributions are extracted and organised according to Eq.\ (\ref{AB}), cf.\  Sec.\ \ref{identim}.
%provides the coupling of ${\rm Re}\,\phi(\epsilon_\vq)$ and $\rho_\phi$.contains 
%where ${\cal B}={\cal B}[\rho_\phi] $ arises from $\rho_\phi$ itself. Its appearance constitutes a (direct) condition for the self-consistency for $\rho_\phi$ in Eq.\ (\ref{AB}). The second term, ${\cal A}={\cal A}[{\rm Re}\,\phi(\epsilon_\vq)]$,  
In the first part of Sec.\ \ref{estAB} the order of magnitude of Im$\,\phi$ will be determined by inserting the known leading order result for ${\rm Re}\,\phi({\tilde \epsilon_\vq},\vq)$ into ${\cal A}$, while the yet unknown term $ {\cal B}$ has to be neglected first. Of course, this estimate for Im$\,\phi$ can be self-consistent only if reinserting it into ${\cal B}$ in Eq.\ (\ref{AB}) only yields sufficiently small corrections to  Im$\,\phi$, i.e. if
\bea\label{selfconsistency1}
{\rm Im}\,\phi \approx {\cal A}\;.
%{\cal B}[{\rm Im}\,\phi]\ll {\cal A}[{\rm Re}\,\phi] \;.
\eea
%This is equivalent to demanding that Im$\,\phi \approx {\cal A}[\rm Re\, \phi(\epsilon_\vq)]$.
This will be analyzed in the second part of Sec.\ \ref{estAB}, where  this estimate for Im$\,\phi$ is used to determine the order of magnitude of ${\cal B}$. It will be shown that Eq.\ (\ref{selfconsistency1}) is indeed fulfilled, i.e. that $ {\cal B}$ constitutes a small correction to ${\cal A}$.
%,and that the assumption $ \rho_\phi(\omega,\vk)\approx{\cal A} (\omega,\vk) $ is selfconsistent to subleading order. 
With these results  the order of magnitude of ${\rm Re}\,\tilde \phi({\epsilon_\vk},\vk)$ is estimated in Sec.\ \ref{hilbert} via the dispersion relation Eq.\ (\ref{disp1}), i.e. by Hilbert transforming $\cal A$ and $\cal B$.
%\bea\label{realphi}
%{\rm Re}\,\tilde\phi(\epsilon_\vk,\vk)&\equiv&-\frac{1}{\pi}\,\mathcal P\!\int\limits _{-\infty}^\infty d\omega \,\frac{{\rm Im}\,\tilde\phi(\omega,\vk)}{\epsilon_\vk -\omega}
%=-\frac{1}{\pi}\,\mathcal P\!\int\limits _{-\infty}^\infty d\omega \,\frac{{\cal A}(\omega,\vk)}{\epsilon_\vk -\omega}-\frac{1}{\pi}\,\mathcal P\!\int\limits _{-\infty}^\infty d\omega \,\frac{{\cal B}(\omega,\vk)}{\epsilon_\vk -\omega}\;.
%\eea 
It is found that ${\rm Re}\,\tilde\phi(\epsilon_\vk,\vk)\sim {\cal H}[{\cal A}]\sim \phi$ for momenta $\vk$ close to the Fermi surface, while ${\cal H}[{\cal B}]\sim g^2\phi$, i.e.\  Im$\,\phi$ contributes beyond subleading order to  ${\rm Re}\,\tilde \phi({\epsilon_\vk},\vk)$ or according to Eq.\ (\ref{assumption1})
\bea\label{confirm1}
{\rm Re}\,\tilde \phi={\rm Re}\,\tilde \phi\,[{\rm Re}\,\phi]\;. 
\eea
In Sec.\ \ref{phi0} it is shown that also $\phi_0(\vk)\sim \phi$ and that the  imaginary part of the gap function contributes to  $\phi_0(\vk)$  at $g^2\phi$, i.e.\ beyond subleading order. Consequently, 
\bea\label{confirm2}
\phi_0=\phi_0[{\rm Re}\, \phi]
\eea
to subleading order. Hereby it is shown that to subleading order Eq.\ (\ref{coupled2Re}) becomes Eq.\ (\ref{assumption1}),
%\bea
%{\rm Re}\,\phi(\epsilon_\vk) = {\rm Re}\,\tilde\phi(\epsilon_\vk)[{\rm Re}\,\phi(\epsilon_\vk)]
%+\phi_0[{\rm Re}\,\phi(\epsilon_\vk)] = {\rm Re}\,\phi(\epsilon_\vk)[{\rm Re}\,\phi(\epsilon_\vk)]\;,
%\eea
which means that the imaginary part of the gap function does not enter the real part at this accuracy. In other words, the real part of the gap equation is decoupled from its imaginary part.
On the other hand, the imaginary part of the gap equation is not decoupled from the real part of the gap equation, cf.\ Eq.\ (\ref{selfconsistency1}). %its self-consistency condition via ${\cal B}[{\rm Im}\,\phi]$ is removed.
It turns out that for momenta close to the Fermi surface Im$\,\phi$ can be calculated self-consistently from Re$\,\phi$ alone, which is performed in Sec.\ \ref{calcimphi}. 
%, which is similar to the coupling of the gap equation to the DSE of the gluon propagator. 
%This ``one-way''-coupling is peculiar to the subleading order solution of the complex gap equation
%\begin{subequations}
%\bea
%{\rm exact:~~}&&{\rm Re}\,\phi(\epsilon_\vk)={\rm Re}\,\phi[{\rm Re}\,\phi(\epsilon_\vk),\rho_\phi]\;,~~\rho_\phi=\rho_\phi[{\rm Re}\,\phi(\epsilon_\vk),\rho_\phi]\\
%{\rm to~subleading~order:~~ }&&{\rm Re}\,\phi(\epsilon_\vk)={\rm Re}\,\phi[{\rm Re}\,\phi(\epsilon_\vk)]\;,~~~~~~\rho_\phi=\rho_\phi[{\rm Re}\,\phi(\epsilon_\vk)]\;.
%\eea
%\end{subequations}

\subsubsection{Performing the Matsubara sum}\label{Matsu}

The complex gap equation (\ref{gapequation1}) reads to subleading order, cf.\ Eq.\ (\ref{gapequation2}),
 \bea \label{gapequation2a}
\phi(K) = \frac{g^2}{3} \, \frac{T}{V} \sum_Q 
{\rm Tr}_s \left( \Lambda_{\bf k}^+ \gamma_\mu 
\Lambda_{\bf q}^- \gamma_\nu \right)
\Delta^{\mu \nu}(K-Q)
\,\frac{\phi(Q)}{[q_0/Z(q_0)]^2 - \epsilon_q^2}\;,
\eea
where the quark wave-function renormalization factor $Z(q_0) \equiv [1+\bar g^2\ln(M^2/q_0^2)]^{-1}$ contains the effect of the normal quark self-energy $\Sigma(Q)$. As explained before, only its real part contributes at subleading order and therefore will be accounted for in the following, while its imaginary part will be neglected. Physically this is equivalent to assume the (normal) quasi-particle excitations to be infinitely long-lived, i.e.\ ignoring their decay due to scattering processes with quarks inside the Fermi sea. Mathematically it amounts to neglecting the cut of the logarithm when performing the Matsubara sum over $q_0$. Consequently, at subleading order the only effect of $Z(q_0)$ in Eq.\ (\ref{gapequation2a}) is a shift of the poles of the propagator
\bea\label{quarkpropcompact}
\tilde\Delta(Q)\equiv \frac{Z^2(q_0)}{q_0^2 - [Z(q_0)\,\epsilon_q]^2} =\frac{1}{2\tilde\epsilon_\vq}\sum\limits_{\sigma=\pm}\frac{\sigma\,Z^2(q_0)}{q_0 -\sigma Z(q_0)\,\epsilon_\vq}
\eea 
to $q_0=\pm Z(\epsilon_\vq)\,\epsilon_\vq \equiv \pm \tilde \epsilon_\vq$.

\begin{figure}[ht]
\centerline{\includegraphics[width=8cm]{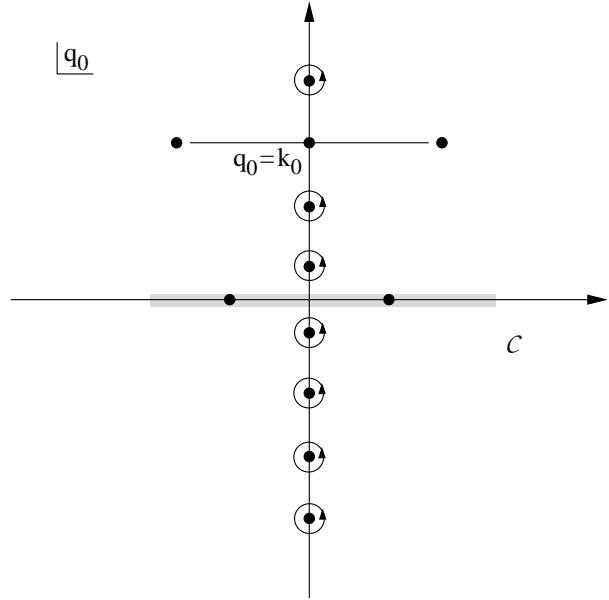}}
\caption[The contour ${\cal C}$ in Eq.\ (\ref{Mdef}).]{The contour ${\cal C}$ in Eq.\ (\ref{Mdef})
encloses the poles of $\tanh [q_0/(2T)]$ on the
imaginary $q_0$ axis. The additional poles and the cut at $q_0=k_0$ arise from the the gluon propagator, while the two poles on the real axis are due to the quasiquarks. The undetermined singularities of the gap function on the real $q_0-$axis are indicated by the shaded area.}
\label{phicontour1}
\end{figure}

\begin{figure}[ht]
\centerline{\includegraphics[width=8cm]{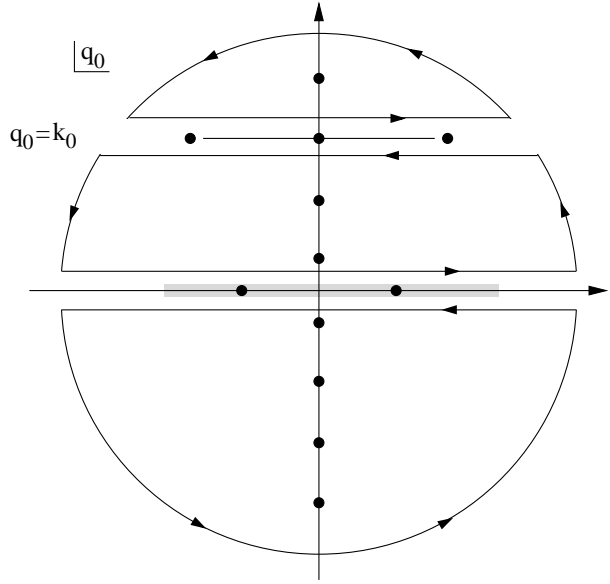}}
\caption[Deforming the contour ${\cal C}$.]{Deforming the contour ${\cal C}$ to introduce the spectral densities of $\Delta^{\ell,t}$ and $\phi$, cf.\ Eq.\ (\ref{M1}).}
\label{phicontour2}
\end{figure}

The Matsubara sum over $q_0$ in Eq.\ (\ref{gapequation1}) is performed via contour integration, cf.\ Fig.\ \ref{phicontour1}, 
\bea\label{Mdef}
{\mathcal M}^{\ell,t}(k_0,\vp,\vq)&\equiv&
 T \sum\limits_{q_0\neq k_0}\Delta^{\ell,t}(Q-K)\tilde\Delta(Q)\phi(Q)\non
&=&\int\limits_{\mathcal C} \frac{dq_0}{2\pi i}\,
\frac{1}{2}\tanh\left(\frac{q_0}{2T} \right)\Delta^{\ell,t}(Q-K)\tilde\Delta(Q)\phi(Q)\non
&\equiv&
 \int\limits_{\mathcal C} \frac{dq_0}{2\pi i}\,{\mathcal K}^{\ell,t} (q_0) \;.
\eea
In order to introduce the spectral densities of the gap function and of the magnetic and longitudinal gluon propagators, the contour ${\cal C}$ in Eq.\ (\ref{Mdef}) is deformed as shown in Fig.\ \ref{phicontour2}, yielding
\bea\label{M}
{\mathcal M}^{\ell,t}(k_0,\vp,\vq)&=&
\int\limits_{\mathcal C_\infty} \frac{dq_0}{2\pi i}\,{\mathcal K}^{\ell,t}(q_0)
+\int\limits_{-\infty}^\infty \frac{dq_0}{2\pi i}\,\left[{\mathcal K}^{\ell,t}(q_0+i\eta)-{\mathcal K}^{\ell,t}(q_0-i\eta)\right]\non
&&+ \int\limits_{-\infty}^\infty \frac{dq_0}{2\pi i}\,\left[{\mathcal K}^{\ell,t}(q_0+k_0+i\eta)-{\mathcal K}^{\ell,t}(q_0+k_0-i\eta)\right] \non
&\equiv& I_\infty + I_0 + I_{k_0}\;.
\label{M1}
\eea
The explicit form of the integral $I_\infty$ reads
\bea
I_\infty \equiv \int\limits_{\mathcal C_\infty} \frac{dq_0}{2\pi i}\,
\frac{1}{2}\tanh\left(\frac{q_0}{2T} \right)\Delta^{\ell,t}(Q-K)\tilde\Delta(Q)\phi(Q)\;.
\eea
For the integration along the contour ${\cal C}_\infty$ one parametrizes $dq_0 = i|q_0|e^{i\theta}d\theta$ and find for $|q_0|\longrightarrow \infty$
\bea
&&
\left|\tanh\left(\frac{q_0}{2T} \right)\right| \rightarrow 1\;,~ \tilde\Delta(Q)\rightarrow \frac{1}{q_0^2}\;,~\phi(Q) \rightarrow \phi_0(\vq)\;,~ 
\non &&
\Delta^{\ell}(Q-K)\rightarrow -\frac{1}{|\vq-\vk|^2}\;,~\Delta^{t}(Q-K)\rightarrow -\frac{1}{q_0^2}\;.
\eea
Consequently,
\bea
dq_0\,{\mathcal K}^{\ell}(q_0)\sim \frac{1}{q_0}\;,~dq_0\,{\mathcal K}^{\ell}(q_0)\sim \frac{1}{q_0^3}
\eea
and $I_\infty$ vanishes, $I_\infty =0\;.$ Using Eq.\ (\ref{quarkpropcompact}) the integral $I_0$ can be decomposed
\bea \label{I00}
I_0&=&\int\limits_{-\infty}^\infty \frac{dq_0}{2\pi i}\,
\frac{1}{2}\tanh\left(\frac{q_0}{2T} \right)\Delta^{\ell,t}(Q-K)
\non
&&\times\left[\tilde\Delta(q_0+i\eta,\vq)\phi(q_0+i\eta,\vq)-\tilde\Delta(q_0-i\eta,\vq)\phi(q_0-i\eta,\vq)\right]
\non
&&\hspace*{-0.8cm}
=\frac{1}{2{\tilde \epsilon_\vq}}\sum\limits_{\sigma=\pm}\sigma{\mathcal P}_{\sigma{\tilde \epsilon_\vq}} \int\limits_{-\infty}^\infty \frac{dq_0}{2\pi i}\, \frac{1}{2}\tanh\left(\frac{q_0}{2T}\right)\Delta^{\ell,t}(Q-K)Z^2(q_0)\,\frac{\phi(q_0+i\eta,\vq)-\phi(q_0-i\eta,\vq)}{q_0-\sigma{\tilde \epsilon_\vq}}\non
&&\hspace*{-0.3cm}
+\frac{1}{2{\tilde \epsilon_\vq}}\sum\limits_{\sigma=\pm}\sigma \int\limits_{\sigma\tilde\epsilon_\vq-\eta}^{\sigma\tilde\epsilon_\vq+\eta} \frac{dq_0}{2\pi i}\, \frac{1}{2}\tanh\left(\frac{q_0}{2T}\right)\Delta^{\ell,t}(Q-K)Z^2(q_0)\,\frac{\phi(q_0+i\eta,\vq)}{q_0-\sigma{\tilde \epsilon_\vq}+i\eta}\non
&&\hspace*{-0.3cm}
+\frac{1}{2{\tilde \epsilon_\vq}}\sum\limits_{\sigma=\pm}\sigma \int\limits_{\sigma\tilde\epsilon_\vq+\eta}^{\sigma\tilde\epsilon_\vq-\eta}\frac{dq_0}{2\pi i}\, \frac{1}{2}\tanh\left(\frac{q_0}{2T}\right)\Delta^{\ell,t}(Q-K)Z^2(q_0)\,\frac{\phi(q_0-i\eta,\vq)}{q_0-\sigma{\tilde \epsilon_\vq}-i\eta}\;,
\eea
where the notation $ {\mathcal P}_{\pm{\tilde \epsilon_\vq}}$ refers to the principal value prescription for the quark pole at $q_0 = \pm {\tilde \epsilon_\vq}$. In the first term on the r.h.s.\ of Eq.\ (\ref{I00}) the difference of the retarded and advanced gap function may be replaced by the spectral density of the gap function, cf.\ Eq.\ (\ref{specphi3}). In the second term one may substitute $q_0^\prime \equiv q_0 -\sigma \tilde\epsilon_\vq$ and deform the contour of integration into an infinitesimally small semicircle around the quark pole at $q^\prime_0 =0$. With $dq^\prime_0 =i\,\eta\, e^{i\theta}d\theta$ this term becomes
\bea
&&\frac{1}{2{\tilde \epsilon_\vq}}\sum\limits_{\sigma=\pm}\sigma \int\limits_{\pi}^0 \frac{i\,\eta\,e^{i\theta}d\theta}{2\pi\,i}\,\frac{1}{\eta\,e^{i\theta}}\, \frac{1}{2}\tanh\left(\frac{\sigma\tilde\epsilon_\vq}{2T}\right)\Delta^{\ell,t}(\sigma\tilde\epsilon_\vq-k_0,\vp)Z^2(\tilde\epsilon_\vq)\,\phi(\sigma\tilde\epsilon_\vq+i\eta,\vq)\non
&=&
-\frac{1}{4{\tilde \epsilon_\vq}}\sum\limits_{\sigma=\pm}\sigma \frac{1}{2}\tanh\left(\frac{\sigma\tilde\epsilon_\vq}{2T}\right)\Delta^{\ell,t}(\sigma\tilde\epsilon_\vq-k_0,\vp)Z^2(\tilde\epsilon_\vq)\,\phi(\sigma\tilde\epsilon_\vq+i\eta,\vq)\;.
\eea
The third term in Eq.\ (\ref{I00}) can be evaluated analogously to the second. The result will differ from the second term only by containing the advanced instead of the retarded gap function. Collecting all three terms one obtains using $\phi(\omega +i \eta)+\phi(\omega-i\eta) = 2{\rm Re}\,\phi(\omega+i\eta)$
\bea\label{I0}
I_0
&=&
\frac{1}{2{\tilde \epsilon_\vq}}\sum\limits_{\sigma=\pm}\sigma{\mathcal P}_{\sigma{\tilde \epsilon_\vq}} \int\limits_{-\infty}^\infty dq_0 \frac{1}{2}\tanh\left(\frac{q_0}{2T}\right)\Delta^{\ell,t}(q_0-k_0,\vp)Z^2(q_0)\frac{\rho_{\phi}(q_0,\vq)}{q_0-\sigma{\tilde \epsilon_\vq}}\non
&&-\frac{1}{2{\tilde \epsilon_\vq}}\frac{1}{2}\tanh\left(\frac{{\tilde \epsilon_\vq}}{2T}\right)Z^2(\tilde\epsilon_\vq)\, {\rm Re}\,\phi({\tilde \epsilon_\vq},\vq) \sum\limits_{\sigma=\pm}\Delta^{\ell,t}(\sigma{\tilde \epsilon_\vq}-k_0,\vp)\;.
\eea
The first term arises from the non-analyticities of the gap function, $\phi(Q)$, and the second from the poles of the quark propagator $\tilde \Delta(Q)$, cf.\ Eq.\ (\ref{quarkpropcompact}), at $q_0 = \pm {\tilde \epsilon_\vq}$. The last integral $I_{k_0}$ reads
\bea\label{Ik00}
&&I_{k_0}=\non
&&{\cal P}_0\int\limits_{-\infty}^\infty \frac{dq_0}{2\pi i}\,
\frac{1}{2}\coth\left(\frac{q_0}{2T} \right)\tilde\Delta(q_0+k_0,\vq)\phi(q_0+k_0,\vq)
\left[\Delta^{\ell,t}(q_0+i\eta,\vp)-\Delta^{\ell,t}(q_0-i\eta,\vp) \right]
\non
&&+\int\limits_{-\eta}^\eta \frac{dq_0}{2\pi i}\,
\frac{1}{2}\coth\left(\frac{q_0+i\eta}{2T} \right)\Delta^{\ell,t}(q_0+i\eta,\vp)
\tilde\Delta(q_0+k_0,\vq)\phi(q_0+k_0,\vq)
\non
&&+\int\limits_{\eta}^{-\eta} \frac{dq_0}{2\pi i}\,
\frac{1}{2}\coth\left(\frac{q_0-i\eta}{2T} \right)\Delta^{\ell,t}(q_0-i\eta,\vp)
\tilde\Delta(q_0+k_0,\vq)\phi(q_0+k_0,\vq)\;.
\eea
In the first term the difference of the retarded and advanced gluon propagator, $\Delta^{\ell,t}(q_0\pm i\eta,\vp)$,  may be replaced by the gluon spectral density. In the second and third term the integrals over $q_0$ can be performed after taking the limit $\coth(q_0/2T)/2\rightarrow T/q_0$ for $\eta \rightarrow 0$ and parametrize the contour as $dq_0 =i\,\eta\, e^{i\theta}d\theta$. One obtains for the second term
\bea
T\int\limits_{\pi}^0 \frac{i\,\eta\, e^{i\theta}d\theta}{2\pi i}\,\frac{1}{\eta\,e^{i\theta}}
\Delta^{\ell,t}(0+i\eta,\vp)
\tilde\Delta(k_0,\vq)\phi(k_0,\vq) = -\frac{T}{2}\Delta^{\ell,t}(0+i\eta,\vp)\tilde\Delta(k_0,\vq)\phi(k_0,\vq)\;.\non
\eea
For the third term the result differs only in the occurence of the advanced gluon propagator instead of the retarded. Combining all three terms in Eq.\ (\ref{Ik00}) yields
\bea\label{Ik0}
I_{k_0}&=&\frac{1}{2{\tilde \epsilon_\vq}}\sum\limits_{\sigma=\pm}\sigma{\mathcal P}_{0} \int\limits_{-\infty}^\infty dq_0 \frac{1}{2}\coth\left(\frac{q_0}{2T}\right)\phi(q_0+k_0,\vq)\,Z^2(q_0+k_0)\,\frac{\rho^{\ell,t}(q_0,\vp)}{q_0-\sigma{\tilde \epsilon_\vq}+k_0}\non
&& -T\, {\rm Re}\,\Delta^{\ell,t}(0+i\eta,\vp) \,\tilde\Delta(k_0,\vq)\,\phi(k_0,\vq)\;.
\eea
The first term is due to the non-analyticities of the longitudinal and magnetic gluon propagators, $\Delta^{\ell,t}$. The second arises from the pole of $\coth(q_0/2T)$ at $q_0=0$ corresponding to the large  occupation number density of gluons in the classical limit, $q_0 \ll T$. The latter contribution has been shown to be beyond subleading order \cite{rdpdhr} and will be discarded in the following.

\subsubsection{The imaginary part of the gap equation}\label{identim}

After having performed the Matsubara sum one considers energies $k_0$ close to the real axis, $k_0 = \omega + i\eta$ in order to employ the Dirac identity to split the complex gap equation (\ref{gapequation2a}) into its real and imaginary part. To this end one extracts the imaginary parts of the contributions $I_0$ and $I_{k_0}$, cf.\ Eqs.\ (\ref{I0},\ref{Ik0}). For the imaginary part of Eq.\ (\ref{gapequation2a}) one finds using that Re$\phi$ and $Z^2$ are even in $q_0$
%with Im$\,\phi(\omega+i\eta, \vk)=-\pi\rho_\phi(\omega,\vk)$ 
\bea\label{phieq}
&&{\rm Im}\,\phi(\omega+i\eta, \vk)=\non
&&\hspace*{-1.5cm}
\frac{g^2}{3}\int\frac{d^3q}{(2\pi)^3} \left[{\rm Tr}_s^\ell ( k,p, q)\,{\rm Im\mathcal M}^{\ell}(\omega+i\eta,\vp,\vq)+{\rm Tr}_s^t ( k,p,q)\,{\rm Im\mathcal M}^{t}(\omega+i\eta,\vp,\vq)\right]\;,
\eea
where
\bea\label{ImM}
&&{\rm Im \mathcal M}^{\ell,t}(\omega+i\eta,\vp,\vq) =\non
&&\frac{\pi}{4{\tilde \epsilon_\vq}}\sum\limits_{\sigma=\pm}\sigma{\mathcal P} \int\limits_{-\infty}^\infty dq_0
\frac{\rho^{\ell,t}(\omega-q_0,\vp)\rho_{\phi}(q_0,\vq)}{q_0-\sigma{\tilde \epsilon_\vq}}
Z^2(q_0)\left[\tanh\left(\frac{q_0}{2T}\right) +\coth\left(\frac{\omega-q_0}{2T}\right) \right] 
\non
&&-\frac{\pi}{4{\tilde \epsilon_\vq}}{\rm Re}\,\phi({\tilde \epsilon_\vq},\vq)\,Z^2(\tilde\epsilon_\vq)\sum\limits_{\sigma=\pm}\sigma\rho^{\ell,t}({\omega-\sigma\tilde \epsilon_\vq},\vp)\left[\tanh\left(\frac{\sigma{\tilde \epsilon_\vq}}{2T}\right) +\coth\left(\frac{{\omega-\sigma\tilde \epsilon_\vq}}{2T}\right) \right]\non
&\equiv& {\rm Im \mathcal M}^{\ell,t}_{\cal B}(\omega+i\eta,\vp,\vq)+ {\rm Im \mathcal M}^{\ell,t}_{\cal A}(\omega+i\eta,\vp,\vq)\,.
\eea
Note, that in the above Eq.\ (\ref{phieq}) $\omega+i\eta$ appears with a minus sign in the argument of the gluon propagators.
The traces over Dirac space ${\rm Tr}_s^{\ell,t} (k,p,q)$ are introduced in Eqs. (\ref{traces})
%\begin{subequations} \label{traces}
%\bea
%{\rm Tr}_s^\ell (k,p,q)
%&\equiv& {\rm Tr}_s \left( \Lambda_{\bf k}^+ \gamma_0 
%\Lambda_{\bf q}^- \gamma_0 \right) 
%= \frac{(k+q)^2 - p^2}{2\, k\,q} \;, \\
%{\rm Tr}_s^t (k,p,q) &\equiv& 
%(\delta^{ij} - \hat{p}^i \hat{p}^j)\, 
%{\rm Tr}_s \left( \Lambda_{\bf k}^+ \gamma_i \Lambda_{\bf q}^-
%\gamma_j \right)  
%=  -2 - \frac{p^2}{2\, k\,q} + \frac{(k^2-q^2)^2}{2\, k\, q\, p^2}\;,
%\eea
%\end{subequations}
%where we used $p^2 \equiv ({\bf k}-{\bf q})^2 = k^2 + q^2 - 2\, k\, q \,
%\hat{\bf k} \cdot \hat{\bf q}$ to eliminate $\hat{\bf k} \cdot
%\hat{\bf q}$ in favor of $p^2$.
The first term on the r.h.s\ of Eq.\ (\ref{ImM}), ${\rm Im \mathcal M}^{\ell,t}_{\cal B}$, is due to the non-analyticities of $\phi(K)$ and has been neglected in all previous treatments. Inserted into Eq.\ (\ref{phieq}) it constitutes the term ${\cal B}[{\rm Im}\,\phi]$ introduced in Eq.\ (\ref{AB}). The second term, ${\rm Im \mathcal M}^{\ell,t}_{\cal A}$,  contains all contributions to $\phi(K)$ that have been considered so far to subleading order. Inserted into Eq.\ (\ref{phieq}) it constitutes the term ${\cal A}[{\rm Re}\,\phi]$. As already announced it contains the real part of the gap function always on the quasiparticle mass-shell. It is instructive to shortly analyze this issue before proceeding further. The anomalous propagator $\Xi(Q) \equiv Z^2(q_0) \,\phi(Q)/(q_0^2 - \tilde\epsilon_q^2)$ has the spectral density
\bea\label{specxi}
\rho_\Xi(\omega,\vq)&\equiv& 
\frac{1}{2\pi i}\left[  \Xi(\omega+i\eta,\vq) - \Xi(\omega-i\eta,\vq) \right]\non
&=&
\frac{1}{2\pi i}\left[ \frac{ Z^2(\omega+i\eta)\,\tilde\phi(\omega+i\eta,\vq) }{\omega^2-[Z(\omega)\epsilon_\vq]^2+{\rm sign}(\omega)i \eta} -
\frac{ Z^2(\omega-i\eta)\,\tilde\phi(\omega-i\eta,\vq) }{\omega^2-[Z(\omega)\epsilon_\vq]^2-{\rm sign}(\omega)i \eta} \right]\non
&\simeq& \frac{1}{2\pi i}\left\{Z^2(\omega){\mathcal P}\; \frac{\tilde\phi(\omega+i\eta,\vq) - \tilde\phi(\omega-i\eta,\vq)}{\omega^2-[Z(\omega)\epsilon_\vq]^2}
\right.\non&&\left.
-i\pi {\rm sign}(\omega)\,Z^2(\tilde\epsilon_\vq)\,\delta\left(\omega^2 -\tilde\epsilon_\vq^2 \right)[\phi(\omega+i\eta,\vq) + \phi(\omega-i\eta,\vq)]\frac{}{}\right\}\non
&=&
- Z^2(\omega){\mathcal P}\; \frac{\rho_{\phi}(\omega,\vq)}{\omega^2-[Z(\omega)\epsilon_\vq]^2} - {\rm sign}(\omega)\,Z^2(\tilde\epsilon_\vq)\,{\rm Re}\,\phi(\omega+i\eta,\vq)\,\delta\!\left(\omega^2 -\tilde\epsilon_\vq^2 \right)\,,\non
\eea
where the cut of $Z(\omega)$ has been neglected.
It follows that $\rho_{\phi}(\omega,\vq)\neq 0$ leads to additional support of $\rho_\Xi(\omega,\vq)$ around the quasiparticle pole $\omega \equiv \tilde\epsilon_q$. Furthermore, neglecting ${\rm Im}\,\phi$ is equivalent to approximating 
\bea\label{ansatz}
\rho_\Xi(\omega,\vq)\simeq  -{\rm sign}(\omega)\,{\rm Re}\,\phi(\tilde\epsilon_\vq+i\eta,\vq)\,Z^2(\tilde\epsilon_\vq)\,\delta\!\left(\omega^2 -\tilde\epsilon_\vq^2 \right)
\eea
as it has been done in \cite{rdpdhr}, cf. Eq.\ (41). As a consequence, the gap function on the r.h.s.\ of the gap equation is always forced onto the quasiparticle mass-shell and the gap equation takes the standard form, cf.\ Eq.\ (\ref{gapequation5}).
The occurrence of the external energy $\tilde\epsilon_k$ on the r.h.s.\ due to the energy-dependent gluon propagators indicates that the solution still would possess some energy dependence although not provided in the ansatz Eq.\ (\ref{ansatz}). This demonstrates explicitly the inherent inconsistency of this ansatz.

It is interesting to note that in Eq.\ (\ref{specxi}) ${\rm Im}\,\phi\neq 0$ does not shift the quasiparticle pole into complex $q_0-$ plane, since $\e_q\equiv \sqrt{(q-\mu)^2+|\phi|^2}$. Therefore, no damping of $\Xi$ is caused. In order to restore the effect of damping due to Im$\,\phi$ one would have to include Im$\,\Sigma$, which is beyond the scope of this work.

In the limit of small temperatures, $T\rightarrow 0$, the hyperbolic functions in Eq.\ (\ref{ImM}) simplify, yielding for $\omega > 0$
\bea\label{ImMT0}
{\rm Im \mathcal M}^{\ell,t}_{T=0}(\omega+i\eta,\vp,\vq) &=&
\frac{\pi}{2{\tilde \epsilon_\vq}}\left[\sum\limits_{\sigma=\pm}\sigma\,{\mathcal P} \int\limits_{0}^{\omega} dq_0
\;\frac{\rho^{\ell,t}(\omega-q_0,\vp)\rho_{\phi}(q_0,\vq)}{q_0-\sigma{\tilde \epsilon_\vq}}\,Z^2(q_0)\right.\non
&&\left.
-\frac{}{}Z^2(\tilde\epsilon_\vq)\,{\rm Re}\,\phi({\tilde \epsilon_\vq},\vq)\,\rho^{\ell,t}(\omega-{\tilde \epsilon_\vq},\vp)\,\theta(\omega-{\tilde \epsilon_\vq})\right]\non
&\equiv& {\rm Im \mathcal M}^{\ell,t}_{{\cal B},T=0}(\omega+i\eta,\vp,\vq)+ {\rm Im \mathcal M}^{\ell,t}_{{\cal A},T=0}(\omega+i\eta,\vp,\vq)\;.\non
\eea
Here, use was made of the oddness of $\rho_\phi$ in $\omega$, cf.\ Eq.\ (\ref{odd}).

\subsection{Estimating the order of magnitude of ${\cal A}$ and ${\cal B}$} \label{estAB}

Following the strategy explained in Sec.\ \ref{solving} it is necessary to know the order of magnitude of the real part of the on-shell gap function Re$\,\phi(\tilde\epsilon_\vq,\vq)$ in order to estimate the magnitude of the parts ${\cal A}$ and ${\cal B}$ of Im$\,\phi$, cf.\ Eq.\ (\ref{AB}). It is known that
$\phi_k \equiv {\rm Re}\,\phi({\epsilon_\vk},\vk)$ is momentum dependent \cite{DHRreview}
\begin{equation} \label{solution}
\phi(x) \equiv \phi_0^{\rm 2SC} \, F(x)\,\, ,
\end{equation}
where  $\phi_0^{\rm 2SC}$ is the value of the gap for momenta on the Fermi surface
\be \label{phi02SC}
\phi_0^{\rm 2SC}=2 \, \tilde{b} \, b_0'\, \mu\, \exp\left(
-\frac{\pi}{2\, \bar{g}}\right) \,\,
\ee
with $\bar g\equiv g/(3\sqrt{2}\pi)$ and the constants (subleading and therefore in principle irrelevant for the present purposes)
\be \label{constants}
\tilde{b} \equiv 256\, \pi^4 \left(\frac{2}{N_f g^2} \right)^{5/2}\,\, , \qquad
b_0' \equiv \exp\left(-\frac{\pi^2+4}{8}\right)
\,\, .
\ee
The variable $x$ is a exponential measure for the distance of $\vk$ from the Fermi surface
\be \label{vartrans}
x \equiv \bar{g} \, \ln\left(\frac{2\tilde b\m}{k-\m+\e_{\vk}}\right)\;.
\ee
For $|k-\mu| \sim \phi$ one finds
$x = \pi/2 +O(\bar g)$, while for $|k-\mu| \sim M$ it is
$x \sim O(\bar g)$. 
The function $F(x)$ is to leading order given by $F(x)=\sin(x)$. 
Hence, for momenta exponentially close to the Fermi surface, $|k-\mu| \sim \phi$,
the on-shell gap function is $\phi(\epsilon_\vk,\vk)=\phi_0^{\rm 2SC}[1+O(\bar g^2)]$, i.e.\ constant to leading order. For momenta $|k-\mu| \sim M$, one has $\phi({\epsilon_\vk},k)=\bar g\,\phi_0^{\rm 2SC}$. Hence, the real part of the on-shell gap function
is sharply peaked around the Fermi surface. This exponential decay off the Fermi surface becomes apparent when considering {\it intermediate} distances  from the Fermi surface (between $\phi$ and $M$) defined by the variable scale 
\bea\label{Lambday}
\Lambda_y \equiv \phi^y M^{1-y}
\eea
 with $0<y<1$. For $|k-\mu| \sim \Lambda_y$ one has $x = y\, \pi/2 +O(\bar g)$ and $F(y) \simeq \sin( y\, \pi/2)\simeq y$. Hence, decreasing $y$ (i.e.\ exponentially admixing the scale $M$ to $\Lambda_y$ and exponentially receding from the Fermi surface) decreases the magnitude of the gap approximately linearly. For $y \sim \bar g$ the gap function has decreased to $\bar g\,\phi_0^{\rm 2SC}$. The intermediate scales may be also used to investigate the generation of the so-called BCS-log. Denoting $\Lambda_1 = \phi$ and $\Lambda_0 = M$ and exploiting $\Lambda_1 \ll \Lambda_{\bar g} \lesssim \Lambda_0$ one may write 
%for a BCS superconductor with $\phi_q \sim g\mu\exp(-
\bea
g^2\int\limits_0^M \frac{d\xi}{\epsilon_\vq} \;\phi_q
&\equiv&
g^2\int\limits_0^{\Lambda_1} \frac{d\xi}{\epsilon_\vq} \;\phi_q
+g^2\int\limits_{\Lambda_1}^{\Lambda_{\bar g}} \frac{d\xi}{\epsilon_\vq} \;\phi_q
+g^2\int\limits_{\Lambda_{\bar g}}^{\Lambda_{0 }} \frac{d\xi}{\epsilon_\vq} \;\phi_q\non
&\simeq& g^2\ln(\sqrt{2}+1) \;\phi_0 +g^2\left[\ln\left(\frac{2\Lambda_{\bar g}}{\Lambda_1}\right) - \ln(\sqrt{2}+1) \right] \phi_0+ g^2 \ln\left( \frac{\Lambda_0}{\Lambda_{\bar g}}\right) g\phi_0\non
&\simeq&
g^2\ln\left( \frac{2M^{1-\bar g}}{\phi^{1-\bar g}}\right)\phi_0 +g^3\ln\left( \frac{M^{\bar g}}{\phi^{\bar g}}\right)\phi_0 \non
&\sim& g \phi_0\;,
\eea
where use was made of Eq.\ (\ref{phi02SC}) yielding the BCS-log, $\ln(M/\phi)\sim 1/g$. It is shown that the BCS-log arises from integrating over intermediate scales, $\Lambda_1 < \xi < \Lambda_{\bar g}$. In the region  $\Lambda_{\bar g} < \xi < M$ the BCS-log does not occur due to the exponential $\bar g$ under the logarithm. In addition the contribution from this region is suppressed by an extra factor $\bar g$ from $\phi_q \sim\bar g \phi_0$. In the QCD gap-equation the gluon propagator has to be added to the integrand. (Then  the region $\Lambda_1 <\xi< \Lambda_{\bar g}$ is enhanced by an additional large logarithm due to almost static, Landau-damped magnetic gluons.) However, the above observation that the BCS-log is generated by  intermediate scales $\Lambda_{1>y>{\bar g}}$ remains valid in the full gap equation.

\subsubsection{Landau-damped gluons contributing to $\cal A$}

For the purpose of power counting the various terms in Eq.\ (\ref{phieq}) one may restrict oneself to the leading contribution of the Dirac traces Eqs.\ (\ref{traces}), which is of order one. The integral over the absolute magnitude of the quark
momentum is $\int dq \, q^2$, while the angular integration is 
$\int d \cos \theta \equiv \int d p \, p / (kq)$. Furthermore,  only the zero-temperature limit, $T=0$ is considered, allowing for the estimate $Z^2(\tilde\epsilon_\vq)\sim 1$. The contribution of ${\cal M}^{\ell,t}_{{\cal A},T=0}$ in Eq.\ (\ref{ImMT0}) arising from $\rho_{\rm cut}^{\ell,t}(\omega-\epsilon_\vq,\vp)$ to $\cal A$ is
\bea\label{Acut}
{\cal A}^{\ell,t}_{\rm cut}(\omega,\vk)&\sim&g^2\int\limits_0^\delta\frac{d\xi}{\epsilon_\vq}\,{\rm Re}\,\phi(\epsilon_\vq,\vq)\int\limits_\lambda^{\Lambda_{\rm gl}} dp\, p \,\rho^{\ell,t}_{\rm cut}(\omega^*,\vp)\;,
\eea
where $\omega^*\equiv\omega-\epsilon_\vq<\omega$, $\delta\equiv{{\rm min}(\omega,\Lambda_{\rm q})}$ and $\lambda\equiv {\max}(|\xi -\zeta|,\omega^*)$ with $\zeta \equiv |k-\mu|$. Due to the condition $\lambda<p<\Lambda_{\rm gl}$ it immediately follows that ${\cal A}^{\ell,t}_{\rm cut}= 0$ for $\omega >\Lambda_{\rm gl}+\Lambda_{\rm q}\sim \mu$. Inserting the approximative forms
\be \label{appcut}
\rho_{\rm cut}^t (\omega^*, {\bf p}) \simeq \frac{M^2}{\pi} \,
\frac{\omega^* \, p}{p^6 + (M^2\, \omega^*)^2}\;,~~~
\rho^{\ell}_{\rm cut}(\omega^*,{\bf p}) \simeq
\frac{2 M^2}{\pi}\, \frac{\omega^*}{p}\,\frac{1}{
( p^2 + 3\, m_g^2 )^2} \;
\ee
the integration over $p$ can be performed analytically. 
%In the region $M<p<\Lambda_{\rm gl}$ and $\omega \lessim p$ the above estimate for $\rho^\ell_{\rm cut}$ slightly underestimates the exact expression, cf.\ Eq.({rhollong}). However, this will not affect the power counting analysis presented. it turns out that the contribution of this momentum regime is highly suppressed. 
For energies $\omega <\Lambda_{\rm gl}$ one finds for the transverse part
\bea\label{t}
{\cal A}^t_{\rm cut}(\omega,\vk)&\sim&g^2\int\limits_0^\omega\frac{d\xi}{\epsilon_\vq}\,{\rm Re}\,\phi(\epsilon_\vq,\vq)
\left[\arctan\left(\frac{\Lambda_{\rm gl}^3}{M^2\omega^*}\right)-\arctan\left(\frac{\lambda^3}{M^2\omega^*}\right)\right]
\;.
\eea
For all $\zeta \leq \Lambda_{\rm q}$ and $\omega <\Lambda_{\rm gl}$ it is $\Lambda_{\rm gl}^3/M^2\omega^* \gg 1$ and the first arctan in the squared brackets may be set equal to $\pi/2$. Considering $\zeta, \omega \ll M$ the argument of the second arctan is $\lambda^3/M^2\omega^*\ll 1$ and the combination of both arctans is  $\sim 1$. Increasing the energy to $\omega \sim M$ one finds $\lambda^3/M^2\omega^*\sim M/(M-\xi)$, which  becomes large only for $\xi\rightarrow M$. However, since the integration over $\xi$ stops here anyway this case does not have to be analyzed further. Consequently, also for  $\omega \sim M$ one may estimate the arctans to be $\sim 1$. For $\omega \gg M$ one has  $\lambda^3/M^2\omega^*\gg 1$ and the arctans finally cancel. Considering $\zeta\lesssim M$ one finds for $\omega \ll M$ that  $\lambda^3/M^2\omega^*\sim M/\omega^* \gg 1$ and the arctans cancel. The same is true for $\omega \gg M$ since then  $\lambda^3/M^2\omega^*\sim (\omega/M)^2 \gg 1$. Only in the region $\omega \sim M$ it is  $\lambda^3/M^2\omega^*\sim M/\omega^* \sim 1$ and the arctans do not cancel. 

One first considers $\zeta\ll M$ and $\omega\sim \phi$.  Because in this special case the integral over $\xi$ does not yield the BCS-log one finds
\bea\label{Aphi}
{\cal A}^t_{\rm cut}(\phi,\vk)&\sim&g^2\phi\;.
%,~~~{\cal A}^\ell_{\rm cut}(\phi,\vk)\sim g^2\phi\,\frac{\phi}{M}\;.
\eea
For larger energies $\omega\sim\Lambda_y$ with $0<y<1$ and $\zeta \ll M$ one substitutes $\xi(y^\prime)\equiv \Lambda_{y^\prime}$, $d\xi/\xi = \ln(\phi/M)\,dy^\prime$, one finds with Eq.\ (\ref{solution})
\bea
{\cal A}^t_{\rm cut}(\omega,\vk)&\sim& g^2\,\ln\left(\frac{\phi}{M}\right) \phi\int\limits_1^y dy\,\sin\left( \frac{\pi\,y}{2}\right)\sim  g\,\phi\,\cos\left( \frac{\pi\,y}{2}\right)\;.\label{Aphi2}
\eea
Following the discussion after Eq.\ (\ref{t}) the limit $y\rightarrow 0$ of Eq.\ (\ref{Aphi2}) is valid also for $\zeta\sim M$. Hence for  $\omega\sim  M$ and all $\zeta < \Lambda_{\rm q}$ it is ${\cal A}^t_{\rm cut}(\omega,\vk)\sim g\,\phi\,.$
For all other values of $\omega$ and $\zeta$ the contribution ${\cal A}^t_{\rm cut}(\omega,\vk)$ is strongly suppressed.

In the longitudinal sector one finds for the integral over the gluon momentum $p$ 
\bea\label{pintlong}
{\cal I}(\lambda)&\equiv& M^2\int\limits_\lambda^{\Lambda_{\rm gl}} \frac{dp}{
( p^2 + X^2 )^2}\sim \frac{1}{X}\left[\arctan\left(\frac{\Lambda_{\rm gl}}{X}\right)-\arctan\left(\frac{\lambda}{X}\right)\right] - \frac{\lambda}{X^2 + \lambda^2}\non
&\sim&
\left\{
\begin{array}{c}
1/X\,,~~{\rm for}~\lambda \leq X\\
1/\lambda\,,~~{\rm for}~\lambda \gg X
\end{array}
\right.
\;,
\eea
where it is abbreviated $X^2\equiv 3m_g^2$. Since $\zeta \leq \Lambda_{\rm q} \sim X$, solely the magnitude of $\omega$ decides whether $\lambda\leq X$ or $\lambda\gg X$ is realized. It follows that in contrast to the transversal case the order of magnitude of ${\cal A}^\ell_{\rm cut}$ is not dependent on the choice of $\zeta$. Energies $\omega \sim \Lambda_y$ with $0\leq y < 1 $ correspond to $\lambda \leq X$, where the r.h.s.\ of Eq.\ (\ref{pintlong}) is $\sim 1/X$. Beginning with the special case $\omega \sim \Lambda_1$ one finds
\bea
{\cal A}^\ell_{\rm cut}(\omega,\vk)&\sim&g^2\int\limits_0^\omega\frac{d\xi}{\epsilon_\vq}\,{\rm Re}\,\phi(\epsilon_\vq,\vq)\,\,\frac{\omega^*}{M} \sim g^2\,\phi\,\frac{\phi}{M}\;.
\eea
For $\omega \sim \Lambda_y$ with $0< y < 1 $ one has similarly
\bea\label{Aphi2b}
{\cal A}^\ell_{\rm cut}(\omega,\vk)&\sim&g\,\phi\int\limits_1^y dy^\prime\,\sin\left( \frac{\pi\,y^\prime}{2}\right)\,\frac{\omega^*}{M} 
\sim g\,\phi\int\limits_1^y dy^\prime\,\sin\left( \frac{\pi\,y^\prime}{2}\right)\,\left[\left(\frac{\phi}{M} \right)^y-\left(\frac{\phi}{M} \right)^{y^\prime}\right]\non
&\sim& g\,\phi\,\cos\left( \frac{\pi\,y}{2}\right) \,\left(\frac{\phi}{M} \right)^y\;.
\eea
Hence, in the considered energy regime ${\cal A}^\ell_{\rm cut}$ is suppressed by the factor $(\phi/M)^y$ as compared to ${\cal A}^t_{\rm cut}$, i.e.\ ${\cal A}^\ell_{\rm cut} \ll {\cal A}^t_{\rm cut}$.
For energies $\omega \sim M$ the longitudinal and the transversal cut contribute at the same order,   ${\cal A}^\ell_{\rm cut} \sim {\cal A}^t_{\rm cut}\sim g\,\phi$.
For much larger energies, $M \ll \omega < \mu$, ($\zeta$ is bounded by $\Lambda_{\rm q}$) it is $\lambda=\omega^* \simeq \omega \gg X$ and the r.h.s.\ of Eq.\ (\ref{pintlong}) is $\sim 1/\lambda \sim 1/\omega$. It follows with $\delta = \Lambda_{\rm q} \sim \Lambda_0$
\bea
{\cal A}^\ell_{\rm cut}(\omega,\vk)&\sim&g\,\phi\int\limits_1^0 dy\,\sin\left( \frac{\pi\,y}{2}\right)\,\frac{\omega^*}{\omega} 
\sim g\,\phi\int\limits_1^0 dy\,\sin\left( \frac{\pi\,y}{2}\right)
\sim g\,\phi\label{Acuthigh}
\eea
and one found that ${\cal A}^\ell_{\rm cut} \gg {\cal A}^t_{\rm cut}$ in this large-energy regime. The results for ${\cal A}^{\ell,t}_{\rm cut}$ are summarized in Tables \ref{tableAcutzeta<M} and \ref{tableAcutzetasimM}.

\begin{table}  
\centerline{\begin{tabular}[t]{|c||c|c|c|c|}
\hline
$\omega$  & $\phi$ & $\Lambda_{1>y>0}$ & ~$M$~ & ~$M\ll\omega<\mu$~
\\ \hline\hline
~ ${\cal A}_{\rm cut}^t$ ~  &  $g^2\phi$	& $g\,\phi\,\cos\left(\frac{\pi\,y}{2}\right)$ & $g\,\phi$ & 0
\\ \hline
${\cal A}_{\rm cut}^\ell$  & ~ $g^2\phi\,\frac{\phi}{M}$~ 	& ~$g\,\phi\left(\frac{\phi}{M}\right)^y\cos\left(\frac{\pi\,y}{2}\right)$ ~ &  $g\,\phi$  & $g\,\phi$
\\ \hline
${\cal A}_{\rm cut}$  &  $g^2\phi$	& $g\,\phi\,\cos\left(\frac{\pi\,y}{2}\right)$ & $g\,\phi$ & $g\,\phi$
\\ \hline
\end{tabular}}
\caption[Estimates for ${\cal A}_{\rm cut}^{\ell,t}$ and  ${\cal A}_{\rm cut}$ at different energy scales and $\zeta\ll M$.]{Estimates for ${\cal A}_{\rm cut}^{\ell,t}$ and  ${\cal A}_{\rm cut}={\cal A}_{\rm cut}^{\ell}+{\cal A}_{\rm cut}^{t}$ at different energy scales and $\zeta\ll M$.}
\label{tableAcutzeta<M}
\end{table}

\begin{table}  
\centerline{\begin{tabular}[t]{|c||c|c|c|c|}
\hline
$\omega$  & $\phi$ & $\Lambda_{1>y>0}$ & ~$M$~ & ~$M\ll\omega<\mu$~
\\ \hline\hline
~ ${\cal A}_{\rm cut}^t$ ~  &  0 & 0  & $g\,\phi$ & 0
\\ \hline
${\cal A}_{\rm cut}^\ell$  & ~ $g^2\phi\,\frac{\phi}{M}$~ 	& ~$g\,\phi\left(\frac{\phi}{M}\right)^y\cos\left(\frac{\pi\,y}{2}\right)$ ~ &  $g\,\phi$  & $g\,\phi$
\\ \hline
${\cal A}_{\rm cut}$  &  $g^2\phi\,\frac{\phi}{M}$	& $g\,\phi\left(\frac{\phi}{M}\right)^y\,\cos\left(\frac{\pi\,y}{2}\right)$ & $g\,\phi$ & $g\,\phi$
\\ \hline
\end{tabular}}
\caption[Estimates for ${\cal A}_{\rm cut}^{\ell,t}$  and  ${\cal A}_{\rm cut}$ at different energy scales and $\zeta\lesssim M$.]{Estimates for ${\cal A}_{\rm cut}^{\ell,t}$  and  ${\cal A}_{\rm cut}={\cal A}_{\rm cut}^{\ell}+{\cal A}_{\rm cut}^{t}$ at different energy scales and $\zeta\lesssim M$.}
\label{tableAcutzetasimM}
\end{table}
 
\subsubsection{Undamped gluons contributing to $\cal A$}

\begin{figure}[ht]
\centerline{\includegraphics[width=10cm]{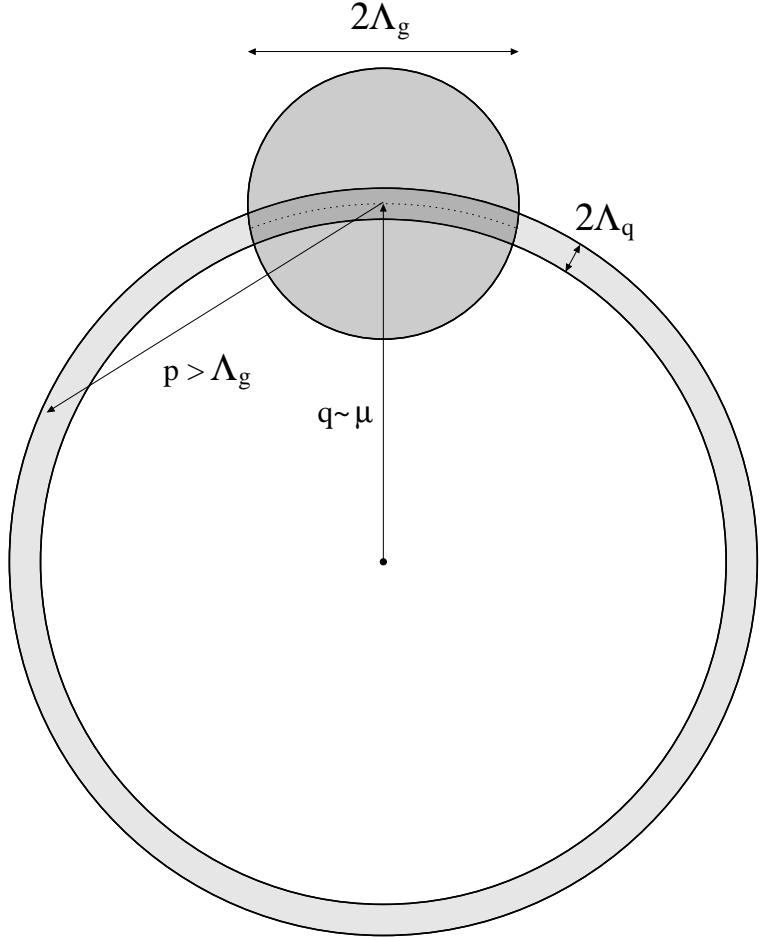}}
\caption[Hard gluon exchange with momentum $p>\Lambda_{\rm gl}\sim \mu$.]{Hard gluon exchange with momentum $p>\Lambda_{\rm gl}\sim \mu$. The quark has to remain within the layer of width $2\Lambda_{\rm q}\sim g\mu$ around the Fermi-surface. This effectively restricts the hard gluon momentum, $p\lesssim \mu+2\Lambda_{\rm q}$.}
\label{Sphere3}
\end{figure}

The contributions from the undamped gluon excitations to $\cal A$ read analogously to Eq.\ (\ref{Acut})
\bea
{\cal A}^{\ell,t}_{\rm pole}(\omega,\vk)&\sim&g^2\int\limits_0^\delta\frac{d\xi}{\epsilon_\vq}\,{\rm Re}\,\phi(\epsilon_\vq,\vq)\int\limits_{|\xi-\zeta|}^{2\mu} dp\, p \,\rho^{\ell,t}_{\rm pole}(\omega^*,\vp)\,\delta[\omega^*-\omega_{\ell,t}(\vp)]\;.\label{Apole}
\eea
Due to the $\delta-$function and $\omega_{\ell,t}(\vp)\geq m_g$ this contribution will be nonzero only for energies $\omega >  m_g \sim M$. The upper boundary $\sim 2\mu$ in the integral over $p$ is due to the constraint on relevant quark momenta near the Fermi surface in the effective theory, $\xi\leq\Lambda_{\rm q}$, where $\Lambda_{\rm q}\sim g\mu$ is the quark cutoff, cf.\ Fig.\ \ref{Sphere3}.
Due to this kinematic limitation it immediately follows that ${\cal A}^{\ell,t}_{\rm pole}=0$ for $\omega>2\mu+\Lambda_{\rm q}\simeq 2\mu$, since then $\omega^*\equiv \omega-\xi > \omega_{\ell,t}$ always and the $\delta-$function on the r.h.s.\ of Eq.\ (\ref{Apole}) is identically zero.

\begin{figure}[ht]
\centerline{\includegraphics[width=8cm]{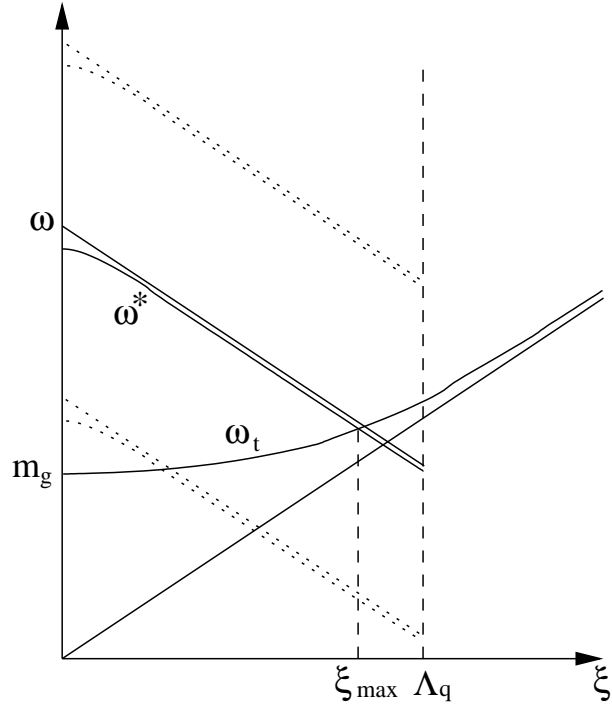}}
\caption[The integration regions of $\xi$ and $\omega_t$ in Eq.\ (\ref{tpole}).]{The integration regions of $\xi$ and $\omega_t$ in Eq.\ (\ref{tpole}). Only those values of $\xi$ contribute, for which  the integral along $\omega_t$, starting at $\omega_t = \sqrt{m_g^2+\xi^2}$ and ending at $\omega_t \simeq 2\mu$, has a point of intersection with $\omega^*=\omega-\xi$,  $\omega_t = \omega^*$. Only for $\omega > m_g+\Lambda_y$ with $y< 1$ the integration region is sufficiently large to generate the BCS-log. The dotted lines correspond to different values of $\omega$.}
\label{xiregion}
\end{figure}

For the transverse sector one may approximate for all momenta $p$
\bea\label{approxt}
\rho^{t}_{\rm pole}(\omega_{t}(\vp),\vp)\simeq-\frac{1}{2\omega_t(\vp)}\;,~~~
%\rho^{\ell}_{\rm pole}(\omega_{\ell}(\vp),\vp)\simeq-\frac{\omega_\ell(\vp)}{2p^2}\;,
\eea
and $\omega_{t}(\vp)\simeq \sqrt{p^2+m_g^2}$. One finds after substituting $dp\,p \simeq d\omega_t\,\omega_t$  and assuming $\zeta\ll M$
\bea\label{tpole}
{\cal A}^{t}_{\rm pole}(\omega,\vk)&\sim&
g^2\int\limits_0^{\Lambda_{\rm q}}\frac{d\xi}{\epsilon_\vq}\,{\rm Re}\,\phi(\epsilon_\vq,\vq)\int\limits_{\sqrt{m_g^2+\xi^2}}^{2\mu}\!\! d\omega_t\,\delta[\omega^*-\omega_{t}]\non
&\sim&g^2\int\limits_0^{\xi_{\rm max}}\frac{d\xi}{\epsilon_\vq}\,{\rm Re}\,\phi(\epsilon_\vq,\vq)\sim g\,\phi\;,
\eea
where $\xi_{\rm max} \equiv  {\rm min}[(\omega^2-m_g^2)/(2\omega),\Lambda_{\rm q}]$. In Fig.\ \ref{xiregion} the integration regions for $\xi$ and $\omega_t$ are visualized. For $\xi >\xi_{\rm max}$ it is $\omega^* = \omega -\xi < \sqrt{m_g^2+\xi^2}<\omega_t$ and the $\delta-$function under the integral over $\omega_t$ is always zero. The last estimate on the r.h.s.\ of Eq.\ (\ref{tpole}) is valid for energies, which are at least $\omega >m_g +\Lambda_y$ with $y < 1$, because then $\xi_{\rm max}$ is at least $\Lambda_y$ and the BCS-log is generated cancelling one power of $g$. 
For $\zeta \lesssim M$ one has $p>|\xi-\zeta|$ and therefore $\omega_t > \sqrt{m_g^2+|\xi-\zeta|^2}$. For $\Lambda_1 < \xi< \Lambda_{\bar g}$ it is  $\omega_t \simeq \sqrt{m_g^2+\zeta^2}\agt \sqrt{2}m_g$, while for $\Lambda_{\bar g} < \xi < \Lambda_0$ it is $\omega_t > m_g$ (independent of the signs of $\xi $ and $\zeta$). It turns out that for $\omega$ exponentially close to $m_g$, i.e.\ for energies of the form $\omega\sim m_g+\Lambda_y$ with $y>0$, $\omega^*\equiv \omega-\xi = \omega_t$ cannot be fulfilled in the whole integration region $\Lambda_1 < \xi< \Lambda_0$. Consequently, ${\cal A}^{t}_{\rm pole}= 0$ in this energy domain. Only for energies $\omega > \sqrt{m_g^2+\zeta^2}\agt \sqrt{2} m_g$ one finds again ${\cal A}^{t}_{\rm pole}\sim g\,\phi$, which remains valid up to $\omega \lesssim 2\mu$.
%Note, that for energies $\omega > m_g +\Lambda_{\rm q}$ the contribution ${\cal A}_{\rm pole}^t$ vanishes as $\omega^* = \omega_t$ cannot be fulfilled. The same is true for large gluon momenta $p\geq m_g + \Lambda_{\rm q}$. 

In the longitudinal gluon sector one may approximate for gluon momenta $p < m_g$
\bea \label{rholapp}
\rho^\ell_{\rm pole}(\omega_\ell(\vp), \vp) \simeq -\frac{\omega_\ell(\vp)}{2p^2}
\eea
and $\omega_{\ell}(\vp)\simeq \sqrt{p^2+m_g^2}$. For $\omega = m_g + \Lambda_y$ with $0<y<1$ one finds after substituting $dp\,p \simeq d\omega_\ell\,\omega_\ell$ and assuming $\zeta\ll M$
\bea\label{lpole}
{\cal A}^{\ell}_{\rm pole}(\omega,\vk)&\sim&
g^2\int\limits_0^{\Lambda_{\rm q}}\frac{d\xi}{\epsilon_\vq}\,{\rm Re}\,\phi(\epsilon_\vq,\vq)\int\limits_{\sqrt{m_g^2+\xi^2}}^{\sqrt{2}\,m_g}\!\!\! d\omega_\ell\,\frac{\omega_\ell^2}{\omega^2_\ell-m_g^2}
\,\delta[\omega^*-\omega_{\ell}]\non
&\sim& g^2 \int\limits_0^{\xi_{\rm max}}\frac{d\xi}{\epsilon_\vq}\,{\rm Re}\,\phi(\epsilon_\vq,\vq)
\frac{(\omega-\epsilon_\vq)^2}{(\omega-\epsilon_\vq)^2-m_g^2}\;,
\eea
%where as in the transverse case the last estimate is valid for energies $\omega  \sim  m_g + \Lambda_y$ with $y < 1$. 
where the upper limit $\xi_{\rm max}\equiv (\omega^2-m_g^2)/(2\omega) \sim \Lambda_y$ is obtained analogously to the transversal case. Since $\xi_{\rm max}\ll \omega$ and $\omega \agt m_g$ one may approximate the energy fraction under the integral by $\omega^2/(\omega^2-m_g^2)\sim\omega/(\omega-m_g)$. Then the integral over $\xi$ may be readily performed yielding
\bea \label{l3}
{\cal A}^{\ell}_{\rm pole}(\omega,\vk)
\sim  g\,\phi\,\frac{\omega}{\omega-m_g}\sim  g\,\phi\,\left(\frac{M}{\phi}\right)^y\;,
\eea
where the BCS-log has cancelled one power of $g$. Analogously to the transversal case, for $\zeta \lesssim M$ one finds ${\cal A}^{\ell}_{\rm pole}=0$ if $\omega$ is exponentially close to $m_g$. For $\omega > \sqrt{m_g^2+\zeta^2}\agt \sqrt{2}m_g$ it follows  ${\cal A}^{\ell}_{\rm pole} \sim g\,\phi$ again.
%To continue we consider two cases for $\omega-m_g \sim \Lambda_y$, the first with $0\leq y \leq \bar g$ and the second with $\bar g \leq y < 1$\;. In the first case with $0\leq y \leq \bar g$ it is $\xi_{\rm max} \sim \Lambda_y$ and one can split the integral over $\xi$ as
%\bea\label{case1}
%{\cal A}^{\ell}_{\rm pole}(\omega,\vk)&\simeq&-\frac{g^2\,\phi}{2}\int\limits_{\Lambda_1}^{\Lambda_{\bar g}}\frac{d\xi}{\xi}\,
%\frac{\omega^2}{\omega^2-m_g^2}-\frac{g^3\,\phi}{2}\int\limits_{\Lambda_{\bar g}}^{\Lambda_0}\frac{d\xi}{\xi}\,
%\frac{(\omega+m_g^2)^2}{(\omega-m_g^2)^2}\non
%&\sim& g\,\phi\,\frac{\omega^2}{\omega^2-m_g^2}\;,
%\eea
%where in the first step the integrand in the second term is estimated by its maximum value. In the second case, $\omega-m_g \sim \Lambda_y$ with $\bar g \leq y < 1$, one finds
%\bea
%{\cal A}(\omega,\vk)&\simeq&-\frac{g^2\,\phi}{2}\int\limits_{\Lambda_1}^{\Lambda_{\bar g}}\frac{d\xi}{\xi}\,\frac{\omega^2}{(\omega-\xi)^2-m_g^2} \sim g\,\phi\,\frac{\omega^2}{\omega^2-m_g^2}\;.
%\eea
%
For much larger energies, $\omega \gg m_g$,
%integral over $\xi$ starts at some value $\xi \lesssim m_g$, cf.\ Fig.\ \ref{xiregion} and the estimate Eq. (\ref{l3}) becomes invalid.
%as well as for large gluon momenta $p\agt\Lambda_{\rm q}$ 
%It follows for the considered momentum regime $p < m_g$ the contribution ${\cal A}_{\rm pole}^\ell$ vanishes because $\omega^* = \omega_l$ cannot be fulfilled, as $\omega^*\equiv \omega -\epsilon_\vq\agt 2m_g$ and $\omega_\ell\leq\sqrt{2}\,m_g$,
gluon momenta $p \gg m_g$ have to be considered in the estimate of ${\cal A}^{\ell}_{\rm pole}$, cf.\ upper dotted lines in Fig.\ \ref{xiregion}. For those the longitudinal gluon spectral density becomes exponentially suppressed
\bea
\rho_{\rm pole}^{\ell}(\omega_l(\vp), \vp) \sim \frac{\exp\left(-\frac{2p^2}{3m_g^2} \right)}{p}
\eea
and one has for  $m_g \ll \omega<2\mu$ with $\omega_l(\vp) \simeq p$
\bea\label{lpole2}
{\cal A}^{\ell}_{\rm pole}(\omega,\vk)&\sim&g^2\int\limits_0^{\Lambda_{\rm q}}\frac{d\xi}{\epsilon_\vq}\,{\rm Re}\,\phi(\epsilon_\vq,\vq)\int\limits_{\sqrt{2}m_g}^{2\mu} d\omega_\ell\,\exp\left(-\frac{2\omega_\ell^2}{3m_g^2} \right)
\,\delta[\omega^*-\omega_{\ell}]\non
&\sim& g^2\phi\int\limits_{\Lambda_1}^{\Lambda_{\bar g}}\frac{d\xi}{\xi}\,\exp\left[-\frac{2(\omega-\xi)^2}{3m_g^2} \right]
\sim g\,\phi \,\exp\left(-\frac{2\omega^2}{3m_g^2} \right)\;,
\eea
which is the continuation of the estimate given in Eq.\ (\ref{l3}) to large energies $\omega\gg m_g$ and is valid for all $\zeta \leq \Lambda_{\rm q}$.  The results for ${\cal A}^{\ell,t}_{\rm pole}$ are summarized in Tables \ref{tableApolezeta<M} and \ref{tableApolezetasimM}.

\begin{table}  
\centerline{\begin{tabular}[t]{|c||c|c|c|c|}
\hline
$\omega$  &~  $<m_g+\Lambda_1~ $ &~  $\sim m_g+\Lambda_{1>y>0}$~ & ~$\agt\sqrt{2}m_g$~ & ~$m_g\ll\omega<2\mu$~
\\ \hline\hline
~ ${\cal A}_{\rm pole}^t$ ~  &  0 & $g\,\phi$ & $g\,\phi$ & $g\,\phi$
\\ \hline
${\cal A}_{\rm pole}^\ell$  & 0 	& ~$g\,\phi\left(\frac{M}{\phi}\right)^y$ ~ &  $g\,\phi$  & ~ $g\,\phi \,\exp\left(-\frac{2\omega^2}{3m_g^2} \right)$~ 
\\ \hline
${\cal A}_{\rm pole}$  &  0	& $g\,\phi\left(\frac{M}{\phi}\right)^y$ & $g\,\phi$ & $g\,\phi$
\\ \hline
\end{tabular}}
\caption[Estimates for ${\cal A}_{\rm pole}^{\ell,t}$ and  ${\cal A}_{\rm pole}$ at different energy scales and $\zeta\ll M$.]{Estimates for ${\cal A}_{\rm pole}^{\ell,t}$ and  ${\cal A}_{\rm pole}={\cal A}_{\rm pole}^{\ell}+{\cal A}_{\rm pole}^{t}$ at different energy scales and $\zeta\ll M$.}
\label{tableApolezeta<M}
\vspace{0.5cm}
\centerline{\begin{tabular}[t]{|c||c|c|c|c|}
\hline
$\omega$  & ~ $<m_g+\Lambda_1$ ~ & ~$\sim m_g+\Lambda_{1>y>0}$~ & ~$\agt\sqrt{2}m_g$~ & ~$m_g\ll\omega<2\mu$~
\\ \hline\hline
~ ${\cal A}_{\rm pole}^t$ ~  &  0 & 0  & $g\,\phi$ & $g\,\phi$
\\ \hline
${\cal A}_{\rm pole}^\ell$  &0 	& 0  &  $g\,\phi$  & ~$g\,\phi \,\exp\left(-\frac{2\omega^2}{3m_g^2} \right)$~ 
\\ \hline
${\cal A}_{\rm pole}$  &  0	& 0 & $g\,\phi$ & $g\,\phi$
\\ \hline
\end{tabular}}
\caption[Estimates for ${\cal A}_{\rm pole}^{\ell,t}$  and  ${\cal A}_{\rm pole}$ at different energy scales and $\zeta\lesssim M$.]{Estimates for ${\cal A}_{\rm pole}^{\ell,t}$  and  ${\cal A}_{\rm pole}={\cal A}_{\rm pole}^{\ell}+{\cal A}_{\rm pole}^{t}$ at different energy scales and $\zeta\lesssim M$.}
\label{tableApolezetasimM}
\end{table}

\subsubsection{Landau-damped gluons contributing to ${\cal B}$ }

In the Landau-damped gluon sector ${\cal M}^{\ell,t}_{{\cal B},T=0}$ in Eq.\ (\ref{ImMT0})  contributes to $\cal B$ as
\bea\label{Blt}
{\cal B}^{\ell,t}_{\rm cut}(\omega,\vk) \sim g^2\int\limits_0^{\Lambda_{\rm q}} \frac{d\xi}{\epsilon_\vq} \int\limits _0^\omega dq_0 \sum\limits_{\sigma=\pm}\frac{\sigma}{q_0-\sigma \epsilon_\vq}\,\rho_\phi(q_0,\vq) \int\limits _\lambda^{\Lambda_{\rm gl}} dp\,p \,\rho^{\ell,t}_{\rm cut}(\omega^\prime,\vp)\;,
\eea
where $\omega^\prime\equiv\omega-q_0<\omega$ and $\lambda\equiv {\max}(|\xi -\zeta|,\omega^\prime)$. Substituting the estimates of ${\cal A}(q_0,\vq)$ for $\rho_\phi(q_0,\vq)$  one estimates ${\cal B}^{\ell,t}_{\rm cut}(\omega,\vk)$ for different domains of $\omega$ and $\zeta$. Due to the condition $\lambda<\Lambda_{\rm gl}$ in Eq.\ (\ref{Blt}) and ${\cal A}(q_0,\vq) = 0$ for $q_0 > 2\mu$ it immediately follows that ${\cal B}^{\ell,t}_{\rm cut}(\omega,\vk) = 0$ for $\omega>\Lambda_{\rm gl}+2\mu\sim 3\mu$.  Inserting the approximative forms (\ref{appcut}) for $\rho^{\ell,t}_{\rm cut}$ into Eq.\ (\ref{Blt}) the integration over $p$ can be performed analogously to Eqs.\ (\ref{t},\ref{pintlong}). In the transverse case one finds
\bea\label{t2}
\hspace*{-0.6cm}
{\cal B}^t_{\rm cut}(\omega,\vk)\sim
 g^2\int\limits_0^{\Lambda_{\rm q}} \frac{d\xi}{\epsilon_\vq} \int\limits _0^\omega dq_0 \sum\limits_{\sigma=\pm}\frac{\sigma\,{\cal A}(q_0,\vq) }{q_0-\sigma \epsilon_\vq}\left[\arctan\left(\frac{\Lambda_{\rm gl}^3}{M^2\omega^\prime}\right)-\arctan\left(\frac{\lambda^3}{M^2\omega^\prime}\right)\right]\;.
%\\ {\cal B}^\ell_{\rm cut}(\omega,\vk)&\sim& g^2\int\limits_0^{\Lambda_{\rm q}} \frac{d\xi}{\epsilon_\vq} \int\limits _0^\omega dq_0 \sum\limits_{\sigma=\pm}\frac{\sigma}{q_0-\sigma \epsilon_\vq}\,\rho_\phi(q_0,\vq) \,\omega^\prime\,\frac{M-\lambda}{M^2}\label{l2}
\eea
The analysis of the domains of $\omega,\,\zeta$ and $\xi$ where the arctans in the squared brackets in Eq.\ (\ref{t2}) do not cancel is analogous the one after Eq.\ (\ref{t}).
For all $\xi,\,\zeta \leq \Lambda_{\rm q}$ and $\omega <3\mu$ it is $\Lambda_{\rm gl}^3/M^2\omega^\prime \gg 1$ and the first arctan in the squared brackets may be set equal to $\pi/2$. Considering $\zeta, \omega \ll M$ the argument of the second arctan is $\lambda^3/M^2\omega^\prime\sim \xi^3/M^2\omega^\prime$. As long as $\xi < (M^2\omega^\prime)^{1/3}$, this is not $\gg 1$ and the combination of both arctans is $\sim 1$. Increasing the energy to $\omega \sim M$ one finds that $\lambda^3/M^2\omega^\prime$ becomes large only for $q_0\rightarrow \omega$, because then $\lambda^3/M^2\omega^\prime\sim \xi^3/M^2(\omega-q_0) \gg 1$. However, since the integration over $q_0$ stops here anyway this case does not have to be analyzed further. Consequently, also for  $\omega \sim M$ one may estimate the arctans to be $\sim 1$. For $\omega \gg M$ one has  $\lambda^3/M^2\omega^\prime \sim [(\omega-q_0)/M]^2$ which is $\lesssim 1$ only for $\omega-M \lesssim q_0 \lesssim \omega$. For values of $\omega$ outside this range, the arctans cancel. Considering $\zeta\lesssim M$ one finds for $\omega \ll M$ that  $\lambda^3/M^2\omega^\prime\sim M/\omega^\prime \gg 1$ and the arctans cancel. (In the region where $|\xi-\zeta|$ is sufficienty small the arctans do not cancel. However, since this requires $\xi \lesssim M$, the BCS-log cannot be generated. Therefore, this special case can always be neglected.) In the region $\omega \sim M$ it is  $\lambda^3/M^2\omega^\prime\sim M/\omega^\prime \sim 1$ and the arctans do not cancel. For $\omega \gg M$ the respective analysis as made for $\zeta \ll M$ applies.

Before proceeding further one proves that the generation of the BCS-log can be prevented if additional logarithmic dependences appear under the integral over $\xi$ in the following form
\bea\label{noBCS1}
\int\limits_{\Lambda_1}^{\Lambda_0}\frac{d\xi}{\xi}\,\ln\left|\frac{\xi + \Lambda_y}{\xi - \Lambda_y}\right|
=\int\limits_{\Lambda_1}^{\Lambda_y}\frac{d\xi}{\xi}\,\ln\left(\frac{\xi + \Lambda_y}{\Lambda_y-\xi }\right)+ \int\limits_{\Lambda_y}^{\Lambda_0}\frac{d\xi}{\xi}\,\ln\left(\frac{\xi + \Lambda_y}{\xi - \Lambda_y}\right)\;,
\eea
where $0\leq y \leq 1$. Introducing the dilogarithm \cite{Nielson}
\bea
{\rm Li}_2(x)\equiv \int\limits_x^0 \frac{d\xi}{\xi}\,\ln(1-\xi)\;,
\eea
which has the values $-\frac{1}{12} \pi^2 \equiv {\rm Li}_2\left(-1\right)\leq {\rm Li}_2\left(x\right)\leq{\rm Li}_2\left(1\right) \equiv  \frac{1}{6} \pi^2$ for $-1 \leq x\leq 1$, one may express the first term on the r.h.s.\ of Eq.\ (\ref{noBCS1}) as
\bea
\int\limits_{\Lambda_1}^{\Lambda_y}\frac{d\xi}{\xi}\,\ln\left(\frac{\xi + \Lambda_y}{\Lambda_y-\xi }\right) &=&
\int\limits_{\Lambda_1/\Lambda_y}^1\frac{d\xi}{\xi}\,\ln\left(\frac{1+\xi}{1-\xi }\right)=
{\rm Li}_2\left(1\right)-{\rm Li}_2\left(-1\right)+{\rm Li}_2\left(-\frac{\Lambda_1}{\Lambda_y}\right)-{\rm Li}_2\left(\frac{\Lambda_1}{\Lambda_y}\right)
\non
&=& \frac{\pi^2}{4}+{\rm Li}_2\left(-\frac{\Lambda_1}{\Lambda_y}\right)-{\rm Li}_2\left(\frac{\Lambda_1}{\Lambda_y}\right)
\;.\label{noBCS2}
\eea
Since $0<\Lambda_1/\Lambda_y \leq 1$ this term is of order 1 and no BCS-log has been generated in this term.
The second term on the r.h.s.\ of Eq.\ (\ref{noBCS1}) is
\bea
 \int\limits_{\Lambda_y}^{\Lambda_0}\frac{d\xi}{\xi}\,\ln\left(\frac{\xi + \Lambda_y}{\xi - \Lambda_y}\right) &=&
\int\limits_1^{\Lambda_0/\Lambda_y}\frac{d\xi}{\xi}\,\ln\left(\frac{1+\xi}{\xi-1}\right)=-\int\limits_1^{\Lambda_y/\Lambda_0}\frac{d\chi}{\chi}\,\ln\left(\frac{1+1/\chi}{1/\chi-1}\right)\non
&=&
\int\limits_{\Lambda_y/\Lambda_0}^1\frac{d\chi}{\chi}\,\ln\left(\frac{1+\chi}{1-\chi }\right)=
{\rm Li}_2\left(1\right)-{\rm Li}_2\left(-1\right)+{\rm Li}_2\left(-\frac{\Lambda_y}{\Lambda_0}\right)-{\rm Li}_2\left(\frac{\Lambda_y}{\Lambda_0}\right)
\non
&=& \frac{\pi^2}{4}+{\rm Li}_2\left(-\frac{\Lambda_y}{\Lambda_0}\right)-{\rm Li}_2\left(\frac{\Lambda_y}{\Lambda_0}\right)
\;.
\eea
%\bea
% \int\limits_{\Lambda_y}^{\Lambda_0}\frac{d\xi}{\xi}\,\ln\left(\frac{\xi + \Lambda_y}{\xi - \Lambda_y}\right)=
%-\frac{\pi^2}{12}-\ln\left( \frac{\Lambda_0}{\Lambda_y}\right)\ln\left( \frac{\Lambda_0-\Lambda_y}{\Lambda_y}\right)-{\rm Li}_2\left(-\frac{\Lambda_0}{\Lambda_y}\right)-{\rm Li}_2\left( \frac{\Lambda_0-\Lambda_y}{\Lambda_y}\right).\label{noBCS2}
%\eea
%For $y \agt 0$ this term is of order 1 (approaches 0 for $y \rightarrow$ 0). For $y \lesssim 1 $ it is $\Lambda_0 \gg \Lambda_y$ or equivalently $x\equiv \Lambda_0/\Lambda_y \gg 1$. Then we may use the identity $L_2(x)+Li_2(1-x) = $ This corresponds to the limit of $x\gg 1$, where ${\rm Li}_2(x) \approx \frac{1}{2}[\ln(x)]^2$. Consequently the two dilogarithms on the r.h.s.\ of Eq.\ (\ref{noBCS2}) cancel the product of the two logarithms and the complete integral is of order 1. 
In the second step one substituted $\chi \equiv 1/\xi$ with $d\chi/\chi = -d\xi/\xi$.
Similarly to Eq.\ (\ref{noBCS2}) it is $0<\Lambda_y/\Lambda_0 \leq 1$. Hence, also this term is of order 1 and no BCS-log has been generated here, either, proving the above statement.

Beginning with energies $\omega\sim \phi$ one may use Eq.\ (\ref{Aphi}) to estimate ${\cal A} \sim{\cal A}^t_{\rm cut}\sim g^2\phi$. Assuming $\zeta \ll M$ one has
\bea
{\cal B}^t_{\rm cut}(\phi,\vk)&\sim& g^4 \phi \int\limits_0^{\Lambda_{\rm q}} \frac{d\xi}{\epsilon_\vq} \int\limits _0^\phi dq_0 \sum\limits_{\sigma=\pm}\frac{\sigma}{q_0-\sigma \epsilon_\vq}\,\left[\arctan\left(\frac{\Lambda_{\rm gl}^3}{M^2\omega^\prime}\right)-\arctan\left(\frac{\lambda^3}{M^2\omega^\prime}\right)\right]\non
&\sim&
g^4 \phi \int\limits_0^{\Lambda_{1/3}} \frac{d\xi}{\epsilon_\vq}\,\ln\left| \frac{\epsilon_\vq-\phi}{\epsilon_\vq + \phi}\right| \sim g^4\phi\;.\label{Bcutphi}
\eea
The arctans in the squared brackets cancel for $\lambda^3 \gg M^2\phi$, which leads to the upper boundary $\Lambda_{1/3}$ in the inegral over $\xi$. As discussed before the generation of the BCS-log was prevented by the additional logarithm under the integral. For $\zeta \lesssim M$ the arctans in Eq.\ (\ref{Bcutphi}) would have cancelled yielding ${\cal B}^t_{\rm cut}(\phi,\vk)\simeq 0$.

For $\omega\sim \Lambda_y$ with $ 0 \leq y < 1$ one conservatively estimates ${\cal A }\sim {\cal A}^t_{\rm cut} \sim g\,\phi$, cf.\ Eq.\ (\ref{Aphi2}), and obtain similarly to Eq.\ (\ref{Bcutphi}), assuming $\zeta \ll M$
\bea\label{est1}
{\cal B}^t_{\rm cut}(\Lambda_y,\vk)&\sim& 
g^3 \phi \int\limits_0^{\Lambda_{\rm q}} \frac{d\xi}{\epsilon_\vq} \int\limits _0^{\Lambda_y} dq_0 \sum\limits_{\sigma=\pm}\frac{\sigma}{q_0-\sigma \epsilon_\vq}\,\left[\arctan\left(\frac{\Lambda_{\rm gl}^3}{M^2\omega^\prime}\right)-\arctan\left(\frac{\lambda^3}{M^2\omega^\prime}\right)\right]\non
&\sim& 
g^3 \phi \int\limits_0^{\Lambda_{y/3}} \frac{d\xi}{\epsilon_\vq}\,\ln\left| \frac{\epsilon_\vq-\Lambda_y}{\epsilon_\vq + \Lambda_y}\right| \sim g^3\phi\;,
\eea
where again no BCS-log was generated. Here, the arctans cancel for $\lambda \gg M^2 \Lambda_y$, leading to the constraint $\xi < \Lambda_{y/3}$. Again, for $\zeta \lesssim M$ the arctans in Eq.\ (\ref{Bcutphi}) would have cancelled yielding ${\cal B}^t_{\rm cut}(\Lambda_y,\vk)\simeq 0$.

For energies $\omega \sim m_g +\Lambda_y$ with $0\leq y <1$ the combination of the arctangents in Eq.\ (\ref{t2}) is $\sim 1$ for all $\zeta\leq \Lambda_{\rm q}$.
 Integrating over $q_0$ from 0 to $m_g$ one finds similarly to Eq.\ (\ref{est1})
\bea
g^3\phi \int\limits_0^{\Lambda_{\rm q}} \frac{d\xi}{\epsilon_\vq}\,\ln\left| \frac{\epsilon_\vq-\Lambda_0}{\epsilon_\vq + \Lambda_0}\right| \sim g^3\phi\;.
\eea
The dominant contribution, however, is generated when integrating over $q_0$ from $m_g$ to $m_g +\Lambda_y$ where one has ${\cal A} \sim {\cal A}^\ell_{\rm pole}\sim g\,\phi \,(M/\phi)^y$, cf.\ Eq.\ (\ref{l3}),
\bea\label{est2}
{\cal B}^t_{\rm cut}(m_g+\Lambda_y,\vk)&\sim& 
g^3 \phi \int\limits_0^{\Lambda_{\rm q}} \frac{d\xi}{\epsilon_\vq} \int\limits _{m_g+\Lambda_1}^{m_g+\Lambda_y} dq_0 \sum\limits_{\sigma=\pm}\frac{\sigma}{q_0-\sigma \epsilon_\vq}\,\frac{q_0}{q_0-m_g}\non
&\sim &
g^2 \phi \int\limits_0^{\Lambda_{\rm q}} \frac{d\xi}{\epsilon_\vq}\,\sum\limits_{\sigma=\pm}\frac{\sigma\,m_g}{\sigma \epsilon_\vq-m_g}\int\limits _1^y dy^\prime
\sim g^2 \phi \int\limits_0^{\Lambda_{\rm q}} d\xi\,\frac{m_g}{\xi^2-m_g^2}\non
&\sim& g^2\,\phi\;.
\eea
For energies $\omega \agt m_g$ but not exponentially close to $m_q$, the integration over $q_0$ from $2m_g$ to $\omega$ may be performed after estimating ${\cal A} \sim g\,\phi$, cf.\ Eqs.\ (\ref{Acuthigh},\ref{lpole}), yielding
\bea\label{est4}
g^3 \phi \int\limits_0^{\Lambda_{\rm q}} \frac{d\xi}{\epsilon_\vq} \int\limits _{2m_g}^{\omega} dq_0\,
\sum\limits_{\sigma=\pm}\frac{\sigma}{q_0-\sigma \epsilon_\vq}&\sim&
g^3 \phi \int\limits_0^{\Lambda_{\rm q}} \frac{d\xi}{\epsilon_\vq} \left[
\ln\left(\frac{\omega-\epsilon_q}{\epsilon_q+\omega} \right)
-\ln\left(\frac{\omega-M+\epsilon_q}{\omega-M-\epsilon_q} \right) \right]\non
&\sim& g^3 \phi\;.
\eea
Hence, one found that generally for energies $\omega \agt m_g$ it is 
\bea
\label{est3}
{\cal B}^t_{\rm cut}(\omega,\vk)&\sim& g^2\phi\;.
\eea
In the limit of very large energies, $\omega \gg M$, it is $\lambda^3/(M^2\omega^\prime) \equiv (\omega -q_0)^2/M^2 \lesssim 1$ only for $\omega -M \lesssim q_0 \lesssim \omega$. For values of $q_0$ outside this range the arctans in Eq.\ (\ref{t2}) cancel. Inside this range one may estimate ${\cal A} \sim g\, \phi$ and find ${\cal B}^t_{\rm cut}$  to be strongly suppressed
\bea\label{est4b}
{\cal B}^t_{\rm cut}(\omega,\vk) \sim g^3 \phi \int\limits_0^{\Lambda_{\rm q}} \frac{d\xi}{\epsilon_\vq} \int\limits _{\omega-M}^{\omega} dq_0\,
\sum\limits_{\sigma=\pm}\frac{\sigma}{q_0-\sigma \epsilon_\vq}\sim
g^3 \phi \int\limits_{\Lambda_1}^{\Lambda_0} d\xi \int\limits _{\omega-M}^{\omega} \frac{dq_0}{q_0^2}
\sim g^3 \phi\left(\frac{M}{\omega}\right)^2 .
\eea
% Since here it is ${\cal A} =0$,  the integrand vanishes and we found  $ {\cal B}^t_{\rm cut}= 0$ for these energies.
%Im$\,\phi = {\cal B}$ we have 
%\bea
%{\cal B}^t_{\rm cut}(\omega,\vk)&\sim& 
%g^2 \int\limits_0^{\Lambda_{\rm q}} \frac{d\xi}{\epsilon_\vq} \int\limits _{\omega-\Lambda_{\rm gl}}^{\omega} dq_0 \sum\limits_{\sigma=\pm}\frac{\sigma\,{\cal B}(q_0,\vq)}{q_0-\sigma \epsilon_\vq}
%\sim 
%g^2  \int\limits_0^{\Lambda_{\rm q}} d\xi \int\limits _{\omega-\Lambda_{\rm gl}}^{\omega} \frac{dq_0}{q_0^2}\,{\cal B}(q_0,\vq)\non
%\eea

In the case of longitudinal gluons one obtains
\bea\label{l2}
{\cal B}^\ell_{\rm cut}(\omega,\vk)&\sim&
 g^2\int\limits_0^{\Lambda_{\rm q}} \frac{d\xi}{\epsilon_\vq} \int\limits _0^\omega dq_0 \sum\limits_{\sigma=\pm}\frac{\sigma\,{\cal A}(q_0,\vq) }{q_0-\sigma \epsilon_\vq}\;\omega^\prime\;{\cal I}(\lambda)\;,
%\\ {\cal B}^\ell_{\rm cut}(\omega,\vk)&\sim& g^2\int\limits_0^{\Lambda_{\rm q}} \frac{d\xi}{\epsilon_\vq} \int\limits _0^\omega dq_0 \sum\limits_{\sigma=\pm}\frac{\sigma}{q_0-\sigma \epsilon_\vq}\,\rho_\phi(q_0,\vq) \,\omega^\prime\,\frac{M-\lambda}{M^2}\label{l2}
\eea
where ${\cal I}(\lambda)$ is defined in Eq.\ (\ref{pintlong}).
Analogously to the analysis of ${\cal A}_{\rm cut}^\ell$ one finds for $\omega \sim \phi$
\bea
{\cal B}^\ell_{\rm cut}(\phi,\vk)&\sim& g^4\phi\int\limits_0^{\Lambda_{\rm q}} \frac{d\xi}{\epsilon_\vq} \int\limits _0^\phi dq_0 \sum\limits_{\sigma=\pm}\frac{\sigma}{q_0-\sigma \epsilon_\vq}\,\frac{\phi}{M} 
\sim
 g^4\phi\,\frac{\phi}{M} \int\limits_0^{\Lambda_{\rm q}} \frac{d\xi}{\epsilon_\vq} \,\ln\left| \frac{\epsilon_\vq-\phi}{\epsilon_\vq + \phi}\right|\non
&\sim&
 g^4\phi\,\frac{\phi}{M} 
\eea
and similarly for $\omega\sim \Lambda_y$ with $ 0\leq y < 1$
\bea%\label{est3}
{\cal B}^\ell_{\rm cut}(\Lambda_y,\vk)&\sim& g^3 \phi \int\limits_0^{\Lambda_{\rm q}} \frac{d\xi}{\epsilon_\vq} \int\limits _0^{\Lambda_y} dq_0 \sum\limits_{\sigma=\pm}\frac{\sigma}{q_0-\sigma \epsilon_\vq} \frac{\Lambda_y}{M}
\sim g^3 \phi \, \frac{\Lambda_y}{M}\int\limits_0^{\Lambda_{\rm q}} \frac{d\xi}{\epsilon_\vq} \,\ln\left| \frac{\epsilon_\vq-\Lambda_y}{\epsilon_\vq + \Lambda_y}\right|
\non
&\sim& g^3\phi\left(\frac{\phi}{M}\right)^{y}\;.
\eea
For $\omega \sim m_g +\Lambda_y$ with $0\leq y < 1$ it is $\omega^\prime\, {\cal I}(\lambda) \sim 1$ and one finds as in Eq.\ (\ref{est2})
\bea
{\cal B}^\ell_{\rm cut}(m_g+\Lambda_y,\vk)&\sim& g^2 \phi \;.
\eea
Also for energies  $\omega \agt M$ the analysis is very similar to the transversal case since $\omega^\prime\, {\cal I}(\lambda) \sim 1$, and one finds again
\bea\label{est7}
{\cal B}^\ell_{\rm cut}(\omega,\vk)\sim g^2\phi\,.
\eea
In the limit of very large energies, $\omega \gg \Lambda_{\rm gl}\sim \mu$, only the range $\omega-\Lambda_{\rm gl}<q_0 < \omega$ contributes, cf. Eq.\ (\ref{Blt}). As soon as $\omega-\Lambda_{\rm gl} > 2m_g$ one may estimate ${\cal A} \sim g\,\phi$ and obtain with  $\omega^\prime\, {\cal I}(\lambda) \sim 1$
\bea\label{large1}
{\cal B}^\ell_{\rm cut}(\omega,\vk)&\sim&
g^3 \phi \int\limits_0^{\Lambda_{\rm q}} \frac{d\xi}{\epsilon_\vq} \int\limits _{\omega-\Lambda_{\rm gl}}^{\omega} dq_0 \sum\limits_{\sigma=\pm}\frac{\sigma}{q_0-\sigma \epsilon_\vq} 
\sim g^3 \phi \int\limits_0^{\Lambda_{\rm q}} d\xi \int\limits _{\omega-\Lambda_{\rm gl}}^{\omega} \frac{dq_0}{q_0^2} \non
&\sim& g^3\phi \,\frac{M\Lambda_{\rm gl}}{\omega^2}\sim  g^3\phi\, \frac{M}{\omega}\;.
\eea
Hence, also ${\cal B}^\ell_{\rm cut}$ becomes small in the limit of large energies, cf.\ Eq.\ (\ref{est4b}).

\begin{table}  
\centerline{\begin{tabular}[t]{|c||c|c|c|c|}
\hline
$\omega$  &~  $\phi $~ &~  $\Lambda_{1>y>0}$~ & ~$\agt m_g$~ & ~$m_g\ll\omega<3\mu$~
\\ \hline\hline
~ ${\cal B}_{\rm cut}^t$ ~  &  $g^4\phi$ & $g^3\phi$ & $g^2\phi$ & $g^3\phi\left(\frac{M}{\omega} \right)^2$
\\ \hline
${\cal B}_{\rm cut}^\ell$  & ~$g^4\phi\,\frac{\phi}{M}$~ & ~$g^3\phi\left(\frac{\phi}{M}\right)^y$ ~ &  $g^2\phi$  & ~ $g^3\phi\left(\frac{M}{\omega} \right) $~ 
\\ \hline
${\cal B}_{\rm cut}$  &  $g^4\phi$	& $g^3\phi$ & $g^2\phi$ & $g^3\phi\left(\frac{M}{\omega} \right)$
\\ \hline
\end{tabular}}
\caption[Estimates for ${\cal B}_{\rm cut}^{\ell,t}$ and  ${\cal B}_{\rm cut}$ at different energy scales and $\zeta\ll M$.]{Estimates for ${\cal B}_{\rm cut}^{\ell,t}$ and  ${\cal B}_{\rm cut}={\cal B}_{\rm cut}^{\ell}+{\cal B}_{\rm cut}^{t}$ at different energy scales and $\zeta\ll M$.}
\label{tableBcutzeta<M}
\vspace{0.5cm} 
\centerline{\begin{tabular}[t]{|c||c|c|c|c|}
\hline
$\omega$  &~  $\phi $~ &~  $\Lambda_{1>y>0}$~ & ~$\agt m_g$~ & ~$m_g\ll\omega<3\mu$~
\\ \hline\hline
~ ${\cal B}_{\rm cut}^t$ ~  &  0 & 0 & $g^2\phi$ & $g^3\phi\left(\frac{M}{\omega} \right)^2$
\\ \hline
${\cal B}_{\rm cut}^\ell$  & ~$g^4\phi\,\frac{\phi}{M}$~ & ~$g^3\phi\left(\frac{\phi}{M}\right)^y$ ~ &  $g^2\phi$  & ~ $g^3\phi\left(\frac{M}{\omega} \right) $~ 
\\ \hline
${\cal B}_{\rm cut}$  &  $g^4\phi\,\frac{\phi}{M}$	& ~$g^3\phi\left(\frac{\phi}{M}\right)^y$~ & $g^2\phi$ & $g^3\phi\left(\frac{M}{\omega} \right)$
\\ \hline
\end{tabular}}
\caption[Estimates for ${\cal B}_{\rm cut}^{\ell,t}$  and  ${\cal B}_{\rm cut}$ at different energy scales and $\zeta\lesssim M$.]{Estimates for ${\cal B}_{\rm cut}^{\ell,t}$  and  ${\cal B}_{\rm cut}={\cal B}_{\rm cut}^{\ell}+{\cal B}_{\rm cut}^{t}$ at different energy scales and $\zeta\lesssim M$.}
\label{tableBcutzetasimM}
\end{table}

\subsubsection{Undamped gluons contributing to $\cal B$}

In the undamped gluon sector the term  ${\cal M}^{\ell,t}_{{\cal B},T=0}$ in Eq.\ (\ref{ImMT0}) gives the contribution 
\bea\label{Bpole}
{\cal B}^{\ell,t}_{\rm pole}(\omega,\vk) \sim g^2\int\limits_0^{\Lambda_{\rm q}} \frac{d\xi}{\epsilon_\vq} \int\limits _0^\omega dq_0 \sum\limits_{\sigma=\pm}\frac{\sigma\,\rho_\phi(q_0,\vq)}{q_0-\sigma \epsilon_\vq} \int\limits _{|\zeta-\xi|}^{2\mu} dp\,p \,\rho^{\ell,t}_{\rm pole}(\omega^\prime,\vp)\,\delta[\omega^\prime-\omega_{\ell,t}(\vp)]\;,
\eea
which is nonzero only for energies $\omega > m_g$. Due to the restriction $p< 2\mu$ in Eq.\ (\ref{Bpole}) it follows with similar arguments as for ${\cal B}^{\ell,t}_{\rm cut}$ that ${\cal B}^{\ell,t}_{\rm pole}(\omega,\vk)=0$ for $\omega >4\mu$. 
For transversal gluons one finds using the same approximations as for ${\cal A}^{t}_{\rm pole}$ beginning with energies $\omega \sim m_g +\Lambda_{1<y\leq 0}$ and assuming $\zeta \ll M$
\bea
{\cal B}^{t}_{\rm pole}(\omega,\vk)&\sim& g^2\int\limits_{\Lambda_1}^{\Lambda_0} \frac{d\xi}{\xi} \int\limits _0^\omega dq_0 \sum\limits_{\sigma=\pm}\frac{\sigma\,{\cal A}(q_0,\vq)}{q_0-\sigma \epsilon_\vq} \int\limits _{\sqrt{\xi^2+m_g^2}}^{2\mu}\!\!\! d\omega_t \,\delta[\omega^\prime-\omega_{t}]\non
&\sim& g^3\phi\int\limits_{\Lambda_1}^{\Lambda_{y/2}} \frac{d\xi}{\xi} \int\limits _0^{\omega-\sqrt{m_g^2+\xi^2}}\!\!\! dq_0 \sum\limits_{\sigma=\pm}\frac{\sigma}{\sigma \xi-q_0}
%\non&\sim& 
\sim g^3\phi\int\limits_{\Lambda_1}^{\Lambda_{y/2}} \frac{d\xi}{\xi} \,
\,\ln\left|\frac{\omega-\sqrt{m_g^2+\xi^2}-\xi}{\omega-\sqrt{m_g^2+\xi^2}+\xi}\right|\non
%&\sim& g^3\phi\int\limits_{\Lambda_{\bar g}}^{\Lambda_0} \frac{d\xi}{\xi} \,
%\,\ln\left|\frac{\omega-\sqrt{m_g^2+\xi^2}-\xi}{\omega-\sqrt{m_g^2+\xi^2}+\xi}\right| 
&\sim& g^3\phi\;,\label{Bcuttrans1}
\eea
where to guarantee $\omega-\sqrt{\xi^2+m_g^2} > 0$ the upper boundary of the integral over $\xi$ is reduced from $\Lambda_{\rm q} \sim \Lambda_0$  to the scale $\sqrt{m_g\Lambda_y}\sim \Lambda_{y/2}$. Furthermore, one estimated ${\cal A}\sim g\,\phi$, which is valid since $q_0<\Lambda_y \leq m_g$ for the considered energies. For $\omega > 2m_g$ the integral over $q_0$ also runs over values $q_0 >m_g$ receiving contributions from ${\cal A}^\ell_{\rm pole}$, cf.\ Eq.\ (\ref{l3}). As a consequence one finds 
\bea\label{est8}
{\cal B}^{t}_{\rm pole}(\omega,\vk)\sim g^2\phi
\eea
 analogously to Eq.\ (\ref{est2}). For $\omega \gg m_g$ the additional contributions from $2m_q<q_0 <\omega$ are only $\sim g^3 \phi$ as can be seen in the same way as in Eq.\ (\ref{est4}) and ${\cal B}^{t}_{\rm pole}\sim g^2\phi$. However, for $\omega\agt 2\mu+2m_g$ the condition $\omega^\prime = \omega_\ell$ can be fulfilled only for $q_0>\omega - 2\mu\agt 2m_g $, where ${\cal A }\sim g\,\phi$, and one may estimate
\bea\label{large2}
{\cal B}^{t}_{\rm pole}(\omega,\vk)&\sim& 
g^3\phi\int\limits_{\Lambda_1}^{\Lambda_0} \frac{d\xi}{\xi} \int\limits _{\omega-2\mu}^\omega dq_0 \sum\limits_{\sigma=\pm}\frac{\sigma}{q_0-\sigma \epsilon_\vq}
\sim g^3\phi\int\limits_{\Lambda_1}^{\Lambda_0} d\xi \int\limits _{\omega-2\mu}^\omega \frac{dq_0}{q_0^2}\non
&\sim& g^3\phi\,\frac{M\,2\mu}{\omega^2}\sim g^3\phi\,\frac{M}{\omega}\;.
\eea
For $\zeta \lesssim M$ and $\omega \sim  m_g +\Lambda_{1<y<0}$ the condition  $\omega- \sqrt{m_g^2+|\xi-\zeta|^2}>0$ leads to $|\zeta| -\Lambda_{y/2}<\xi<|\zeta|+\Lambda_{y/2}$. Then the integral over $\xi$ finally yields
${\cal B}^{t}_{\rm pole} \sim g^3\phi\,(\Lambda_{y/2}/M)\sim g^3\phi\,(\phi/M)^{y/2}$.  For $\omega >2 m_g$ and $\omega \gg m_g$ the same analyses as in the case of $\zeta \ll M$ apply.

In the longitudinal sector starting with energies $\omega =m_g +\Lambda_y$ with $0< y<1$ and momenta $p \lesssim m_g$ one may employ Eq.\ (\ref{rholapp}). Assuming $\zeta \ll M$ one obtains
\bea
{\cal B}^{\ell}_{\rm pole}(\omega,\vk)
&\sim& g^2\int\limits_{\Lambda_1}^{\Lambda_0} \frac{d\xi}{\xi}\int\limits _0^\omega dq_0 \sum\limits_{\sigma=\pm}\frac{\sigma\,{\cal A}(q_0,\vq) }{q_0-\sigma \epsilon_\vq}
\int\limits _{\sqrt{\xi^2+m_g^2}}^{\omega} d\omega_\ell\,\frac{\omega_\ell^2}{\omega^2_\ell-m_g^2} 
\,\delta[\omega^\prime-\omega_{\ell}]\non
&\sim&g^2\int\limits_{\Lambda_1} ^{\Lambda_{y/2}} \frac{d\xi}{\xi}\int\limits _0^{\omega-\sqrt{\xi^2+m_g^2}} \!\!\!dq_0\sum\limits_{\sigma=\pm}\frac{\sigma\,{\cal A}(q_0,\vq)}{q_0-\sigma \epsilon_\vq}\;\frac{(\omega-q_0)^2}{(\omega-q_0)^2-m_g^2} \;,\label{est5}
\eea
where upper boundaries of the integrals over $\xi$ and $q_0$ are analogous to the transversal case, cf.\ Eq.\ (\ref{Bcuttrans1}). % $\omega_\ell$ is given by $\omega$, cf.\ Fig.\ \ref{xiregion}.
%, which in the considered regime for $\omega$ is smaller than $\sqrt{m_g^2 + p^2_{\rm max}} = \sqrt{2}m_g$. 
Since $q_0\leq\omega-\sqrt{\xi^2+m_g^2}$ is much smaller than $\omega \sim m_g+ \Lambda_y$, one may neglect $q_0$ against $\omega$ on the r.h.s\ of Eq.\ (\ref{est5}). Therefore, one may estimate ${\cal A}\sim g\,\phi$ and finally obtain
\bea
{\cal B}^{\ell}_{\rm pole}(\omega,\vk)
&\sim&g^3\phi\,\frac{\omega^2}{\omega^2-m_g^2}\int\limits_{\Lambda_1} ^{\Lambda_{y/2}} \frac{d\xi}{\xi}\,\ln\left|\frac{\omega-\sqrt{m_g^2+\xi^2}-\xi}{\omega-\sqrt{m_g^2+\xi^2}+\xi}\right|\non
&\sim&  g^3\phi\,\frac{\omega}{\omega-m_g}\sim  g^3\phi\,\left(\frac{M}{\phi}\right)^y\;.
\eea
%The first term in the  integral over $\omega_\ell$ has a pole at $\omega_\ell = m_g$, which is reached only for $\xi = 0$. The integral over $\omega_\ell$ is therefore logarithmic divergent for $\xi \rightarrow 0$. This logarithmic divergence in $\xi$ gives a finite result when integrating over $\xi$. Furthermore there is a pole at $\omega_\ell = \omega-\epsilon_\vq$ in the second term in the integral over $\omega_\ell$. Since it is located inside the integration region of $\omega_\ell$ the integral over $\omega_\ell$ remains finite. The integration over $\Lambda_1\leq \xi \leq \Lambda_{\bar g}$ finally gives to leading order
%\bea
%{\cal B}^{\ell}_{\rm pole}(\omega,\vk)
%\sim  g^3\,\phi\,\frac{\omega}{\omega-m_g}\sim  g^3\,\phi\,\left(\frac{M}{\phi}\right)^y\;.
%\eea
For larger energies $\omega \agt 2m_g$ the upper boundary of the integral over $q_0$ will just exceed $m_g$, where it is ${\cal A} \sim {\cal A}^\ell_{\rm pole}$, cf.\ Eq.\ (\ref{l3}). One finds analogously to ${\cal B}^t_{\rm pole}$ that this gives the main contribution
\bea
{\cal B}^\ell_{\rm pole}(\omega,\vk)&\sim&g^2\int\limits_{\Lambda_1}^{\Lambda_0} \frac{d\xi}{\xi}\int\limits _{m_g}^\omega dq_0 \sum\limits_{\sigma=\pm}\frac{\sigma\,{\cal A}(q_0,\vq) }{q_0-\sigma \epsilon_\vq}
\int\limits _{\sqrt{\xi^2+m_g^2}}^{\omega} d\omega_\ell\,\frac{\omega_\ell^2}{\omega^2_\ell-m_g^2}
\,\delta[\omega^\prime-\omega_{\ell}]\non
&\sim&g^3\phi\int\limits_{\Lambda_1} ^{\Lambda_0} \frac{d\xi}{\xi}\int\limits _{m_g}^{\omega-\sqrt{\xi^2+m_g^2}} \!\!\!dq_0\sum\limits_{\sigma=\pm}\frac{\sigma}{q_0-\sigma \epsilon_\vq}\,\frac{q_0}{q_0-m_g}\;\frac{(\omega-q_0)^2}{(\omega-q_0)^2-m_g^2}\non
&\sim & g^2\phi \;,\label{est6}
\eea
where one exploited ${(\omega-q_0)^2}/[{(\omega-q_0)^2-m_g^2}]\sim 1$ (since always $\omega -q_0 \sim m_g$) and estimated the integral over $q_0$ as in Eq.\ (\ref{est2}).
For energies $\omega\gg 2m_g$ one finds analogously to Eq.\ (\ref{lpole2}) for the contribution from the integration region $2m_g<q_0<\omega$, where one may approximate ${\cal A} \sim g^2\phi$,
\bea\label{lpole3}
&& g^3\phi\int\limits_0^{\Lambda_{\rm q}}\frac{d\xi}{\epsilon_\vq}\int\limits _{2m_g}^\omega dq_0 \sum\limits_{\sigma=\pm}\frac{\sigma}{q_0-\sigma \epsilon_\vq}\int\limits_{m_g}^{2\mu} d\omega_\ell\,\exp\left(-\frac{2\omega_\ell^2}{3m_g^2} \right)
\,\delta[\omega^\prime-\omega_{\ell}]\non
&\sim& g^3\phi\int\limits_{\Lambda_1}^{\Lambda_0}\frac{d\xi}{\xi}\,\int\limits _{2m_g}^\omega dq_0 \sum\limits_{\sigma=\pm}\frac{\sigma}{q_0-\sigma \epsilon_\vq}\,\exp\left[-\frac{2(\omega-q_0)^2}{3m_g^2} \right]\non
&\sim& g^3\phi\int\limits_0^{\Lambda_{\rm q}}d\xi\int\limits _{2m_g}^\omega \frac{dq_0}{q_0^2}\exp\left(-\frac{2\omega_\ell^2}{3m_g^2} \right)
\sim g^3\phi\left(\frac{M}{\omega}\right)^2\;.
\eea
Hence, in the considered large energy regime, $\omega\gg 2m_g$ the main contribution to ${\cal B}^\ell_{\rm pole}$ comes from Eq.\ (\ref{est6}), and it is ${\cal B}^\ell_{\rm pole}\sim g^2\phi$. Similarly to the transversal case, this holds up to energies $\omega > 2\mu +2m_g$ since then it is always $q_0 > 2m_g$ leading to
\bea
{\rm B}^\ell_{\rm pole}(\omega,\vk) \sim g^3\,\phi\, \left(\frac{M}{\omega}\right)^2\;.
\eea
Analogously to the transversal case one finds that for $\zeta \lesssim M$ and $\omega \sim  m_g +\Lambda_{1<y<0}$ the condition  $\omega- \sqrt{m_g^2+|\xi-\zeta|^2}>0$ leads to $|\zeta| -\Lambda_{y/2}<\xi<|\zeta|+\Lambda_{y/2}$. Then the integral over $\xi$ finally yields
${\cal B}^{\ell}_{\rm pole} \sim g^3\phi\,(M/\phi)^y\,(\Lambda_{y/2}/M)\sim g^3\phi\,(M/\phi)^{y/2}$.  For $\omega >2 m_g$ and $\omega \gg m_g$ the same analyses as in the case of $\zeta \ll M$ apply. The results of this subsection are summarized in Tables \ref{tableBpolezeta<M} and \ref{tableBpolezetasimM}.

\begin{table}  
\centerline{\begin{tabular}[ht]{|c||c|c|c|c|}
\hline
$\omega$  &~  $<m_g+\Lambda_1~ $ &~  $\sim m_g+\Lambda_{1>y\geq 0}$~ & ~$\agt 2m_g$~ & ~$m_g\ll\omega<4\mu$~
\\ \hline\hline
~ ${\cal B}_{\rm pole}^t$ ~  &  0 & $g^3\phi$ & $g^2\phi$ & $g^3\phi\,\frac{M}{\omega}$
\\ \hline
${\cal B}_{\rm pole}^\ell$  & 0 	& ~$g^3\phi\left(\frac{M}{\phi}\right)^y$ ~ &  $g^2\phi$  & ~ $g^3\phi \,\left(\frac{M}{\omega}\right)^2$~ 
\\ \hline
${\cal B}_{\rm pole}$  &  0	& $g^3\phi\left(\frac{M}{\phi}\right)^y$ & $g^2\phi$ & $g^3\phi\,\frac{M}{\omega}$
\\ \hline
\end{tabular}}
\caption{Estimates for ${\cal B}_{\rm pole}^{\ell,t}$ and  ${\cal B}_{\rm pole}$ at different energy scales and $\zeta\ll M$.}
\label{tableBpolezeta<M}
\vspace{1cm}
\centerline{\begin{tabular}[ht]{|c||c|c|c|c|}
\hline
$\omega$  & ~ $<m_g+\Lambda_1$ ~ & ~$\sim m_g+\Lambda_{1>y\geq0}$~ & ~$\agt2m_g$~ & ~$m_g\ll\omega<4\mu$~
\\ \hline\hline
~ ${\cal B}_{\rm pole}^t$ ~  &  0 & $g^3\phi\,\left(\frac{\phi}{M}\right)^{y/2}$ & $g^2\phi$ &  $g^3\phi\,\frac{M}{\omega}$
\\ \hline
${\cal B}_{\rm pole}^\ell$  & 0 	& ~$g^3\phi\left(\frac{M}{\phi}\right)^{y/2}$ ~ &  $g^2\phi$  & ~ $g^3\phi \,\left(\frac{M}{\omega}\right)^2$~ 
\\ \hline
${\cal B}_{\rm pole}$  &  0	& ~$g^3\phi\left(\frac{M}{\phi}\right)^{y/2}$ ~ & $g^2\phi$ &  $g^3\phi\,\frac{M}{\omega}$
\\ \hline
\end{tabular}}
\caption[Estimates for ${\cal B}_{\rm pole}^{\ell,t}$  and  ${\cal B}_{\rm pole}$ at different energy scales and $\zeta\lesssim M$.]{Estimates for ${\cal B}_{\rm pole}^{\ell,t}$  and  ${\cal B}_{\rm pole}={\cal B}_{\rm pole}^{\ell}+{\cal B}_{\rm pole}^{t}$ at different energy scales and $\zeta\lesssim M$.}
\label{tableBpolezetasimM}
\end{table}

\subsection{Estimating ${\cal H}[{\cal A}]$ and  ${\cal H}[{\cal B}]$ and $\phi_0$}\label{hilbert}

\subsubsection{Hilbert transforming ${\cal A}$ and  ${\cal B}$}

Having determined the order of magnitude of Im$\,\phi$ at various characteristic energy scales one is now in the position to (qualitatively) Hilbert-transform Im$\,\phi$ for $\zeta \ll M$ and to find the order of magnitude of Re$\,\tilde \phi$ and the corrections due to $\cal B$. To this end one splits the integral over $\omega$ in the dispersion relation Eq.\ (\ref{disp1}) as
\bea\label{split}
{\rm Re}\,\tilde\phi(\epsilon_\vk,\vk)&=& \mathcal P\int\limits _{-\infty}^\infty d\omega \,\frac{\rho_\phi(\omega,\vk)}{\omega-\epsilon_\vk} =
\mathcal P\int\limits _{0}^\infty d\omega \,\sum\limits_{\sigma=\pm}\frac{\rho_\phi(\omega,\vk)}{\omega-\sigma\epsilon_\vk}\non
&=&\mathcal P\left[\int\limits _{0}^{\Lambda_1}+\int\limits _{\Lambda_1}^{\Lambda_{\bar g}}+\int\limits _{\Lambda_{\bar g}}^{\Lambda_0}+\int\limits _{m_g+\Lambda_1}^{2m_g}+\int\limits _{2m_g}^{2\mu}+\int\limits _{2\mu}^{4\mu}\right] d\omega \,\sum\limits_{\sigma=\pm}\frac{\rho_\phi(\omega,\vk)}{\omega-\sigma\epsilon_\vk} \;.
\eea
Since the integral is additive, one may choose for each region the leading contribution. At the first scale $0\leq\omega \leq \Lambda_1$ it is $\rho_\phi \sim {\cal A}^t_{\rm cut} \sim g^2\phi$, cf.\ Eq. (\ref{Aphi}), and one has the contribution 
\bea
&&\mathcal P\int\limits _0^{\Lambda_1} d\omega \,\sum\limits_{\sigma=\pm}\frac{{\cal A}^t_{\rm cut}(\omega,\vk)}{\omega-\sigma\epsilon_\vk} \sim g^2 \phi\, \ln\left(\frac{\phi}{\epsilon_\vk}\right)\sim g^2\, \phi\;.
\eea
At the scale $\Lambda_1\leq \omega \leq \Lambda_{\bar g}$ one has  $\rho_\phi \sim {\cal A}^t_{\rm cut} \sim g\,\phi$, cf. Eq.\ (\ref{Aphi2}). With the substitution $d\omega/\omega = \ln(\phi/M)\,dy$ one finds
\bea\label{hil1}
&&\mathcal P\int\limits _{\Lambda_1}^{\Lambda_{\bar g}} d\omega \,\sum\limits_{\sigma=\pm}\frac{{\cal A}^t_{\rm cut}(\omega,\vk)}{\omega-\sigma\epsilon_\vk}\sim g\,\phi\int\limits _{\Lambda_1}^{\Lambda_{\bar g}}\frac{d\omega}{\omega}  \sim g\, \phi \ln\left(\frac{\phi}{M}\right)\int\limits _{1}^{{\bar g}}dy \sim \phi\;.\label{magcon}
\eea
The contribution from ${\cal A}^\ell_{\rm cut} \sim g\,\phi\, (\phi/M)^y$ at the same scale,
%$\omega \sim \Lambda_y$ with $1\leq y \leq \bar g$
cf. Eq.\ (\ref{Aphi2b}),  can be shown to be much smaller. One has 
\bea
&&\mathcal P\int\limits _{\Lambda_1}^{\Lambda_{\bar g}} d\omega \,\sum\limits_{\sigma=\pm}\frac{{\cal A}^\ell_{\rm cut} (\omega,\vk)}{\omega-\sigma\epsilon_\vk}\sim g\,\phi\,\ln\left(\frac{\phi}{M}\right)\int\limits _{1}^{\bar g} dy \,\left(\frac{\phi}{M}\right)^y  \sim g\,\phi\,\frac{\phi}{M}\;.
\eea
At the scale $\Lambda_{\bar g}\leq \omega \leq \Lambda_0$ one has  $\rho_\phi \sim {\cal A}^t_{\rm cut} \sim {\cal A}^\ell_{\rm cut} \sim g\,\phi$, cf. Eq.\ (\ref{Aphi2}, \ref{Aphi2b}),  and finds
\bea
&&\mathcal P\int\limits_{\Lambda_{\bar g}}^{\Lambda_0} d\omega \,\sum\limits_{\sigma=\pm}\frac{{\cal A}_{\rm cut}(\omega,\vk)}{\omega-\sigma\epsilon_\vk}\sim g\,\phi\int\limits_{\Lambda_{\bar g}} ^{\Lambda_0} \frac{d\omega}{\omega}  \sim \phi\int\limits_{\bar g}^0 dy\sim g\,\phi\;.
\eea
For energies $\omega \agt m_g + \Lambda_y$ with $0\leq y < 1$ one has $\rho_\phi \sim {\cal A}^\ell_{\rm pole} \sim g\,\phi\, (M/\phi)^y$, cf. Eq.\ (\ref{l3}),  and finds with $d\omega = \ln(\phi/M) \,\Lambda_y\,dy$
\bea
\mathcal P\int\limits_{m_g+\Lambda_1}^{m_g+\Lambda_0} d\omega \,\sum\limits_{\sigma=\pm}\frac{{\cal A}^\ell_{\rm pole}(\omega,\vk)}{\omega-\sigma\epsilon_\vk}
&\sim&
 \frac{g\,\phi}{M}\ln\left(\frac{\phi}{M}\right)\int\limits_{1}^{0}  dy \,\Lambda_{y}\left(\frac{M}{\phi}\right)^y  
%\non&\sim&
\sim\frac{\phi}{M}\int\limits_{1}^{0}  dy \,M  \sim \phi\;.\label{elcon}
\eea
Integrating over $2m_g < \omega < 2\mu$ with $\rho_\phi \sim {\cal A}^t_{\rm pole}\sim g\,\phi$, cf.\ Eq.\ (\ref{tpole}), one obtains
\bea
\mathcal P\int\limits_{2m_g}^{2\mu} d\omega \,\sum\limits_{\sigma=\pm}\frac{{\cal A}^t_{\rm pole}(\omega,\vk)}{\omega-\sigma\epsilon_\vk}
\sim g\,\phi\int\limits_{2m_g}^{2\mu} \frac{d\omega}{\omega} \sim g\,\phi \,\ln\left(\frac{\mu}{M} \right)\sim {g\,\phi}\;.\label{conlargeomega}
\eea
Finally, integrating over $2\mu < \omega < 4\mu$ with $\rho_\phi \sim {\cal B}^t_{\rm pole}\sim g^3\phi\,(M/\omega)$, cf.\ Eqs.\ (\ref{large1},\ref{large2}), one obtains
\bea
\mathcal P\int\limits_{2\mu}^{4\mu} d\omega \,\sum\limits_{\sigma=\pm}\frac{{\cal B}^t_{\rm pole}(\omega,\vk)}{\omega-\sigma\epsilon_\vk}
\sim g^3\phi\,M\int\limits_{2\mu}^{4\mu} \frac{d\omega}{\omega^2} \sim g^3\phi \,\frac{M}{\mu} \sim g^4\phi\;.\label{converylargeomega}
\eea
As expected,  Re$\,\tilde \phi\sim \phi$. Furthermore, I find several potential sources for sub-subleading order correction from to ${\cal H}[{\cal B}]$ to Re$\,\phi$. The first comes from ${\cal B}^t_{\rm cut}$, which is $\sim g^2{\cal A}^t_{\rm cut}$  for $\omega \sim \Lambda_y$ with $0<y<1$, cf.\ Eq.\ (\ref{est1}). After Hilbert transformation it yields a contribution of order $g^2\phi$ to  Re$\,\tilde \phi$, cf.\ Eq.\ (\ref{hil1}), and is therefore of sub-subleading order. At this point,  also from ${\cal B}^\ell_{\rm pole}\sim g^2{\cal A}^\ell_{\rm pole}$ a sub-subleading order contribution seems possible, since the corresponding contribution from ${\cal A}^\ell_{\rm pole}$ is $\sim \phi$, cf.\ (\ref{elcon}). As the latter, however, combines with $\phi_0$ to a subleading order term, cf.\ Sec.\ \ref{repro}, it would be interesting to investigate if also ${\cal B}^\ell_{\rm pole}$ finds an analogous partner to cancel similarly. If not, it would contribute at sub-subleading order to Re$\,\tilde \phi$. Furthermore, one found that ${\cal B}\sim g {\cal A} \sim g^2\phi$ for $\omega \agt M$, cf.\ Eqs.\ (\ref{est3},\ref{est7},\ref{est8},\ref{est6}). From the estimate in Eq.\ (\ref{conlargeomega}) one deduces that the corresponding contribution of ${\cal H}[{\cal B}]$ to Re$\,\phi$ is of sub-subleading order. In the next section it will be analyzed at which order Im$\,\phi$ contributes to the local part of the gab function, $\phi_0$.

\subsubsection{The contribution of Im$\,\phi$ to $\phi_0$}\label{phi0}

The gap equation for $\phi_0(\vk)$ is obtained by considering the integrals $I_0$ and $I_{k_0}$, cf.\ Eqs.\ (\ref{I0},\ref{Ik0}), in the limit $k_0 \rightarrow \infty$. Since $p\lesssim 2\mu$ the gluon spectral densities $\rho^{\ell,t}(q_0,\vp)$ are nonzero only for $q_0 \lesssim 2\mu$. Consequently, the integral over $q_0$ in Eq.\ (\ref{Ik0}) is effectively bounded by $\pm 2\mu$. Then, due to the energy denominator under the integral, $I_{\rm k_0}$ tends to zero as $k_0 \rightarrow \infty$. In the second term on the r.h.s.\ of Eq.\ (\ref{I0})  one has ${\tilde \epsilon_\vq}<\Lambda_{\rm q} \sim g\mu$. Therefore, one may neglect the imaginary part of the gluon propagator $\Delta^{\ell,t}(\pm\tilde\epsilon_\vq-k_0,\vp)$, cf.\ Eqs. (\ref{propHDL}), for $k_0 \gg 2\mu>p$ . In the transverse case the gluon propagator becomes $\Delta^t\sim 1/k_0^2$ and the respective contribution vanishes in the limit $k_0 \rightarrow \infty$. In the longitudinal sector one has $\Delta^\ell \rightarrow -1/p^2$. Hence, the longitudinal contribution of the considered term does not vanish. In the first term on the r.h.s.\ of Eq.\ (\ref{I0}) the integral over $q_0$ only runs to values $q_0 \sim \mu$ due to the presence of $\rho_\phi$. In the limit $k_0 \rightarrow \infty$ again only the contribution from the static electric gluon propagator $\Delta^\ell \rightarrow -1/p^2$ remains. Consequently, one finds for $\phi_0(\vk)$ to subleading order
\bea
\phi_0(\vk) &=& -\frac{g^2}{3(2\pi)^2} \int\limits _{\Lambda_1}^{\Lambda_0}\frac{d\xi}{\xi}\int\limits _{\xi}^{2\mu}\frac{dp}{p}\,{\rm Tr}_s^\ell (k,p,q)
\left\{ \left[\phi_0(\vq)+{\rm Re}\,\tilde\phi({\tilde \epsilon_\vq}+i\eta,\vq)\right]Z^2(\tilde\epsilon_\vq)\tanh\left(\frac{{\tilde \epsilon_\vq}}{2T} \right)\right.
\non&&
\left. -{\mathcal P}\int\limits _{-\infty}^\infty d\omega \,
\frac{\rho_{\phi}(\omega,\vq)}{{\tilde \epsilon_\vq}-\omega}\,Z^2(\omega)
\tanh\left(\frac{\omega}{2T} \right)\right\}\;.
\eea
%\bea
%\phi_0(\vk) &=& \frac{g^2}{3} \int\limits _{\Lambda_1}^{\Lambda_0}\frac{d\xi}{\xi}\int\limits _{\xi}^{2\Lambda_{\rm gl}}\frac{dp}{p}
%\left\{ \left[\phi_0(\vq)+{\rm Re}\,\tilde\phi({\tilde \epsilon_\vq}+i\eta,\vq)\right]\tanh\left(\frac{{\tilde \epsilon_\vq}}{2T} \right)\right.
%\non&&
%\left. -{\mathcal P}\int\limits _{-\infty}^\infty d\omega \,
%\frac{\rho_{\phi}(\omega,\vq)}{{\tilde \epsilon_\vq}-\omega}
%\tanh\left(\frac{\omega}{2T} \right)\right\}\;.
%\eea
In the limit $T\rightarrow 0$ the hyperbolic functions simplify, yielding after performing the integral over $p$
and with ${\rm Tr}_s^\ell (k,p,q)\sim 1$
\bea
\phi_0(\vk) &\sim& g^2 \int\limits _{\Lambda_1}^{\Lambda_0}\frac{d\xi}{\xi}\ln\left(\frac{2\mu}{\xi}\right)
\left\{ \left[\phi_0(\vq)+{\rm Re}\,\tilde\phi({\tilde \epsilon_\vq}+i\eta,\vq)\right]
\right.\non&&\left.
-{\mathcal P}\int\limits _{0}^\infty d\omega \,
\sum\limits_{\sigma=\pm}\rho_{\phi}(\omega,\vq)\,\frac{\sigma}{\sigma{\tilde \epsilon_\vq}-\omega}\right\}\;.\label{phi02}
\eea
Using ${\rm Re}\,\tilde\phi({\tilde \epsilon_\vq}+i\eta,\vq)\sim \phi$ one finds for the respective contribution in Eq.\ (\ref{phi02})
\bea
 g^2\phi \int\limits _{\Lambda_1}^{\Lambda_0}\frac{d\xi}{\xi}\ln\left(\frac{2\mu}{\xi}\right)
\sim \phi\;,
\eea
and hence $\phi_0(\vk)\sim \phi$. Consequently, the integration over the first term in the squared brackets, $\phi_0(\vq)$, also gives a contribution of order $\phi$ to $\phi_0(\vk)$. The remaining term is the contribution from $\rho_\phi$ and is identical to Eq.\ (\ref{split}) up to an extra sign $\sigma$ arising from the hyperbolic tangent. One can conservatively estimate this term by appoximating $\rho_\phi \sim g\,\phi$ for $0<\omega<4\mu$ and all $\Lambda_1\leq\xi\leq\Lambda_0$ and adding $\rho_\phi\sim g\,\phi\,(M/\phi)^y$ in the range $\omega\sim m_g+\Lambda_y\,,~ 1>y>0$ and $\Lambda_1<\xi<\Lambda_{\bar{g}}$. 
One obtains
%In the integration region from $\Lambda_1\leq \omega \leq \Lambda_0$ one may approximate $\rho_\phi \sim g\,\phi$ and finds
\bea
&&g^3\phi \int\limits _{\Lambda_1}^{\Lambda_0}\frac{d\xi}{\xi}\ln\left(\frac{2\mu}{\xi}\right) {\mathcal P}\int\limits _{0}^{4\mu} d\omega \sum\limits_{\sigma=\pm}\frac{\sigma}{\omega-\sigma\xi}\sim
g^3\phi \int\limits _{\Lambda_1}^{\Lambda_0}\frac{d\xi}{\xi}\ln\left(\frac{2\mu}{\xi}\right)\,\ln\left(\frac{\xi+4\mu}{\xi-4\mu}\right)\non
&&\sim g^3\phi\,\frac{1}{\mu} \int\limits _{\Lambda_1}^{\Lambda_0}d\xi\,\ln\left(\frac{2\mu}{\xi}\right)
\sim g^3\phi\,\frac{M}{\mu}
\sim g^4\phi
\eea
%In the first term one may exploit $\xi > \Lambda_1$, expand the respective logarithm in $\Lambda_1/\xi$ and find
and
%\bea
% g^3\phi \int\limits _{\Lambda_1}^{\Lambda_0}\frac{d\xi}{\xi}\ln\left(\frac{2\mu}{\xi}\right)
%\ln\left(\frac{\xi-\Lambda_1}{\xi+\Lambda_1}\right)
%&\sim&
% g^3\phi\,\Lambda_1 \int\limits _{\Lambda_1}^{\Lambda_0}\frac{d\xi}{\xi^2}\ln\left(\frac{2\mu}{\xi}\right)
%\sim  g^3\phi\ln\left(\frac{2\mu}{\Lambda_1}\right)\sim g^2 \phi\;.
%\eea
%In the second term one has  $\xi < \Lambda_0$ and expanding the respective logarithm in $\xi/\Lambda_0$ yields
%\bea
% g^3\phi \int\limits _{\Lambda_1}^{\Lambda_0}\frac{d\xi}{\xi}\ln\left(\frac{2\mu}{\xi}\right)
%\ln\left(\frac{\Lambda_0-\xi}{\xi+\Lambda_0}\right)
%&\sim&
% g^3\phi\,\frac{1}{\Lambda_0} \int\limits _{\Lambda_1}^{\Lambda_0}d\xi\,\ln\left(\frac{2\mu}{\xi}\right)\sim g^3 \phi\;.
%\eea
%In the region $m_g+\Lambda_1 \leq \omega \leq m_g +\Lambda_0$ it is with $\rho_\phi\sim g\,\phi\,(M/\phi)^y$ and $d\omega = \ln(\phi/M)\,\Lambda_y\,dy$
\bea
&& g^2 \int\limits _{\Lambda_1}^{\Lambda_{\bar g}}\frac{d\xi}{\xi}\ln\left(\frac{2\mu}{\xi}\right)
{\mathcal P}\int\limits _{m_g+\Lambda_1}^{m_g+\Lambda_0} d\omega\sum\limits_{\sigma=\pm}\frac{\sigma\,\rho_{\phi}(\omega,\vq)}{\omega-\sigma\xi}
\sim g^2 \frac{\phi}{M^2} \int\limits _{\Lambda_1}^{\Lambda_{\bar g}}d\xi\,\ln\left(\frac{2\mu}{\xi}\right)
\int\limits _{1}^{0} dy \,\Lambda_y\, \left(\frac{M}{\phi} \right)^y\non
&&\sim g^2 \frac{\phi}{M} \int\limits _{\Lambda_1}^{\Lambda_{\bar g}}d\xi\,\ln\left(\frac{2\mu}{\xi}\right)
\sim g^2\phi\;,
\eea
where one has estimated $\sum_\sigma\sigma/(\omega-\sigma\xi)\sim \xi/M$, since $\xi\ll\omega\sim M$. It follows that the contributions from $\rho_\phi$ to $\phi_0$ are of order $g^2\phi$ and hence of sub-subleading order, cf.\ discussion after Eq.\ (\ref{converylargeomega}).

Finally, one has found that the contribution of Im$\,\phi$ to ${\rm Re}\,\phi(\epsilon_\vk, \vk) = {\rm Re}\,\tilde\phi(\epsilon_\vk, \vk)+\phi_0(\vk)$ are in total beyond subleading order.

\subsection{Reproducing  Re$\,\phi(\epsilon_\vk+i\eta,\vk)$ to subleading order}\label{repro}

Although  $\rho_\phi(\omega,\vk)$ and  $\phi_0(\vk)$  have not been calculated (in principle possible) but only estimated in magnitude, the main goal of this analysis has been achieved. The imaginary part of the gap function is shown to be at most of order $g\, {\rm Re}\,\phi(\epsilon_\vk,\vk)$ for $\epsilon_\vk < \Lambda_{\rm q} \sim g \mu$ (then it is $|\phi| =  {\rm Re}\,\phi$, cf.\ Eq.\ (\ref{gapexcite})) and it enters ${\rm Re}\,\phi(\epsilon_\vk,\vk)$ beyond subleading order. 

Neglecting all terms that have been identified to originate from $\rho_\phi$, namely Im$\,{\cal M_B}(\omega+i\eta,\vq,\vp)$ in   Eq.\ (\ref{ImM}), and Hilbert transfoming the remaining term  Im$\,{\cal M_A}(\omega+i\eta,\vq,\vp)$ gives the gap equation for  Re$\,\tilde\phi(\epsilon_\vk,\vk)$ to subleading order
\bea
{\rm Re}\,\tilde\phi({ \epsilon_\vk}+i\eta,\vk)&=&
-\frac{g^2}{3} \int\limits\frac{d^3\vq}{(2\pi)^3}\,\frac{Z^2(\tilde\epsilon_\vq)}{2\tilde\epsilon_\vq}\,{\rm Re}\,\phi({\tilde \epsilon_\vq}+i\eta,\vq)
\non
&&\times \,{\cal P} \!\int\limits_{-\infty}^\infty d\omega\,\sum\limits_{\sigma=\pm}\frac{\sigma\left[{\rm Tr}_s^\ell (k,p,q)\rho^{\ell}(\omega,\vp)+ {\rm Tr}_s^t (k,p,q) \rho^{t}(\omega,\vp)\right]}{\omega-\epsilon_\vk+\sigma\tilde\epsilon_\vq}\non
&&\times\,\frac{1}{2}\left[\tanh\left(\frac{\sigma\tilde\epsilon_\vq}{2T} \right) +\coth\left( \frac{\omega}{2T}\right) \right]
\;.\label{rephitilde}
\eea
Adding
\bea\label{phi03}
\phi_0(\vk) = -\frac{g^2}{3} \int\limits\frac{d^3\vq}{(2\pi)^3}\,\frac{Z^2(\tilde\epsilon_\vq)}{2\tilde\epsilon_\vq}\,{\rm Tr}_s^\ell (k,p,q)\left(-\frac{2}{p^2}\right)\,
{\rm Re}\,\phi({\tilde \epsilon_\vq}+i\eta,\vq)\frac{1}{2}\tanh\left(\frac{{\tilde \epsilon_\vq}}{2T} \right)
\eea
to Eq.\ (\ref{rephitilde}) one reproduces the well known gap equation for  Re$\,\phi(\epsilon_\vk,\vk)$ to subleading order, cf.\ \cite{rdpdhr}. %The contributions from Im$\,{\cal M_B}(\omega+i\eta,\vq,\vp)$ give contributions, which are 
To see explicitly how $\phi_0\sim \phi$ and ${\cal H}[{\cal A}_{\rm pole}^\ell]\sim \phi$ combine to a subleading order contribution, one first notes that all terms $\sim \coth$ are at most of sub-subleading order, cf. Sec.\ \ref{solving}. Then Eq.\ (\ref{rephitilde}) simplifies to
\bea
{\rm Re}\,\tilde\phi({ \epsilon_\vk}+i\eta,\vk)&=&
-\frac{g^2}{3} \int\limits\frac{d^3\vq}{(2\pi)^3}\,\frac{Z^2(\tilde\epsilon_\vq)}{2\tilde\epsilon_\vq}\,{\rm Re}\,\phi({\tilde \epsilon_\vq}+i\eta,\vq)\,\frac{1}{2}\tanh\left(\frac{\tilde\epsilon_\vq}{2T} \right)
\non &&\times\,
{\cal P} \int\limits_{-\infty}^\infty d\omega\sum\limits_{\sigma=\pm}\frac{{\rm Tr}_s^\ell (k,p,q)\rho^{\ell}(\omega,\vp)+ {\rm Tr}_s^t (k,p,q) \rho^{t}(\omega,\vp)}{\omega-\epsilon_\vk+\sigma\tilde\epsilon_\vq}\non
&&\hspace{-3.8cm}
= -\frac{g^2}{3(2\pi)^2} \int\limits_0^{\Lambda_{\rm q}}d\xi\frac{Z^2(\tilde\epsilon_\vq)}{2\tilde\epsilon_\vq}\,{\rm Re}\,\phi({\tilde \epsilon_\vq}+i\eta,\vq)\,\tanh\left(\frac{\tilde\epsilon_\vq}{2T} \right)
\non
&&\hspace{-3.4cm}
\times\sum\limits_{\sigma=\pm}\left[
\int\limits_{|\xi-\zeta|}^{\Lambda_{\rm gl}}\!\!dp\,p\left\{{\rm Tr}_s^\ell (k,p,q)\left[\frac{1}{p^2}+\Delta_{\rm HDL}^{\ell}(\epsilon_\vk-\sigma\tilde\epsilon_\vq,\vp)\right]+ {\rm Tr}_s^t (k,p,q)\Delta^{t}_{\rm HDL}(\epsilon_\vk-\sigma\tilde\epsilon_\vq,\vp)\right\}\right.\non
&&\hspace{-2.2cm}
+\left.
\int\limits_{\Lambda_{\rm gl}}^{2\mu}dp\,p\,{\rm Tr}_s^t (k,p,q)\, \Delta^{t}_{0,22}(\epsilon_\vk-\sigma\tilde\epsilon_\vq,\vp)
\right]\!,\label{rephitilde2}
\eea
where one used $\rho^\ell(\omega, \vp) \equiv 0$ for $p>\Lambda_{\rm gl}$ in the effective theory, cf.\ Eq.\ (\ref{hardspecl}).  Adding Eqs.\ (\ref{phi03},\ref{rephitilde2}), the $1/p^2$-term from the soft electric gluon propagator in Eq.\ (\ref{rephitilde2}) effectively restricts the $p-$integral in $\phi_0$ from $\Lambda_{\rm gl}$ to $2\mu$. After aproximating the hard magnetic gluon propagator as $\Delta^{t}_{0,22}(\epsilon_\vk-\sigma\tilde\epsilon_\vq,\vp)= 1/p^2 +O(\Lambda_{\rm q}/\Lambda_{\rm gl})$, one can combine it with the remaining contribution from $\phi_0$. Using ${\rm Tr}_s^\ell (k,p,q)-{\rm Tr}_s^t(k,p,q) =4 +O(\Lambda_{\rm q}/\Lambda_{\rm gl})$, one finally arrives at Eq.\ (124) of \cite{qwdhr}. Hence, indeed $\phi_0 +{\cal H}[{\cal A}_{\rm pole}^\ell] \sim g\,\phi$, i.e.\ contribute at subleading order to Re$\,\phi$.

\subsection{Calculating ${\rm Im}\,\phi(\epsilon_\vk+i\eta,\vk)$ near the Fermi surface}\label{calcimphi}

In Sec.\ \ref{estAB} the contributions $\cal A$ and $\cal B$ to ${\rm Im}\,\phi$ have been estimated for different regimes of $\omega$ and $\zeta$, cf.\ Tab.\ \ref{tableAcutzeta<M}-\ref{tableBpolezetasimM}. In the case that both $\omega$ and $\zeta$ are $\sim \Lambda_y$ with some $y$ satisfying $0<y\leq 1$ one found ${\cal A}^t_{\rm cut}$ to be the dominant contribution to ${\rm Im}\,\phi$, while in all other regimes of  $\omega$ and $\zeta$ different gluon sectors contribute at the respective leading order and therefore mix. Furthermore, for $\omega <2m_g$ also the contributions from $\cal B$ have to be considered, since there ${\cal B}$ is suppressed relative to ${\cal A}$ only by one power of $g$. 

In the following the imaginary part of the on-shell gap function ${\rm Im}\,\phi(\epsilon_\vk+i\eta,\vk)$ near the Fermi surface, $\zeta\sim \Lambda_y$ with $1<y<0$,  at zero temperature, $T=0$, is calculated.
%Furthermore, gluon momenta $p\leq M$ give the main contribution and we may approximate ${\rm Tr}_s^t (k,p,q)\approx -2$.
It follows
\bea\label{Imphiexact}
{\rm Im}\,\phi({\epsilon_\vk}+i\eta,\vk)&\simeq&
\frac{g^2\,\pi}{3(2\pi)^2}\int\limits_{\Lambda_1}^{\epsilon_\vk}\frac{d\xi}{\xi}\,Z^2(\tilde\epsilon_\vq)\,{\rm Re}\,\phi({\tilde \epsilon_\vq}+i\eta,\vq)\int\limits_\lambda^{\Lambda_{\rm gl}} dp\,p \,\rho_{\rm cut}^t(\omega^*,\vp)\,{\rm Tr}_s^t (k,p,q)\non
&\simeq&
\frac{g^2\,\pi}{3(2\pi)^2}\int\limits_{\Lambda_1}^{\epsilon_\vk}\frac{d\xi}{\xi}\,Z^2(\tilde\epsilon_\vq)\,{\rm Re}\,\phi({\tilde \epsilon_\vq}+i\eta,\vq)\int\limits_\lambda^{\Lambda_{\rm gl}} dp \,\frac{2M^2\,\omega^*}{\pi}\,\frac{p^2}{p^6+(M^2\omega^*)^2}\non
&\simeq&\frac{g^2\,\pi}{9(2\pi)^2}\int\limits_{\Lambda_1}^{\epsilon_\vk}\frac{d\xi}{\xi}\,Z^2(\tilde\epsilon_\vq)\,{\rm Re}\,\phi({\tilde \epsilon_\vq}+i\eta,\vq) \non
&\simeq& \frac{g^2\,\pi}{9(2\pi)^2}\ln\left(\frac{\phi}{M}\right)\,\phi \int\limits_{1}^{y}dy^\prime\,\sin\left(\frac{\pi\,y^\prime}{2} \right) \,\left(1-\frac{\bar g\,\pi\,y^\prime}{2} \right)
\non
&=&-\frac{g^2}{18\,\pi^2}\ln\left(\frac{\phi}{M}\right)\,\phi\,\cos\left(\frac{\pi\,y}{2} \right)+ {\cal O}(\bar g^2)\non
&=&\bar{g}\,\phi \;\frac{\pi}{2}\cos\left(\frac{\pi\,y}{2} \right)+ {\cal O}(\bar g^2)\;,
\eea
where one substituted $\epsilon_\vk = \Lambda_y$ and $d\xi/\xi = dy^\prime\,\ln(\phi/M)$ and used $\ln(\phi/M)=-3\pi^2/(\sqrt{2}\,g)$ and $\bar g\equiv g/(3\sqrt{2}\pi)$. Furthermore, it was sufficient to approximate ${\rm Tr}_s^t (k,p,q)\simeq -2$. The corrections due to the term $-p^2/2\mu^2$ are easily shown to be suppressed by a factor $\sim (M/\mu)^2\sim g^2$ compared to the final result in Eq.\ (\ref{Imphiexact}). To estimate the contributions from the term $-(k^2-q^2)^2/(2\, k\, q\, p^2)\simeq -2(\xi-\zeta)^2/p^2$ one first conservatively sets $\zeta =0$. At the lower boundary of the $p-$integral, $p\geq \lambda$, this term is $\sim 1$.  One finds for the integral over $p$
\bea
\int\limits_\lambda^{\Lambda_{\rm gl}} dp \,\frac{M^2\,\omega^*\,\xi^2}{p^6+(M^2\omega^*)^2}<\left(\frac{\Lambda_{3y/5}}{\Lambda_{y/3}}\right)^5\sim \left(\frac{\phi}{M}\right)^{\frac{4y}{15}}\;.
\eea
The integral over $\xi$ then can be estimated to be at most of order
\bea
\int\limits_{1}^{y}dy^\prime\,\sin\left(\frac{\pi\,y^\prime}{2} \right) \left(\frac{\phi}{M}\right)^{\frac{4y^\prime}{15}}<\int\limits_{1}^{y}dy^\prime \left(\frac{\phi}{M}\right)^{\frac{4y^\prime}{15}}
\sim \frac{\left(\frac{\phi}{M}\right)^{\frac{4}{15}}-\left(\frac{\phi}{M}\right)^{\frac{4y}{15}}}{\ln\left(\frac{\phi}{M}\right)}\lesssim g\left(\frac{\phi}{M}\right)^{\frac{4y}{15}}\;.
\eea
Hence, also this contribution is suppressed by a factor of $g$ compared to the result given in  Eq.\ (\ref{Imphiexact}).

		\chapter{Summary and Outlook}\label{V}

In this work I have presented a formal derivation of a general
effective action for non-Abelian gauge theories, Eq.\ (\ref{Seff}). 
This was motivated by the occurence of well-separated momentum scales
in hot and/or dense quark matter. To this end
I first introduced cut-offs in momentum space for
quarks, $\Lambda_{\rm q}$, and gluons, $\Lambda_{\rm gl}$. These cut-offs
separate relevant from irrelevant quark modes and
soft from hard gluon modes. I then 
explicitly integrated out irrelevant quark and hard gluon modes.
The effective action (\ref{Seff}) is completely general and, as
shown explicitly in Sec.\ \ref{IIIA}, after appropriately
choosing $\Lambda_{\rm q}$ and $\Lambda_{\rm gl}$, it comprises
well-known effective actions as special cases, for instance,
the ``Hard Thermal Loop'' (HTL) and
``Hard Dense Loop'' (HDL) effective actions. I also demonstrated,
cf.\ Sec.\ \ref{IIIB}, that the high-density effective theory introduced
by Hong and others \cite{hong,hong2,HLSLHDET,schaferefftheory,NFL,others}
is contained in the effective action (\ref{Seff}).

In Sec.\ \ref{toeft} it is argued that in order to obtain a complete effective theory, in principle, all diagrams appearing in the effective action (\ref{Seff}) have to be power-counted. This remains as a future project. A first step towards this goal, however, is performed for the special case of an effective theory for cold and dense quark matter, i.e.\ for quarks with momenta $|k-\mu| \sim \phi$ around the Fermi surface and soft gluons with momenta of order $\Lambda_{\rm gl}^s \ll \mu$. To this end, one specific class of diagrams occurring in the effective action (\ref{Seff}), loops of irrelevant quarks with $N$ external soft gluon legs, is power-counted for the  cutoff parameters fulfilling $\phi \ll \Lambda_{\rm q} \ll \Lambda_{\rm gl} \simeq \mu$ and the projection operators given in Eqs.\ (\ref{P12}) and (\ref{Q12}).
It is shown that for the considered momentum regime these loops are of the order of the corresponding bare vertices (bare gluon propagator and gluon vertices) times the factor $(g\mu)^2/(\Lambda_{\rm gl}^s)^2$. They are therefore ``relevant'' operators in the effective action, as their magnitude increases when the soft gluon scale $\Lambda_{\rm gl}^s$ decreases.  They contain Debye screening and Landau damping (as known from HDLs), whereas the Meissner effect for the magnetic gluons is negligible as long as $\Lambda_{\rm gl} ^s \gg \phi$. 
 A possible field of application could be the precursory effects to color superconductivity at temperatures above the (non-perturbative) onset of Cooper pairing, $T \agt T_c \sim \phi$. 

As it is known from the HTL/HDL power-counting scheme there is another important class of diagrams, i.e.\ those with 2 external quark and $N-2$ external gluon legs. It would be interesting to analyze how their orders of magnitude change as the cutoffs are shifted. Especially the 1-loop corrections to the quark-gluon vertex could contribute to the gap parameter at sub-subleading order.

In Sec.\ \ref{IV} I showed how the QCD gap equation can be derived from
the effective action (\ref{Seff}). The gap equation is a
Dyson-Schwinger equation for the anomalous part of the
quark self-energy. It has to be solved
self-consistently, which is feasible only after truncating
the set of all possible diagrams
contributing to the Dyson-Schwinger equation. Such truncations
can be derived in a systematic way within the general
Cornwall-Jackiw-Tomboulis (CJT) formalism \cite{CJT}.
Here, I only include diagrams of the sunset-type, cf.\ 
Fig.\ \ref{Gamma2eff}, in the
CJT effective action, which gives rise to one-loop diagrams
(with self-consistently determined quark and gluon propagators)
in the quark and gluon self-energies.

In principle, the CJT effective action (\ref{Gamma}) for
the effective theory contains the same information as the
corresponding one for QCD. However, while in full QCD self-consistency
is maintained for {\em all\/} momentum modes via the solution of the
stationarity condition (\ref{statcond}), in the effective theory
self-consistency is only required for the {\em relevant\/} quark 
and {\em soft\/} gluon modes. These are the {\em only\/} 
dynamical degrees of freedom in the CJT effective action; the
irrelevant fermion and hard gluon modes, which were integrated out,
only appear in the vertices of the {\rm tree-level\/}
action (\ref{Seff}). In this sense, the effective theory 
provides a simplification of {\it any} physical problem that has to be solved
self-consistently. How this facilitates the actual computation is
exemplified explicitly with the solution of the color-superconducting gap equation.

Usually, the advantage of an effective theory is that 
the degree of importance of various operators can be estimated
(via power counting) at the level of the effective action, i.e.,
{\em prior\/} to the actual calculation of a physical quantity.
This tremendously simplifies the computation of quantities which
are accessible within a perturbative framework.
On the other hand, the requirement of self-consistency for the
solution of the Dyson-Schwinger equation invalidates any such
power-counting scheme on the level of the effective action. 
For instance, perturbatively,
the right-hand side of the gap equation (\ref{gapeq}) is proportional
to $g^2$. However, self-consistency generates additional
large logarithms $\sim \ln (\mu/\phi) \sim 1/g$ which
cancel powers of $g$.

% combination with the CJT formalism the  derived effective action (\ref{Seff})
%has a distinct advantage for the derivation and the solution of Dyson-Schwinger 
%equations for quantities which have to be determined self-consistently, 
%such as the color-superconducting gap function in QCD. 
%This advantage originates from 
%the introduction of the two cut-offs which separate various
%regions in momentum space. They allow for a {\em rigorous\/} power
%counting of different contributions to the Dyson-Schwinger equation.
%I explicitly demonstrated this in Sec.\ \ref{IV},
%where I reviewed the calculation of the color-superconducting
%gap parameter to subleading order and in Sec.\ \ref{imsec}, where I
%considered the imaginary part of the gap function. 

Nevertheless, it turns out that there is still a 
distinct advantage in using the CJT effective action 
(\ref{Gamma}) of the effective theory
for the derivation and the solution of Dyson-Schwinger 
equations for quantities which have to be determined self-consistently,
such as the color-superconducting gap function in QCD. This advantage 
originates from the introduction of the cut-offs which separate various
regions in momentum space. Choosing $\Lambda_{\rm q} \ll \Lambda_{\rm gl}$
they allow for a {\em rigorous\/} power
counting of different contributions to the Dyson-Schwinger equation after expanding 
in $\Lambda_{\rm q}/\Lambda_{\rm gl} \ll 1$.
I explicitly demonstrated this in Sec.\ \ref{IV},
where I reviewed the calculation of the color-superconducting
gap parameter to subleading order and in Sec.\ \ref{imsec}, where I
considered the imaginary part of the gap function.

\begin{figure}[ht]
\centerline{\includegraphics[width=8cm]{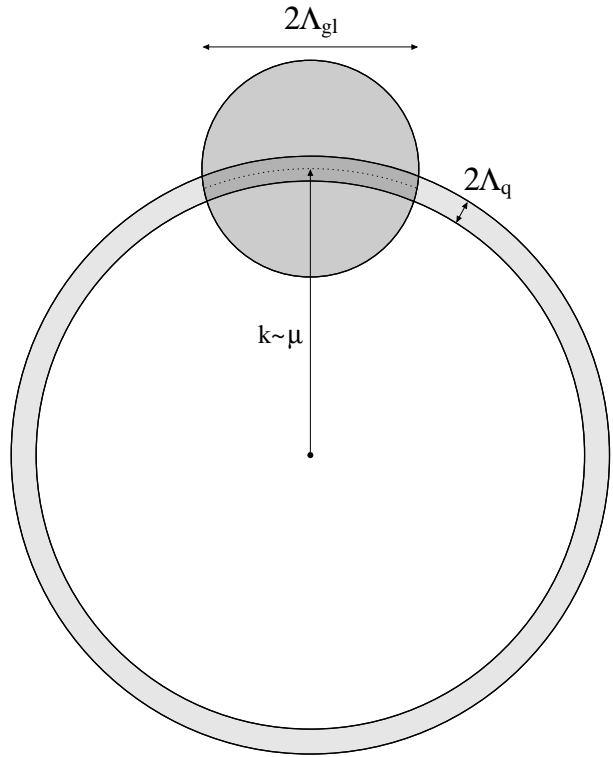}}
\caption[Momentum regime of quarks near the Fermi surface and soft gluons.]{Momentum regime of quarks near the Fermi surface, $|k-\mu|\ll\Lambda_{\rm q}$, and soft gluons, $p\ll \Lambda_{\rm gl}$.}
\label{spheressum}
\end{figure}

In order to obtain the standard result (\ref{gapsol}), it was mandatory to
choose $\Lambda_{\rm q}  \alt g \mu \ll \Lambda_{\rm gl} \alt \mu$, cf.\ Fig.\ \ref{spheressum}. 
This is in contrast
to previous statements in the literature 
\cite{hong2,HLSLHDET,schaferefftheory} 
that a consistent power-counting scheme requires $\Lambda_{\rm q} \sim \Lambda_{\rm gl}$.
In particular, the choice $\Lambda_{\rm q} \ll \Lambda_{\rm gl}$ has the consequence
that the gluon energy in the QCD gap equation is restricted
to values $p_0 \alt \Lambda_{\rm q}$, while the gluon momentum can be much larger,
$p \alt \Lambda_{\rm gl}$. This naturally explains why it is permissible to use
the low-energy limit (\ref{lel}) of the HDL gluon propagators
in order to extract the dominant contribution 
to the gap equation (which arises from soft magnetic gluons).
In previous calculations of the gap within the framework
of an effective theory \cite{hong2,HLSLHDET,schaferefftheory}, the
low-energy limit for the HDL propagators was used 
without further justification,
even though for the choice $\Lambda_{\rm q} \sim \Lambda_{\rm gl}$ 
the gluon energy can be of the same order as the gluon momentum.
The physical picture which arises from the choice $\Lambda_{\rm q} \alt g
\mu \ll \Lambda_{\rm gl} \alt \mu$ is summarized in Fig.\ \ref{spheressum}.
Relevant quarks are located within a thin layer of width
$\sim \Lambda_{\rm q}$ around the Fermi surface. Soft gluon exchange
mediates between quarks within a ``patch'' of size $\sim \Lambda_{\rm gl}$
inside this layer. The area of the patch is much larger than its
thickness. Hard gluon exchange mediates between quark states 
inside and outside of the patch.

Obviously, this picture, as well as all power-counting arguments,
are rigorously valid only at asymptotically large values of
the quark chemical potential, where $g \ll 1$. In the physically relevant
region, $\mu \alt 1$ GeV and
$g \sim 1$, the scale hierarchy $\Lambda_{\rm q} \alt g\mu \ll \Lambda_{\rm gl}
\alt \mu$ breaks down. When all scales are of the same order,
the patch on the Fermi surface becomes a sphere of the size
of the Fermi sphere. 

In the course of the calculation, I was able to
identify various potential contributions of sub-subleading order.
However, I argued that, at this order, a self-consistent solution
of the gap equation must take into account the {\em off-shell\/}
behavior of the gap function. To this end, I investigated the imaginary part of the 
gap function, which (loosely speaking) generates the energy dependence of the gap function. 
Due to the energy dependent gluon propagator in the gap equation a self-consistent solution 
requires a energy dependent and therefore complex ansatz. Consequently, I considered a complex gap equation, which is pair of two coupled gap equations, one for the real and one for the imaginary part of the gap function. 

To solve it self-consistently, I first inserted the known leading order result of the real part of the gap function into the imaginary gap equation. In this way I estimated the order of magnitude of the imaginary part of the gap function for various energy and momentum regimes, cf.\ Tables I-III. The self-consistency of this approach was controlled by reinserting these estimates into the imaginary gap equation, cf.\ Tables IV-VI. Using these results it was checked at which order  the imaginary part contributes to the real part of the gap function. It was found that it enters at sub-subleading order. Furthermore, several sub-subleading order terms in the real part of the gap equation arising from the imaginary part of the gap function have been identified. 

Finally, I calculated the imaginary part of the on-shell gap function for momenta exponentially close to the Fermi surface, cf.\ Eq. (\ref{Imphiexact}). In this regime only Landau damped transversal gluons had to be considered. Also for the complex gap equation the two momentum cutoffs $\Lambda_{\rm q}$ and $\Lambda_{\rm gl}$ proved to be a powerful means for the rigorous power counting of innumerable terms.

For a complete sub-subleading order calculation it also appears to be necessary to include
2PI diagrams beyond those of sunset topology 
in $\Gamma_2$, cf.\ Eq.\ (\ref{EqGamma2}) and Fig.\ \ref{Gamma2eff}. As a first step it would be interesting to include the vertex corrections into the gap equation.

The general effective action (\ref{Seff}) derived in this work 
puts the known effective theories for hot and 
dense matter on a more formal basis and gives better insight and control 
over the different approximations made. It connects previous results and 
puts them on a common footing. Thereby, it has the potential to provide 
a framework for systematic studies of various problems in the field of hot 
and dense systems. 
Besides an improvement of the result for the color-superconducting
gap parameter beyond subleading order, I believe that it can serve as a 
convenient starting point to investigate other 
interesting problems pertaining to hot and/or dense 
quark matter.

%Outlook: application to CSC: sub-subleading order (->beyond mean-field, vertex corrections), integrate out mixed scales, other applications on damping rates?, Ward identies, RG-analysis (my work with Qun..., connection to above vertexcorrection, $e^I$...), continuation of power counting analysis (quarks as external legs transform towards the Fermi surface is different makes a different from standard HTL/HDL analysis and my own I gave here...),
\begin{appendix}
    \chapter{Matsubara sums in quark loops} \label{matsum}

In the following the Matsubara sum and the $\t-$integrals in quark loops with three and four external gluon fields is performed, always starting from the general Eq.\ (\ref{gentauint}).

\subsubsection{Three external glouns}

For three external glouns ($N=3$) one has $m=1,2$ and \eqrf{gentauint} becomes 
\bea
&&{\cal J} ^{\mu_1\mu_2\mu _3}(P_{1},P_2,P_{3})=\non
&=&
- \int \limits \frac{d^{3}\mathbf{k}}{(2\pi )^{3}}
\sum_{\ve,\vs}\mathcal{T}_\ve^{\mu_1\mu_2\mu_3}\Theta_\vs^\ve 
\sum_{m=1}^{2}(-1)^m
\prod _{i=1}^{2}
\left[\int \limits _{0}^{\beta }d\tau _{i}
%\non&&\times
\ft^{e_i}_{s_i}  e^{ \Omega_i \tau _{i}} \right]
\times\non&& \times
 \ft^{e_3}_{s_3}
e^{ -s_3\e_3 m\beta}
%\times 
%\non &&
%\times
\Theta \left( m\beta -\sum _{j=1}^{2}\tau _{j}\right) \Theta \left( \sum _{j=1}^{2}\tau _{j}-(m-1)\beta \right)
\non
&=&
- \int \limits \frac{d^{3}\mathbf{k}}{(2\pi )^{3}}
\sum_{\ve,\vs}\mathcal{T}_\ve^{\mu_1\mu_2\mu_3}\Theta_\vs^\ve 
\times \non
&&\times
\left\{
\prod _{i=1}^{2}
\left[\int \limits _{0}^{\beta }d\tau _{i}
%\non&&\times
 \ft^{e_i}_{s_i}  e^{ \Omega_i \tau _{i}} \right]
%\times\non
%&& \times
\ft^{e_3}_{s_3}
e^{ -2s_3\e_3 \beta}
%\times 
%\non &&
%\times
\Theta \left( \t_1 + \t_2-\beta \right)-
\right.
\non
&&
\left.
-\prod _{i=1}^{2}
\left[\int \limits _{0}^{\beta }d\tau _{i}
%\non&&\times
 \ft^{e_i}_{s_i}  e^{ \Omega_i \tau _{i}} \right]
%\times\non&& \times
 \ft^{e_3}_{s_3}
e^{ -s_3\e_3 \beta}
%\times 
%\non &&
%\times
 \Theta \left( \beta -\tau _{1}-\t_2  \right)
\right\}
\eea
Using \eqrftw{rel1}{rel2} one finds
\bea
&&\ft^{e_3}_{s_3}e^{ -s_3\e_3 \beta}\Theta \left( \beta -\tau _{1}-\t_2\right)-
\ft^{e_3}_{s_3}e^{ -2s_3\e_3 \beta}\Theta \left( \tau _{1}+\t_2-\b\right)\non
=&&
\ft^{e_3}_{-s_3}\Theta \left( \beta -\tau _{1}-\t_2\right)+
\left[\ft^{e_3}_{-s_3}-e^{ -s_3\e_3 \beta}\right]\Theta \left( \tau _{1}+\t_2-\b\right)\non
=&&
\ft^{e_3}_{-s_3}
-e^{ -s_3\e_3 \beta}\Theta \left( \tau _{1}+\t_2-\b\right)\;,
\eea
which leads to
\bea
&&{\cal J} ^{\mu_1\mu_2\mu _3}(P_{1},P_2,P_{3})=\non
&=&
-\int \limits \frac{d^{3}\mathbf{k}}{(2\pi )^{3}}
\sum_{\ve,\vs}\mathcal{T}_\ve^{\mu_1\mu_2\mu_3}\Theta_\vs^\ve 
%\Theta^{e_1}_{s_1}\Theta^{e_2}_{s_2} \Theta^{e_3}_{s_3}
\times
\non
&&\times
\left\{
e^{ -s_3\e_3 \beta}
 \int \limits _{0}^{\beta }d\tau_1 \int \limits _{\b-\t_1}^{\beta }d\tau _2
\ft^{e_1}_{s_1}  
e^{ \Omega_1 \tau _{1}}
\ft^{e_2}_{s_2}  
e^{ \Omega_2 \tau _{2}}
-
\prod _{i=1}^{2}
\left[\int \limits _{0}^{\beta }d\tau _{i}
%\non&&\times
 \ft^{e_i}_{s_i}  e^{ \Omega_i \tau _{i}} \right]
\ft^{e_3}_{-s_3}
\right\}\;.
\eea
After having performed the integral over $\t_2$ one can combine three of the four remaining $\t_1-$integrals by using $\ft_{-s_2}^{e_2}\ft_{s_3}^{e_3}-\ft_{s_2}^{e_2}\ft_{-s_3}^{e_3}-\ft_{-s_2}^{e_2} =-\ft_{-s_3}^{e_3}$ and obtains
\bea
&&{\cal J} ^{\mu_1\mu_2\mu _3}(P_{1},P_2,P_{3})=\non
&=&
-\int \limits \frac{d^{3}\mathbf{k}}{(2\pi )^{3}}
\sum_{\ve,\vs}\mathcal{T}_\ve^{\mu_1\mu_2\mu_3}\Theta_\vs^\ve 
%\Theta^{e_1}_{s_1}\Theta^{e_2}_{s_2} \Theta^{e_3}_{s_3}
\times
\non
&&\times
\left\{\frac{1}{\Omega_2}
\int \limits _{0}^{\beta }d\tau _{1}
  e^{ \Omega_1 \tau _{1}}
 \ft^{e_1}_{s_1}  \ft^{e_3}_{-s_3}
-
\frac{1}{\Omega_2} \int \limits _{0}^{\beta }d\tau_1
e^{( \Omega_1-\Omega_2) \tau _{1}}
\ft^{e_1}_{s_1}  
\ft^{e_2}_{-s_2}  
\right\}\non
&=&
-\int \limits \frac{d^{3}\mathbf{k}}{(2\pi )^{3}}
\sum_{\ve,\vs}\mathcal{T}_\ve^{\mu_1\mu_2\mu_3}\Theta_\vs^\ve 
%\Theta^{e_1}_{s_1}\Theta^{e_2}_{s_2} \Theta^{e_3}_{s_3}
\times
\non
&&\times
\frac{1}{\Omega_2}
\left\{\frac{ 1}{\Omega_1}\left[\ft^{e_1}_{-s_1}  \ft^{e_3}_{s_3}- \ft^{e_1}_{s_1}  \ft^{e_3}_{-s_3}\right]
-
\frac{1 }{ \Omega_1-\Omega_2}\left[\ft^{e_1}_{-s_1}  \ft^{e_2}_{s_2}- \ft^{e_1}_{s_1}  \ft^{e_2}_{-s_2}\right]
\right\}\non
&=&
-\int \limits \frac{d^{3}\mathbf{k}}{(2\pi )^{3}}
\sum_{\ve,\vs}\mathcal{T}_\ve^{\mu_1\mu_2\mu_3}\Theta_\vs^\ve 
%\Theta^{e_1}_{s_1}\Theta^{e_2}_{s_2} \Theta^{e_3}_{s_3}
\;\frac{1} {p_2^{0}-p_3^0 - s_2\e_{2} +s_3\e_{3}}
\times
\non
&&\times
\left[
\frac{\frac{s_3-s_1}{2} +s_1N(\e_1) - s_3 N(\e_3)} {p_1^{0}-p_3^0 - s_1\e_{1} +s_3\e_{3}}
-
\frac{\frac{s_2-s_1}{2} +s_1N(\e_1) - s_2 N(\e_2)} {p_1^{0}-p_2^0 - s_1\e_{1} +s_2\e_{2} }
\right]\;.\label{3gluons}
\eea
In the limit $\mu \rightarrow 0$ and after using Eq.\ (\ref{limtheta}) this reproduces the corresoponding formula (A.17) in \cite{braatenpisarski}.

\subsubsection{Four external Gluons}

In the case of four external gluons one has $m=1,2,3$ and  \eqrf{gentauint} becomes 
\bea
&&{\cal J} ^{\mu_1\cdots\mu _4}(P_{1},\cdots , P_{4})=\non
&=&
- \int \limits \frac{d^{3}\mathbf{k}}{(2\pi )^{3}}
\sum_{\ve,\vs}\mathcal{T}_\ve^{\mu_1\cdots\mu_4}\Theta_\vs^\ve 
\sum_{m=1}^{3}(-1)^m
\prod _{i=1}^{3}
\left[\int \limits _{0}^{\beta }d\tau _{i}
%\non&&\times
\ft^{e_i}_{s_i}  e^{ \Omega_i \tau _{i}} \right]
\times\non&& \times
 \ft^{e_4}_{s_4}
e^{ -s_4\e_4 m\beta}
%\times 
%\non &&
%\times
\Theta \left( m\beta -\sum _{j=1}^{3}\tau _{j}\right) \Theta \left( \sum _{j=1}^{3}\tau _{j}-(m-1)\beta \right)
\non
&&
\hspace*{-1.4cm} = - \int \limits \frac{d^{3}\mathbf{k}}{(2\pi )^{3}}
\sum_{\ve,\vs}\mathcal{T}_\ve^{\mu_1\cdots\mu_4}\Theta_\vs^\ve 
\prod _{i=1}^{3}
\left[\int \limits _{0}^{\beta }d\tau _{i} \ft^{e_i}_{s_i}  e^{ \Omega_i \tau _{i}} \right]
%\times \non
%&&\times
\left\{
-
\ft^{e_4}_{s_4}
e^{ -s_4\e_4 \beta}
%\times 
%\non &&
%\times
\Theta \left(\beta -\sum _{j=1}^{3}\tau _{j} \right) +
\right.
\non
&&
\hspace*{-1cm}\left.+
 \ft^{e_3}_{s_4}
e^{ -2s_4\e_4 \beta}
%\times 
%\non &&
%\times
\Theta \left(2\beta -\sum _{j=1}^{3}\tau _{j}\right) \Theta \left(\sum _{j=1}^{3}\tau _{j} -\beta \right)
- \ft^{e_4}_{s_4}
e^{ -3s_4\e_4 \beta}
%\times 
%\non &&
%\times
 \Theta \left(\sum _{j=1}^{3}\tau _{j} -2\beta \right)
\right\}\;.
\eea
Using \eqrftw{rel1}{rel2} and respecting $0\leq\t_i\leq\b$ one finds 
\bea
&&-
\ft^{e_4}_{s_4}
e^{ -s_4\e_4 \beta}
%\times 
%\non &&
%\times
\Theta \left(\beta -\sum _{j=1}^{3}\tau _{j} \right) +
 \ft^{e_4}_{s_4}
e^{ -2s_4\e_4 \beta}
%\times 
%\non &&
%\times
\Theta \left(2\beta -\sum _{j=1}^{3}\tau _{j}\right) \Theta \left(\sum _{j=1}^{3}\tau _{j} -\beta \right)-
\non
&&
- \ft^{e_4}_{s_4}
e^{ -3s_4\e_4 \beta}
%\times 
%\non &&
%\times
 \Theta \left(\sum _{j=1}^{3}\tau _{j} -2\beta \right)=
\non
&&
= - \ft^{e_4}_{-s_4}
+e^{ -s_4\e_4 \beta}
\left\{\Theta \left(\tau_1+\t_2 -\beta \right)+ 
\Theta \left( \sum _{j=1}^{3}\tau _{j} -\beta \right) \Theta \left(\beta-\t_1-\t_2 \right)   \right\}-
\non&&
-e^{ -2s_4\e_4 \beta}\Theta \left( \sum _{j=1}^{3}\tau _{j} -2\beta \right)\;.
\eea
Hence, we have the following $\t$-integrals
\bea
&&-\prod _{i=1}^{3}
\left[\int \limits _{0}^{\beta }d\tau _{i} \ft^{e_i}_{s_i}  e^{ \Omega_i \tau _{i}} \right]\ft^{e_4}_{-s_4}
+
\ft^{e_1}_{s_1}\ft^{e_2}_{s_2}\ft^{e_3}_{s_3}e^{ -s_4\e_4 \beta}
\int \limits _{0}^{\beta }d\tau _1\int \limits _{\b-\t_1}^{\beta }d\tau _{2}\int \limits _{0}^{\beta }d\tau _{3}
 e^{ \Omega_1 \tau _{1}} e^{ \Omega_2 \tau _{2}} e^{ \Omega_3 \tau _{3}}+
\non
&&
+
\ft^{e_1}_{s_1}\ft^{e_2}_{s_2}\ft^{e_3}_{s_3}e^{ -s_4\e_4 \beta}
\int \limits _{0}^{\beta }d\tau _1\int \limits _0^{\b-\t_1}d\tau _{2}
\int \limits _{\b-\t_1-\t_2}^{\beta }d\tau _{3}
 e^{ \Omega_1 \tau _{1}} e^{ \Omega_2 \tau _{2}} e^{ \Omega_3 \tau _{3}}
-\non
&&-
\ft^{e_1}_{s_1}\ft^{e_2}_{s_2}\ft^{e_3}_{s_3}e^{ -2s_4\e_4 \beta}
\int \limits _{0}^{\beta }d\tau _1\int \limits _{\b-\t_1}^{\beta }d\tau _{2}\int \limits _{2\b-\t_1-\t_2}^{\beta }d\tau _{3}
 e^{ \Omega_1 \tau _{1}} e^{ \Omega_2 \tau _{2}} e^{ \Omega_3 \tau _{3}}
\eea
Performing the $\t$-integrals one by one  and using \eqrftw{rel1}{rel2} after each integration to combine the corresponding terms one finally obtains the result
\bea
&&{\cal J} ^{\mu_1\cdots\mu _4}(P_{1},\cdots , P_{4})=\non
&=&
- \int \limits \frac{d^{3}\mathbf{k}}{(2\pi )^{3}}
\sum_{\ve,\vs}\mathcal{T}_\ve^{\mu_1\cdots\mu_4}\Theta_\vs^\ve \times \non
&&\frac{1}{\Omega_3}\left\{
\frac{1}{\Omega_2}
\left[ 
\frac{\ft^{e_1}_{-s_1}  \ft^{e_2}_{s_2}- \ft^{e_1}_{s_1}  \ft^{e_2}_{-s_2}}{\Omega_1-\Omega_2} -
\frac{\ft^{e_1}_{-s_1}  \ft^{e_4}_{s_4}- \ft^{e_1}_{s_1}  \ft^{e_4}_{-s_4}}{\Omega_1}
\right]+ \right. \non
&&\left.
\frac{1}{\Omega_2-\Omega_3}\left[
\frac{\ft^{e_1}_{-s_1}  \ft^{e_3}_{s_3}- \ft^{e_1}_{s_1}  \ft^{e_3}_{-s_3}}{\Omega_1-\Omega_3}-\frac{\ft^{e_1}_{-s_1}  \ft^{e_2}_{s_2}- \ft^{e_1}_{s_1}  \ft^{e_2}_{-s_2}}{\Omega_1-\Omega_2} \right]
\right\}=\non
&=&
- \int \limits \frac{d^{3}\mathbf{k}}{(2\pi )^{3}}
\sum_{\ve,\vs}\mathcal{T}_\ve^{\mu_1\cdots\mu_4}\Theta_\vs^\ve 
\;\frac{1}{p_3^{0}-p_4^0 - s_3\e_{3} +s_4\e_{4}}\times \non
&&\left\{
\frac{1}{p_2^{0}-p_4^0 - s_2\e_{2} +s_4\e_{4}}
\left[ 
\frac{\frac{s_2-s_1}{2} +s_1N(\e_1) - s_2 N(\e_2)}{p_1^{0}-p_2^0 - s_1\e_{1} +s_2\e_{2}} -
\frac{\frac{s_4-s_1}{2} +s_1N(\e_1) - s_4 N(\e_4)}{p_1^{0}-p_4^0 - s_1\e_{1} +s_4\e_{4}}
\right]+ \right. \non
&&\left.
+\frac{1}{p_2^{0}-p_3^0 - s_2\e_{2} +s_3\e_{3}}\left[
\frac{\frac{s_3-s_1}{2} +s_1N(\e_1) - s_3 N(\e_3)}{p_1^{0}-p_3^0 - s_1\e_{1} +s_3\e_{3}}
-\frac{\frac{s_2-s_1}{2} +s_1N(\e_1) - s_2 N(\e_2)}{p_1^{0}-p_2^0 - s_1\e_{1} +s_2\e_{2}} \right]
\right\}\non\label{4gluons}
\eea
In the limit $\mu \rightarrow 0$ and after using Eq.\ (\ref{limtheta}) this reproduces the corresoponding formula (A.19) in \cite{braatenpisarski}.
%Combining the first with the last lerm yields
%\bea
%&&{\cal J} ^{\mu_1\cdots\mu _4}(P_{1},\cdots , P_{4})=\non
%&=&
%- \int \limits \frac{d^{3}\mathbf{k}}{(2\pi )^{3}}
%\sum_{\ve,\vs}\mathcal{T}_\ve^{\mu_1\cdots\mu_4}\Theta_\vs^\ve 
%\times \non
%&&\left\{
%-\frac{1}{p_3^{0}-p_4^0 - s_3\e_{3} +s_4\e_{4}}
%\,\frac{1}{p_2^{0}-p_4^0 - s_2\e_{2} +s_4\e_{4}} 
%\,\frac{\frac{s_4-s_1}{2} +s_1N(\e_1) - s_4 N(\e_4)}{p_1^{0}-p_4^0 - s_1\e_{1} +s_4\e_{4}}
% \right. \non
%&&\left.
%+\frac{1}{p_3^{0}-p_4^0 - s_3\e_{3} +s_4\e_{4}}
%\,\frac{1}{p_2^{0}-p_3^0 - s_2\e_{2} +s_3\e_{3}}
%\,\frac{\frac{s_3-s_1}{2} +s_1N(\e_1) - s_3 N(\e_3)}{p_1^{0}-p_3^0 - s_1\e_{1} +s_3\e_{3}}
% \right. \non
%&&\left.
%-\frac{1}{p_2^{0}-p_3^0 - s_2\e_{2} +s_3\e_{3}}
%\,\frac{1}{p_2^{0}-p_4^0 - s_2\e_{2} +s_4\e_{4}}
%\,\frac{\frac{s_2-s_1}{2} +s_1N(\e_1) - s_2 N(\e_2)}{p_1^{0}-p_2^0 - s_1\e_{1} +s_2\e_{2}}
%\right\}
%\eea

    %Quark loops
%\germanTeX
%\include{zusf} 			%deutsche Zusammenfassung
%\originalTeX
\end{appendix}
 		%Bibliography%
%\germanTeX
%\include {TABLEBEN}  	%Lebenslauf
\chapter*{Acknowledgements}

I would like to express my gratitude to Dirk Rischke for giving me the opportunity to do research in this interesting and stimulating field. I am thankful to Dirk Rischke and to Qun Wang for the fruitful and friendly collaboration and for many valueable discussions. Especially at the beginning, as I freshly entered this field, your help was absolutely indispensable. Thanks a lot!

I highly appreciate the instructive and lively discussions with Andreas Schmitt and Igor Shovkovy, who were always ready to clarify all kinds of open questions. 

I would like to thank all members and visitors of this institute for their 
openness and for the congenial atmosphere. In particular I would like to thank Mei Huang, Hossein Malekzadeh, Amruta Mishra, Dirk Rischke, Andreas Schmitt, Igor Shovkovy, and Qun Wang for their friendship and for a great time! Thanks to the system administrators Alexander Achenbach, Manuel Reiter and Gebhard Zeeb, and to the secretaries Mrs.\ Veronika Palade and Mrs.\ Daniela Radulescu for their support.

Special thanks are due to  Andreas Schmitt for the happy time -- in the shared office as well as on our memorable tours -- and for hundreds of seminal jokes!

% for being so well-grounded,

Innumerable thanks go to my family and to Aline. Thanks for your support and your love!

\end{document}